\documentclass[11pt]{article}
\pdfoutput=1

\usepackage[usenames,dvipsnames,table]{xcolor}
\usepackage[utf8]{inputenc}

\usepackage{jheppub} 

\usepackage{tikz-cd} 
\usepackage{circuitikz}
\usepgfmodule{shapes}
\usetikzlibrary{decorations.pathmorphing}
\usetikzlibrary{decorations.pathreplacing,decorations.markings,snakes}
\usetikzlibrary{backgrounds, arrows,calc,shapes,decorations.pathreplacing, automata,positioning}

\usepackage{cancel} 
\usepackage{xcolor}

\usetikzlibrary{shapes.misc}

\tikzset{cross/.style={cross out, draw=black, minimum size=5*(#1-\pgflinewidth), inner sep=0pt, outer sep=0pt},
cross/.default={2pt}}

\tikzset{snake it/.style={decorate, decoration=snake}}

\tikzset{mid arrow/.style={postaction={decorate,decoration={
        markings,
        mark = at position .55 with {\arrow[#1]{Straight Barb[width=5pt]}}
      }}}}
      
\tikzset{mid arrowsm/.style={postaction={decorate,decoration={
        markings,
        mark = at position .55 with {\arrow[#1]{Straight Barb[width=3pt]}}
      }}}}
      
\tikzset{middx arrowsm/.style={postaction={decorate,decoration={
        markings,
        mark = at position .7 with {\arrow[#1]{Straight Barb[width=3pt]}}
      }}}}
      
\tikzset{midsx arrowsm/.style={postaction={decorate,decoration={
        markings,
        mark = at position .4 with {\arrow[#1]{Straight Barb[width=3pt]}}
      }}}}

\newcommand{\wigE}{\draw[snake=zigzag]}

\newcommand{\wigM}{\draw[snake=coil,,segment aspect=0,segment length=6pt, double]}

\newcommand{\wigT}{\draw[dashed]}
\newcommand{\NSbrane}{\draw[ultra thick, blue]}
\newcommand{\Dbrane}{\draw[dashed, ultra thick, gray]}
\newcommand{\Dthree}{\draw[thick, red]}
\newcommand{\hyper}{\draw[-,double distance=3]}
\newcommand{\chir}{\draw[-,mid arrow]}

\newenvironment{proposal}[1][Proposal]{\begin{trivlist}
\item[\hskip \labelsep {\bfseries #1}]}{\end{trivlist}}

\newcommand{\M}{\mathfrak{M}}

\usepackage{graphicx}
\usepackage{amsmath, amssymb}
\usepackage{youngtab}
\usepackage{changepage}
\usepackage{mathabx} 

\newcommand{\be}{\begin{eqnarray}}
\newcommand{\ee}{\end{eqnarray}}
\newcommand{\ba}{\begin{array}}
\newcommand{\ea}{\end{array}}
\newcommand{\bea}{\begin{eqnarray}}
\newcommand{\eea}{\end{eqnarray}}
\newcommand{\bpic}{\begin{tikzpicture}}
\newcommand{\epic}{\end{tikzpicture}}

\newcommand{\nn}{\nonumber}
\newcommand{\bn}{\begin{enumerate}}
\newcommand{\en}{\end{enumerate}}

\def\Ge{\Gamma_e}

\def\CF{{\cal F}}

\def\CI{{\cal I}}

\def\cN{{\cal N}}
\def\CN{{\cal N}}

\def\CS{{\cal S}}

\def\CW{{\cal W}}
\def\cW{{\cal W}}

\def\Tr{\mathop{\text{Tr}}\nolimits}

\def\U{\mathrm{U}}

\def\b{\beta}

\def\d{\delta}

\def\r{\rho}

\def\s{\sigma}

\def\D{\Delta}

\def\Qt{\tilde{Q}}
\def\Pt{\tilde{P}}

\title{Mirror dualities with four supercharges}

\author[a]{Sergio Benvenuti}
\author[b,c]{, Riccardo Comi}
\author[b,c]{, Sara Pasquetti}

\affiliation[a]{Istituto Nazionale di Fisica Nucleare, sezione di Trieste}
\affiliation[b]{Dipartimento di Fisica, Università di Milano-Bicocca,Piazza della Scienza 3, I-20126 Milano, Italy}
\affiliation[c]{INFN, sezione di Milano-Bicocca, Piazza della Scienza 3, I-20126 Milano, Italy}

\emailAdd{benve79@gmail.com, r.comi2@campus.unimib.it, sara.pasquetti@gmail.com}

\abstract{We consider $3d$ $\mathcal{N}=2$ non-abelian Hanany-Witten brane setups with chiral flavor symmetry. We propose that the associated field theories are quivers with \emph{improved bifundamentals}, instead of standard bifundamentals. The improved bifundamental is a strongly coupled SCFT that carries one more $U(1)$ global symmetry than the standard bifundamental. As a consequence, our proposal overcomes the long standing challenge of associating to each $\mathcal{N}=2$ brane setup a gauge theory with the full rank global symmetry, allowing the study of all the usual supersymmetric observables, such as superconformal index, sphere partition function, chiral ring and moduli space. The construction passes many non-trivial tests, for instance we algorithmically prove that any two improved quivers associated to $\mathcal{S}$-dual brane setups are infrared dual.  The $3d$ $\mathcal{N}=2$ mirror dualities can be uplifted to $4d$ dualities with  $4d$ improved bifundamentals connecting $USp(2N)$  nodes. }

\begin{document} 

\maketitle
\flushbottom

\newpage

\section{Introduction}

One of the important offshoots of the second superstring revolution is the brane construction of gauge theories. Hanany-Witten brane setups \cite{Hanany:1996ie} engineer $3d$ $\cN=4$ linear quiver gauge theories. 
An immediate but extremely profound consequence of this construction is the observation that mirror dualities relating $3d$ $\cN=4$ theories \cite{Intriligator:1996ex} are inherited from $\CS$-duality in Type IIB string theory, which swaps $NS$ and $D5$ branes.

In the abelian case, it is possible to prove $3d$ mirror symmetry using purely field theory arguments. \cite{Kapustin:1999ha} showed how to piecewise dualize a general $\cN=2,3,4$ abelian QFT.
The proof uses as basic ingredient only the duality between $U(1)$ with 1 flavor and a free hypermultiplet.\footnote{There is compelling evidence for the validity of this duality: one can prove that in the gauge theory there is a free sector using the monopole R-charges and the unitarity bounds \cite{Borokhov:2002cg}. The matching of the $S_b^3$ partition function of the gauge theory with the one of the free hyper implies that there is nothing on top of the free sector.} Let us mention that the abelian $3d$ mirror symmetry is related to $3d$ abelian non supersymmetric bosonization \cite{Kachru:2016rui,Karch:2016aux}.

$3d$ mirror symmetry led to many advances in our understanding of the quantum dynamics of gauge theories. Theories with $8$ supercharges in $d=4,5,6$, whose Higgs branches are not corrected upon circle reduction \cite{Argyres:1996eh},  admit a $3d$ mirror also known as magnetic quiver  (see for instance \cite{Benini:2010uu, Xie:2012hs, DelZotto:2014kka, Cremonesi:2015lsa, Ferlito:2017xdq, Hanany:2018uhm, Hanany:2018vph, Cabrera:2018jxt, Cabrera:2019izd, Bourget:2019rtl, Cabrera:2019dob}), which is often crucial in uncovering the quantum dynamics of QFT’s which do not admit a Lagrangian description.

In light of the above, it would clearly be desirable to extend our understanding of non-abelian $3d$ mirror symmetry to theories with less than 8 supercharges. 
A recent advance comes to our help in this direction.  A couple of years ago \cite{Hwang:2021ulb,Comi:2022aqo} was able to \emph{prove} non-abelian $3d$ $\cN=4$ mirror symmetry via the {\it dualization algorithm}. 

The idea  of the algorithm originates from the observation   \cite{Gaiotto:2008ak,Gulotta:2011si,Assel:2014awa} that on linear or circular brane setups, we can think of $\CS$-duality  as acting locally on each 5-brane, creating an $\CS$-duality  wall on its left and an $\CS^{-1}$-duality  wall on its right:
$D5= \CS \!\cdot\! NS \!\cdot\! \CS^{-1}$ and $\widebar{NS}= \CS \!\cdot\! D5 \!\cdot\! \CS^{-1}.$\\
The intersection of the  $\CS$-duality  wall  with the $N$ $D3$ branes was argued to be captured by the $3d$ $\mathcal{N}=4$
 $\CS$-duality wall  $FT[SU(N)]$ theory introduced in  \cite{Gaiotto:2008ak}, represented by the quiver below.\footnote{The  $FT[SU(N)]$ theory  we use here differs from the  $T[SU(N)]$  introduced in \cite{Gaiotto:2008ak} only by the adjoint singlet flipping the meson operator.}
\be
\resizebox{.8\hsize}{!}{
\begin{tikzpicture}[thick,node distance=3cm,gauge/.style={circle,draw,minimum size=5mm},flavor/.style={rectangle,draw,minimum size=5mm}] 
 
\begin{scope}
	
	\path (0,0) node[gauge] (g1) {$\!\!\!1\!\!\!$} -- (1.5,0) node[gauge] (g2)	{$\!\!\!2\!\!\!$} 
		-- (4,0) node[gauge] (g3) {\!\!\tiny{$N$-$1$}\!\!\!} -- (5.5,0) node[flavor] (g4) {$\!N\!$};
		
	\draw[-, shorten >= 6, shorten <= 8, shift={(-0.05,0.07)}, mid arrowsm] (0,0) -- (1.5,0);
	\draw[-, shorten >= 6, shorten <= 8, shift={(0.05,-0.07)}, mid arrowsm] (1.5,0) -- (0,0);
	
	\draw[-, shorten >= 5, shorten <= 5, shift={(0.05,0.07)}, mid arrowsm] (1.5,0) -- (2.5,0);
	\draw[-, shorten >= 4, shorten <= 6, shift={(0.1,-0.07)}, mid arrowsm] (2.5,0) -- (1.5,0);
	
	\draw (2.73,0) node {$\cdots$};	
	
	\draw[-, shorten >= 5.5, shorten <= 2.5, shift={(-0.1,0.07)}, mid arrowsm] (3,0) -- (4,0);
	\draw[-, shorten >= 1.5, shorten <= 7, shift={(-0.05,-0.07)}, mid arrowsm] (4,0) -- (3,0);
	
	\draw[-, shorten >= 6, shorten <= 9, shift={(-0.05,0.07)}, mid arrowsm] (4,0) -- (5.5,0);
	\draw[-, shorten >= 6.5, shorten <= 8.5, shift={(0.05,-0.07)}, mid arrowsm] (5.5,0) -- (4,0);
	
	\draw[-] (g1) to[out=60,in=0] (0,0.5) to[out=180,in=120] (g1);
	\draw[-] (g2) to[out=60,in=0] (1.5,0.5) to[out=180,in=120] (g2);
	\draw[-] (g3) to[out=60,in=0] (4,0.55) to[out=180,in=120] (g3);
	\draw[-] (g4) to[out=60,in=0] (5.5,0.6) to[out=180,in=120] (g4);
	
\end{scope}

	\draw (6.75,0) node {$=$};

\begin{scope}[shift={(9,0.25)}]

	\path (0,0) node[flavor] (x) {$\!N\!$} -- (2,0) node[flavor] (y) {$\!N\!$};
	
	\wigT (x) -- (y);
	
	\draw (1,-1) node {\small{$SU(N) \times SU(N)$}};

\end{scope}
	 
\end{tikzpicture}}
\ee
The $\CS$-wall theory has IR $SU(N)\times SU(N)$ symmetry, where one of the two  $SU(N)$ factors arises from the IR
enhancement of the $U(1)^{N-1}$ topological symmetry, we denote it in compact form by a dashed line connecting two square blocks  to indicate its non-abelian IR symmetries.
The chiral ring of the $FT[SU(N)]$ theory  is generated by the moment map operators in the adjoint of the two $SU(N)$ symmetries.
In \cite{Bottini:2021vms} the $\CS$-wall theory was shown to satisfy the $SL(2,\mathbb{Z})$ relations and in particular the
\emph{fusion to identity} property $\CS \!\cdot\! \CS^{-1}=1$, a relation which is going to play an important role:
\be
\begin{tikzpicture}[thick,node distance=3cm,gauge/.style={circle,draw,minimum size=5mm},flavor/.style={rectangle,draw,minimum size=5mm}]
	
	\path (0,0) node[gauge] (g1) {$\!\!\!N\!\!\!$} -- (-1.5,0) node[flavor] (f1) {$\!N\!$} -- (1.5,0) node[flavor] (f2) {$\!N\!$};
     
    \draw[-] (g1) to[out=300,in=0] (0,-0.6) to[out=180,in=240] (g1);
	\wigT (g1) -- (f1); 	\wigT (g1) -- (f2); 
	
	\draw (3,0) node {$\Longleftrightarrow$};
	
	\draw (4,0) node[right] {\LARGE{$\mathbb{I}$}-wall};
	
\end{tikzpicture}
\label{Swallprop}
\ee
On the r.h.s. we have the Identity wall theory $\mathbb{I}$-wall, whose partition function is a delta function identifying the Cartans of the two $U(N)$ global symmetry groups.

The dualization algorithm  (similar in spirit to \cite{Kapustin:1999ha}) basically implements  in field theory the local action $\CS$-duality.  
Roughly speaking, the $\cN=4$ algorithm consists in ungauging a linear quiver in two types of basic matter blocks: the bifundamental matter and the flavor matter. At the level of brane setups the bifundamental is associated to a $NS$ brane, the flavor to a $D5$ brane. 
Each block is then locally dualized by a basic duality move which implements at the field theory level the local action of the $\CS$-dualization of  each 5-brane. 
The basic duality moves are in turn genuine IR dualities which can be proven assuming only the basic Seiberg-like Aharony duality \cite{Aharony:1997gp}.

In this paper we generalize this strategy to the mirror dualities with four supercharges which can be realized in setups with $NS$ and $D5'$ branes.
The salient feature of our proposal is that in such setups the bifundamentals in the quiver, associated to the presence of a $NS$ brane over which $D3-D3$ strings stretch, is not a standard bifundamental, but an \emph{improved bifundamental}.
The improved bifundamental is a strongly coupled $3d$ $\cN=2$ CFT, the  $FM[U(N)]$ theory introduced in \cite{Pasquetti:2019tix}, with the UV completion given by the quiver below:
\be
\resizebox{.85\hsize}{!}{
\begin{tikzpicture}[thick,node distance=3cm,gauge/.style={circle,draw,minimum size=5mm},flavor/.style={rectangle,draw,minimum size=5mm}] 
 
\begin{scope}
	
	\path (0,0) node[gauge] (g1) {$\!\!\!1\!\!\!$} -- (1.5,0) node[gauge] (g2)	{$\!\!\!2\!\!\!$} 
		-- (4,0) node[gauge] (g3) {\!\!\tiny{$N$-$1$}\!\!\!} -- (5.5,0) node[flavor] (g4) {$\!N\!$}
		-- (-0.75,-1.25) node[flavor] (x1) {$\!1\!$} -- (0.75,-1.25) node[flavor] (x2) {$\!1\!$} -- (4.75,-1.25) node[flavor] (x3) {$\!1\!$};
		
	\draw[-, shorten >= 6, shorten <= 8, shift={(-0.05,0.07)}, middx arrowsm] (0,0) -- (1.5,0);
	\draw[-, shorten >= 6, shorten <= 8, shift={(0.05,-0.07)}, midsx arrowsm] (1.5,0) -- (0,0);
	\draw (0.5,0) node[cross] {};
	
	\draw[-, shorten >= 5, shorten <= 5, shift={(0.05,0.07)}, mid arrowsm] (1.5,0) -- (2.5,0);
	\draw[-, shorten >= 4, shorten <= 6, shift={(0.1,-0.07)}, mid arrowsm] (2.5,0) -- (1.5,0);
	
	\draw (2.75,-0.5) node {$\cdots$};	
	
	\draw[-, shorten >= 5.5, shorten <= 2.5, shift={(-0.1,0.07)}, mid arrowsm] (3,0) -- (4,0);
	\draw[-, shorten >= 1.5, shorten <= 7, shift={(-0.05,-0.07)}, mid arrowsm] (4,0) -- (3,0);
	
	\draw[-, shorten >= 6, shorten <= 9, shift={(-0.05,0.07)}, middx arrowsm] (4,0) -- (5.5,0);
	\draw[-, shorten >= 6.5, shorten <= 8.5, shift={(0.05,-0.07)}, midsx arrowsm] (5.5,0) -- (4,0);
	\draw (4.5,0) node[cross] {};
	
	\draw[-] (g2) to[out=60,in=0] (1.5,0.5) to[out=180,in=120] (g2);
	\draw[-] (g3) to[out=60,in=0] (4,0.55) to[out=180,in=120] (g3);
	\draw[-] (g4) to[out=60,in=0] (5.5,0.6) to[out=180,in=120] (g4);
	
	\draw[-, shorten >= 5.5, shorten <= 8, shift={(-0.1,0.02)}, middx arrowsm] (-0.75,-1.25) -- (0,0);
	\draw[-, shorten >= 8.5, shorten <= 8, shift={(0.05,0)}, midsx arrowsm] (0,0) -- (-0.75,-1.25);
	\draw (-0.485,-0.75) node {\rotatebox{-30}{\LARGE{$\times$}}};
	
	\draw[-, shorten >= 7.5, shorten <= 8.5, shift={(-0.07,0.02)}, mid arrowsm] (0,0) -- (0.75,-1.25);
	\draw[-, shorten >= 5.5, shorten <= 8, shift={(0.1,0)}, mid arrowsm] (0.75,-1.25) -- (0,0);
	
	\draw[-, shorten >= 5.5, shorten <= 8, shift={(-0.1,0.02)}, middx arrowsm] (0.75,-1.25) -- (1.5,0);
	\draw[-, shorten >= 8.5, shorten <= 8, shift={(0.05,0)}, midsx arrowsm] (1.5,0) -- (0.75,-1.25);
	\draw (1.015,-0.75) node {\rotatebox{-30}{\LARGE{$\times$}}};
	
	\draw[-, shorten >= 0, shorten <= 8.5, shift={(-0.07,0.02)}, middx arrowsm] (1.5,0) -- (2,-0.85);
	\draw[-, shorten >= 5.5, shorten <= 3, shift={(0.1,0)}, midsx arrowsm] (2,-0.85) -- (1.5,0);
	
	\draw[-, shorten >= 7, shorten <= 2, shift={(-0.1,0.02)}, midsx arrowsm] (3.5,-0.85) -- (4,0);
	\draw[-, shorten >= 0, shorten <= 9, shift={(0.05,0)}, middx arrowsm] (4,0) -- (3.5,-0.85);
	
	\draw[-, shorten >= 7.5, shorten <= 9.5, shift={(-0.07,0.02)}, mid arrowsm] (4,0) -- (4.75,-1.25);
	\draw[-, shorten >= 6.5, shorten <= 8, shift={(0.1,0)}, mid arrowsm] (4.75,-1.25) -- (4,0);
	
	\draw[-, shorten >= 8, shorten <= 8, shift={(-0.1,0.02)}, middx arrowsm] (4.75,-1.25) -- (5.5,0);
	\draw[-, shorten >= 8.5, shorten <= 8.5, shift={(0.05,0)}, midsx arrowsm] (5.5,0) -- (4.75,-1.25);
	\draw (5.015,-0.75) node {\rotatebox{-30}{\LARGE{$\times$}}};
	
\end{scope}

	\draw (6.5,-0.5) node {$=$};

\begin{scope}[shift={(9,-0.25)}]

	\path (0,0) node[flavor] (x) {$\!N\!$} -- (2,0) node[flavor] (y) {$\!N\!$};
	
	\wigM (x) -- (y);
	
	\draw (1,-1) node {\small{$S[U(N) \times U(N)] \times U(1)^2$}};

\end{scope}
	 
\end{tikzpicture}}
\ee
The superpotential of the $FM[U(N)]$ quiver includes a  term  linear in the fundamental monopoles at each node.
This term breaks all the topological symmetries and the manifest global symmetry is $U(N)\times U(1)^N\times U(1)^2$.
In the IR the  $U(1)^N$ UV global symmetry  acting on the fields of the {\it saw} enhances to $U(N)$. 
The IR spectrum of the improved bifundamental includes two adjoint operators and two  bifundamental $(N, \bar N)$, $(\bar N, N)$ operators $\Pi,\tilde \Pi$ of the two $U(N)$ symmetries. 
We use the compact notation on the right with two wiggle lines connecting the two square nodes to visualize the two non-abelian symmetries.
In addition to them, the improved bifundamental has two $U(1)$ global  symmetries, so one extra  symmetry with respect to the standard bifundamental. As we will see, improved quiver theories built with improved bifundamentals have an interesting pattern of symmetry enhancement and we can introduce a notion of {\it balanced} nodes leading to non-abelian symmetry enhancement.
For example, in a  string of $k$ consecutive  improved bifundamentals  linking $k-1$ $U(N)$ gauge  nodes, $k$   $U(1)$ symmetries rotating each improved bifundamental and the $(k-1)$ $U(1)$ topological symmetries will enhance to $U(k)^2/U(1)$.\\

The local dualization of improved quivers requires a new set of $\mathcal{N}=2$ basic duality moves implementing the $\CS$-dualization of  generalized flavors into generalized  bifundamentals  and viceversa. 
The \emph{$3d$ basic move} corresponding to the dualization  $\CS \!\cdot\! D5 \!\cdot\! \CS^{-1} =\widebar{NS}$ 
is given by:
\be\label{fig:3dbasicmove}
\begin{tikzpicture}[thick,node distance=3cm,gauge/.style={circle,draw,minimum size=5mm},flavor/.style={rectangle,draw,minimum size=5mm}]

	\path (0,0) node[gauge] (g1) {$\!\!\!N\!\!\!$} -- (-1.5,0) node[flavor] (f1) {$\!N\!$} -- (1.5,0) node[flavor] (f2) {$\!N\!$} 
		-- (0,1.5) node[flavor] (f3) {$\!1\!$};
	
	\wigT (g1) -- (f1); 	\wigT (g1) -- (f2); 
		
	\draw[-, shorten >= 6, shorten <= 9, shift={(-0.07,-0.05)}, mid arrowsm] (0,0) -- (0,1.5);
	\draw[-, shorten >= 6.5, shorten <= 8.5, shift={(0.07,0.05)}, mid arrowsm] (0,1.5) -- (0,0);
	
    \draw[-] (g1) to[out=300,in=0] (0,-0.6) to[out=180,in=240] (g1);
	
	\draw (3.25,0.75) node {$\Longleftrightarrow$};
	
	\path (5,0.75) node[flavor] (f1) {$\!N\!$} -- (7,0.75) node[flavor] (f2) {$\!N\!$};
	
	\wigM (f1) -- (f2);	
	
\end{tikzpicture}\ee
This move is a genuine IR duality which can be obtained via compactification and real mass deformation from the $4d$ $\cN=1$ \emph{star-triangle} or \emph{braid} duality, which can be proven by induction in $N$ using only the basic Seiberg-like Intriligator-Pouliot duality as shown in \cite{BCP1}.
In the move \eqref{fig:3dbasicmove}, on the l.h.s., the adjoint chiral couples to the moment maps of the $\CS$-walls to its left and to its right while the flavor is not coupled to the adjoint (as it would be in the $\cN=4$ case).

The opposite transformation, the dualization of an improved bifundantal into a generalized flavor, corresponding to the dualization  $\CS \!\cdot\! NS \!\cdot\! \CS^{-1} =D5$ 
can be easily obtained by  combining \eqref{fig:3dbasicmove} and \eqref{Swallprop} and is given by:
\be\label{fig:3dbasicmoveB}
\begin{tikzpicture}[thick,node distance=3cm,gauge/.style={circle,draw,minimum size=5mm},flavor/.style={rectangle,draw,minimum size=5mm}]
	
	\path (-1,-4) node[flavor] (x1) {$\!N\!$} -- (-1,-2.5) node[flavor] (v1) {$\!1\!$} 
		-- (2,-3.25) node {$\Longleftrightarrow$}
		-- (4,-3.25) node[flavor] (y1) {$\!N\!$} -- (5.5,-3.25) node[gauge] (y2) {$\!\!\!N\!\!\!$} -- (7,-3.25) node[gauge] (y3) {$\!\!\!N\!\!\!$} 
		-- (8.5,-3.25) node[flavor] (y4) {$\!N\!$}; 
	
	\draw[-, shorten >= 6, shorten <= 9, shift={(-0.07,-0.05)}, mid arrowsm] (-1,-4) -- (-1,-2.5);
	\draw[-, shorten >= 6.5, shorten <= 8.5, shift={(0.07,0.05)}, mid arrowsm] (-1,-2.5) -- (-1,-4);
	\draw (-1+0.4,-4) node[right] {\LARGE{$\mathbb{I}$}-wall};
	\wigT (y1) -- (y2); 
	\wigM (y2) -- (y3); 
	\wigT (y3) -- (y4); 
    \draw[-] (y2) to[out=300,in=0] (5.5,-3.85) to[out=180,in=240] (y2);
    \draw[-] (y3) to[out=300,in=0] (7,-3.85) to[out=180,in=240] (y3);
    
\end{tikzpicture}\ee
So the Braid duality is the fundamental move and since it can be demonstrated by induction by assuming only the elementary Seiberg-like dualities,
it follows that all the $\cN=2$ mirror dualities following from the algorithm are demonstrated to be consequence of Seiberg-like dualities only. \\

The $\mathcal{N}=2$ dualization algorithm based on the above basic duality moves allows us to work out the $3d$ mirror dual of linear quivers corresponding to Hanany-Witten brane setups with four supercharges formed by a sequence of $NS$ and $D5’$ branes. In doing so we propose how to read the associated gauge theories. 
More precisely we focus on $\cN=2$ brane setup made of an arbitrary sequence of $NS$ and $D5'$ branes in the case that the number of $D3$ branes is constant along the brane setup. The $D3$/$NS$/$D5'$ branes extend along $0126$/$012345$/$012457$, respectively. We propose that the IR QFT associated to such setups is given by an {\it improved} linear quiver with $U(N)$ adjoint nodes, joined by improved bifundamentals links with flavors distributed among the nodes, according to the position of the $D5'$ branes. The superpotential couples the adjoint of each $U(N)$ nodes to the adjoint operators of the nearby improved bifundamentals. The flavors do not enter the superpotential.
Crucially, we will show that this proposal is consistent with $\mathcal{S}$-duality, that is two improved quivers corresponding to $\CS$-dual brane setups are mirror dual as $3d$ $\cN=2$ QFT's.

Basically, our proposal differs from the the \emph{naive} reading of the brane setup (see for instance \cite{deBoer:1997ka, Aharony:1997ju, Giacomelli:2017vgk, Benvenuti:2018bav, Giacomelli:2019blm}) in that instead of a standard bifundamental hypermultiplet, we use the $FM$ theory, which carries an additional $U(1)$ global symmetry.\footnote{Let us remind that in the abelian case the improved and standard bifundamentals coincide, hence our proposed quivers and the naive quivers are the same.} The naive quiver  indeed has the problem that the UV theory sees a global symmetry with rank strictly smaller than the rank of the IR global symmetry, hence it is impossible to use the naive quiver to compute observables in the IR SCFT, even supersymmetric localized partition functions and chiral rings are not accessible. One interesting comment is that it is possible to turn on an holomorphic deformation that turns the improved bifundamental theory into a standard bifundamental, but these operators are generically trivial in the chiral ring of the quiver, hence \emph{chiral ring stability} of \cite{Benvenuti:2017lle}  imples that these deformations do not lead to a new IR SCFT. In other words, for each brane setup there is only one IR SCFT, whose properties can be explored using the UV quiver with improved bifundamental, but not using the naive UV quiver.

Let us provide a concrete example of how to associate a quiver and how to prove the mirror duality. Consider the sequence  $NS - D5'- NS - (D5')^3 - NS$, part of a longer brane setup, together with its $\CS$-dual setup depicted below.\footnote{For convenience in the picture we present the action of $\CS$-duality combined with the rotation acting by $NS' \to NS$ and $D5 \to D5'$.}
\be \label{MSexample}
\resizebox{\hsize}{!}{
 \bpic[thick,node distance=3cm,gauge/.style={circle,draw,minimum size=5mm},flavor/.style={rectangle,draw,minimum size=5mm}] 
	 
\begin{scope}[shift={(0,0)}]	
	\draw (0,0) node {$\cdots$};
	\NSbrane (1, 1.5) -- (1, -1.5);
	\Dbrane (1.8, 1.5) -- (2.2, -1.5);
	\NSbrane (3, 1.5) -- (3, -1.5);
	\Dbrane (3.2, 1.5) -- (3.6, -1.5);
	\Dbrane (3.8, 1.5) -- (4.2, -1.5);
	\Dbrane (4.4,1.5) -- (4.8, -1.5);
	\NSbrane (5,1.5) -- (5, -1.5);
	\draw (6,0) node {$\cdots$};
	
	\Dthree (0.5,0.1) -- (5.5,0.1);
	\Dthree (0.5,-0.1) -- (5.5,-0.1);
\end{scope}	 

	\draw (8.5,0.4) node {$\CS$-duality};
	\draw (8.5,0) node {$\Longleftrightarrow$};
	\draw (8.5,-4.2) node {Mirror};
	\draw (8.5,-4.6) node {symmetry};
	\draw (8.5,-5) node {$\Longleftrightarrow$};

\begin{scope}[shift={(11,0)}]
	\draw (-1,0) node {$\cdots$};
    \Dbrane (-0.2,1.5) -- (0.2,-1.5);
	\NSbrane (1,1.5) -- (1,-1.5);
	\Dbrane (1.8,1.5) -- (2.2,-1.5);
	\NSbrane (3,1.5) -- (3,-1.5);
	\NSbrane (5,1.5) -- (5,-1.5);
	\NSbrane (7,1.5) -- (7,-1.5);
	\Dbrane (7.8,1.5) -- (8.2,-1.5);
	\draw (9,0) node {$\cdots$};
	
	\Dthree (-0.5,0.1) -- (8.5,0.1);
	\Dthree (-0.5,-0.1) -- (8.5,-0.1);	
\end{scope}

\begin{scope}[shift={(0,-5)}]	
	\path (-0.75,0) node {$\cdots$} -- (0,0) node[gauge](g0) {$\!\!\!N\!\!\!$} -- (2,0) node[gauge](g1) {$\!\!\!N\!\!\!$} 
		-- (4,0) node[gauge](g2) {$\!\!\!N\!\!\!$} -- (6,0) node[gauge](g3) {$\!\!\!N\!\!\!$} -- (6.75,0) node {$\cdots$}
		-- (1.5,1.5) node[flavor] (x1) {$\!1\!$} -- (2.5,1.5) node[flavor] (x2) {$\!1\!$} 
		-- (3.5,1.5) node[flavor] (y1) {$\!3\!$} -- (4.5,1.5) node[flavor] (y2) {$\!3\!$}; 
		
	\wigM (g1) -- (g0); \wigM (g1) -- (g2);\wigM (g3) -- (g2);\chir (g1) -- (x1); \chir (x2) -- (g1); \chir (g2) -- (y1); \chir (y2) -- (g2);
    \draw[-] (g0) to[out=300,in=0] (0,-0.6) to[out=180,in=240] (g0);
    \draw[-] (g1) to[out=300,in=0] (2,-0.6) to[out=180,in=240] (g1);
    \draw[-] (g2) to[out=300,in=0] (4,-0.6) to[out=180,in=240] (g2);
    \draw[-] (g3) to[out=300,in=0] (6,-0.6) to[out=180,in=240] (g3);
\end{scope}	 	
	
\begin{scope}[shift={(11,-5)}]	
	 \path  (-0.75,0) node {$\cdots$} -- (0,0) node[gauge](g0) {$\!\!\!N\!\!\!$} --  (2,0) node[gauge](g1) {$\!\!\!N\!\!\!$} 
	 	-- (4,0) node[gauge](g2) {$\!\!\!N\!\!\!$} -- (6,0) node[gauge] (g3) {$\!\!\!N\!\!\!$}-- (8,0) node[gauge] (g4) {$\!\!\!N\!\!\!$} 
	 	-- (8.75,0) node {$\cdots$}
	 	-- (-0.5,1.5) node[flavor] (x1) {$\!1\!$} -- (0.5,1.5) node[flavor] (x1p) {$\!1\!$} 
	 	-- (1.5,1.5) node[flavor] (x2) {$\!1\!$} -- (2.5,1.5) node[flavor] (x2p) {$\!1\!$} 
	 	-- (7.5,1.5) node[flavor] (z) {$\!1\!$} -- (8.5,1.5) node[flavor] (zp) {$\!1\!$} ;
	 
	 \chir (g0) -- (x1); \chir (x1p) -- (g0);
	 \chir (g1) -- (x2); \chir (x2p) -- (g1);
 	 \chir (g4) -- (z); \chir (zp) -- (g4);
	 \wigM (g0) -- (g1); \wigM (g1) -- (g2); \wigM (g2) -- (g3); \wigM (g3) -- (g4); 
     \draw[-] (g0) to[out=300,in=0] (0,-0.6) to[out=180,in=240] (g0);
     \draw[-] (g1) to[out=300,in=0] (2,-0.6) to[out=180,in=240] (g1);
     \draw[-] (g2) to[out=300,in=0] (4,-0.6) to[out=180,in=240] (g2);
     \draw[-] (g3) to[out=300,in=0] (6,-0.6) to[out=180,in=240] (g3);
     \draw[-] (g4) to[out=300,in=0] (8,-0.6) to[out=180,in=240] (g4);
\end{scope}	 	

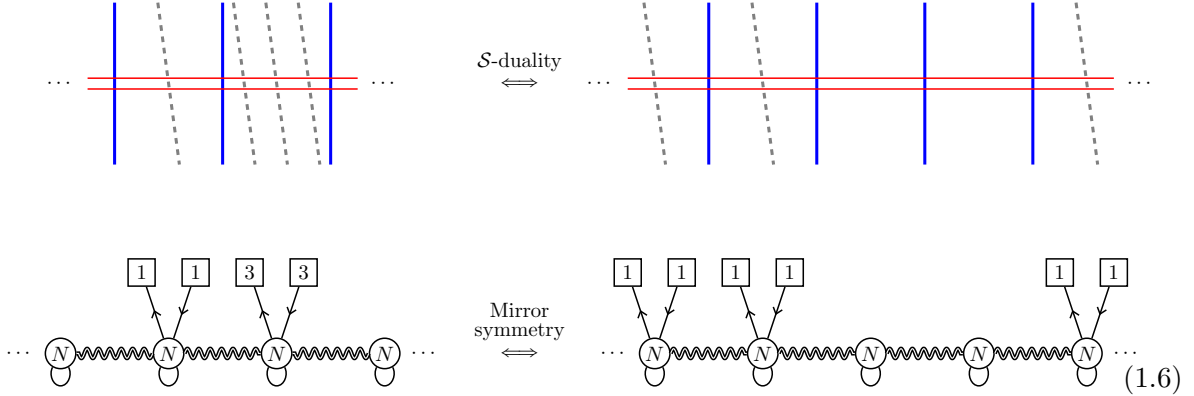
\epic}\ee
According to our proposal the part of the quiver associated to this sequence, given in the bottom left corner, contains three improved bifundamentals and four flavors which don't enter the superpotential.
The quiver corresponding to the $\CS$-dual section of the brane setup, given in the bottom right corner, contains instead four improved bifundamentals and three flavors. Notice that in the left quiver, the global symmetry will include a non-abelian $U(3)^2/U(1)$ factor associated to the 
3 consecutive $D5'$ branes. In the right quiver this symmetry appears in the IR via the enhancement mentioned above of the string of 3 improved bifundamentals associated to the 3 consecutive $NS$ branes. \\
We can then prove that the two improved quivers associated to the $\CS$-dual brane setups in \eqref{MSexample} are mirror dual, by running the $\mathcal{N}=2$ algorithm.
Let us start from the quiver on the l.h.s., we freeze the gauge interactions, breaking up the theory into the two types of generalized matter blocks:
\be
\resizebox{.8\hsize}{!}{
\begin{tikzpicture}[thick,node distance=3cm,gauge/.style={circle,draw,minimum size=5mm},flavor/.style={rectangle,draw,minimum size=5mm}]

	\path (-0.75,0) node (g0) {$\cdots$} --  (0,0) node[flavor] (f1) {$\!N\!$} -- (1.5,0) node[flavor] (f2) {$\!N\!$} ;	
	\wigM (f1) -- (f2);
	
\begin{scope}[shift={(2.3,0)}]
	\path (0,0) node[flavor] (g) {$\!N\!$} -- (-0.5,1.5) node[flavor] (y1) {$\!1\!$} -- (0.5,1.5) node[flavor] (y2) {$\!1\!$}; 
	\chir (g) -- (y1); \chir (y2) -- (g); 
	\draw (0.2,0) node[right] {\large{$\mathbb{I}$-wall}};
\end{scope}	 	

	\path  (4,0) node[flavor] (f1) {$\!N\!$} -- (5.5,0) node[flavor] (f2) {$\!N\!$} ;	
	\wigM (f1) -- (f2);

\begin{scope}[shift={(6.5,0)}]
	\path (0,0) node[flavor] (g) {$\!N\!$} -- (-0.5,1.5) node[flavor] (y1) {$\!1\!$} -- (0.5,1.5) node[flavor] (y2) {$\!1\!$}; 
	\chir (g) -- (y1); \chir (y2) -- (g); 
	\draw (0.2,0) node[right] {\large{$\mathbb{I}$-wall}};
\end{scope}	 	

\begin{scope}[shift={(8.5,0)}]
	\path (0,0) node[flavor] (g) {$\!N\!$} -- (-0.5,1.5) node[flavor] (y1) {$\!1\!$} -- (0.5,1.5) node[flavor] (y2) {$\!1\!$}; 
	\chir (g) -- (y1); \chir (y2) -- (g); 
	\draw (0.2,0) node[right] {\large{$\mathbb{I}$-wall}};
\end{scope}	 	
	
\begin{scope}[shift={(10.3,0)}]
	\path (0,0) node[flavor] (g) {$\!N\!$} -- (-0.5,1.5) node[flavor] (y1) {$\!1\!$} -- (0.5,1.5) node[flavor] (y2) {$\!1\!$}; 
	\chir (g) -- (y1); \chir (y2) -- (g); 
	\draw (0.2,0) node[right] {\large{$\mathbb{I}$}-wall};
\end{scope}	 	

	\path (12,0) node[flavor] (f3) {$\!N\!$} -- (13.5,0) node[flavor] (f4) {$\!N\!$}  -- (14.25,0) node (g15) {$\cdots$}; 	
	\wigM (f3) -- (f4);

\end{tikzpicture}}
\ee
Now we dualize each block, using \eqref{fig:3dbasicmove} and \eqref{fig:3dbasicmoveB}, transforming generalized flavors into improved bifundamentals and viceversa and glue back:
\be
\resizebox{\hsize}{!}{
\begin{tikzpicture}[thick,node distance=3cm,gauge/.style={circle,draw,minimum size=5mm},flavor/.style={rectangle,draw,minimum size=5mm}]

	\path (-0.75,0) node (g0) {$\cdots$} -- (0,0) node[gauge] (g1) {$\!\!\!N\!\!\!$} -- (1,0) node[gauge] (g2) {$\!\!\!N\!\!\!$}
		-- (-0.5,1.5) node[flavor] (x1) {$\!1\!$} -- (0.5,1.5) node[flavor] (x1p) {$\!1\!$};
	\wigT (g1) -- (g2);	\chir (g1) -- (x1); \chir (x1p) -- (g1); 

	\path  (2,0) node[gauge] (gg2) {$\!\!\!N\!\!\!$} -- (3.5,0) node[gauge] (gg3) {$\!\!\!N\!\!\!$} -- (4.5,0) node[gauge] (gg4) {$\!\!\!N\!\!\!$}; 
	\wigT (g2) -- (gg2);	\wigM (gg2) -- (gg3);	\wigT (gg3) -- (gg4);
	
	\path (5.5,0) node[gauge] (g3) {$\!\!\!N\!\!\!$} -- (6.5,0) node[gauge] (g4) {$\!\!\!N\!\!\!$}
		-- (5,1.5) node[flavor] (y1) {$\!1\!$} -- (6,1.5) node[flavor] (y2) {$\!1\!$};
	\chir (g3) -- (y1); \chir (y2) -- (g3); \wigT (gg4) -- (g3);  \wigT (g4) -- (g3);
 
	\path (7.5,0) node[gauge] (g5) {$\!\!\!N\!\!\!$} -- (9,0) node[gauge] (g6) {$\!\!\!N\!\!\!$} -- (10,0) node[gauge] (g7) {$\!\!\!N\!\!\!$};
	\wigT (g4) -- (g5); \wigM (g5) -- (g6); \wigT (g6) -- (g7);

	\path (11,0) node[gauge] (g8) {$\!\!\!N\!\!\!$} -- (12.5,0) node[gauge] (g9) {$\!\!\!N\!\!\!$} -- (13.5,0) node[gauge] (g10) {$\!\!\!N\!\!\!$};
	\wigT (g7) -- (g8); \wigM (g8) -- (g9); \wigT (g9) -- (g10);

	\path (14.5,0) node[gauge] (g11) {$\!\!\!N\!\!\!$} -- (16,0) node[gauge] (g12) {$\!\!\!N\!\!\!$} -- (17,0) node[gauge] (g13) {$\!\!\!N\!\!\!$};
	\wigT (g10) -- (g11); \wigM (g11) -- (g12); 	\wigT (g12) -- (g13);

	\path  (18,0) node[gauge] (g14) {$\!\!\!N\!\!\!$} -- (18.75,0) node (g15) {$\cdots$}
		-- (17.5,1.5) node[flavor] (z) {$\!1\!$} -- (18.5,1.5) node[flavor] (zp) {$\!1\!$};
	\wigT (g13) -- (g14); \chir (g14) -- (z); \chir (zp) -- (g14); 
	
	\draw[-] (g1) to[out=300,in=0] (0,-0.6) to[out=180,in=240] (g1);
	\draw[-] (g2) to[out=300,in=0] (1,-0.6) to[out=180,in=240] (g2);
	\draw[-] (gg2) to[out=300,in=0] (2,-0.6) to[out=180,in=240] (gg2);
	\draw[-] (gg3) to[out=300,in=0] (3.5,-0.6) to[out=180,in=240] (gg3);
	\draw[-] (gg4) to[out=300,in=0] (4.5,-0.6) to[out=180,in=240] (gg4);
	\draw[-] (g3) to[out=300,in=0] (5.5,-0.6) to[out=180,in=240] (g3);
	\draw[-] (g4) to[out=300,in=0] (6.5,-0.6) to[out=180,in=240] (g4);
	\draw[-] (g5) to[out=300,in=0] (7.5,-0.6) to[out=180,in=240] (g5);
	\draw[-] (g6) to[out=300,in=0] (9,-0.6) to[out=180,in=240] (g6);
	\draw[-] (g7) to[out=300,in=0] (10,-0.6) to[out=180,in=240] (g7);
	\draw[-] (g8) to[out=300,in=0] (11,-0.6) to[out=180,in=240] (g8);
	\draw[-] (g9) to[out=300,in=0] (12.5,-0.6) to[out=180,in=240] (g9);
	\draw[-] (g10) to[out=300,in=0] (13.5,-0.6) to[out=180,in=240] (g10);
	\draw[-] (g11) to[out=300,in=0] (14.5,-0.6) to[out=180,in=240] (g11);
	\draw[-] (g12) to[out=300,in=0] (16,-0.6) to[out=180,in=240] (g12);
	\draw[-] (g13) to[out=300,in=0] (17,-0.6) to[out=180,in=240] (g13);
	\draw[-] (g14) to[out=300,in=0] (18,-0.6) to[out=180,in=240] (g14);

\end{tikzpicture}}
\ee
Now we implement the \emph{fusion to Identity}  property  of the $\CS$-walls \eqref{Swallprop}, with the effect of removing all the  $\CS$-wall theories from the improved quiver, to obtain:
\be
\begin{tikzpicture}[thick,node distance=3cm,gauge/.style={circle,draw,minimum size=5mm},flavor/.style={rectangle,draw,minimum size=5mm}]
	
	\path  (-0.75,0) node {$\cdots$} -- (0,0) node[gauge](g0) {$\!\!\!N\!\!\!$} --  (2,0) node[gauge](g1) {$\!\!\!N\!\!\!$} 
	 	-- (4,0) node[gauge](g2) {$\!\!\!N\!\!\!$} -- (6,0) node[gauge] (g3) {$\!\!\!N\!\!\!$}-- (8,0) node[gauge] (g4) {$\!\!\!N\!\!\!$} 
	 	-- (8.75,0) node {$\cdots$}
	 	-- (-0.5,1.5) node[flavor] (x1) {$\!1\!$} -- (0.5,1.5) node[flavor] (x1p) {$\!1\!$} 
	 	-- (1.5,1.5) node[flavor] (x2) {$\!1\!$} -- (2.5,1.5) node[flavor] (x2p) {$\!1\!$} 
	 	-- (7.5,1.5) node[flavor] (z) {$\!1\!$} -- (8.5,1.5) node[flavor] (zp) {$\!1\!$} ;
	 
	\chir (g0) -- (x1); \chir (x1p) -- (g0);
	\chir (g1) -- (x2); \chir (x2p) -- (g1);
 	\chir (g4) -- (z); \chir (zp) -- (g4);
	\wigM (g0) -- (g1); \wigM (g1) -- (g2); \wigM (g2) -- (g3); \wigM (g3) -- (g4); 
    \draw[-] (g0) to[out=300,in=0] (0,-0.6) to[out=180,in=240] (g0);
    \draw[-] (g1) to[out=300,in=0] (2,-0.6) to[out=180,in=240] (g1);
    \draw[-] (g2) to[out=300,in=0] (4,-0.6) to[out=180,in=240] (g2);
    \draw[-] (g3) to[out=300,in=0] (6,-0.6) to[out=180,in=240] (g3);
    \draw[-] (g4) to[out=300,in=0] (8,-0.6) to[out=180,in=240] (g4);
    
\end{tikzpicture}\ee
which is precisely the quiver  associated to the $\CS$-dual brane setup in fig. \eqref{MSexample}.
\vspace{0.5cm}

There are various  natural generalization of this result. 
We can easily describe $(p,q)$-webs  of rectangular shape formed by an arbitrary number of $D5'$ branes and one $NS$. Using the algorithm we can obtain the QFT description of the $\CS$-dual $(p,q)$-web containing many $NS$'s sitting on top of a single $D5'$. \\
We can turn on real mass deformations in our quivers to generate Chern-Simons interactions and/or theories with chiral matter (different number of fundamentals vs anti-fundamentals). The corresponding brane setup might include $(p,q)$ 5-branes and non-rectangular $(p,q)$-webs. 
We will discuss these theories in \cite{BCP3}, using the chiral improved bifundamental introduced in \cite{BCP1}.

We still don't know how to  describe more generic $3d$ $\cN=2$ setups involving all four types of $5$-branes ($NS$, $NS'$, $D5$, $D5'$) and a non-constant number of $D3$ branes along the brane setup. For such setups we need a new object: an asymmetric improved bifundamental with non-abelian global symmetry $S[U(N_1) \times U(N_2)]$. We plan to investigate this in the future. \\

Our results can  play a role also in the study of non-perturbative properties of $4d$ $\mathcal{N}=1$ QFT’s. For instance in some cases the $3d$ mirror of a $4d$ SCFT, defined through a stringy or higher dimensional construction, might be a quiver containing improved bifundamentals. As illustrative examples, we show that $4d$ $SU(N)$ adjoint SQCD with $F$ flavors possesses a $3d$ mirror which is a linear quiver with $F-1$ $U(N)$, one $U(1)$ gauge nodes and $F-2$ improved bifundamentals, and we work out the $3d$ mirror of $4d$ $\cN=1$ $SU(N)$ quivers associated to linear Type IIA brane setups. \\

We also present a family of $4d$ $\mathcal{N}=1$ improved quivers related by mirror-like dualities.
The $4d$ improved quivers contain $4d$ improved bifundamental links which we identify with the $FE[USp(2N)]$ theory introduced  in \cite{Pasquetti:2019hxf}. 
$4d$ mirror dualities can be demonstrated via a $4d$ dualization algorithm. The basic move, dualizing an improved bifundamental block into a generalized flavor block is given by the $4d$ $\cN=1$ \emph{star-triangle} or \emph{braid} duality.
As a simple example, we present the mirror dual of the $4d$ $\mathcal{N}=1$ antisymmetric $USp(2N)$ SQCD with $2F+4$ flavors  which reduces to the $3d$ $\mathcal{N}=2$ mirror pair for the $U(N)$ adjoint SQCD, upon a suitable dimensional reduction limit. The proposed duality generalizes the self-dual case with eight flavors, called CSST duality \cite{Csaki:1997cu}.

\vspace{0.5cm}

The paper is organized as follows.
In section \ref{sqcdmirror} we present the mirror of the adjoint $\mathcal{N}=2$ SQCD, we work out the operator map, study various deformations and perform
several consistency checks.
In section \ref{sec:3d_algorithm} we introduce the $\mathcal{N}=2$ dualization algorithm and we apply it to the derivation of the SQCD mirror dual.
In section \ref{sec:branes} we formulate our proposal to associate improved quivers to brane setups with four supercharges and discuss various examples.
In section \ref{sec:pqwebs} we study $(p,q)$-webs, while section \ref{3dmirror} studies $3d$ mirrors of $4d$ $SU(N)$ quivers.
Finally in section \ref{4dsec} we discuss  $4d$ $\mathcal{N}=1$ mirror dualities.

\section{$3d$ $\cN=2$ adjoint $U(N)$ SQCD and its mirror}\label{sqcdmirror}

\begin{figure}
\centering
\begin{tikzpicture}[thick,node distance=3cm,gauge/.style={circle,draw,minimum size=5mm},flavor/.style={rectangle,draw,minimum size=5mm}] 

\begin{scope}[shift={(1,0)}]
	\path (0,0) node[gauge](g) {$\!\!\!N\!\!\!$} -- (-0.5,1.5) node[flavor] (x1) {$\!F\!$} -- (0.5,1.5) node[flavor] (x2) {$\!F\!$} 
		-- (2,0) node{$\Longleftrightarrow$};
	
	\chir (g) -- (x1); \draw (-0.3,0.6) node[left] {$Q$};
	\chir (x2) -- (g); \draw (0.3,0.6) node[right] {$\tilde{Q}$};
	\draw[-] (g) to[out=-60,in=0] (0,-0.6) to[out=180,in=-120] (g); \draw (0,-0.6) node[right] {$A$};
	
	\draw (0,-1.5) node{$\cW = 0$};
\end{scope}
	 
	\path (5,0) node[gauge](g1) {$\!\!\!N\!\!\!$} -- (7,0) node[gauge](g2) {$\!\!\!N\!\!\!$} -- (10,0) node[gauge](g3) {$\!\!\!N\!\!\!$} -- (12,0) node[gauge](g4) {$\!\!\!N\!\!\!$} -- (5,1.5) node[flavor](y1) {$\!1\!$} -- (12,1.5) node[flavor](y2) {$\!1\!$};
	 
	\wigM (g1) -- (g2); \draw (6,0.4) node {$\Pi_2$};
	\wigM (g2) -- (8,0); 
	\wigM (g3) -- (9,0); 
	\wigM (g3) -- (g4); \draw (11,0.4) node {$\Pi_{F-1}$};
	\draw[-, shorten >= 9, shorten <= 6, shift={(-0.05,0.05)}, mid arrowsm] (5,0) -- (5,1.5); \draw (5,0.75) node[left] {$V_1$};
	\draw[-, shorten >= 4, shorten <= 11, shift={(0.05,0.15)}, mid arrowsm] (5,1.5) -- (5,0);
	\draw[-, shorten >= 9, shorten <= 6, shift={(0.05,0.05)}, mid arrowsm] (12,0) -- (12,1.5); \draw (12,0.75) node[right] {$V_2$};
	\draw[-, shorten >= 4, shorten <= 11, shift={(-0.05,0.15)}, mid arrowsm] (12,1.5) -- (12,0); 
	\draw (5,0.45) node[cross]{}; \draw (12,0.45) node[cross]{};	
	\draw (8.5,0) node {$\cdots$};
	\draw[-] (g1) to[out=-60,in=0] (5,-0.6) to[out=180,in=-120] (g1); \draw (5,-0.7) node[right] {$A_1$};
	\draw[-] (g2) to[out=-60,in=0] (7,-0.6) to[out=180,in=-120] (g2); \draw (7,-0.7) node[right] {$A_2$};
	\draw[-] (g3) to[out=-60,in=0] (10,-0.6) to[out=180,in=-120] (g3); \draw (10,-0.7) node[right] {$A_{F-2}$};
	\draw[-] (g4) to[out=-60,in=0] (12,-0.6) to[out=180,in=-120] (g4); \draw (12,-0.7) node[right] {$A_{F-1}$};
	
	\draw (4,-1.5) node[right]{$\cW = \sum_{j=0}^{N-1} ( Flip[V_1 A_1^j \tilde{V}_1] + Flip[V_2 A_{F-1}^j \tilde{V}_2] )+\mathcal{W}_{\text{gluing}}$  };
	 
\end{tikzpicture}
\caption{Mirror duality for the $\CN=2$ adjoint SQCD. Each node, round or square, denotes a $U(N)$ group, gauge or flavor respectively. Lines with an ingoing or outgoing arrow are fields in the fundamental or antifundamental representation of the group to whom is linked. Arches denotes field in the traceless adjoint representation. In the mirror theory there are also double wiggly-lines that represent a $FM[U(N)]$ theory and crosses denoting flipping fields. In the pictures we also give the name of the fields beside the line that represent it, the names of flipping fields are omitted in the picture but their presence can be read from the superpotential given below the theory. To avoid cluttering, whenever we have a double line, straight or wiggly, we just give the name of one field, it is implied the presence of a second field, distinguished by a tilde, that is in the conjugate representation.}
\label{fig:SQCD_Dual} 
\end{figure}
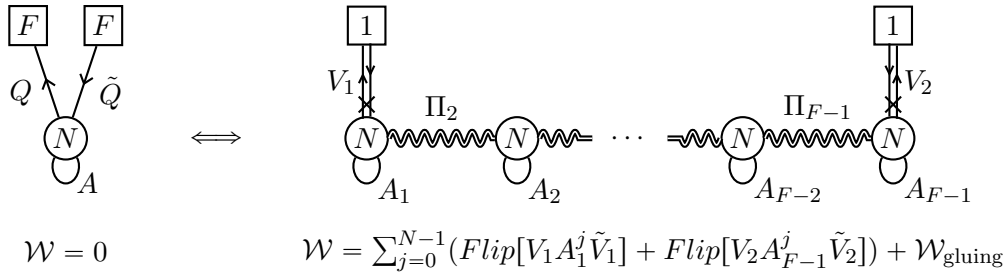

In this section we present the mirror duality for the $\CN=2$ adjoint SQCD, which is depicted in figure \ref{fig:SQCD_Dual}. The electric theory is an $\CN=2$ $U(N)$ gauge theory with a chiral field $A$ in the traceless adjoint representation and $F$ fundamental chirals $Q$ and antifundamental chirals $\tilde{Q}$, with zero superpotential. The global symmetry group of the theory is:
 \begin{align}
	SU(F)_U \times SU(F)_W \times U(1)_m \times U(1)_\tau \times U(1)_Y\,,
\end{align}
where we denote by  $U_j$ with $\sum_{j=1}^{N_f} U_j=0$, $W_j$ with $\sum_{j=1}^{N_f} W_j=0$, $m$  the real masses for the fundamental chirals,  $\tau$ is the real mass for the adjoint chiral and $Y$ is the FI parameter.
The  charges and representations for all the fields is given in table \ref{tab:SQCD_ChargesE}.\footnote{We recall that in $\CN=2$ theories the R-symmetry group is abelian and can mix with other abelian symmetries along the RG-flow and the value of the superconformal R-charge can be fixed via $F$-extremization \cite{Jafferis:2010un}.  In the table we give  trial $U(1)_{R_0}$-charges. }

\begin{table}[h]
\centering
\renewcommand{\arraystretch}{1.15}
\begin{tabular}{|c|c|c|c|c|c|c|}
	\hline
		&  $U(1)_{R_0}$ & $U(1)_\tau$ & $U(1)_m$ & $SU(F)_U \times SU(F)_W$  \\
	\hline
	$Q$ & $1$ & $0$ & $-1$ & $ \bar{\textbf{F}}\times\textbf{1} $ \\
	$\tilde{Q}$ &  $1$ & $0$ & $-1$ & $ \textbf{1}\times\textbf{F} $ \\
	$A$ &  $0$ & $1$ & $0$ & $ \textbf{1}\times\textbf{1}$ \\
	\hline
\end{tabular} 
\caption{List of the charges and representation of the fields in the electric theory.}
	\label{tab:SQCD_ChargesE}
\end{table}

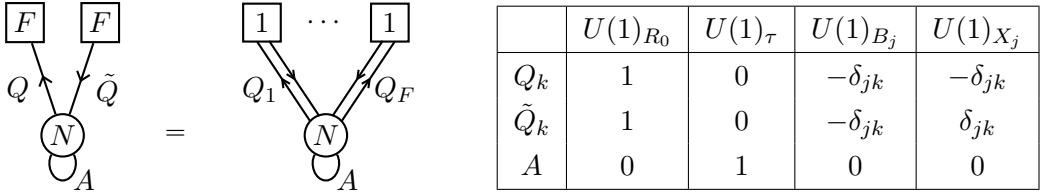
\begin{figure}
\centering
\begin{tikzpicture}[thick,node distance=3cm,gauge/.style={circle,draw,minimum size=5mm},flavor/.style={rectangle,draw,minimum size=5mm}] 
 
\begin{scope}[shift={(0.5,0)}]
	\path (0,0) node[gauge](g) {$\!\!\!N\!\!\!$} -- (-0.5,1.5) node[flavor] (x1) {$\!F\!$} -- (0.5,1.5) node[flavor] (x2) {$\!F\!$} -- (1.5,0) node{$=$};
	
	\chir (g) -- (x1); \draw (-0.3,0.6) node[left] {$Q$};
	\chir (x2) -- (g); \draw (0.3,0.6) node[right] {$\tilde{Q}$};
	\draw[-] (g) to[out=-60,in=0] (0,-0.6) to[out=180,in=-120] (g); \draw (0,-0.6) node[right] {$A$};
\end{scope}

	\path (4,0) node[gauge](g) {$\!\!\!N\!\!\!$} -- (3.15,1.5) node[flavor](x1) {$\!1\!$} -- (4.85,1.5) node[flavor] (x2) {$\!1\!$};
	
	\draw[-, shorten >= 7, shorten <= 8, shift={(-0.05,-0.05)}, mid arrowsm] (4,0) -- (3,1.5); \draw (3.5,0.6) node[left] {$Q_1$};
	\draw[-, shorten >= 7, shorten <= 10, shift={(0.05,0.05)}, mid arrowsm] (3,1.5) -- (4,0);
	\draw[-, shorten >= 7, shorten <= 8, shift={(0.05,-0.05)}, mid arrowsm] (4,0) -- (5,1.5); \draw (4.5,0.6) node[right] {$Q_F$};
	\draw[-, shorten >= 7, shorten <= 10, shift={(-0.05,0.05)}, mid arrowsm] (5,1.5) -- (4,0); 
	\draw[-] (g) to[out=-60,in=0] (4,-0.6) to[out=180,in=-120] (g); \draw (4,-0.6) node[right] {$A$};
	\draw (4,1.5) node{$\cdots$};
	
	\draw (6,0.5) node[right] {
	\renewcommand{\arraystretch}{1.2}
	$\begin{tabular}{|c|c|c|c|c|c|}
		\hline
				&  $U(1)_{R_0}$ & $U(1)_\tau$ & $U(1)_{B_j}$ & $U(1)_{X_j}$ \\
		\hline
		$Q_k$ &  $1$ & $0$ & $- \delta_{jk}$ & $ -\delta_{jk}$ \\
		$\tilde{Q}_k$ & $1$ & $0$ & $- \delta_{jk} $ & $\delta_{jk}$ \\
		$A$ & $0$ & $1$ & $0$ & $0$  \\
		\hline
	\end{tabular}$};
	
\end{tikzpicture}
\caption{Reparameterization of the electric theory. On the right of the picture we also listed the abelian charges of all the fields of the reparameterized theory. The convention is to take the fields $Q_j$ in the fundamental and $\tilde{Q}_j$ antifundamental representation of the gauge group.}
\label{fig:SQCD_Manifest} 
\end{figure}

It will also be useful to parameterize the SQCD theory so that its manifest symmetry matches that of the mirror theory
by combining pairs of fundamental/anti-fundamental chirals into flavors with axial-like mass $B_j$ and vector-like mass $X_j$. The reparameterized theory is depicted in figure \ref{fig:SQCD_Manifest}, along with a table with all the representation and charges of the fields after the reparameterization.

The set of real masses for the vector-like symmetries can be taken such that: $\sum_{j=1}^F X_j=0$, since the gauge group is $U(N)$. The $U(1)_m$ symmetry of the theory before the reparameterization is related to the axial masses as:
\begin{align}
	m = \frac{1}{F} \sum_{j=1}^F B_j \,.
\label{axdef}
\end{align}
We can also define a new set of axial masses as: $\tilde{B}_j = B_j - m$, so that $\sum_{j=1}^F \tilde{B}_j=0$. The real masses of the two parameterization are related as:
\begin{align}
	& U_j = X_j - \tilde{B}_j \nonumber \\
	& W_j = X_j + \tilde{B}_j \,,
	\label{uvdef}
	\end{align}
while the symmetries recombine as:
\begin{align}
	\prod_{j=1}^{F} U(1)_{B_j}  \times S\left[\prod_{j=1}^{F} U(1)_{X_j}\right] = &	
	S\left[\prod_{j=1}^{F} U(1)_{\tilde B_j}\right] \times S\left[\prod_{j=1}^{F} U(1)_{X_j}\right] 
		 \times U(1)_m \nn \\ 
	\to & \, SU(N)_U\times SU(N)_W \times U(1)_m	\,.
\end{align}

\paragraph{Dual quiver}
Let's now discuss the mirror theory. The main ingredient is the {\it improved bifundamental} which is identified with the $FM[U(N)]$ theory, a $3d$ $\CN=2$ SCFT introduced in \cite{Pasquetti:2019tix} which we describe in appendix \ref{app:FM}.
We denote this theory compactly by  two wiggle lines connecting the two $U(N)$ nodes to visualize the two non-abelian  $U(N)$ IR symmetries.
In addition to them, the improved bifundamental has a $U(1)_\tau \times U(1)_\D$ abelian symmetry. The IR spectrum of the improved bifundamental includes two traceless adjoint operators and two  bifundamental $(N, \bar N)$, $(\bar N, N)$ operators $\Pi,\tilde \Pi$ of the two $U(N)$ symmetries and a matrix of singlets $\mathsf{B}_{n, m}$, with charges given in \ref{tab:FM_operators}. In particular the bifundamental
operators carry charge one under the {\it axial} $ U(1)_\D$ symmetry.

Our adjoint SQCD mirror dual is given by a linear quiver with $F-2$ improved bifundamental links 
 and at each end of the quiver we have the the  flavors $V_1,\tilde{V}_1$ and  $V_2,\tilde{V}_2$:
\be
\begin{tikzpicture}[thick,node distance=3cm,gauge/.style={circle,draw,minimum size=5mm},flavor/.style={rectangle,draw,minimum size=5mm}] 	 
\begin{scope}
	
	\path (5,0) node[gauge](g1) {$\!\!\!N\!\!\!$} -- (7,0) node[gauge](g2) {$\!\!\!N\!\!\!$} -- (10,0) node[gauge](g3) {$\!\!\!N\!\!\!$} 
		-- (12,0) node[gauge](g4) {$\!\!\!N\!\!\!$} -- (5,1.5) node[flavor](y1) {$\!1\!$} -- (12,1.5) node[flavor](y2) {$\!1\!$};	 
		
	\wigM (g1) -- (g2); \draw (6,0.4) node {$\Pi_2$};
	\wigM (g2) -- (8,0); 
	\wigM (g3) -- (9,0); 
	\wigM (g3) -- (g4); \draw (11,0.4) node {$\Pi_{F-1}$};
	\draw[-, shorten >= 9, shorten <= 6, shift={(-0.05,0.05)}, mid arrowsm] (5,0) -- (5,1.5); \draw (5,0.75) node[left] {$V_1$};
	\draw[-, shorten >= 4, shorten <= 11, shift={(0.05,0.15)}, mid arrowsm] (5,1.5) -- (5,0);
	\draw[-, shorten >= 9, shorten <= 6, shift={(0.05,0.05)}, mid arrowsm] (12,0) -- (12,1.5); \draw (12,0.75) node[right] {$V_2$};
	\draw[-, shorten >= 4, shorten <= 11, shift={(-0.05,0.15)}, mid arrowsm] (12,1.5) -- (12,0); 
	\draw (5,0.45) node[cross]{}; \draw (12,0.45) node[cross]{};	
	\draw (8.5,0) node {$\cdots$};
	\draw[-] (g1) to[out=-60,in=0] (5,-0.6) to[out=180,in=-120] (g1); \draw (5,-0.7) node[right] {$A_1$};
	\draw[-] (g2) to[out=-60,in=0] (7,-0.6) to[out=180,in=-120] (g2); \draw (7,-0.7) node[right] {$A_2$};
	\draw[-] (g3) to[out=-60,in=0] (10,-0.6) to[out=180,in=-120] (g3); \draw (10,-0.7) node[right] {$A_{F-2}$};
	\draw[-] (g4) to[out=-60,in=0] (12,-0.6) to[out=180,in=-120] (g4); \draw (12,-0.7) node[right] {$A_{F-1}$};
	
	\draw (4,-1.5) node[right]{$\cW = \sum_{j=0}^{N-1} ( Flip[V_1 A_1^j \tilde{V}_1] + Flip[V_2 A_{F-1}^j \tilde{V}_2] )+\mathcal{W}_{\text{gluing}}$  };
	
\end{scope} 
\end{tikzpicture}
\ee
The list of charges and representations for all the fields and  bifundamental operators  is given in table \ref{tab:SQCD_Charges}.
\begin{table}[h]
\centering
\renewcommand{\arraystretch}{1.2}
\begin{tabular}{|c|c|c|c|c|c|}
	\hline
		&  $U(1)_{R_0}$ & $U(1)_\tau$ & $U(1)_{B_j}$ & $U(1)_Y$ \\
	\hline
	$V_1 ,\tilde{V}_1 $ &  $0$ & $\frac{1-N}{2}$ & $ \delta_{j1}$ & $\mp 1$ \\
	$V_2,\tilde{V}_2$ &   $0$ & $\frac{1-N}{2}$ & $ \delta_{jF}$ & $0$ \\
	$\Pi_k, \tilde \Pi_k$ & $0$ & $0$ & $ \delta_{jk}$ & $0$ \\
	$A_k$ &  $0$ & $1$ & $0$ & $0$ \\
	\hline
\end{tabular}
\caption{
List of charges and representations. The $\Pi_j$ operators  are in the fundamental representation of the $(j-1)$-th and antifundamental of the $j$-th gauge groups. The fields $V_1$ and $V_2$ are in fundamental representation of the first and last gauge group, respectively.
 }
 \label{tab:SQCD_Charges}
\end{table}

The manifest UV global symmetry is:
\begin{align}
	\prod_{j=1}^{F} U(1)_{B_j}  \times \prod_{j=1}^{F-1} U(1)_{X_j-X_{j-1}}  \times U(1)_\tau \times U(1)_Y \,,
\end{align}
Where $X_{j+1}-X_j$ is the FI parameter related to the $U(1)$ topological symmetry of the $j$-th gauge node. The parameters $B_j$ for $j=2,\cdots F-1$ are the real  axial mass associated to the $U(1)_{B_j}$ symmetries of each improved bifundamental, while $B_1$ and $B_F$ are the axial symmetries for the left and right vertical flavors, respectively. 
Notice that we can re-absorbe a  $U(1)$ vector-like symmetry by a gauge transformation since all nodes are $U(N)$.
We have the following  symmetry enhancement in the IR 
\begin{align}\label{eq:mirror_symenh}
	\prod_{j=1}^{F} U(1)_{B_j}  \times \prod_{j=1}^{F-1} U(1)_{X_{j+1} - X_{j}}  \to
	SU(N)_U\times SU(N)_V \times U(1)_m\,,
\end{align}
so that in the IR, the mirror dual theory has exactly  the same global symmetry group of the electric theory.

The theory also contains singlets. To avoid introducing too many names for the singlet fields, we will adopt the following notation.
Given an operator $X$ and a gauge singlet elementary field $O_X$ we denote a superpotential term of the form $\mathcal{W}= O_X X$ as $Flip[X]$. In addition we will refer to the flipper singlet $O_X$ as $\mathcal{F}[X]$. In our mirror theory we have singlets $\mathcal{F}[V_1A^j \tilde{V}_1]$ and $\mathcal{F}[V_2 A^j \tilde{V}_2]$, which flip the dressed mesons constructed with the left and right flavors, they are represented as crosses in the picture \ref{fig:SQCD_Dual}. \\

In the mirror theory we have a string of consecutive improved bifundamentals  which are glued by gauging a diagonal combination of the two $U(N)$ symmetries with the addition of an adjoint field $A$. More precisely we couple the  adjoint operator $\mathsf{A}_L$ of the improved bifundamental on the left and the adjoint operator $\mathsf{A}_R$ of the improved bifundamental on the right to the extra adjoint field $A$ as: $\cW = A ( \mathsf{A}_L - \mathsf{A}_R )$. We will also use the short-hand notation $\cW_{\text{gluing}}$ to collect all the superpotential terms coming from this procedure \footnote{Notice that here we glue improved bifundamentals 
by turning on only $\cW_{\text{gluing}}$.
If one adds also a monopole superpotential of the type $\cW = \mathfrak{M}^+ + \mathfrak{M}^-$, then two  improved bifundamental theories fuse to an $\mathbb{I}$-wall  as explained in appendix \ref{app:FM}.}. Notice that when we glue a string of improved bifundamentals, all the $U(1)_\tau$ symmetries are identified while the $U(1)_{B_j}$ symmetries are all preserved. These symmetries then recombine with the topological symmetries and enhance to match the global symmetry group of the electric theory as in \eqref{eq:mirror_symenh}.\\
At the level of $S_b^3$ partition functions, the duality in figure \ref{fig:SQCD_Dual} implies the  identity:\footnote{We follow the conventions summarized in appendix \ref{inpaconv}.}
\begin{align}\label{eq:SQCD_Zidentity}
	Z_{\text{SQCD}} \left(\tau, \vec{B}, \vec{X}, Y \right) = Z_{\widecheck{\text{SQCD}}} \left(\tau, \vec{B}, \vec{X}, Y \right)\,.
\end{align}
On the l.h.s. we have the partition function of the $U(N)$ adjoint SQCD parameterized as in figure \ref{fig:SQCD_Manifest}, which is given as:
\begin{align}\label{eq:SQCD_parfun}
	Z_{\text{SQCD}} \left(\tau,\vec{B},\vec{X}, Y\right) = & \int d\vec{Z}_N \D_N (\vec{Z},\tau) e^{2\pi i Y \sum_{j=1}^N Z_j} \prod_{j=1}^N \prod_{a=1}^F s_b\left( B_a \pm (Z_j - X_a) \right) \,.
\end{align}
The partition function of the mirror dual theory, given on the r.h.s.\!\! of figure \ref{fig:SQCD_Dual}, is instead given as:
\begin{align}\label{eq:SQCDmirr_parfun}
	Z_{\widecheck{\text{SQCD}}} \left(\tau, \vec{B}, \vec{X}, Y\right) & = e^{2 \pi i N Y X_1} \int \prod_{a=1}^{F-1}  d\vec{Z}_N^{(a)}  \D_N (\vec{Z}^{(a)},\tau) e^{2\pi i (X_{a+1}-X_{a}) \sum_{j=1}^N Z_j } \nonumber \\ 
	& \prod_{j=1}^N \big[ s_b\left( \frac{iQ}{2} - \frac{1-N}{2}\tau - B_1 \pm (Z_j^{(1)} - Y) \right) s_b( -\frac{iQ}{2} + (j-N)\tau + 2B_1)  \nn \\
	& s_b \left( \frac{iQ}{2} - \frac{1-N}{2}\tau - B_F \pm Z_j^{(F-1)} \right) s_b( -\frac{iQ}{2} + (j-N)\tau + 2B_F) \big] \nn \\
	& \prod_{a=1}^{F-2} Z_{FM}^{(N)} \left( \vec{Z}^{(a)},\vec{Z}^{(a+1)},\tau,B_{a+1} \right) \,.
\end{align}
The $S_b^3$ partition function of the $FM[U(N)]$ theory is defined in appendix \ref{app:FM}.


\subsection{Comments on the $F=1$ and $F=2$ cases}
The  cases $F=1,2$ were already discussed in literature, in this section we wish to comment on how our result reconciles with these known results. \\

Let us start with the $F=2$ case. In this case our mirror pair in figure \ref{fig:SQCD_Dual} has no improved bifundamental links and it reduces to a self-duality modulo flips:
\be\label{CSST3d}
\begin{tikzpicture}[thick,node distance=3cm,gauge/.style={circle,draw,minimum size=5mm},flavor/.style={rectangle,draw,minimum size=5mm}] 
 
	\path (0,0) node[gauge](g) {$\!\!\!N\!\!\!$} -- (-1,1.5) node[flavor] (x1) {$\!1\!$} -- (1,1.5) node[flavor] (x2) {$\!1\!$} 
		-- (2.5,0) node{$\Leftrightarrow$};
	
	\draw[-, shorten >= 7, shorten <= 8, shift={(-0.05,-0.05)}, mid arrowsm] (0,0) -- (-1,1.5); \draw (-0.5,0.6) node[left] {$Q_1$};
	\draw[-, shorten >= 7, shorten <= 10, shift={(0.05,0.05)}, mid arrowsm] (-1,1.5) -- (0,0);
	\draw[-, shorten >= 7, shorten <= 8, shift={(0.05,-0.05)}, mid arrowsm] (0,0) -- (1,1.5); \draw (0.5,0.6) node[right] {$Q_2$};
	\draw[-, shorten >= 7, shorten <= 10, shift={(-0.05,0.05)}, mid arrowsm] (1,1.5) -- (0,0); 
	\draw[-] (g) to[out=-60,in=0] (0,-0.6) to[out=180,in=-120] (g); \draw (0,-0.6) node[right] {$A$};
	
	\draw (0,-1.5) node {$\CW = 0$};
	
\begin{scope}[shift={(2,0)}]
	\path (4,0) node[gauge](g) {$\!\!\!N\!\!\!$} -- (3.15,1.5) node[flavor](x1) {$\!1\!$} -- (4.85,1.5) node[flavor] (x2) {$\!1\!$};
	
	\draw[-, shorten >= 7, shorten <= 8, shift={(-0.04,-0.04)}, middx arrowsm] (4,0) -- (3,1.5); \draw (3.5,0.6) node[left] {$P_1$};
	\draw[-, shorten >= 7, shorten <= 10, shift={(0.04,0.04)}, midsx arrowsm] (3,1.5) -- (4,0);
	\draw[-, shorten >= 7, shorten <= 8, shift={(0.04,-0.04)}, middx arrowsm] (4,0) -- (5,1.5); \draw (4.5,0.6) node[right] {$P_2$};
	\draw[-, shorten >= 7, shorten <= 10, shift={(-0.04,0.04)}, midsx arrowsm] (5,1.5) -- (4,0); 
	\draw (3.7,0.45) node[rotate={30}] {\LARGE{$\times$}}; \draw (4.3,0.45) node[rotate={-30}] {\LARGE{$\times$}};
	\draw[-] (g) to[out=-60,in=0] (4,-0.6) to[out=180,in=-120] (g); \draw (4,-0.6) node[right] {$C$};
	
	\draw (4,-1.5) node {$\CW = \sum_{j=0}^{N-1} \big( Flip[P_1 C^j \tilde{P}_1] +$};
	\draw (4,-2.2) node {$ + Flip[P_2 C^j \tilde{P}_2] \big)$};
\end{scope}
	
\end{tikzpicture}\ee

\begin{table}[h]
\centering
\renewcommand{\arraystretch}{1.2}
\begin{tabular}{|c|c|c|c|c|c|c|}
	\hline
		&  $U(1)_{R_0}$ & $U(1)_\tau$ & $U(1)_{B_1}$ &  $U(1)_{B_2}$ &  $U(1)_{X}$ & $U(1)_Y$ \\
	\hline
	$Q_1 ,\tilde{Q}_1 $ &  $1$ & $0$ & $ -1$ & $0$  & $\pm 1/2$ & $0$ \\
	$Q_2 ,\tilde{Q}_2 $ &  $1$ & $0$ & $ 0$ & $-1$  & $\mp 1/2$ & $0$ \\
	$P_1 ,\tilde{P}_1 $ &  $0$ & $\frac{1-N}{2}$ & $ 1$ & $0$  & $0$ & $\mp 1/2$ \\
	$P_2 ,\tilde{P}_2 $ &  $0$ & $\frac{1-N}{2}$ & $ 0$ & $1$  & $0$ & $\pm 1/2$ \\
	$A,C $ &  $0$ & $1$ & $ 0$ & $0$  & $0$ & $0$ \\
	\hline
\end{tabular}
\caption{
Charges for the fields in the mirror $F=2 $ self-duality. 
In the electric theory the FI parameter for the topological symmetry is $Y$, while in the dual it is $X$. 
}
 \label{f2table}
\end{table}
The duality \eqref{CSST3d} was  interpreted as a mirror symmetry in  \cite{Benvenuti:2018bav}, which obtained it reducing the CSST self-duality modulo flips of $4d$ $\cN=1$ $Usp(2N)$ with antisymmetric and $8$ fundamentals \cite{Csaki:1997cu}.\footnote{A similar self-duality with $6N$ instead of $2N$ flipping fields on the r.h.s. can  be  obtained via  sequential deconfinement, \cite{Benvenuti:2020gvy}. As shown in \cite{Benvenuti:2018bav}, \eqref{CSST3d} is the $3d$ reduction of the CSST self-duality of $Usp(2N)$ with antisymmetric and $8$ fundamentals, while the sequential deconfinement method \cite{Benvenuti:2020gvy, Bajeot:2022lah} proves the IP-like self-duality of $Usp(2N)$ with antisymmetric and $8$ fundamentals and its $3d$ reductions.}

The $F=1$  case can not be  directly read from  the mirror pair proposed in figure \ref{fig:SQCD_Dual}, which is defined only for $F\geq 2$.
However our dualization algorithm, which as we will see  in section \ref{sec:3d_algorithm} allows us to prove the $F\geq 2$ duality, can be run also in the $F=1$ case and the result produced is consistent with the earlier  duality shown in figure
\eqref{fig:SQCD_1flav} which  was discussed in  \cite{Benvenuti:2018bav} and  derived via sequential deconfinement in \cite{Pasquetti:2019uop, Bajeot:2022lah}. 
 \be
\begin{tikzpicture}[thick,node distance=3cm,gauge/.style={circle,draw,minimum size=5mm},flavor/.style={rectangle,draw,minimum size=5mm}] 
 
	\path (0,0) node[gauge](g) {$\!\!\!N\!\!\!$} -- (0,1.5) node[flavor] (x1) {$\!1\!$} -- (1.75,0.5) node{$\Leftrightarrow$};
	
	\draw[-, shorten >= 6, shorten <= 9, shift={(-0.05,-0.05)}, mid arrowsm] (0,0) -- (0,1.5); \draw (0,0.6) node[left] {$Q$};
	\draw[-, shorten >= 6, shorten <= 9, shift={(0.05,0.05)}, mid arrowsm] (0,1.5) -- (0,0);
	\draw[-] (g) to[out=-60,in=0] (0,-0.6) to[out=180,in=-120] (g); \draw (0,-0.6) node[right] {$A$}; \draw (0,-0.6) node[cross] {};
	
	\draw (-3,0) node {$\CW = \sum_{j=2}^N Flip[A^j] $};
	
	\draw (3,0.8) node[right] {$3N$ chirals: $R_j,S_j,T_j$ with:};
	\draw (3,0.2) node[right] {$ \CW = \sum_{j,k,l=1}^N \d_{j+k+l,N+2} R_j S_l T_k$};
	
\end{tikzpicture}
\label{fig:SQCD_1flav}
\ee
This duality relates the SQCD with one flavor where we flipped the  tower of powers of the adjoint (which would  all be below the unitarity bound), to a Wess-Zumino model. Notice that the superpotential on the Wess-Zumino side contains a number of terms growing quadratically with $N$. This superpotential was proposed in \cite{Benvenuti:2018bav} and tested  using sequential deconfinement in \cite{Bajeot:2022lah}. 

Starting from the the SQCD on the l.h.s of figure \eqref{fig:SQCD_1flav}, the algorithm yields on the dual side a collection of  $3N$ chiral fields  with a charge assignment which is indeed  compatible with  the superpotential given \eqref{fig:SQCD_1flav}.

It will be  also useful to consider a flipped version of this duality where on the electric side  we flip   the tower of dressed mesons:
\be
\begin{tikzpicture}[thick,node distance=3cm,gauge/.style={circle,draw,minimum size=5mm},flavor/.style={rectangle,draw,minimum size=5mm}] 
 
	\path (0,0) node[gauge](g) {$\!\!\!N\!\!\!$} -- (0,1.5) node[flavor] (x1) {$\!1\!$} -- (2,0.5) node{$\Leftrightarrow$};
	
	\draw[-, shorten >= 6, shorten <= 9, shift={(-0.05,-0.05)}, middx arrowsm] (0,0) -- (0,1.5); \draw (0,0.6) node[left] {$Q$};
	\draw[-, shorten >= 6, shorten <= 9, shift={(0.05,0.05)}, midsx arrowsm] (0,1.5) -- (0,0); \draw (0,0.5) node[cross] {};
	\draw[-] (g) to[out=-60,in=0] (0,-0.6) to[out=180,in=-120] (g); \draw (0,-0.6) node[right] {$A$}; \draw (0,-0.6) node[cross] {};
	
	\draw (-4,0.5) node {$\CW = \sum_{j=0}^{N-1} Flip[QA^j\tilde{Q}] +$};
	\draw (-4,-0.2) node {$ + \sum_{j=2}^N Flip[A^j] $};
	
	\draw (3,0.5) node[right] {$ (\text{Free Hyper} )^N $};

\end{tikzpicture}
\label{flipf1}
\ee
In this case on the dual side we have  $N$ free hypers mapping to the tower of dressed monopoles $\M^\pm_{A^j}$ in the  electric SQCD side.

\subsection{Operator Map}\label{sec:3d_SQCD_map}
We now show how the gauge invariant operators of the SQCD theory are mapped into the mirror dual.
\begin{itemize}
\item In the electric theory we have the meson operator $Q \tilde{Q}$ in the $\bar{\textbf{F}} \times \textbf{F}$ representation of $SU(F)_U \times SU(F)_W$, with R-charge 2, $m$-charge -2 and zero $\tau$-charge.
 This operator is mapped into the following collection of operators of the mirror theory:
\begin{itemize}
	\item Singlets $B^{(k)}_{1,1}$ of the $(k)$-th improved bifundamental theory, with $k=2,\cdots F-1$, and  singlets $\mathcal{F}[V_1A_1^{N-1} \tilde{V}_1]$ and $\mathcal{F}[V_2 A_{F-1}^{N-1} \tilde{V}_2]$. These are $F$ operators, each with R-charge 2 and $m$-charge -2.

	\item  $F(F-1)$ monopole  operators with topological charge given by strings of contiguous $+1$ (or $-1$) under the topological symmetries $U(1)_{X_{j+1} - X_j}$. As discussed in the appendix \ref{app:FMmonopoles}, it is possible to prove that they carry R-charge 2 and
	$m$-charge -2. 
\end{itemize}
 We can arrange all these operator into a matrix transforming in the $\textbf{F} \times \bar{\textbf{F}}$ representation of $SU(F)_U \times SU(F)_W$ which is naturally mapped to the electric meson. For example, for a theory with $F=4$, meaning that the mirror theory has 3 gauge nodes, this matrix is given as:
\begin{align}\label{eq:meson_F=4_mirror}
	\begin{pmatrix}
		\mathcal{F}[V_1 A_1^{N-1} \tilde{V}_1] & \mathfrak{M}^{(+,0,0)} & \mathfrak{M}^{(+,+,0)} &
		 \mathfrak{M}^{(+,+,+)} \\
		\mathfrak{M}^{(-,0,0)} & \mathsf{B}_{1,1}^{(2)} & \mathfrak{M}^{(0,+,0)} & \mathfrak{M}^{0,+,+)} \\
		\mathfrak{M}^{(-,-,0)} & \mathfrak{M}^{(0,-,0)} & \mathsf{B}_{1,1}^{(3)} & \mathfrak{M}^{(0,0,+)} \\
		\mathfrak{M}^{(-,-,-)} & \mathfrak{M}^{(0,-,-)} & \mathfrak{M}^{(0,0,-)} & \mathcal{F}[V_2 A^{N-1}_3 \tilde{V}_2]
	\end{pmatrix}
\end{align}
Where we denote with $\mathfrak{M}^{(i,j,k)}$ a monopole with topological charges $i,j,k$ under the three topological symmetries. 

We can also consider dressed mesons with powers of the adjoint in the electric theory.
If we parameterize the SQCD as in figure \ref{fig:SQCD_Manifest} the map works very intuitively as:
\begin{align}\label{tab:SQCD_Explicit_Map}
\renewcommand{\arraystretch}{1.25}
\begin{tabular}{c|c}
	SQCD & Mirror \\
	\hline
	$Q_1 A^l \tilde{Q}_1$ & $\mathcal{F}[V_1 A^{N-1-l}_1 \tilde{V}_1]$ \\
	$Q_F  A^l \tilde{Q}_F$ & $\mathcal{F}[V_2 A^{N-1-l}_{F-1} \tilde{V}_2]$ \\
	$Q_j A^l\tilde{Q}_j$ for $j=2,\ldots,F-1$ & $\mathsf{B}_{1,1+l}^{(j)}$ \\
	$Q_j A^l \tilde{Q}_k$ for $j \neq k$ & $\mathfrak{M}_{A^l }^{(0,\ldots,0,+,\ldots,+,0,\ldots,0)}$ non-null entries: $j$ to $k-1$ 
\end{tabular}
\end{align}

\item  In the electric theory we then have the traces of powers of the adjoint field $A^j$, for $j=2,\ldots,N$, that are only charged under the $U(1)_\tau$ symmetry with a charge of $j$. These operators are mapped into similar operators that can be built in the mirror theory. In the mirror we have an adjoint field for each gauge node, all with a charge $1$ under the $U(1)_\tau$ symmetry. However, quantum effects relate the traces of powers of all these operators such that they are all identified, leaving only one independent set of operators with charges $j$ under the $U(1)_\tau$ symmetry for $j=2,\ldots,N$. \\

\item Lastly, we also have monopoles in the SQCD theory. The lowest charged monopoles, with $\pm 1$ charge under $U(1)_Y$, have $m$-charge $F$, $\tau$-charge $1-N$ and are singlets under all the other symmetries. These are mapped into long mesons $\tilde{V}_1 \Pi_2 \ldots \Pi_{F-1} V_2$ and $V_1 \tilde{\Pi}_2 \ldots \tilde{\Pi}_{F-1} \tilde{V}_2$ in the mirror theory. These have $\tau$-charge $N-1$ and charge $+1$ under all the $U(1)_{B_j}$ symmetries, which implies that under the $U(1)_m$ symmetry it has charge $F$. Also, they have charges $\pm 1$ under $U(1)_Y$, which we recall is mapped into the topological symmetry of the SQCD theory. \\
Dressed monopoles of the SQCD theory will be mapped similarly into dressed long mesons with the same level of dressing.

\end{itemize}


To conclude, let us mention that not all the holomorphic gauge invariant operators in the quiver side are mapped to the electric theory. Notable absent from the list of mapped operators are  the gauge singlets $\mathsf{B}^{(j)}_{n,m}$ for $n \neq 1$. We claim that the holomorphic operators $\mathsf{B}^{(j)}_{n \neq 1,m}$ are trivial in the chiral ring of the magnetic theory, since there is no operator in the electric theory chiral ring with the correct global symmetries. 

In particular, the triviality of the $\mathsf{B}^{(j)}_{2,1}$'s has interesting consequences. The $\mathsf{B}^{(j)}_{2,1}$'s, if turned on in the superpotential, would \emph{iron} an improved bifundamental into a standard one (see \eqref{fig:FM_c=1-t/2}), providing an RG flow to the putative IR SCFT associated to the naive reading of the magnetic brane setup. Chiral ring stability tells us that adding to the superpotential chiral-ring-trivial operators has trivial consequences to the IR SCFT. Hence chiral ring stability tells us that our mirror and the naive mirror flow in the IR to the same SCFT. See section \ref{oldprop} for more comments about the relation between our and the old proposals of $3d$ mirror symmetry with $4$ supercharges.

\subsection{Deformations and consistency checks}\label{sec:3d_SQCD_deformations}

In this section we  study the effect of some interesting deformations of our dual pair, providing non-trivial consistency checks of our duality.\\
Before discussing deformations we notice that in the magnetic theory, thanks to two {\it swapping} dualities, we are allowed to shuffle and reorder all the improved bifundamentals and the two vertical flavors. \\

The first duality in figure \eqref{fig:3d_starstar} allows us to \emph{swap} two consecutive improved bifundamentals, that is under the duality the two $U(1)$ symmetries rotating the bifundamentals are exchanged. Using this duality we get:\footnote{We list operators and their R-charge in a table beside the quiver representation of a theory. The R-charge is expressed as $R_0 + \sum_{E} q_E E$, where $R_0$ is the trial R-charge. Also, the sum runs over all the $U(1)_E$ global symmetries and we denote by $E$ the mixing coefficient for $U(1)_E$ and by $q_E$ the charge of the operator under the group. So with a slight abuse of notation we denote by $E$ both the real mass and the mixing coefficient for $U(1)_E$. }
\be
 \bpic[thick,node distance=3cm,gauge/.style={circle,draw,minimum size=5mm},flavor/.style={rectangle,draw,minimum size=5mm}]  
    
    \path (-6,0) node[gauge](x1) {$\!\!\!N\!\!\!$} -- (-4,0) node[gauge](x2) {$\!\!\!N\!\!\!$} -- (-2,0) node[gauge](x3) {$\!\!\!N\!\!\!$} 
    	 --  (0,0) node{$\Longleftrightarrow$};
    
    \wigM (x1) -- (-7,0);
    \wigM (x1) -- (x2); \draw (-5,0.4) node {$\Pi_j$};
    \wigM (x3) -- (x2); \draw (-3,0.4) node {$\Pi_{j+1}$};
    \wigM (x3) -- (-1,0);
    \draw [-] (x1) to[out=-60, in=0] (-6,-0.6) to[out=180,in=-120] (x1); 
    \draw [-] (x2) to[out=-60, in=0] (-4,-0.6) to[out=180,in=-120] (x2); 
    \draw [-] (x3) to[out=-60, in=0] (-2,-0.6) to[out=180,in=-120] (x3); 
    
	
	\path (-4,-1.5) node {\begin{tabular}{c|c}
							$\Pi_j$ & $B_j$ \\
							$\Pi_{j+1}$ & $B_{j+1}$
						\end{tabular}};
    
    \path (2,0) node[gauge](x1) {$\!\!\!N\!\!\!$} -- (4,0) node[gauge](x2) {$\!\!\!N\!\!\!$} -- (6,0) node[gauge](x3) {$\!\!\!N\!\!\!$};
    	
    \wigM (x1) -- (1,0);
    \wigM (x1) -- (x2); \draw (3,0.4) node {$\Pi_j'$};
    \wigM (x3) -- (x2); \draw (5,0.4) node {$\Pi_{j+1}'$};
    \wigM (x3) -- (7,0);
    \draw [-] (x1) to[out=-60, in=0] (2,-0.6) to[out=180,in=-120] (x1); 
    \draw [-] (x2) to[out=-60, in=0] (4,-0.6) to[out=180,in=-120] (x2); 
    \draw [-] (x3) to[out=-60, in=0] (6,-0.6) to[out=180,in=-120] (x3); 
  
	
	\path (4,-1.5) node {\begin{tabular}{c|c}
							$\Pi_j'$ & $B_{j+1}$ \\
							$\Pi_{j+1}'$ & $B_{j}$
						\end{tabular}};
	 \label{fig:mirror_swapBB}					
\epic\ee
Notice that under this duality, the matrix  $\mathsf{B}_{n,m}^{(j)}$ is mapped to $\mathsf{B}_{n,m}^{'(j+1)}$ while
$\mathsf{B}_{n,m}^{(j+1)}$ is mapped to $\mathsf{B}_{n,m}^{'(j)}$. For more details see \eqref{fig:3d_starstar}. \\

A specialisation of the previous duality, given in figure \eqref{fig:3d_starstar_specialised}, allows us to exchange an improved bifundamental with a flavor. For example, we can exchange the left vertical flavor with the first improved bifundamental:
\be
 \bpic[thick,node distance=3cm,gauge/.style={circle,draw,minimum size=5mm},flavor/.style={rectangle,draw,minimum size=5mm}]  
    
    \path (-4,0) node[gauge](x2) {$\!\!\!N\!\!\!$} -- (-2,0) node[gauge](x3) {$\!\!\!N\!\!\!$}
   		-- (-4,1.5) node[flavor](y1) {$\!1\!$} --  (0.5,0) node{$\Longleftrightarrow$};

    \draw[-, shorten >= 9, shorten <= 6, shift={(-0.05,0.05)}, mid arrowsm] (-4,0) -- (-4,1.5); \draw (-4,0.75) node[left] {$V_1$};
	\draw[-, shorten >= 4, shorten <= 11, shift={(0.05,0.15)}, mid arrowsm] (-4,1.5) -- (-4,0);
	\draw (-4,0.45) node[cross]{};	
    \wigM (x3) -- (x2); \draw (-3,0.4) node {$\Pi_2$};
    \wigM (x3) -- (-1,0);
    \draw [-] (x2) to[out=-60, in=0] (-4,-0.6) to[out=180,in=-120] (x2); 
    \draw [-] (x3) to[out=-60, in=0] (-2,-0.6) to[out=180,in=-120] (x3); 
    
	
	\path (-4.5,-1.5) node[right] {\begin{tabular}{c|c}
							$V_1$ & $\frac{1-N}{2}\tau + B_1$ \\
							$\Pi_2$ & $B_{2}$
						\end{tabular}};
    
    \path (2,0) node[gauge](x1) {$\!\!\!N\!\!\!$} -- (4,0) node[gauge](x2) {$\!\!\!N\!\!\!$} -- (2,1.5) node[flavor](y1) {$\!1$};
    
    \draw[-, shorten >= 9, shorten <= 6, shift={(-0.05,0.05)}, mid arrowsm] (2,0) -- (2,1.5); \draw (2,0.75) node[left] {$V'_1$};
	\draw[-, shorten >= 4, shorten <= 11, shift={(0.05,0.15)}, mid arrowsm] (2,1.5) -- (2,0);
	\draw (2,0.45) node[cross]{};
    \wigM (x1) -- (x2); \draw (3,0.4) node {$\Pi_2'$};
    \wigM (x2) -- (5,0);
    \draw [-] (x1) to[out=-60, in=0] (2,-0.6) to[out=180,in=-120] (x1); 
    \draw [-] (x2) to[out=-60, in=0] (4,-0.6) to[out=180,in=-120] (x2); 
  

	\path (1.5,-1.5) node[right] {\begin{tabular}{c|c}
							$V_1'$ & $\frac{1-N}{2}\tau + B_{2}$ \\
							$\Pi_2'$ & $B_{1}$ 
						\end{tabular}};
		\label{fig:mirror_swapBF}				
\epic\ee
Notice that under this duality the tower of flipping singlets $\CF[V_1 A_1^k \tilde{V}_1]$ is mapped into part of the matrix of singlets of the improved bifundamental theory $\mathsf{B}^{(2)}_{1,k}$, and vice-versa. Instead, the rest of the singlet matrix is not mapped under this duality, this is consistent with our claim that the singlets that are not mapped are not in the chiral ring. \\

One can combine the two moves \eqref{fig:mirror_swapBB} and \eqref{fig:mirror_swapBF} to rearrange all the bifundamentals in any desired way. This property will be important to discuss the deformations, as we will show in detail below.

\subsubsection{Shortening}
The first deformation that we consider is given by complex mass terms in the electric theory. 
This means that we turn on a superpotential term as: $\delta \cW = Q_j \tilde{Q}_j$ for any $j=1,\ldots,F$. For $j=1,F$ this deformation is mapped respectively into $\mathcal{F}[V_1 A^{N-1}_1 \tilde{V}_1]$ and   $\mathcal{F}[V_2 A^{N-1}_{F-1} \tilde{V}_2]$ while for $j=2,\cdots F-1$ it is mapped into $\mathsf{B}_{1,1}^{(j)}$, as it can be read from table \eqref{tab:SQCD_Explicit_Map}. \\
By means of the freedom of rearranging the improved bifundamentals, using the duality \eqref{fig:mirror_swapBB} we see that all the deformations for $j=2,\ldots,F-1$ are on the same footing. Also using the duality \eqref{fig:mirror_swapBF} the case $j=1$ is equivalent to $j=2$ and, analogously, $j=F$ is equal to $j=F-1$. Therefore we conclude that this deformation can be implemented on any improved bifundamental without any loss of generality.
Let us then consider $j=3$ for simplicity, the superpotential term $\d \CW = Q_3 \tilde{Q}_3$ is mapped to $\d \CW = \mathsf{B}_{1,1}^{(3)}$.
This deformation has the effect of transforming the improved bifundamental in an $\mathbb{I}-wall$, as explained in appendix \ref{app:FM}  (eq. \eqref{FMid}) which identifies the two $U(N)$ symmetry groups which  is connecting:
\be
 \bpic[thick,node distance=3cm,gauge/.style={circle,draw,minimum size=5mm},flavor/.style={rectangle,draw,minimum size=5mm}]  
    
    \path (-4,0) node[gauge](x2) {$\!\!\!N\!\!\!$} -- (-2,0) node[gauge](x3) {$\!\!\!N\!\!\!$} 
    	 --  (0,0) node{$\Longrightarrow$};
    
    \wigM (x2) -- (-5,0);
    \wigM (x3) -- (x2); \draw (-3,0.4) node {$\Pi_3$};
    \wigM (x3) -- (-1,0);
    \draw [-] (x2) to[out=-60, in=0] (-4,-0.6) to[out=180,in=-120] (x2); 
    \draw [-] (x3) to[out=-60, in=0] (-2,-0.6) to[out=180,in=-120] (x3); 
    
    \draw (-3,-1.2) node {$\CW = \CW_{\text{gluing}} + \mathsf{B}^{(3)}_{1,1}$};
	
    
    \path (3,0) node[gauge](x1) {$\!\!\!N\!\!\!$};
    	
    \wigM (x1) -- (2,0);
    \wigM (x1) -- (4,0);
    \draw [-] (x1) to[out=-60, in=0] (3,-0.6) to[out=180,in=-120] (x1);
  
    \draw (3,-1.2) node {$\cW = \CW_{\text{gluing}} $};
	
						
\epic\ee
Intuitively, one can think that the linear superpotential term $\d \CW = \mathsf{B}^{(3)}_{1,1}$ has the effect of giving a VEV to the $\Pi_3$ and $\tilde{\Pi}_3$ operators. The effect of this VEV is to Higgs the $U(N)\times U(N)$ gauge symemtry under which is charged down to the diagonal $U(N)$. This is supported from the fact that imposing the $\mathsf{B}^{(3)}_{1,1}$ operator to have R-charge 2 implies that the $\Pi_3$ operator has R-charge 0, as one would expect from an operator acquiring a VEV. The net effect of this deformation is then to shorten the sequence of improved bifundamentals by one unit. This is consistent with our duality, indeed the electric SQCD after the mass deformation has $F-1$ flavors and its mirror has one less  improved bifundamental links.\\

In  the electric theory we can also turn on a non-diagonal mass term of the type: $\d \CW = Q_j \tilde{Q}_k$, with $j \neq k$ which has also the effect of just removing one flavor. From the operator map \eqref{tab:SQCD_Explicit_Map}, we see that this deformation is mapped into monopole superpotentials in the magnetic theory. Let us consider the case $\d \CW = Q_2\tilde{Q}_3$ without any loss of generality. This superpotential term is mapped to $\d \CW = \mathfrak{M}^{(0,+,0,\ldots,0)}$ in the magnetic theory. We can now use the duality \eqref{fig:FM_fusion} where two improved bifundamentals glued with a monopole superpotential $\CW =\CW_{gluing}+ \mathfrak{M}^+$ (and equivalently for $\cW = \mathfrak{M}^-$) are shown to be dual to a single improved bifundamental.
\be
 \bpic[thick,node distance=3cm,gauge/.style={circle,draw,minimum size=5mm},flavor/.style={rectangle,draw,minimum size=5mm}]  
    
    \path (-6,0) node[gauge](x1) {$\!\!\!N\!\!\!$} -- (-4,0) node[gauge](x2) {$\!\!\!N\!\!\!$} -- (-2,0) node[gauge](x3) {$\!\!\!N\!\!\!$} 
    	 -- (-6,1.5) node[flavor](y1) {$\!1\!$} --  (0,0) node{$\Longrightarrow$};
    
    \wigM (x1) -- (x2); \draw (-5,0.4) node {$\Pi_2$};
    \wigM (x3) -- (x2); \draw (-3,0.4) node {$\Pi_3$};
    \wigM (x3) -- (-1,0);
    \draw[-, shorten >= 9, shorten <= 6, shift={(-0.05,0.05)}, mid arrowsm] (-6,0) -- (-6,1.5); 
	\draw[-, shorten >= 4, shorten <= 11, shift={(0.05,0.15)}, mid arrowsm] (-6,1.5) -- (-6,0);
	\draw (-6,0.45) node[cross]{};
    \draw [-] (x1) to[out=-60, in=0] (-6,-0.6) to[out=180,in=-120] (x1); 
    \draw [-] (x2) to[out=-60, in=0] (-4,-0.6) to[out=180,in=-120] (x2); 
    \draw [-] (x3) to[out=-60, in=0] (-2,-0.6) to[out=180,in=-120] (x3); 
    
    \draw (-6,-1.2) node[right]{$\cW = \CW_{\text{gluing}} + \mathfrak{M}^{(0,+,0\ldots,0)} $};
	
    
    \path (2,0) node[gauge](x1) {$\!\!\!N\!\!\!$} -- (4,0) node[gauge](x2) {$\!\!\!N\!\!\!$} -- (2,1.5) node[flavor](y1) {$\!1\!$};
    	
    \wigM (x1) -- (x2); \draw (3,0.4) node {$\Pi'_2$};   
    \wigM (x2) -- (5,0);
    \draw[-, shorten >= 9, shorten <= 6, shift={(-0.05,0.05)}, mid arrowsm] (2,0) -- (2,1.5); 
	\draw[-, shorten >= 4, shorten <= 11, shift={(0.05,0.15)}, mid arrowsm] (2,1.5) -- (2,0);
	\draw (2,0.45) node[cross]{};
    \draw [-] (x1) to[out=-60, in=0] (2,-0.6) to[out=180,in=-120] (x1); 
    \draw [-] (x2) to[out=-60, in=0] (4,-0.6) to[out=180,in=-120] (x2); 
  
    \draw (2,-1.2) node[right]{$\cW = \CW_{\text{gluing}} $};
	
						
\epic\ee
We can then conclude that this deformation has the effect of shortening the sequence of improved bifundamentals by one unit. So our duality passes also this consistency check. \\
Notice that instead  the deformation $\d \CW = Q_2\tilde{Q}_3+\tilde{Q}_2 Q_3 $ which corresponds to integrating out two flavors
in the electric theory, maps to  $\d \CW = \mathfrak{M}^{(0,+,0,\ldots,0)}+\mathfrak{M}^{(0,-,0,\ldots,0)}$ in the magnetic theory. We can now use 
the fact that two improved bifundamentals  glued with a monopole superpotential $\CW =\CW_{gluing}+ \mathfrak{M}^+ + \mathfrak{M}^-$ fuse to an $\mathbb{I}-wall$ (see \eqref{FMdelta}), to conclude that this deformation has the effect of shortening the sequence of improved bifundamentals by two units, as expected.

\subsubsection{Ironing}\label{ironing}

The second set of deformations that we consider are cubic terms for the flavors in the SQCD: $\d \cW= Q_j A \tilde{Q}_j$. Following the operator map \eqref{tab:SQCD_Explicit_Map} we see that those are mapped in either $\mathcal{F}[V_1 A_1^{N-2} \tilde{V}_1]$, $\mathcal{F}[V_2 A_{F-1}^{N-2} \tilde{V}_2]$ for $j=1,F$ and into 
$\mathsf{B}_{1,2}^{(j)}$ for $j=2,\ldots F-1$.
As before, using the swapping dualities \eqref{fig:mirror_swapBB} and \eqref{fig:mirror_swapBF} to rearrange improved bifundamentals and flavors, we can focus on  the case $j=3$ for simplicity without any loss of generality where the deformation $\d \CW = Q_3 A \tilde{Q}_3$ is mapped to $\d \CW = \mathsf{B}_{1,2}^{(3)}$. As shown in figure
\eqref{ironduality}, this deformation has the effect of ironing an improved bifundamental into an ordinary bifundamental of charge $\tau/2$ coupled to two extra adjoint fields. These extra adjoint fields give mass to the adjoint fields in $\CW_{gluing}$ to its left and right and the bifundamental is then coupled to adjoint operators inside the improved bifundamentals. We summarise graphically this picture below:
\be
 \bpic[thick,node distance=3cm,gauge/.style={circle,draw,minimum size=5mm},flavor/.style={rectangle,draw,minimum size=5mm}]  
    
    \path (-4,0) node[gauge](x2) {$\!\!\!N\!\!\!$} -- (-2,0) node[gauge](x3) {$\!\!\!N\!\!\!$} 
    	 --  (0,0) node{$\Longrightarrow$};
    
    \wigM (x2) -- (-5,0);
    \wigM (x3) -- (x2); \draw (-3,0.4) node {$\Pi_3$};
    \wigM (x3) -- (-1,0);
    \draw [-] (x2) to[out=-60, in=0] (-4,-0.6) to[out=180,in=-120] (x2); 
    \draw [-] (x3) to[out=-60, in=0] (-2,-0.6) to[out=180,in=-120] (x3); 
    
    \draw (-3,-1) node {$\CW = \CW_{\text{gluing}} + B^{(3)}_{1,2}$};
	
    
    \path (2,0) node[gauge](x1) {$\!\!\!N\!\!\!$} -- (4,0) node[gauge](x2) {$\!\!\!N\!\!\!$};
    	
    \wigM (x1) -- (1,0);
    \draw[-, shorten >= 5, shorten <= 11, shift={(-0.1,0.05)}, middx arrowsm] (2,0) -- (4,0); \draw (2.7,0) node[cross] {}; \draw (3,0.4) node {$\Pi'_3$};
	\draw[-, shorten >= 5, shorten <= 11, shift={(0.1,-0.05)}, midsx arrowsm] (4,0) -- (2,0);
    \wigM (x2) -- (5,0);
  
    \draw (3,-1) node {$\cW = \Pi'_3( \mathsf{A}_L + \mathsf{A}_R ) \tilde{\Pi}'_3 +$};
    \draw (3,-1.6) node {$+ Flip[\Pi'_3 \tilde{\Pi}'_3] $};
	
						
\epic\ee 

One can consider also a non diagonal superpotential term: $\d \CW = Q_j A \tilde{Q}_k$ which is mapped to a superpotential term for the magnetic theory given by dressed monopole operators. Again by consistency  this deformation should result into the ironing of an improved bifundamental link. Indeed using the  swapping dualities \eqref{fig:mirror_swapBB} and \eqref{fig:mirror_swapBF}, without any loss of generality, we can consider the case $\d \CW = Q_2 A \tilde{Q}_3$. This superpotential term is mapped to the superpotential $\d \CW =  \mathfrak{M}_{A_2}^{(0,+,0,\ldots,0)}$ involving  a dressed monopole. In this case we can use the duality \eqref{fig:FM_dressedmonopole} under which two improved bifundamentals glued with $\cW = \CW_{gluing}+ \mathfrak{M}_A^+$ (and analogously for $\CW =  \mathfrak{M}_A^-$) are dual to an improved bifundamental glued to an ordinary bifundamental coupled to the adjoint operators  of the improved bifundamental theories. We summarise graphically this picture below.
\be 
 \bpic[thick,node distance=3cm,gauge/.style={circle,draw,minimum size=5mm},flavor/.style={rectangle,draw,minimum size=5mm}]  
    
    \path (-6,0) node[gauge](x1) {$\!\!\!N\!\!\!$} -- (-4,0) node[gauge](x2) {$\!\!\!N\!\!\!$} -- (-2,0) node[gauge](x3) {$\!\!\!N\!\!\!$} 
    	 -- (-6,1.5) node[flavor](y1) {$\!1\!$} --  (0,0) node{$\Longrightarrow$};
    
    \wigM (x1) -- (x2); \draw (-5,0.4) node {$\Pi_2$};
    \wigM (x3) -- (x2); \draw (-3,0.4) node {$\Pi_3$};
    \wigM (x3) -- (-1,0);
    \draw[-, shorten >= 9, shorten <= 6, shift={(-0.05,0.05)}, mid arrowsm] (-6,0) -- (-6,1.5); \draw (-6,0.75) node[left] {$V_1$};
	\draw[-, shorten >= 4, shorten <= 11, shift={(0.05,0.15)}, mid arrowsm] (-6,1.5) -- (-6,0);
	\draw (-6,0.45) node[cross]{};
    \draw [-] (x1) to[out=-60, in=0] (-6,-0.6) to[out=180,in=-120] (x1); \draw (-6,-0.7) node[right] {$A_1$};
    \draw [-] (x2) to[out=-60, in=0] (-4,-0.6) to[out=180,in=-120] (x2); \draw (-4,-0.7) node[right] {$A_2$};
    \draw [-] (x3) to[out=-60, in=0] (-2,-0.6) to[out=180,in=-120] (x3); \draw (-2,-0.7) node[right] {$A_3$};
    
    \draw (-4,-1.5) node {$\cW = \CW_{\text{gluing}} + \mathfrak{M}^{(0,+,0\ldots,0)}_{A_2} $};
	
    
    \path (2,0) node[gauge](x1) {$\!\!\!N\!\!\!$} -- (4,0) node[gauge](x2) {$\!\!\!N\!\!\!$} -- (6,0) node[gauge](x3) {$\!\!\!N\!\!\!$}
    	-- (2,1.5) node[flavor](y1) {$\!1\!$};
    	
    \wigM (x1) -- (x2); \draw (3,0.4) node {$\Pi'_2$};
	\draw[-, shorten >= 5, shorten <= 11, shift={(-0.1,0.05)}, middx arrowsm] (4,0) -- (6,0); \draw (4.7,0) node[cross] {}; \draw (5,0.4) node {$\Pi'_3$};
	\draw[-, shorten >= 5, shorten <= 11, shift={(0.1,-0.05)}, midsx arrowsm] (6,0) -- (4,0);    
    \wigM (x3) -- (7,0);
    \draw[-, shorten >= 9, shorten <= 6, shift={(-0.05,0.05)}, mid arrowsm] (2,0) -- (2,1.5); \draw (2,0.75) node[left] {$V_1$};
	\draw[-, shorten >= 4, shorten <= 11, shift={(0.05,0.15)}, mid arrowsm] (2,1.5) -- (2,0);
	\draw (2,0.45) node[cross]{};
    \draw [-] (x1) to[out=-60, in=0] (2,-0.6) to[out=180,in=-120] (x1); \draw (2,-0.7) node[right] {$A_1$};
  
    \draw (4,-1.5) node {$\cW = \Pi'_3 (\mathsf{A}_L + \mathsf{A}_R) \tilde{\Pi}'_3 $};
    \draw (4,-2.1) node {$+ Flip[\Pi'_3 \tilde{\Pi}'_3] $};
	
						
\epic\ee

We can then conclude that any generic cubic superpotential term  $\d \CW = Q_j A \tilde{Q}_k$ in the electric theory leads to the ironing of a single improved bifundamental link.

\subsubsection{Flow to the $\cN=4$ mirror symmetry}
If we turn on the superpotential  $\d\CW = \sum_{j=1}^F Q_j A \tilde{Q}_j$, that is we couple all flavors to the adjoint chiral, we reach the $\cN=4$ $U(N)$ SQCD. It is then an interesting consistency check to show how our mirror dual reduces to the known $\cN=4$ of the SQCD for $F\geq 2N$ \footnote{It is possible to show that for $F< 2N$ our dual reproduces also the results for the bad SQCD found in \cite{Giacomelli:2023zkk}. However, this analysis is beyond the scope of this work.}. The effect of the deformation on the mirror theory is to iron all the improved bifundamentals. 
Keeping track of extra adjoints appearing in the ironing duality \eqref{ironduality}, we observe that each  node has an adjoint of charge $2-\tau$ which couples to the bifundamentals to its right and its left and also all the bifundamentals are flipped. We denote these couplings as $\CW^{partial}_{\cN=4}$. On the first and last node the adjoint has become massive and now the vertical flavors are coupled to an adjoint operator built from the square of the bifundamentals. On the mirror side  we also turn on linear superpotentials for the flipping fields $\CF[V_1 (\Pi_2\tilde{\Pi}_2)^{N-2} \tilde{V}_1]$ and $\CF[V_2 (\Pi_{F-1}\tilde{\Pi}_{F-1})^{N-2} \tilde{V}_2]$. The resulting duality is depicted below.
\be\label{fig:SQCD_Dual_N=4_vev}
\resizebox{.9\hsize}{!}{ 
\bpic[thick,node distance=3cm,gauge/.style={circle,draw,minimum size=5mm},flavor/.style={rectangle,draw,minimum size=5mm}] 
 
	\path (0,0) node[gauge](g) {$\!\!\!N\!\!\!$} -- (0,1.5) node[flavor] (x1) {$\!F\!$} -- (1.5,0) node{$\Longleftrightarrow$};
	
	\draw[-, shorten >= 6, shorten <= 9, shift={(-0.05,-0.05)}, mid arrowsm] (0,0) -- (0,1.5); \draw (0,0.6) node[left] {$Q$};
	\draw[-, shorten >= 6, shorten <= 9, shift={(0.05,0.05)}, mid arrowsm] (0,1.5) -- (0,0);
	\draw[-] (g) to[out=-60,in=0] (0,-0.6) to[out=180,in=-120] (g); \draw (0,-0.6) node[right] {$A$};
	 
	\draw (0,-1.3) node {$\CW = Q A \tilde{Q}$}; 
	
\begin{scope}[shift={(1,0)}]
	\path (3,0) node[gauge](g1) {$\!\!\!N\!\!\!$} -- (5,0) node[gauge](g2) {$\!\!\!N\!\!\!$} -- (8,0) node[gauge](g3) {$\!\!\!N\!\!\!$} 
		-- (10,0) node[gauge](g4) {$\!\!\!N\!\!\!$} -- (3,1.5) node[flavor](y1) {$\!1\!$} -- (10,1.5) node[flavor](y2) {$\!1\!$};
	 
	\draw[-, shorten >= 5, shorten <= 11, shift={(-0.1,0.05)}, middx arrowsm] (3,0) -- (5,0); \draw (3.7,0) node[cross] {}; \draw (4,0.35) node {$\Pi_3$};
	\draw[-, shorten >= 5, shorten <= 11, shift={(0.1,-0.05)}, midsx arrowsm] (5,0) -- (3,0);
	
	\draw[-, shorten >= 5, shorten <= 8, shift={(0,0.05)}, mid arrowsm] (5,0) -- (6.25,0);
	\draw[-, shorten >= 8, shorten <= 5, shift={(0,-0.05)}, mid arrowsm] (6.25,0) -- (5,0);
	
	\draw[-, shorten >= 8, shorten <= 5, shift={(0,0.05)}, mid arrowsm] (6.77,0) -- (8,0);
	\draw[-, shorten >= 5, shorten <= 8, shift={(0,-0.05)}, mid arrowsm] (8,0) -- (6.75,0);
	 
	\draw[-, shorten >= 5, shorten <= 11, shift={(-0.1,0.05)}, middx arrowsm] (8,0) -- (10,0); \draw (8.7,0) node[cross] {}; \draw (9,0.35) node {$\Pi_{F-1}$};
	\draw[-, shorten >= 5, shorten <= 11, shift={(0.1,-0.05)}, midsx arrowsm] (10,0) -- (8,0);
	
	\draw[-, shorten >= 9, shorten <= 6, shift={(-0.05,0.05)}, mid arrowsm] (3,0) -- (3,1.5); \draw (3,0.75) node[left] {$V_1$};
	\draw[-, shorten >= 4, shorten <= 11, shift={(0.05,0.15)}, mid arrowsm] (3,1.5) -- (3,0);
	\draw[-, shorten >= 9, shorten <= 6, shift={(-0.05,0.05)}, mid arrowsm] (10,0) -- (10,1.5); \draw (10,0.75) node[right] {$V_2$};
	\draw[-, shorten >= 4, shorten <= 11, shift={(0.05,0.15)}, mid arrowsm] (10,1.5) -- (10,0);
	\draw (3,0.45) node[cross]{}; \draw (10,0.45) node[cross]{};
	\draw (6.5,0) node {$\cdots$};
	\draw[-] (g2) to[out=-60,in=0] (5,-0.6) to[out=180,in=-120] (g2); \draw (5,-0.7) node[right] {$A_2$};
	\draw[-] (g3) to[out=-60,in=0] (8,-0.6) to[out=180,in=-120] (g3); \draw (8,-0.7) node[right] {$A_{F-2}$};
	
	\draw (6.5,-1.5) node {$\CW = \sum_{j=0}^{N-1} ( Flip[V_1 (\Pi_2\tilde{\Pi}_2)^j \tilde{V}_1] + Flip[V_2 (\Pi_{F-1}\tilde{\Pi}_{F-1})^j \tilde{V}_2] )$};
	\draw (6.5,-2.2) node {$ + \CF[V_1 (\Pi_2\tilde{\Pi}_2)^{N-2} \tilde{V}_1] + \CF[V_2 (\Pi_{F-1}\tilde{\Pi}_{F-1})^{N-2} \tilde{V}_2] + \CW^{\text{partial}}_{\cN=4} $ };
	
\end{scope}

\epic}\ee
The EOMs for the two singlets $ \mathcal{F}[V_1 (\Pi_2\tilde{\Pi}_2)^{N-2} \tilde{V}_1]$ and $\mathcal{F}[V_2 (\Pi_{F-1}\tilde{\Pi}_{F-1})^{N-2}\tilde{V}_2]$ yield VEVs for the dressed mesons. By  carefully studying the effect of sequential Higgsing triggered by these VEVs (see for example \cite{Comi:2022aqo}), one can show that on each side of the quiver a tail of gauge nodes with increasing ranks from 1 to $N$ is reconstructed. We also have a plateau of $F-2N-1$ gauge nodes of rank $N$ with two flavors on the two ends. The result is depicted below and indeed coincides with the known mirror dual of the $\CN=4$ $U(N)$ SQCD \cite{Hanany:1996ie}:
\be\label{fig:SQCD_Dual_N=4} 
 \bpic[thick,node distance=3cm,gauge/.style={circle,draw,minimum size=5mm},flavor/.style={rectangle,draw,minimum size=5mm}] 
 
	\path (0,0) node[gauge](g) {$\!\!\!N\!\!\!$} -- (0,1.5) node[flavor] (x1) {$\!F\!$} -- (1.5,0) node{$\Longleftrightarrow$};
	
	\draw[-, shorten >= 6, shorten <= 9, shift={(-0.05,-0.05)}, mid arrowsm] (0,0) -- (0,1.5); 
	\draw[-, shorten >= 6, shorten <= 9, shift={(0.05,0.05)}, mid arrowsm] (0,1.5) -- (0,0);
	\draw[-] (g) to[out=-60,in=0] (0,-0.6) to[out=180,in=-120] (g); 
	
	\draw (0,-1.3) node {$\CW = \CW_{\CN=4}$}; 
	 
	\path (3,0) node[gauge](t1) {$\!\!1\!\!$} -- (5,0) node[gauge](t2) {\!\!\tiny{$N$-$1$}\!\!\!} -- (6.5,0) node[gauge](g2) {$\!\!N\!\!$} 
		-- (8.5,0) node[gauge](g3) {$\!\!N\!\!$} -- (10,0) node[gauge](t3) {\!\!\tiny{$N$-$1$}\!\!\!} -- (12,0) node[gauge](t4) {$\!\!1\!\!$}
		-- (6.5,1.5) node[flavor](y1) {$\!1\!$} -- (8.5,1.5) node[flavor](y2) {$\!1\!$};
	 
	\draw[-, shorten >= 0, shorten <= 7, shift={(0,0.05)}, middx arrowsm] (3,0) -- (3.7,0);
	\draw[-, shorten >= 7, shorten <= 0, shift={(0,-0.05)}, midsx arrowsm] (3.7,0) -- (3,0);
	
	\draw (4,0) node {$\cdots$};
	
	\draw[-, shorten >= 9, shorten <= 0, shift={(0,0.05)}, midsx arrowsm] (4.3,0) -- (5,0);
	\draw[-, shorten >= 0, shorten <= 9, shift={(0,-0.05)}, middx arrowsm] (5,0) -- (4.3,0);
	
	\draw[-, shorten >= 6, shorten <= 11, shift={(-0.1,0.05)}, mid arrowsm] (5,0) -- (6.5,0); 
	\draw[-, shorten >= 6, shorten <= 11, shift={(0.1,-0.05)}, mid arrowsm] (6.5,0) -- (5,0);
	
	\draw[-, shorten >= 0, shorten <= 9, shift={(0,0.05)}, middx arrowsm] (6.5,0) -- (7.2,0);
	\draw[-, shorten >= 9, shorten <= 0, shift={(0,-0.05)}, midsx arrowsm] (7.2,0) -- (6.5,0);
	 
	\draw (7.5,0) node {$\cdots$};
	
	\draw[-, shorten >= 9, shorten <= 0, shift={(0,0.05)}, midsx arrowsm] (7.8,0) -- (8.5,0);
	\draw[-, shorten >= 0, shorten <= 9, shift={(0,-0.05)}, middx arrowsm] (8.5,0) -- (7.8,0);
	 
	\draw[-, shorten >= 6, shorten <= 11, shift={(-0.1,0.05)}, mid arrowsm] (8.5,0) -- (10,0); 
	\draw[-, shorten >= 6, shorten <= 11, shift={(0.1,-0.05)}, mid arrowsm] (10,0) -- (8.5,0);
	
	\draw[-, shorten >= 0, shorten <= 9, shift={(0,0.05)}, middx arrowsm] (10,0) -- (10.7,0);
	\draw[-, shorten >= 9, shorten <= 0, shift={(0,-0.05)}, midsx arrowsm] (10.7,0) -- (10,0);
	
	\draw (11,0) node {$\cdots$};
	
	\draw[-, shorten >= 7, shorten <= 0, shift={(0,0.05)}, midsx arrowsm] (11.3,0) -- (12,0);
	\draw[-, shorten >= 0, shorten <= 7, shift={(0,-0.05)}, middx arrowsm] (12,0) -- (11.3,0);

	\draw[-, shorten >= 9, shorten <= 7, shift={(-0.05,0.05)}, mid arrowsm] (6.5,0) -- (6.5,1.5);
	\draw[-, shorten >= 4, shorten <= 11, shift={(0.05,0.15)}, mid arrowsm] (6.5,1.5) -- (6.5,0);
	
	\draw[-, shorten >= 9, shorten <= 7, shift={(-0.05,0.05)}, mid arrowsm] (8.5,0) -- (8.5,1.5);
	\draw[-, shorten >= 4, shorten <= 11, shift={(0.05,0.15)}, mid arrowsm] (8.5,1.5) -- (8.5,0);
	
	\draw[-] (t1) to[out=-60,in=0] (3,-0.55) to[out=180,in=-120] (t1);
	\draw[-] (t2) to[out=-60,in=0] (5,-0.6) to[out=180,in=-120] (t2);
	\draw[-] (g2) to[out=-60,in=0] (6.5,-0.6) to[out=180,in=-120] (g2);
	\draw[-] (g3) to[out=-60,in=0] (8.5,-0.6) to[out=180,in=-120] (g3);
	\draw[-] (t3) to[out=-60,in=0] (10,-0.6) to[out=180,in=-120] (t3);
	\draw[-] (t4) to[out=-60,in=0] (12,-0.55) to[out=180,in=-120] (t4);
	
	\draw (7.5,-1.3) node {$\CW = \CW_{\CN=4}$};
	 
\epic\ee
Notice that in the picture above we have rearranged the singlets flipping the bifundamentals so that all the adjoint chirals are tracefull.

\subsubsection{Higgsing $\leftrightarrow$ massive flavor and sequential confinement}\label{sec:seqdef}
We now consider another deformation which consists in giving a mass to any of the two vertical flavors in the mirror theory (\ref{mthe}).

Actually if we want to turn on a mass term we first need to {\it move} the flippers\footnote{The operation of moving the flippers, simply means 
that on the magnetic side where we have flipping terms of the type $\mathcal{W}=Flip[X]=O_X X$ we add a mass deformation for the flipper $\delta \mathcal{W}=O_X \tilde{O}_X $ which removes the flipping terms. The effect of this mass deformation on the electric side is then worked out using the operator map.}, meaning that  we consider  a modified version of the mirror pair where
on the electric side we have the superpotential $\mathcal{W}= Flip[Q_1 A_1^j \tilde{Q}_1]$
and on the mirror side the first vertical flavor has no flippers. It is also convenient to introduce $N-1$ singlets that on the l.h.s. \!\! flip the traces of powers of the adjoint chiral. On the r.h.s. we recall that the traces of powers of all the adjoint chirals are identified via quantum relations, therefore we can pick any adjoint, say the last one $A_{F-1}$, and flip its traces with the effect of flipping all the traces of powers of the adjoint chirals.
\be\label{flippedmipa}
\resizebox{.95\hsize}{!}{ 
\begin{tikzpicture}[thick,node distance=3cm,gauge/.style={circle,draw,minimum size=5mm},flavor/.style={rectangle,draw,minimum size=5mm}] 
 
	\path (0,0) node[gauge](g) {$\!\!\!N\!\!\!$} -- (-1.5,1.5) node[flavor] (x1) {$\!1\!$} -- (0,1.5) node[flavor] (x3) {$\!1\!$} -- (1.5,1.5) node[flavor] (x2) {$\!1\!$} 
		-- (2.7,0) node{$\Longleftrightarrow$};
	
	\draw[-, shorten >= 6, shorten <= 11, shift={(-0.07,0.02)}, midsx arrowsm] (-1.5,1.5) -- (0,0);
	\draw[-, shorten >= 6, shorten <= 10, shift={(0.1,0)}, middx arrowsm] (0,0) -- (-1.5,1.5); 
	\draw (-0.5,0.53) node[rotate={30}] {\LARGE{$\times$}}; \draw (-0.75,0.6) node[left] {$P_1$};
	
	\draw[-, shorten >= 9, shorten <= 6, shift={(-0.05,0.05)}, mid arrowsm] (0,0) -- (0,1.5); 
	\draw[-, shorten >= 4, shorten <= 11, shift={(0.05,0.15)}, mid arrowsm] (0,1.5) -- (0,0);
	\draw (0,0.9) node[left] {$P_2$};
	
	\draw (0.75,1.5) node {$\cdots$};
	
	\draw[-, shorten >= 6, shorten <= 11, shift={(0.07,0.02)}, mid arrowsm] (1.5,1.5) -- (0,0);
	\draw[-, shorten >= 6, shorten <= 10, shift={(-0.1,0)}, mid arrowsm] (0,0) -- (1.5,1.5);
	\draw (0.75,0.6) node[right] {$P_3$};
	
	\draw[-] (g) to[out=-60,in=0] (0,-0.6) to[out=180,in=-120] (g); \draw (0,-0.6) node[right] {$A$}; \draw (0,-0.6) node[cross] {};
	
	\draw (0,-1.5) node{$\cW = \sum_{j=0}^{N-1} Flip[P_1 A^j \tilde{P}_1 ] + $};
	\draw (0.5,-2.1) node {$ + \sum_{j=2}^{N} Flip[\Tr A^j] $};
	 
	\path (5,0) node[gauge](g1) {$\!\!\!N\!\!\!$} -- (7,0) node[gauge](g2) {$\!\!\!N\!\!\!$} -- (10,0) node[gauge](g3) {$\!\!\!N\!\!\!$} -- (12,0) node[gauge](g4) {$\!\!\!N\!\!\!$} -- (5,1.5) node[flavor](y1) {$\!1\!$} -- (12,1.5) node[flavor](y2) {$\!1\!$};
	 
	\wigM (g1) -- (g2); \draw (6,0.4) node {$\Pi_2$};
	\wigM (g2) -- (8,0); 
	\wigM (g3) -- (9,0); 
	\wigM (g3) -- (g4); \draw (11,0.4) node {$\Pi_{F-1}$};
	\draw[-, shorten >= 9, shorten <= 6, shift={(-0.05,0.05)}, mid arrowsm] (5,0) -- (5,1.5); \draw (5,0.75) node[left] {$V_1$};
	\draw[-, shorten >= 4, shorten <= 11, shift={(0.05,0.15)}, mid arrowsm] (5,1.5) -- (5,0);
	\draw[-, shorten >= 9, shorten <= 6, shift={(0.05,0.05)}, mid arrowsm] (12,0) -- (12,1.5); \draw (12,0.75) node[right] {$V_2$};
	\draw[-, shorten >= 4, shorten <= 11, shift={(-0.05,0.15)}, mid arrowsm] (12,1.5) -- (12,0); 
	\draw (12,0.45) node[cross]{};	
	\draw (8.5,0) node {$\cdots$};
	\draw[-] (g1) to[out=-60, in=0] (5,-0.6) to[out=180,in=-120] (g1); \draw (5,-0.7) node[right] {$A_1$};
	\draw[-] (g2) to[out=-60, in=0] (7,-0.6) to[out=180,in=-120] (g2); \draw (7,-0.7) node[right] {$A_2$};
	\draw[-] (g3) to[out=-60, in=0] (10,-0.6) to[out=180,in=-120] (g3); \draw (10,-0.7) node[right] {$A_{F-2}$};
	\draw[-] (g4) to[out=-60, in=0] (12,-0.6) to[out=180,in=-120] (g4); \draw (12,-0.7) node[right] {$A_{F-1}$}; \draw (12,-0.6) node[cross] {};
	
	\draw (8.5,-1.5) node {$\cW = \mathcal{W}_{\text{gluing}} + \sum_{j=0}^{N-1} (Flip[V_2 A_{F-1}^j \tilde{V}_2] ) + $  };
	\draw (8.5,-2.2) node {$ + \sum_{j=2}^N Flip[\Tr A_{F-1}^j] $};
	 
\end{tikzpicture}}
\ee

We can now can turn on the mass term $\d \CW = V_1 \tilde{V}_1$ on the r.h.s. of \eqref{flippedmipa}. After this deformation we are left with the following theory:
\be\label{fig:3d_mirror_mass}
\begin{tikzpicture}[thick,node distance=3cm,gauge/.style={circle,draw,minimum size=5mm},flavor/.style={rectangle,draw,minimum size=5mm}] 
	 
	\path (5,0) node[gauge](g1) {$\!\!\!N\!\!\!$} -- (7,0) node[gauge](g2) {$\!\!\!N\!\!\!$} -- (10,0) node[gauge](g3) {$\!\!\!N\!\!\!$} -- (12,0) node[gauge](g4) {$\!\!\!N\!\!\!$} -- (12,1.5) node[flavor](y2) {$\!1\!$};
	 
	\wigM (g1) -- (g2); \draw (6,0.4) node {$\Pi_2$};
	\wigM (g2) -- (8,0); 
	\wigM (g3) -- (9,0); 
	\wigM (g3) -- (g4); \draw (11,0.4) node {$\Pi_{F-1}$};
	\draw[-, shorten >= 9, shorten <= 6, shift={(0.05,0.05)}, mid arrowsm] (12,0) -- (12,1.5); \draw (12,0.75) node[right] {$V_2$};
	\draw[-, shorten >= 4, shorten <= 11, shift={(-0.05,0.15)}, mid arrowsm] (12,1.5) -- (12,0); \draw (12,0.45) node[cross]{};	
	\draw (8.5,0) node {$\cdots$};
	\draw[-] (g1) to[out=-60, in=0] (5,-0.6) to[out=180,in=-120] (g1); \draw (5,-0.7) node[right] {$A_1$};
	\draw[-] (g2) to[out=-60, in=0] (7,-0.6) to[out=180,in=-120] (g2); \draw (7,-0.7) node[right] {$A_2$};
	\draw[-] (g3) to[out=-60, in=0] (10,-0.6) to[out=180,in=-120] (g3); \draw (10,-0.7) node[right] {$A_{F-2}$};
	\draw[-] (g4) to[out=-60, in=0] (12,-0.6) to[out=180,in=-120] (g4); \draw (12,-0.7) node[right] {$A_{F-1}$}; \draw (12,-0.6) node[cross] {};
	
	\draw (5.5,-1.5) node[right]{$\cW = \mathcal{W}_{\text{gluing}} + \sum_{j=0}^{N-1} Flip[V_2 A_{F-1}^j \tilde{V}_2]$  };
	\draw (6.5,-2.1) node[right] {$ + \sum_{j=2}^{N} Flip[\Tr A_{F-1}^j] $};
	 
\end{tikzpicture}
\ee
Now we use the fact that an improved bifundamental gauged on one side  {\it confines} to $N$ free hypers (see \eqref{fmconf}):
\be
\begin{tikzpicture}[thick,node distance=3cm,gauge/.style={circle,draw,minimum size=5mm},flavor/.style={rectangle,draw,minimum size=5mm}] 
	 
	\path (0,0) node[gauge](g) {$\!\!\!N\!\!\!$} -- (2,0) node[flavor](x) {$\!N\!$} -- (3,0) node {$\Longleftrightarrow$};
	 
	\wigM (g) -- (x);
	\draw[-] (g) to[out=60,in=0] (0,0.6) to[out=180,in=120] (g);
	
	\draw (1,-1) node {$\cW = \CW_{\text{gluing}}$};
	\draw (5,0) node {$( \text{Free Hyper} )^N$};
	 
\end{tikzpicture}
\ee
Using this fact we can sequentially confine all the improved bifundamentals in \eqref{fig:3d_mirror_mass} into a total of $(F-2) \times N$ hypers.
We are then left just with a $U(N)$ adjoint SQCD with one flipped flavor  ($\mathcal{W}=\sum_{j=0}^{N-1} Flip[V_2 A_{F-1}^j \tilde{V}_2]$) and   $(F-2) \times N$ free hypers.
  
Using the duality \eqref{flipf1} we claim that  that the  $U(N)$ adjoint SQCD with one flipped flavor is dual to $N$ free hypers.
So in conclusion on the mirror side of the duality in \eqref{flippedmipa}, after the mass deformation for the first flavor  we have just $(F-1)\times N $ free hypers. 

Now let's go back to the electric theory in \eqref{flippedmipa}.
Using the operator map we see that in the electric theory the mass term $\CW = V_1\tilde{V}_1$ maps to 
$\mathcal{F}[Q_1 A^{N-1} \tilde{Q}_1]$ inducing a VEV for $Q_1 A^{N-1} \tilde{Q}_1$. This is a VEV  for a meson dressed $N-1$ times which Higgses
completely the theory leaving $(F-1)\times N $ free hypers. So also this consistency check is passed.

\subsection{Flowing to $U(N)$ SQCD without adjoint}\label{sec:noadjQCDduality}
The last deformation we consider is turning on a mass term for the adjoint $A$ in the electric $U(N)$ SQCD. As discussed in sec. \ref{sec:3d_SQCD_map}, $\Tr(A^j)$ maps to $\Tr(A_I^j)$ in the mirror quiver. In the mirror quiver, for each $j$, there are $F-1$ holomorphic operators  $\Tr(A_I^j)$, one for each node. However only one combination is non-zero in the chiral ring.\footnote{This follows from the superpotential $\CW_{gluing}$ and of the chiral ring relation $\Tr(\mathsf{A}_L^k) = \Tr(\mathsf{A}_R^k), k=2,\ldots,N$ in the $FM[U(N)]$ theory, see \cite{BCP1}.} Hence, we turn on masses for all the adjoints
\be \delta \CW = \Tr(A^2) \qquad \Longleftrightarrow \qquad \delta\CW = \sum_{I=1}^{F-1} \Tr(A_I^2) \ee
This deformation breaks the $U(1)_{\tau}$ symmetry and triggers an RG flow to a new dual pair.

On the left hand side the effect is to simply remove the adjoint. On the right hand side we remove the $F-1$ adjoints $A_I$, and the superpotential 
now includes the adjoint operators of the improved bifundamentals $A^{(I)}_{R/L}$.

\be
\resizebox{.95\hsize}{!}{
 \bpic[thick,node distance=3cm,gauge/.style={circle,draw,minimum size=5mm},flavor/.style={rectangle,draw,minimum size=5mm}] 
 
\begin{scope}[shift={(-3,-4.5)}]
    \path (4,0) node[gauge](g) {$\!\!\!N\!\!\!$} -- (3.15,1.5) node[flavor](x1) {$\!1\!$} -- (4.85,1.5) node[flavor] (x2) {$\!1\!$};
	
	\draw[-, shorten >= 7, shorten <= 8, shift={(-0.05,-0.05)}, mid arrowsm] (4,0) -- (3,1.5); \draw (3.5,0.6) node[left] {$Q_1$};
	\draw[-, shorten >= 7, shorten <= 10, shift={(0.05,0.05)}, mid arrowsm] (3,1.5) -- (4,0);
	\draw[-, shorten >= 7, shorten <= 8, shift={(0.05,-0.05)}, mid arrowsm] (4,0) -- (5,1.5); \draw (4.5,0.6) node[right] {$Q_F$};
	\draw[-, shorten >= 7, shorten <= 10, shift={(-0.05,0.05)}, mid arrowsm] (5,1.5) -- (4,0); 
	\draw (4,1.5) node{$\cdots$};
	
	\draw (4,-1) node {$\CW = 0$};
	
\end{scope}

	\draw (3,-4.2) node {$\Longleftrightarrow$};
	
\begin{scope}[shift={(5,-4.5)}]	
	\path (0,0) node[gauge](g1) {$\!\!\!N\!\!\!$} -- (1.5,0) node[gauge](g2) {$\!\!\!N\!\!\!$} -- (3,0) node(gi) {$\ldots$}
		-- (4.5,0) node[gauge](g3) {$\!\!\!N\!\!\!$} -- (6,0) node[gauge](g4) {$\!\!\!N\!\!\!$} 
		-- (0,1.5) node[flavor](y1) {$\!1\!$} -- (6,1.5) node[flavor](y2) {$\!1\!$};
	\wigM (g1) -- (g2); \draw (0.75,0.35) node {$\Pi_2$};
	\wigM (g2) -- (gi);
	\wigM (g3) -- (gi);
	\wigM (g3) -- (g4); \draw (5.25,0.35) node {$\Pi_{F-1}$};
	\draw[-, shorten >= 3, shorten <= 12, shift={(-0.05,-0.15)}, middx arrowsm] (0,0) -- (0,1.5);
	\draw[-, shorten >= 8, shorten <= 7, shift={(0.05,0)}, midsx arrowsm] (0,1.5) -- (0,0);
	\draw (0,0.5) node[cross]{}; \draw (0,0.6) node[left] {$V_1$};
	\draw[-, shorten >= 3, shorten <= 12, shift={(-0.05,-0.15)}, middx arrowsm] (6,0) -- (6,1.5);
	\draw[-, shorten >= 8, shorten <= 7, shift={(0.05,0)}, midsx arrowsm] (6,1.5) -- (6,0);
	\draw (6,0.5) node[cross]{}; \draw (6,0.6) node[right] {$V_2$};
	\draw (3,-1) node {\small{$\cW = \sum_{j=0}^{N-1} ( Flip[V_1 (\mathsf{A}_{L}^{(2)})^j \tilde{V}_1] + Flip[V_2 (\mathsf{A}_{R}^{(F-1)})^j \tilde{V}_2] )+$} };
	\draw (3,-1.6) node {\small{$+\sum_{I=2}^{F-2} \mathsf{A}_{R}^{(I)}\mathsf{A}_{L}^{(I+1)}$} };
\end{scope}

\epic}\label{noadjsqcd}
\ee
The list of charges is given in table \ref{tab:noadjqcd_charges}.
\begin{table}[h]
\centering
\renewcommand{\arraystretch}{1.2}
\begin{tabular}{|c|c|c|c|c|}
	\hline
		&  $U(1)_{R_0}$ & $U(1)_{B_j}$ &  $U(1)_{X_j}$ & $U(1)_Y$ \\
	\hline
	$Q_k ,\tilde{Q}_k $ &  $1$ & $- \d_{j,k}$ & $ \mp \d_{j,k}$ & $0$ \\
	\hline
	\hline 
	$V_1 ,\tilde{V}_1 $ &  $\frac{1-N}{2}$ & $ \d_{1,j} $ & $0$ & $ \mp 1$ \\
	$V_2 ,\tilde{V}_2 $ &  $\frac{1-N}{2}$ & $ \d_{F,j} $ & $0$  & $0$ \\
	$\Pi_k, \tilde{\Pi}_k $ &  $0$ & $ \d_{j,k} $ & $0$  & $0$ \\
	\hline
\end{tabular}
\caption{
Charges of the fields in the mirror duality for the $U(N)$ SQCD in  \eqref{noadjsqcd}. In the first block are listed the fields in the SQCD, while in the second block are listed those of the mirror description. 
}
 \label{tab:noadjqcd_charges}
\end{table}
The global symmetry on the l.h.s. inf \eqref{noadjsqcd} is given by:
\begin{align}\label{eq:noadjsqcd_symm}
	SU(F)_U \times SU(F)_W \times U(1)_m \times U(1)_Y \,,
\end{align}
where the two $SU(F)$ and the $U(1)_m$ global symmetries are obtained from the $U(1)_{B_j} \times U(1)_{X_j}$ as usual with the  redefinitions in eqs. \eqref{axdef}, \eqref{uvdef}.\\
In the mirror theory, on the r.h.s. in \eqref{noadjsqcd}, the UV global symmetry is:
\begin{align}
	\prod_{j=1}^F U(1)_{B_j} \times \prod_{j=1}^{F-1} U(1)_{X_{j+1}-X_j} \times U(1)_Y \,,
\end{align}
where    $U(1)_{X_{j+1}-X_j}$ are the topological symmetries of the $F-1$ gauge nodes.
In the IR the global symmetry enhances to the group in \eqref{eq:noadjsqcd_symm}. 

The chiral ring generators in the SQCD side are the $F^2$ mesons $Q\tilde{Q}$, with  R-charge  $2-2m$ and the $2$ monopoles $\M^\pm$, with R-charge  $Fm-N+1$.  The mapping of the mesons is very similar to  \eqref{eq:meson_F=4_mirror}, for instance if $F=4$:
\begin{align}\label{eq:meson_F=4_mirrorB}
	\Tr( Q \tilde{Q}) 
\Longleftrightarrow
	\begin{pmatrix}
		\mathcal{F}[V_1 (\mathsf{A}_L^{(2)})^{N-1} \tilde{V}_1] & \mathfrak{M}^{(+,0,0)} & \mathfrak{M}^{(+,+,0)} &
		 \mathfrak{M}^{(+,+,+)} \\
		\mathfrak{M}^{(-,0,0)} & \mathsf{B}_{1,1}^{(2)} & \mathfrak{M}^{(0,+,0)} & \mathfrak{M}^{0,+,+)} \\
		\mathfrak{M}^{(-,-,0)} & \mathfrak{M}^{(0,-,0)} & \mathsf{B}_{1,1}^{(3)} & \mathfrak{M}^{(0,0,+)} \\
		\mathfrak{M}^{(-,-,-)} & \mathfrak{M}^{(0,-,-)} & \mathfrak{M}^{(0,0,-)} & \mathcal{F}[V_2 (\mathsf{A}_R^{(3)})^{N-1} \tilde{V}_2]
	\end{pmatrix} \,,
\end{align}
The SQCD monopoles $\M^\pm$ map to long mesons $V_1 \Pi_2 \ldots \Pi_{F-1} \tilde{V}_2$ and $V_2 \tilde{\Pi}_{F-1} \ldots \tilde{\Pi}_2 \tilde{V}_1$.
It is easy to check that the charge assignements given in table \ref{tab:noadjqcd_charges} are consistent with the mapping.

We now comment about the fate of the operators on the quiver side that were mapping to dressed mesons in the duality with adjoint. These operators are the $\mathcal{F}[V_1 (\mathsf{A}_L^{(2)})^{j} \tilde{V}_1]$, $\mathsf{B}_{1,n}^{(a)}$ $\mathcal{F}[V_2 (\mathsf{A}_R^{(F-1)})^{j} \tilde{V}_2]$ and the dressed monopoles and dressed long mesons 
 (monopoles and  long mesons, in the quiver side,  cannot be dressed by the explicit adjoint fields, since they are massive, but we can consider dressing with the adjoints inside the improved bifundamental theories that is the $\mathsf{A}$'s). Such  operators do not exist in the SQCD side of \eqref{noadjsqcd}, while candidate dressed operators appear in the quiver side, so the duality implies that the dressed operators in the quiver are holomorphic operators set to zero in the chiral ring. We would like to understand this feature directly in the quiver side without invoking the duality. 

We can explain why the quiver operators along the diagonal in \eqref{eq:meson_F=4_mirrorB} are set to zero in the quiver chiral ring using the logic of \cite{Benvenuti:2017lle}, where it is shown that when a flipper field is flipping an operator below the unitarity bound (hence the flipper has $R>\frac{3}{2}$), it is zero on the chiral ring as a consequence of quantum effects, e.g. giving a VEV to such a flipper leads to a theory with no supersymmetric vacuum.

On the electric side, we know that the superconformal R-charge is such that \footnote{We are considering the region $F \geq N$.}
\be \label{doublein} \frac{1}{2}<R[Q\tilde{Q}]<1 \,.\ee 
The left inequality is the unitarity bound, the right inequality follows from the fact in absence of superpotential the interactions decrease the R-charge with respect to the free theory, where $R[Q]=\frac{1}{2}$. The inequality \eqref{doublein} implies that on the quiver side,  $\frac{1}{2}<R[\mathcal{F}[ \mathsf{A}^{N-1} \tilde{V}]<1$, while all the operators  $\CF[V \mathsf{A}^h \tilde{V}], h=0,1,\ldots,N-2$, have R-charge greater than $\frac{3}{2}$ (recall $R[\mathsf{A}]=1$). Following the logic of \cite{Benvenuti:2017lle}, we learn such flippers are holomorphic operators which are zero in the chiral ring of the quiver theory. 

The same argument works for the $\mathsf{B}_{1,k}$ operators with $k=2,\ldots,N-1$, which were mapping to dressed mesons in the duality with the adjoint. Such operators, when viewed in the Lagragian UV completion of the improved bifundamental \ref{fig:FM_quiver}, are flippers, see table \ref{tab:FM_operators}, hence they are flippers with $R>\frac{3}{2}$, so they must be zero in the chiral ring.

\section{Derivation via the $\CN=2$ algorithm}\label{sec:3d_algorithm}
In this section we generalize the  dualization algorithm introduced in \cite{Hwang:2021ulb,Comi:2022aqo}  for $\CN=4$ linear quivers to the $\CN=2$ case and show how to construct the mirror dual of a generic $\CN=2$ linear quiver with $U(N)$ gauge nodes, improved bifundamentals and  generalized flavors. 

The idea  of the algorithm builds on the observation   \cite{Gaiotto:2008ak,Gulotta:2011si,Assel:2014awa} that on linear or circular brane setups, $\CS$-duality  can act locally on each 5-brane creating an $\CS$-duality  wall on its right and an $\CS^{-1}$-duality  wall on its left:
$D5= \CS \!\cdot\! NS \!\cdot\! \CS^{-1}$ and $\widebar{NS}= \CS \!\cdot\! D5 \!\cdot\! \CS^{-1}.$
The dualization algorithm implements in field theory this local action of $\CS$-duality. 

We first define the basic $\CN=2$ QFT blocks and  the basic $\CN=2$ duality moves. We then explain the steps of the algorithm and apply them to the example of the $\CN=2$ adjoint SQCD.

\subsection{Generalized QFT blocks and basic moves}\label{sec:blocks_moves}

\subsubsection*{The generalized QFT blocks}
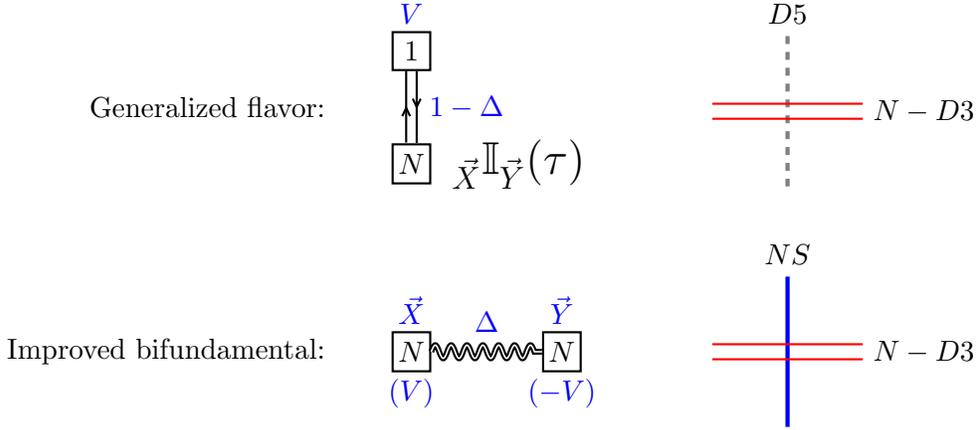
\begin{figure}
\begin{tikzpicture}[thick,node distance=3cm,gauge/.style={circle,draw,minimum size=5mm},flavor/.style={rectangle,draw,minimum size=5mm}]
\begin{scope}
	\path (-1,0.75) node[left] {Generalized flavor:} -- (0,0) node[flavor] (x) {$\!N\!$} -- (0,1.5) node[flavor] (v) {$\!1\!$};
	
	\Dbrane (5,1.7) -- (5,-0.3); \draw (5,2) node {$D5$};
	\Dthree (4,0.8) -- (6,0.8); \Dthree (4,0.6) -- (6,0.6); \draw (6,0.7) node[right] {$N - D3$};
	
	\draw[-, shorten >= 6, shorten <= 9.5, shift={(-0.07,-0.05)}, mid arrowsm] (0,0) -- (0,1.5);
	\draw[-, shorten >= 6.5, shorten <= 9, shift={(0.07,0.05)}, mid arrowsm] (0,1.5) -- (0,0);  
	\draw[blue] (0.1,0.75) node[right] {$1-\D$}; \draw[blue] (0,2) node {$V$};
	
	\draw (0.4,0) node[right] {\LARGE{${}_{\vec{X}}\mathbb{I}_{\vec{Y}}(\tau)$}};
\end{scope}
\begin{scope}[shift={(0,-2.5)}]
	\path (-1,0) node[left] {Improved bifundamental:} -- (0,0) node[flavor] (f1) {$\!N\!$} -- (2,0) node[flavor] (f2) {$\!N\!$};
	
	\NSbrane (5,1) -- (5,-1); \draw (5,1.3) node {$NS$};
	\Dthree (4,0.1) -- (6,0.1); \Dthree (4,-0.1) -- (6,-0.1); \draw (6,0) node[right] {$N - D3$};
	
	\wigM (f1) -- (f2); \draw[blue] (1,0.4) node {$\D$}; 
	\draw[blue] (0,0.55) node {$\vec{X}$}; \draw[blue] (0,-0.55) node {$(V)$}; 
	\draw[blue] (2,0.55) node {$\vec{Y}$}; \draw[blue] (2,-0.55) node {$(-V)$};
\end{scope}
\end{tikzpicture}
\caption{Definition of the generalized blocks. In the picture we write in blue the parameterization of the two theories.
To the generalized flavor block we assign trial R-charge $1$ and charge $-1$ under the axial symmetry $U(1)_\D$, while
$V$ denotes the real mass parameter for its vector-like symmetry. $\vec X, \vec Y$ denote the Cartans of two $U(N)$ flavor groups.
The improved bifundamental block, with trial R-charge $0$ and $\D$-charge $1$,  is defined with background FI couplings for the two $U(N)$ groups. The  FI parameters are denoted by the $(\pm V)$.
}
\label{fig:3d_genblocks}
\end{figure}
The generalized  matter blocks  are depicted in figure \ref{fig:3d_genblocks}.
To a $D5$ brane with $N$ $D3$ branes stretching on the left and right we associate a generalized flavor block, which consist in a flavor with
$U(1)_\D \times U(1)_V$ symmetry together with the identity operator ${}_{\vec{X}}\mathbb{I}_{\vec{Y}}(\tau)$ which 
identifies the Cartans  $\vec{X}$ and $\vec{Y}$ of two $U(N)$ symmetries. \\
To a $NS$ brane with $N$ $D3$ branes stretching on the left and right we associate  an improved bifundamental block given by an $FM[U(N)]$ theory  with background FI couplings for the two $U(N)$ global symmetries. \\
The $S_b^3$ partition functions of the QFT blocks are given by:
\begin{align}
	& Z_{D5}^{(N)} (\vec{X},\vec{Y},\tau,\D,V) = \prod_{j=1}^N s_b(\D \pm (X_j - V)) {}_{\vec{X}}\mathbb{I}_{\vec{Y}}(\tau) \,, \nn \\
	& Z_{NS}^{(N)} (\vec{X},\vec{Y},\tau,\D,V) = e^{2\pi i V \sum_{j=1}^N (X_j - Y_j)} Z_{FM}^{(N)}(\vec{X},\vec{Y},\tau,\D) \,,
\end{align}
where $Z_{FM}^{(N)}$ is defined in appendix \ref{app:FM}, equation \eqref{eq:FM_parfun}. The identity operator instead is defined as follows:
\begin{align}\label{eq:3d_idoperator_main}
	{}_{\vec{X}}\mathbb{I}_{\vec{Y}}(\tau) = \frac{1}{\D_N(\vec{X},\tau)} \sum_{\s \in S_N} \prod_{j=1}^N \d(X_j - Y_{\s(j)}) \,.
\end{align}
The convention used for the $S_b^3$ partition function is given in appendix \ref{inpaconv}.

\subsection*{The $\CS$-wall}
The $3d$ $\mathcal{S}$-wall theory is realized in field theory as the $FT[U(N)]$ theory (see \ref{app:FT})  as explained in \cite{Gaiotto:2008ak}.
The  $FT[SU(N)]$ theory  we use here differs from the  $T[SU(N)]$  introduced in \cite{Gaiotto:2008ak}  by the adjoint singlet flipping the meson operator. In addition here we work in the $\mathcal{N}=2^*$ parameterization.
Together with the $T$ generator it satisfies the $SL(2,\mathbb{Z})$ relations $(\CS T)^3$, $\CS^2 = -1$ and $\CS\CS^{-1} = 1$ \cite{Bottini:2021vms,Comi:2022aqo}. The partition functions of  $\CS$ and $\CS^{-1}$ differ only by a sign:
\begin{align}
	Z_{\CS^{\pm}}^{(N)} (\vec{X}, \vec{Y}, \tau) = 
	Z_{FT}^{(N)}( \vec{X}, \mp \vec{Y}, \tau) = 
	Z_{FT}^{(N)}(\mp \vec{X}, \vec{Y},\tau ) \,.
\end{align}
Graphically we represent an $\CS$-wall by a dashed line connecting two $U(N)$ flavor symmetries, the $\CS$ and $\CS^{-1}$ walls  are then distinguished by the $\pm$ sign over the dashed line.  The $\CS \CS^{\pm} = \mp 1$ relations
\be
\begin{tikzpicture}[thick,node distance=3cm,gauge/.style={circle,draw,minimum size=5mm},flavor/.style={rectangle,draw,minimum size=5mm}]

	\path (0,0) node[flavor](x) {$\!N\!$} -- (2,0) node[gauge](z) {$\!\!\!N\!\!\!$} -- (4,0) node[flavor](y) {$\!N\!$};
	
	\wigT (x) -- (z); \draw (1,0.3) node {$+$};
	\wigT (z) -- (y); \draw (3,0.3) node {$\pm$};
	\draw[-] (z) to[out=60,in=0] (2,0.6) to[out=180,in=120] (z); \draw[blue] (2,0.8) node {\scriptsize{$\tau$}};
	\draw[blue] (0,0.5) node {\scriptsize{$\vec{X}$}}; \draw[blue] (4,0.5) node {\scriptsize{$\vec{Y}$}};
	
	\draw (5,0) node {$=$};
	\draw (6,0) node[right] {\large{${}_{\vec{X}} \mathbb{I}_{\mp\vec{Y}} (\tau)$}};
	
	\draw (2,-1) node {$\cW = \CW_{\text{gluing}}$};

\end{tikzpicture}
\label{ftid}
\ee
correspond to the following partition function identity:
\begin{align}\label{eq:Swall_relations}
	\int d\vec{Z}_N \D_N(\vec{Z},\tau) Z_{\CS}^{(N)} (\vec{X},\vec{Z},\tau) Z_{\CS^{\pm}} (\vec{Z}, \vec{Y}, \tau) =
	{}_{\vec{X}}\mathbb{I}_{\mp \vec{Y}} (\tau) \,,
\end{align}
where the identity operator is defined as in \eqref{eq:3d_idoperator_main} .
It was shown in \cite{Bottini:2021vms} that these relations can be proved by iterating Seiberg-like dualities.

\subsubsection*{Basic duality moves}
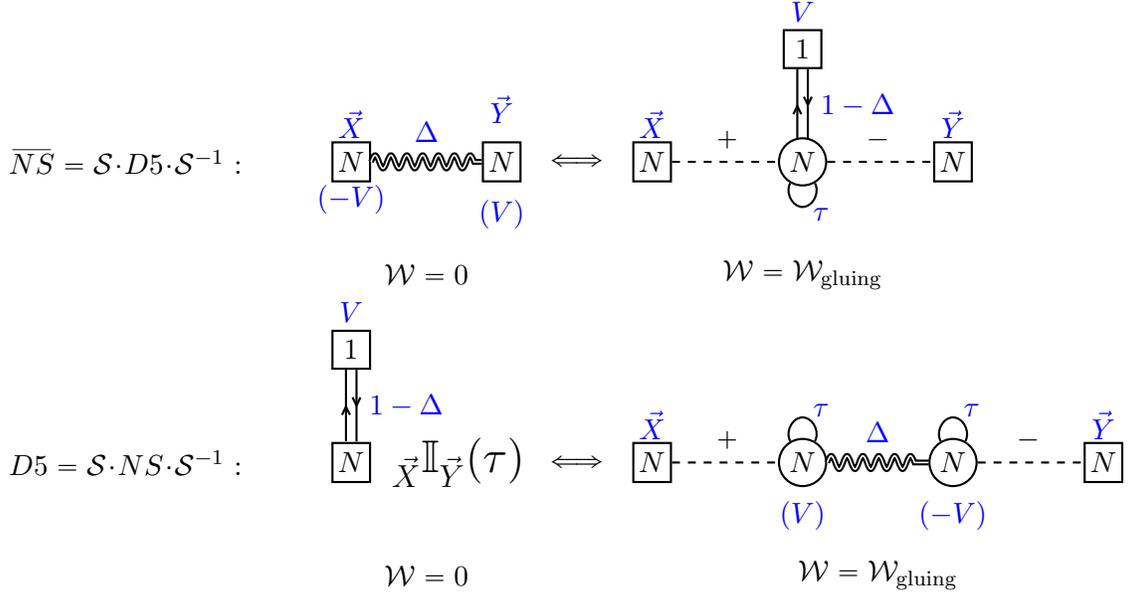
\begin{figure}\label{fig:algo}
\centering
\begin{tikzpicture}[thick,node distance=3cm,gauge/.style={circle,draw,minimum size=5mm},flavor/.style={rectangle,draw,minimum size=5mm}]

\begin{scope}
	\path (0,0) node[flavor] (x1) {$\!N\!$} -- (2,0) node[flavor] (x2) {$\!N\!$} -- (3,0) node {$\Longleftrightarrow$} 
		-- (4,0) node[flavor] (y1) {$\!N\!$} -- (6,0) node[gauge] (y2) {$\!\!N\!\!$} -- (8,0) node[flavor] (y3) {$\!N\!$} 
		-- (6,1.5) node[flavor] (v1) {$\!1\!$} -- (-3,0) node { $\widebar{NS}=\CS \!\cdot\!D5 \!\cdot\! \CS^{-1} $ :};
		
	\wigM (x1) -- (x2); \draw[blue] (1,0.4) node {$\D$};
	\draw[blue] (0,-0.5) node {$(-V)$}; \draw[blue] (2,-0.7) node {$(V)$};
	\draw[blue] (0,0.5) node {$\vec{X}$}; \draw[blue] (2,0.7) node {$\vec{Y}$};
	
	\wigT (y1) -- (y2); \draw (5,0.3) node {$+$};
	\wigT (y2) -- (y3); \draw (7,0.3) node {$-$};
	
	\draw[-, shorten >= 6, shorten <= 10, shift={(-0.07,-0.05)}, mid arrowsm] (6,0) -- (6,1.5);
	\draw[-, shorten >= 7, shorten <= 9, shift={(0.07,0.05)}, mid arrowsm] (6,1.5) -- (6,0);
	\draw[blue] (6.1,0.75) node[right] {$1-\D$}; \draw[blue] (6,2) node {$V$};
	
	\draw[-] (y2) to[out=-60,in=0] (6,-0.6) to[out=180,in=-120] (y2); \draw[blue] (6,-0.7) node[right] {$\tau$};
	
	\draw[blue] (4,0.5) node {$\vec{X}$}; \draw[blue] (8,0.5) node {$\vec{Y}$};
	
	\draw (1,-1.5) node {$\CW = 0$};
	\draw (6,-1.5) node {$\CW = \CW_{\text{gluing}} $}; 

\end{scope}
\begin{scope}[shift={(0,0)}]					
	\path (0,-4) node[flavor] (x1) {$\!N\!$} -- (0,-2.5) node[flavor] (v1) {$\!1\!$} -- (3,-4) node {$\Longleftrightarrow$}
		-- (4,-4) node[flavor] (y1) {$\!N\!$} -- (6,-4) node[gauge] (y2) {$\!\!N\!\!$} -- (8,-4) node[gauge] (y3) {$\!\!N\!\!$} 
		-- (10,-4) node[flavor] (y4) {$\!N\!$} -- (-3,-4) node {$D5=\CS \!\cdot\! NS \!\cdot\! \CS^{-1} $ :};
		
	\draw[-, shorten >= 6, shorten <= 9.5, shift={(-0.07,-0.05)}, mid arrowsm] (0,-4) -- (0,-2.5);
	\draw[-, shorten >= 6.5, shorten <= 9, shift={(0.07,0.05)}, mid arrowsm] (0,-2.5) -- (0,-4); 
	\draw[blue] (0.1,-3.25) node[right] {$1-\D$}; \draw[blue] (0,-2) node {$V$};
	
	\draw (0.4,-4) node[right] {\LARGE{${}_{\vec{X}}\mathbb{I}_{\vec{Y}} (\tau)$}};
	
	\wigT (y1) -- (y2); \draw (5,-3.7) node {$+$};
	\wigM (y2) -- (y3); \draw[blue] (7,-3.6) node {$\D$};
	\wigT (y3) -- (y4); \draw (9,-3.7) node {$-$};
	\draw[-] (y2) to[out=60,in=0] (6,-3.4) to[out=180,in=120] (y2); \draw[blue] (6,-3.3) node[right] {$\tau$};
	\draw[-] (y3) to[out=60,in=0] (8,-3.4) to[out=180,in=120] (y3); \draw[blue] (8,-3.3) node[right] {$\tau$};
	
	\draw[blue] (4,-3.5) node {$\vec{X}$}; \draw[blue] (10,-3.5) node {$\vec{Y}$};
	\draw[blue] (6,-4.7) node {$(V)$}; \draw[blue] (8,-4.7) node {$(-V)$};
	
	\draw (1,-5.5) node {$\CW = 0$};
	\draw (7,-5.5) node {$\CW = \CW_{\text{gluing}} $};
	
\end{scope}
\end{tikzpicture}
\caption{Basic $\CS$-duality moves for the $\CN=2$ QFT blocks.
In the first line a flavor block acted by an $\CS$-wall on on the left and by an
 $\CS^{-1}$-wall on the right  is dualized to an improved bifundamental. On the r.h.s. $\CW_{\text{gluing}}$ couples the adjoint chiral to the adjoint operators of the two $\CS$-walll theories.
 Similarly  in the second line we have the $\CS$-dualization of the improved bifundamental into a flavor block. On the r.h.s $\CW_{\text{gluing}}$ couples the adjoint chirals to the adjoint operators of the improved bifundamental and of the $\CS$-wall theories.}
\label{fig:3d_basic_duality_moves}
\end{figure}

The last ingredient necessary for the definition of the algorithm is given by the \emph{basic duality moves}.  They realize at the field theory level  the local action of  $\CS$-duality on each 5-brane. The two basic moves are given in figure \ref{fig:3d_basic_duality_moves}.

In the duality  move in the first line of  figure \ref{fig:3d_basic_duality_moves}, we see how by acting with an  $\CS$-wall on the left and $\CS^{-1}$-wall on the right of a generalized fundamental block we obtain an improved bifundamental block. The superpotential $\CW_{\text{gluing}}$ couples the adjoint chiral to the two adjoint moment map present in the two $\CS$-wall theories, $\mathsf{A}_L, \mathsf{A}_R$, as $\CW_{\text{gluing}}= a(\mathsf{A}_L + \mathsf{A}_R)$. The flavor does not enter the superpotential and indeed is rotated by a $U(1)_V \times U(1)_\Delta$ symmetry. \\
One can recover the $\CN=4$ basic moves of  \cite{Hwang:2021ulb,Comi:2022aqo}, by adding on the r.h.s. a cubic superpotential coupling the flavor $f$ to the moment maps  as $\d \CW = f ( \mathsf{A}_L - \mathsf{A}_R) \tilde{f}$, therefore making the theory $\CN=4$. This deformation is mapped on the l.h.s. to the $\mathsf{B}_{2,1}$ singlet of the improved bifundamental  theory which has the effect of ironing it to a $U(N) \times U(N)$ bifundamental hypermultiplet, as shown \eqref{fig:FM_c=1-t/2}.\\
In the duality  move in the second line of  figure \ref{fig:3d_basic_duality_moves}, instead the $\CS$-dualization of an improved bifundamental block gives the generalized fundamental block. The superpotential $\CW_{\text{gluing}}$ couples the two adjoint chirals to the adjoint operators inside the $\CS$-walls and
 of improved bifundamentals.

The first $\cN=2$ duality  move can be derived by taking suitable real mass deformations of the $3d$ braid duality \eqref{fig:3dBraid} as shown in \eqref{fig:3d_basicmove}.
The second $\cN=2$  duality move can actually be obtained by acting on the left and right hand side of the the first duality move  with  $\CS$ and $\CS^{-1}$ and using the fusion to identity property $\CS \CS^{-1}=1$ \eqref{ftid},  hence the braid duality is  the fundamental duality move.\footnote{Notice that  also the fusion to identity property  \eqref{ftid}  follows from 
 the first move which can be regarded as an S-confining duality, similar  to $4d$ $\cN=1$ $SU(N)$ SQCD with $N+1$ flavors. Turning on a mass for a flavor one flow to $SU(N)$ SQCD with $N$ flavors whose low energy dynamics is well known to be governed by a quantum deformed moduli space, over which a part of the global symmetry is spontaneously broken. In the same way we can obtain \eqref{ftid}  by  giving a mass to the flavor  in the first duality move to go from a confining duality to a quantum deformed moduli space where the $U(N) \times U(N)$ global symmetry  is broken to the diagonal.}
Moreover it has been shown in \cite{BCP1} that the braid duality can be demonstrated by induction by assuming only the elementary Seiberg-like dualities. Hence all the $\cN=2$ mirror dualities following from the algorithm are demonstrated to be consequence Seiberg-like dualities only. \\

As partition function identities the basic moves are:\footnote{As discussed in  \cite{Comi:2022aqo} the partition function of the  $\widebar{NS}$   block differs from 
 the one of the ${NS}$ block only for the flip of the sign of the parameter $V$.}
\begin{align} \label{eq:basic_moves}
	Z_{NS}^{(N)}(\vec{X},\vec{Y},\tau,\D,-V) = & \int \prod_{a=1}^2 \big( d\vec{Z}^{(a)}_N \D_N(\vec{Z}^{(a)},\tau) \big) Z_{S}^{(N)}(\vec{X},\vec{Z}^{(1)},\tau) \nn \\
	& Z_{D5}^{(N)} (\vec{Z}^{(1)},\vec{Z}^{(2)},\tau,\D,V)  Z_{S^{-1}}^{(N)}(\vec{Z}^{(2)},\vec{Y},\tau) \,, \\
	Z_{D5}^{(N)} (\vec{X},\vec{Y},\tau,\D,V) = & \prod_{a=1}^2 \big( d\vec{Z}^{(a)}_N \D_N(\vec{Z}^{(a)},\tau) \big) Z_{S}^{(N)}(\vec{X},\vec{Z}^{(1)},\tau) \nn \\
	& Z_{NS}^{(N)}(\vec{Z}^{(1)},\vec{Z}^{(2)},\tau,\D,V) Z_{S^{-1}}^{(N)}(\vec{Z}^{(2)},\vec{Y},\tau) \,.
\end{align}
It is useful to regard  the matter blocks and the  $\CS$ generator as  matrices with two indexes $\vec{X}$ and $\vec{Y}$ for their two $U(N)$ symmetries. 
Multiplying these matrices corresponds to gauging $U(N)$ symmetries using the integration measure $\D_N(\vec{Z},\tau)$, defined in equation \eqref{adjme} of appendix \ref{inpaconv}, containing  both the contribution of a $\CN=2$ vector multiplet and an extra adjoint chiral with +1 charge under a $U(1)_\tau$ symmetry. 
Notice that  the $U(1)_\tau$ symmetries in the matter blocks and in the $\CS$-duality walls are all identified, this is because when we gauge $U(N)$ nodes
we always turn on $\CW_{\text{gluing}}$.

Focusing on the first duality only, notice that on the r.h.s. the $U(1)_V$ symmetry can be reabsorbed by a $U(1)$ gauge transformation and therefore it acts trivially on the theory. In fact, on the l.h.s. the $V$ parameter appears just as a background FI term and therefore it is not associated to any symmetry acting on the theory. This feature will recur many times, we find useful to give the dualities writing explicitly also the redundant parameters because they become physical when the duality is used as a local dualization inside a bigger theory. \\

So far we have only considered improved bifundamentals with $U(N)\times U(N)$ non-abelian symmetry. To describe more general theories, corresponding to  brane setups with non-constant number of $D3$ branes,  we would need an improved bifundamentals with  $U(N)\times U(M)$ non-abelian symmetry and  its $\CS$-dual which we do not have at the moment. We plan to focus on this generalization in future works.

In this work we will only need the $M=0$ case. The $U(N)\times U(0)$  bifundamental is a trivial theory consisting only of a background FI term for a $U(N)$ global symmetry, its $\CS$-dualization to a trivial flavor block acted by a trivial $\CS$-wall on its left and an asymmetric $\CS^{-1}$-wall on its right
is shown  in figure \ref{fig:3d_asymm_basicmove}. 
The definition of the asymmetric $\CS$-wall is given  in appendix \ref{app:FT}. 
\begin{figure}
\centering
\begin{tikzpicture}[thick,node distance=3cm,gauge/.style={circle,draw,minimum size=5mm},flavor/.style={rectangle,draw,minimum size=5mm}]
	
	\path (0,0) node[flavor] (g1) {$\!0\!$} -- (1.5,0) node[flavor] (g2) {$\!N\!$} -- (2.75,0) node {$\Longleftrightarrow$};
	
	\draw[-, shorten >= 4, shorten <= 10, shift={(-0.1,0.05)}, mid arrowsm] (0,0) -- (1.5,0); 
	\draw[-, shorten >= 4, shorten <= 10, shift={(0.1,-0.05)}, mid arrowsm] (1.5,0) -- (0,0);
	 
	\draw[blue] (1.5,0.5) node {\scriptsize{$\vec{X}$}}; \draw[blue] (1.5,-0.5) node {\scriptsize{$(-V)$}};
	
	\path (4,0) node[flavor] (f1) {$\!0\!$} -- (5.5,0) node[gauge] (g) {$\!\!\!0\!\!\!$} -- (7,0) node[flavor] (f2) {$\!N\!$} 
		--	(5.5,1.5) node[flavor] (x1) {$\!1\!$};
	
	\wigT (f1) -- (g); \draw (4.75,-0.3) node {$+$};
	\wigT (g) -- (f2); \draw (6.25,-0.3) node {$-$};
	\wigT (x1) -- (6.25,0);
	
	\draw[-, shorten >= 6, shorten <= 9, shift={(-0.07,-0.05)}, mid arrowsm] (5.5,0) -- (5.5,1.5);
	\draw[-, shorten >= 6, shorten <= 9, shift={(0.07,0.05)}, mid arrowsm] (5.5,1.5) -- (5.5,0);
	\draw[blue] (5.5,2) node {\scriptsize{$V$}};
	
	\draw (g) to[out=-60,in=0] (5.5,-0.6) to[out=180,in=-120] (g);
	\draw[blue] (7,0.5) node {\scriptsize{$\vec{X}$}}; \draw (7.5,0) node[right] {$\times \text{ singlets}$};
	
\end{tikzpicture}
\caption{Asymmetric basic duality move relating a trivial $U(N)\times U(0)$ bifundamental  to trivial flavor block.}
\label{fig:3d_asymm_basicmove}
\end{figure}
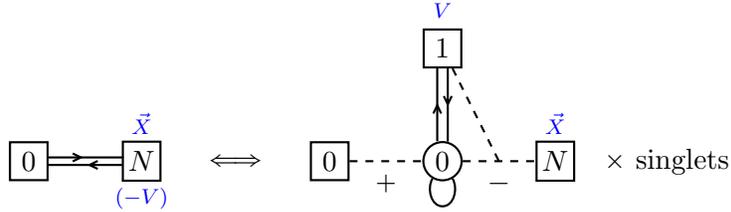
This duality move  corresponds to the dualization of a $D5$ brane into an $NS$ brane, with 0 $D3$ branes  on the left and $N$ on the right.
The  partition function identity associated to this duality is given by:
\begin{align}\label{eq:basic_moves_asymm}
	e^{-2 \pi i V \sum_{j=1}^N X_j} = Z_{\CS^{-1}}^{(N)}\big( \vec{X},\{ \frac{N-1}{2}\tau + V, \ldots, \frac{1-N}{2}\tau + V \}, \tau \big) 
	\prod_{j=2}^N s_b( \frac{iQ}{2} - j \tau) \,.
\end{align}
Where the partition function of the trivial $\CS$-wall and of the trivial matter is equal to one.


\subsubsection*{Useful combined moves}
It is convenient to also define some combined  duality moves that are not fundamental and are obtained by composing  several basic moves. This corresponds to the idea of acting on a set of many 5-branes at the same time, instead of acting on a single one.

For example, it can be useful to consider the $\CS$-dualization of a block of $F$ $\CN=2$ flavors as schematically shown  in figure \ref{fig:3d_flavors_basicmove}.
\begin{figure}
\centering
\resizebox{.95\hsize}{!}{
\begin{tikzpicture}[thick,node distance=3cm,gauge/.style={circle,draw,minimum size=5mm},flavor/.style={rectangle,draw,minimum size=5mm}]
	
	\path (0,0) node[flavor] (g) {$\!N\!$} -- (-0.5,1.5) node[flavor] (f1) {$\!F\!$} -- (0.5,1.5) node[flavor] (f2) {$\!F\!$} 
		-- (2,0) node {$=$};
	
	\chir (g) -- (f1);
	\chir (f2) -- (g);
	\draw (0.2,0) node[right] {\large{${}_{\vec{X}}\mathbb{I}_{\vec{Y}} (\tau)$}};

\begin{scope}[shift={(0.5,0)}]

	\path (3.5,0) node[flavor] (g) {$\!N\!$} -- (2.75,1.5) node[flavor] (f1) {$\!1\!$} -- (4.25,1.5) node[flavor] (f2) {$\!1\!$}
		-- (5.5,0) node {$=$};

	\draw[-, shorten >= 8, shorten <= 9, shift={(-0.07,0.02)}, mid arrowsm] (2.75,1.5) -- (3.5,0);
	\draw[-, shorten >= 8, shorten <= 8, shift={(0.1,0)}, mid arrowsm] (3.5,0) -- (2.75,1.5); 
	\draw[blue] (3.2,0.65) node[left] {\scriptsize{$1-B_1$}}; \draw[blue] (2.75,2) node {\scriptsize{$V_1$}};
	
	\draw[-, shorten >= 8, shorten <= 7.5, shift={(-0.1,0.02)}, mid arrowsm] (3.5,0) -- (4.25,1.5);
	\draw[-, shorten >= 8, shorten <= 8, shift={(0.05,0)}, mid arrowsm] (4.25,1.5) -- (3.5,0); 
	\draw[blue] (3.85,0.65) node[right] {\scriptsize{$1-B_F$}}; \draw[blue] (4.25,2) node {\scriptsize{$V_F$}};
	
	\draw (3.5,1.5) node {$\cdots$};
	
	\draw (3.7,0) node[right] {\large{${}_{\vec{X}}\mathbb{I}_{\vec{Y}} (\tau)$}};
	
\end{scope}
	
\begin{scope}[shift={(0.5,0)}]
	\path (6.5,0) node[flavor] (f1) {$\!N\!$} -- (6.5,1.5) node[flavor] (x1) {$\!1\!$};
	
	\draw[-, shorten >= 6.5, shorten <= 9.5, shift={(-0.07,-0.05)}, mid arrowsm] (6.5,0) -- (6.5,1.5);
	\draw[-, shorten >= 6.5, shorten <= 9, shift={(0.07,0.05)}, mid arrowsm] (6.5,1.5) -- (6.5,0);
	\draw[blue] (6.6,0.75) node[right] {\scriptsize{$1-B_1$}}; \draw[blue] (6.5,2) node {\scriptsize{$V_1$}};
	\draw (6.7,0) node[right] {\large{${}_{\vec{X}}\mathbb{I}_{\vec{Z}_2} (\tau)$}};

\end{scope}

\begin{scope}[shift={(3,0)}]
	\path (6.5,0) node[flavor] (f1) {$\!N\!$} -- (6.5,1.5) node[flavor] (x1) {$\!1\!$};
	
	\draw[-, shorten >= 6.5, shorten <= 9.5, shift={(-0.07,-0.05)}, mid arrowsm] (6.5,0) -- (6.5,1.5);
	\draw[-, shorten >= 6.5, shorten <= 9, shift={(0.07,0.05)}, mid arrowsm] (6.5,1.5) -- (6.5,0);
	\draw[blue] (6.6,0.75) node[right] {\scriptsize{$1-B_2$}}; \draw[blue] (6.5,2) node {\scriptsize{$V_2$}};
	\draw (6.7,0) node[right] {\large{${}_{\vec{Z}_2}\mathbb{I}_{\vec{Z}_3} (\tau)$}};

\end{scope}

	\draw (11.75,0) node {$\cdots$};

\begin{scope}[shift={(6,0)}]
	\path (6.5,0) node[flavor] (f1) {$\!N\!$} -- (6.5,1.5) node[flavor] (x1) {$\!1\!$};
	
	\draw[-, shorten >= 6.5, shorten <= 9.5, shift={(-0.07,-0.05)}, mid arrowsm] (6.5,0) -- (6.5,1.5);
	\draw[-, shorten >= 6.5, shorten <= 9, shift={(0.07,0.05)}, mid arrowsm] (6.5,1.5) -- (6.5,0);
	\draw[blue] (6.6,0.75) node[right] {\scriptsize{$1-B_F$}}; \draw[blue] (6.5,2) node {\scriptsize{$V_F$}};
	\draw (6.7,0) node[right] {\large{${}_{\vec{Z}_F}\mathbb{I}_{\vec{Y}} (\tau)$}};

\end{scope}

	\draw (14.75,0) node {$=$};
	
	\path (1,-2) node[flavor] (f1) {$\!N\!$} -- (2.5,-2) node[gauge] (g1) {$\!\!\!N\!\!\!$} -- (4,-2) node[gauge] (g2) {$\!\!\!N\!\!\!$} --
			(5.5,-2) node[gauge] (g3) {$\!\!\!N\!\!\!$} -- (7,-2) node[gauge] (g4) {$\!\!\!N\!\!\!$} -- (8.5,-2) node[gauge] (g5) {$\!\!\!N\!\!\!$} -- 
			(10.5,-2) node[gauge] (g6) {$\!\!\!N\!\!\!$} -- (12,-2) node[gauge] (g7) {$\!\!\!N\!\!\!$} --
			(13.5,-2) node[flavor] (f2) {$\!N\!$} -- (0,-2) node {$=$} -- (14.5,-2) node {$=$};
	
	\wigT (f1) -- (g1); \draw (1.75,-1.7) node {$+$};
	\draw (g1) to[out=60,in=0] (2.5,-1.4) to[out=180,in=120] (g1); \draw[blue] (2.5,-1.2) node {\scriptsize{$\tau$}};
	\wigM (g1) -- (g2); \draw[blue] (3.25,-1.6) node {\scriptsize{$B_1$}};
	\draw (g2) to[out=60,in=0] (4,-1.4) to[out=180,in=120] (g2); \draw[blue] (4,-1.2) node {\scriptsize{$\tau$}};
	\wigT (g2) -- (g3); \draw (4.75,-1.7) node {$-$}; 
	\draw (g3) to[out=60,in=0] (5.5,-1.4) to[out=180,in=120] (g3); \draw[blue] (5.5,-1.2) node {\scriptsize{$\tau$}};
	\wigT (g3) -- (g4); \draw (6.25,-1.7) node {$+$};
	\draw (g4) to[out=60,in=0] (7,-1.4) to[out=180,in=120] (g4); \draw[blue] (7,-1.2) node {\scriptsize{$\tau$}};
	\wigM (g4) -- (g5); \draw[blue] (7.75,-1.6) node {\scriptsize{$B_2$}};
	\draw (g5) to[out=60,in=0] (8.5,-1.4) to[out=180,in=120] (g5); \draw[blue] (8.5,-1.2) node {\scriptsize{$\tau$}};
	\wigT (g5) -- (9.15,-2);
	\draw (9.5,-2) node {$\cdots$};
	\wigT (9.85,-2) -- (g6);
	\draw (g6) to[out=60,in=0] (10.5,-1.4) to[out=180,in=120] (g6); \draw[blue] (10.5,-1.2) node {\scriptsize{$\tau$}};
	\wigM (g6) -- (g7); \draw[blue] (11.25,-1.6) node {\scriptsize{$B_F$}};
	\draw (g7) to[out=60,in=0] (12,-1.4) to[out=180,in=120] (g7); \draw[blue] (12,-1.2) node {\scriptsize{$\tau$}};
	\wigT (g7) -- (f2); \draw (12.75,-1.7) node {$-$};
	\draw[blue] (1,-1.5) node {\scriptsize{$\vec{X}$}}; \draw[blue] (13.5,-1.5) node {\scriptsize{$\vec{Y}$}};
	\draw[blue] (2.5,-2.5) node {\scriptsize{$(V_1)$}}; \draw[blue] (4,-2.5) node {\scriptsize{$(-V_1)$}};
	\draw[blue] (7,-2.5) node {\scriptsize{$(V_2)$}}; \draw[blue] (8.5,-2.5) node {\scriptsize{$(-V_2)$}};
	\draw[blue] (10.5,-2.5) node {\scriptsize{$(V_F)$}}; \draw[blue] (12,-2.5) node {\scriptsize{$(-V_F)$}};
	
	\path (1,-4) node[flavor] (f1) {$\!N\!$} -- (2.5,-4) node[gauge] (g1) {$\!\!\!N\!\!\!$} -- (4,-4) node[gauge] (g2) {$\!\!\!N\!\!\!$} --
			(5.5,-4) node[gauge] (g3) {$\!\!\!N\!\!\!$} -- (7.5,-4) node[gauge] (g4) {$\!\!\!N\!\!\!$} -- (9,-4) node[gauge] (g5) {$\!\!\!N\!\!\!$} -- 
			(10.5,-4) node[flavor] (f2) {$\!N\!$} -- (0,-4) node {$=$};
			
	\wigT (f1) -- (g1); \draw (1.75,-3.7) node {$+$};
	\draw (g1) to[out=60,in=0] (2.5,-3.4) to[out=180,in=120] (g1); \draw[blue] (2.5,-3.2) node {\scriptsize{$\tau$}};
	\wigM (g1) -- (g2); \draw[blue] (3.25,-3.6) node {\scriptsize{$B_1$}};
	\draw (g2) to[out=60,in=0] (4,-3.4) to[out=180,in=120] (g2); \draw[blue] (4,-3.2) node {\scriptsize{$\tau$}};
	\wigM (g2) -- (g3); \draw[blue] (4.75,-3.6) node {\scriptsize{$B_2$}};
	\draw (g3) to[out=60,in=0] (5.5,-3.4) to[out=180,in=120] (g3); \draw[blue] (5.5,-3.2) node {\scriptsize{$\tau$}};
	\wigM (g3) -- (6.25,-4);
	\draw (6.5,-4) node {$\cdots$}; 
	\wigM (g4) -- (6.75,-4);
	\draw (g4) to[out=60,in=0] (7.5,-3.4) to[out=180,in=120] (g4); \draw[blue] (7.5,-3.2) node {\scriptsize{$\tau$}};
	\wigM (g4) -- (g5); \draw[blue] (8.25,-3.6) node {\scriptsize{$B_F$}};
	\draw (g5) to[out=60,in=0] (9,-3.4) to[out=180,in=120] (g5); \draw[blue] (9,-3.2) node {\scriptsize{$\tau$}};
	\wigT (g5) -- (f2); \draw (9.75,-3.7) node {$-$};
	\draw[blue] (1,-3.5) node {\scriptsize{$\vec{X}$}}; \draw[blue] (10.5,-3.5) node {\scriptsize{$\vec{Y}$}};
	\draw[blue] (2.5,-4.5) node {\scriptsize{$(V_1)$}}; \draw[blue] (4,-4.5) node {\scriptsize{$(V_2-V_1)$}};
	\draw[blue] (5.5,-4.5) node {\scriptsize{$(V_3-V_2)$}}; \draw[blue] (7.5,-4.5) node {\scriptsize{$(V_F-V_{F-1})$}};
	 \draw[blue] (9,-4.5) node {\scriptsize{$(-V_F)$}};
	
\end{tikzpicture}}
\caption{$\CS$-dualization of a block of $F$ $\CN=2$ flavors. In the first step we reparameterize  the  $U(F)\times U(F)$ flavors 
as a set of $F$ fundamental anti-fundamental pairs of flavors. In the second step we cut the block of $F$ flavours into $F$ generalized flavor  blocks. We then dualize each block to an improved bifundamental block and glue together the results to reach the theory in the second line. Implementing the fusion to identity 
$\CS \CS^{-1}=1$ we reach the final frame which is given by a string of $F$ improved bifundamentals with an $\CS$-wall on the left  and an $\CS^{-1}$-wall on the right.}
\label{fig:3d_flavors_basicmove}
\end{figure}
Similarly it can be useful to dualize a string of consecutive  improved bifundamentals \ref{fig:3d_bifunds_basicmove}. This second move can be obtained starting from the duality \ref{fig:3d_flavors_basicmove} by acting on the left and right with a $\CS$ and $\CS^{-1}$ operators and using the fact that $\CS \CS^{-1}=1$.
\begin{figure}
\centering
\begin{tikzpicture}[thick,node distance=3cm,gauge/.style={circle,draw,minimum size=5mm},flavor/.style={rectangle,draw,minimum size=5mm}]
	
	\path (0,0) node[flavor] (g1) {$\!N\!$} -- (1.5,0) node[gauge] (g2) {$\!\!\!N\!\!\!$} -- (3,0) node (gi) {$\ldots$} 
		-- (4.5,0) node[gauge] (g3) {$\!\!\!N\!\!\!$} -- (6,0) node[flavor] (g4) {$\!N\!$};
	
	\wigM (g1) -- (g2); \draw[blue] (0.75,0.4) node {\scriptsize{$B_1$}};
	\wigM (g2) -- (gi);
	\wigM (g3) -- (gi);
	\wigM (g4) -- (g3); \draw[blue] (5.25,0.4) node {\scriptsize{$B_F$}};
	\draw[-] (g2) to[out=60,in=0] (1.5,0.6) to[out=180,in=120] (g2); \draw[blue] (1.5,0.8) node {\scriptsize{$\tau$}};
	\draw[-] (g3) to[out=60,in=0] (4.5,0.6) to[out=180,in=120] (g3); \draw[blue] (4.5,0.8) node {\scriptsize{$\tau$}};
	\draw[blue] (0,0.5) node {\scriptsize{$\vec{X}$}}; \draw[blue] (6,0.5) node {\scriptsize{$\vec{Y}$}};
	\draw[blue] (0,-0.5) node {\scriptsize{$(-V_1)$}}; \draw[blue] (1.5,-0.5) node {\scriptsize{$(V_1-V_2)$}};
	\draw[blue] (4.5,-0.5) node {\scriptsize{$(V_{F-1}-V_F)$}}; \draw[blue] (6,-0.5) node {\scriptsize{$(V_F)$}};
	
	\path (7,0) node {$=$} -- (8,0) node[flavor] (g1) {$\!N\!$} -- (9.5,0) node[gauge] (g2) {$\!\!\!N\!\!\!$} 
		-- (11,0) node[flavor] (g3) {$\!N\!$}
		-- (8.75,1.5) node[flavor] (x1) {$\!1\!$} -- (10.25,1.5) node[flavor] (x2) {$\!1\!$};
		
	\wigT (g1) -- (g2); \draw (8.75,-0.3) node {$+$};
	\wigT (g2) -- (g3); \draw (10.25,-0.3) node {$-$};
	
	\draw[-, shorten >= 6, shorten <= 9, shift={(-0.07,0.02)}, mid arrowsm] (8.75,1.5) -- (9.5,0);
	\draw[-, shorten >= 8, shorten <= 9, shift={(0.1,0)}, mid arrowsm] (9.5,0) -- (8.75,1.5); 
	\draw[blue] (9,0.9) node[left] {\scriptsize{$1-B_1$}}; \draw[blue] (8.75,2) node {\scriptsize{$V_1$}};
	
	\draw[-, shorten >= 8, shorten <= 8, shift={(-0.1,0.02)}, mid arrowsm] (9.5,0) -- (10.25,1.5);
	\draw[-, shorten >= 7, shorten <= 8, shift={(0.05,0)}, mid arrowsm] (10.25,1.5) -- (9.5,0); 
	\draw[blue] (10,0.9) node[right] {\scriptsize{$1-B_F$}}; \draw[blue] (10.25,2) node {\scriptsize{$V_F$}};
	
	\draw[blue] (8,0.5) node {\scriptsize{$\vec{X}$}}; \draw[blue] (11,0.5) node {\scriptsize{$\vec{Y}$}};
	
	\draw (9.5,1.5) node {$\cdots$};

	\draw[-] (g2) to[out=-60,in=0] (9.5,-0.6) to[out=180,in=-120] (g2); \draw[blue] (9.5,-0.8) node {\scriptsize{$\tau$}};
	
\end{tikzpicture}
\caption{$\CS$-dualization of a block of $F$ improved bifundamentals.}
\label{fig:3d_bifunds_basicmove}
\end{figure}
As partition function identities, the two combined duality moves corresponds to:
\begin{align} \label{eq:basic_moves_F}
	 Z_{F-D5}^{(N)} (\vec{X},\vec{Y},\tau,\vec{B},\vec{V}) = 	
	 &\int \prod_{a=1}^{F+1} \big( d\vec{Z}^{(a)}_N \D_N(\vec{Z}^{(a)},\tau) \big) Z_{\CS}^{(N)}(\vec{X},\vec{Z}^{(1)},\tau) \nn \\
	 & \prod_{a=1}^F Z_{NS}^{(N)}(\vec{Z}^{(a)},\vec{Z}^{(a+1)},\tau,B_a,V_a) Z_{\CS^{-1}}^{(N)}(\vec{Z}^{(F+1)},\vec{Y},\tau) \,,
\end{align}
\begin{align}
	& \int \prod_{a=1}^{F-1} \big( d\vec{Z}^{(a)}_N \D_N(\vec{Z}^{(a)},\tau) \big) Z_{NS}^{(N)}(\vec{X},\vec{Z}^{(1)},\tau,B_1,-V_1) \nn \\
	& \prod_{a=2}^{F-1} Z_{NS}^{(N)}(\vec{Z}^{(a-1)},\vec{Z}^{(a)},\tau,B_a,-V_a) Z_{NS}(\vec{Z}^{(F-1)},\vec{Y},\tau,B_F,-V_F)  = \nn \\
	&\qquad = \int \prod_{a=1}^{2} \big( d\vec{Z}^{(a)}_N \D_N(\vec{Z}^{(a)},\tau) \big) Z_{\CS}^{(N)}(\vec{X},\vec{Z}^{(1)},\tau) 
	Z_{F-D5}^{(N)}(\vec{Z}^{(1)},\vec{Z}^{(2)},\tau,\vec{B},\vec{V}) Z_{\CS^{-1}}^{(N)}(\vec{Z}^{(2)},\vec{Y},\tau) \,,
\end{align}
with
\begin{align}\label{eq:flavorsblock_def}
	Z_{F-D5}^{(N)} (\vec{X},\vec{Y},\tau, \vec{B}, \vec{V}) = \prod_{j=1}^{N} \prod_{a=1}^F s_b(B_a \pm (X_j - V_a)) {}_{\vec{X}}\mathbb{I}_{\vec{Y}}(\tau) \,.
\end{align}

\subsection{$\CN=2$ dualization algorithm}\label{algsec}
Now that we have introduced all the necessary ingredients we are ready to present the dualization algorithm. This consists in the following steps:
\begin{itemize}
	\item Ungauge the gauge nodes to cut the quiver theory into QFT matter blocks that can be either improved bifundamental or generalized flavor blocks.
	\item Dualize each block using the two basic duality moves.
	\item Glue back  all the dualized blocks implementing the fusion to identity $\CS \CS^{-1}=1$.
	\item If some operator has acquired a VEV, follow the RG flow triggered by this VEV.  
\end{itemize}
To illustrate this procedure   we will  now implement the algorithm  to derive the mirror dual of the  adjoint SQCD. Another example is given in appendix \ref{app:Quiver_algorithm}.\\

\subsection{Dualization of the $U(N)$ adjoint SQCD}\label{alsqcd}

We start by taking the SCQD parameterized as  in figure \ref{fig:SQCD_Manifest}, we ungauge the $U(N)$ gauge group and chop  the theory into a block of $F$ generalized flavors and two (trivial) bifundamental blocks:
\be\label{fig:step1}
\begin{tikzpicture}[thick,node distance=3cm,gauge/.style={circle,draw,minimum size=5mm},flavor/.style={rectangle,draw,minimum size=5mm}]
	
	\path (0,0) node[gauge] (g) {$\!\!\!N\!\!\!$} -- (-0.75,1.5) node[flavor] (f1) {$\!1\!$} -- (0.75,1.5) node[flavor] (f2) {$\!\!1\!\!$} 
	-- (1.75,0) node {$=$};
	
	\draw[-, shorten >= 6, shorten <= 9, shift={(-0.07,0.02)}, mid arrowsm] (-0.75,1.5) -- (0,0);
	\draw[-, shorten >= 8, shorten <= 9, shift={(0.1,0)}, mid arrowsm] (0,0) -- (-0.75,1.5); 
	\draw[blue] (-0.5,0.9) node[left] {\scriptsize{$1-B_1$}}; \draw[blue] (-0.75,2) node {\scriptsize{$X_1$}};
	
	\draw[-, shorten >= 8, shorten <= 8, shift={(-0.1,0.02)}, mid arrowsm] (0,0) -- (0.75,1.5);
	\draw[-, shorten >= 7, shorten <= 8, shift={(0.05,0)}, mid arrowsm] (0.75,1.5) -- (0,0); 
	\draw[blue] (0.5,0.9) node[right] {\scriptsize{$1-B_F$}}; \draw[blue] (0.75,2) node {\scriptsize{$X_F$}};
	
	\draw (0,1.5) node {$\ldots$};
	
	\draw (g) to[out=150,in=90] (-0.6,0) to[out=-90,in=-150] (g); \draw[blue] (-0.6,0) node[left] {\scriptsize{$\tau$}};
	\draw[blue] (0,-0.5) node[left] {\scriptsize{$(Y_1-Y_2)$}};
	
	\path (3,0) node[flavor] (f1) {$\!0\!$} -- (4.5,0) node[flavor] (f2) {$\!N\!$} 
		-- (6,0) node[flavor] (g) {$\!N\!$} -- (5.25,1.5) node[flavor] (y1) {$\!1\!$} -- (6.75,1.5) node[flavor] (y2) {$\!1\!$}
		-- (8.5,0) node[flavor] (f3) {$\!N\!$} -- (10,0) node[flavor] (f4) {$\!0\!$};
	
	\draw[-, shorten >= 4, shorten <= 10, shift={(-0.1,0.05)}, mid arrowsm] (3,0) -- (4.5,0); 
	\draw[-, shorten >= 4, shorten <= 10, shift={(0.1,-0.05)}, mid arrowsm] (4.5,0) -- (3,0);
	\draw[blue] (4.5,0.5) node {\scriptsize{$\vec{Z}$}}; \draw[blue] (4.5,-0.5) node {\scriptsize{$(Y_1)$}};
	
	\draw[-, shorten >= 8, shorten <= 9, shift={(-0.07,0.02)}, mid arrowsm] (5.25,1.5) -- (6,0);
	\draw[-, shorten >= 8, shorten <= 8, shift={(0.1,0)}, mid arrowsm] (6,0) -- (5.25,1.5); 
	\draw[blue] (5.5,0.95) node[left] {\scriptsize{$1-B_1$}}; \draw[blue] (5.25,2) node {\scriptsize{$X_1$}};
	
	\draw[-, shorten >= 8, shorten <= 8, shift={(-0.1,0.02)}, mid arrowsm] (6,0) -- (6.75,1.5);
	\draw[-, shorten >= 8, shorten <= 8, shift={(0.05,0)}, mid arrowsm] (6.75,1.5) -- (6,0); 
	\draw[blue] (6.55,0.95) node[right] {\scriptsize{$1-B_F$}}; \draw[blue] (6.75,2) node {\scriptsize{$X_1$}};
	
	\draw (6,1.5) node {$\ldots$};
	
	\draw (6.2,0) node[right] {\large{${}_{\vec{Z}}\mathbb{I}_{\vec{W}} (\tau)$}};
	
	\draw[-, shorten >= 4, shorten <= 10, shift={(-0.1,0.05)}, mid arrowsm] (8.5,0) -- (10,0); 
	\draw[-, shorten >= 4, shorten <= 10, shift={(0.1,-0.05)}, mid arrowsm] (10,0) -- (8.5,0);
	\draw[blue] (8.5,0.5) node {\scriptsize{$\vec{W}$}}; \draw[blue] (8.5,-0.5) node {\scriptsize{$(-Y_2)$}};
	
\end{tikzpicture}\ee
For later convenience we have redefined the FI parameter of \ref{fig:SQCD_Manifest} as $Y\to Y_1-Y_2$.
At the level of the partition function this first step consist in the following rewriting:
\begin{align}\label{eq:step_1}
	Z_{\text{SQCD}}=&\int d\vec{Z}_N \D_N(\vec{Z},\tau) e^{2 \pi i (Y_1-Y_2) \sum_{j=1}^N Z_j} \prod_{j=1}^N \prod_{a=1}^F s_b( B_a \pm (Z_j - X_a) ) = \nn \\
	= & \int d\vec{Z}_N \D_N(\vec{Z},\tau) d\vec{W}_N \D_N(\vec{W},\tau) e^{2 \pi i Y_1 \sum_{j=1}^N Z_j}  \nn \\
	& Z_{F-D5}^{(N)}(\vec{Z},\vec{W},\tau,\vec{B},\vec{V}) 
	e^{-2 \pi i Y_2 \sum_{j=1}^N W_j} \,=Z_{\text{step 1}} \,,
\end{align}
where we have isolated the contributions of the two trivial bifundamental blocks corresponding to the  FI couplings and the $F$-flavors  block. The identity between the first and second line follows from the fact that the identity operator ${}_{\vec{Z}}\mathbb{I}_{\vec{W}}(\tau)$, contained in the definition of the $F$-flavor block \eqref{eq:flavorsblock_def}, behaves as a delta function identifying the $\vec{Z}$ and $\vec{W}$ parameters and is normalized as:
\begin{align}
	\int d\vec{W}_N \D_N(\vec{W},\tau) {}_{\vec{Z}}\mathbb{I}_{\vec{W}}(\tau) = 1 \,.
\end{align}

We use the combined duality move in \ref{fig:3d_flavors_basicmove} to dualize the generalized QFT blocks. We also use the asymmetric duality \ref{fig:3d_asymm_basicmove} to dualize the trivial bifundamental blocks. Then we glue back all the dualized blocks:
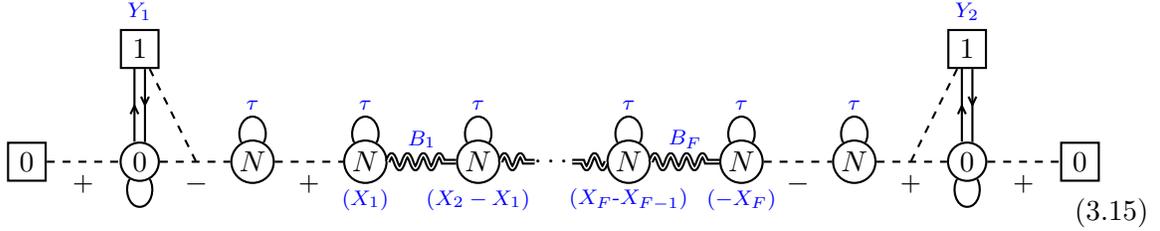
\be\label{fig:step2}
\begin{tikzpicture}[thick,node distance=3cm,gauge/.style={circle,draw,minimum size=5mm},flavor/.style={rectangle,draw,minimum size=5mm}]
	
	\path (0,0) node[flavor] (f1) {$\!0\!$} -- (1.5,0) node[gauge] (g1) {$\!\!\!0\!\!\!$} -- (1.5,1.5) node[flavor] (x1) {$\!1\!$} --
			(3,0) node[gauge] (g2) {$\!\!\!N\!\!\!$} -- (4.5,0) node[gauge] (g3) {$\!\!\!N\!\!\!$} -- (6,0) node[gauge] (g4) {$\!\!\!N\!\!\!$} -- 
			(8,0) node[gauge] (g5) {$\!\!\!N\!\!\!$} -- (9.5,0) node[gauge] (g6) {$\!\!\!N\!\!\!$} -- (11,0) node[gauge] (g7) {$\!\!\!N\!\!\!$} --
			(12.5,0) node[gauge] (g8) {$\!\!\!0\!\!\!$} -- (12.5,1.5) node[flavor] (x2) {$\!1\!$} -- (14,0) node[flavor] (f2) {$\!0\!$};
						
	\wigT (f1) -- (g1); \draw (0.75,-0.3) node {$+$};
	
	\draw[-, shorten >= 6, shorten <= 9, shift={(-0.07,-0.05)}, mid arrowsm] (1.5,0) -- (1.5,1.5);
	\draw[-, shorten >= 6, shorten <= 9, shift={(0.07,0.05)}, mid arrowsm] (1.5,1.5) -- (1.5,0);
	\draw[blue] (1.5,2) node {\scriptsize{$Y_1$}};
	
	\wigT (g1) -- (g2); \draw (2.25,-0.3) node {$-$};
	\wigT (x1) -- (2.25,0); 
	\wigT (g2) -- (g3); \draw (3.75,-0.3) node {$+$};
	\wigM (g3) -- (g4); \draw[blue] (5.25,0.3) node {\scriptsize{$B_1$}};
	\wigM (g4) -- (6.75,0);
	\draw (7,0) node {$\cdots$};
	\wigM (g5) -- (7.25,0);
	\wigM (g5) -- (g6); \draw[blue] (8.75,0.3) node {\scriptsize{$B_F$}};
	\wigT (g6) -- (g7); \draw (10.25,-0.3) node {$-$};
	\wigT (g7) -- (g8); \draw (11.75,-0.3) node {$+$};
	\wigT (x2) -- (11.75,0);
	
	\draw[-, shorten >= 6, shorten <= 9, shift={(-0.07,-0.05)}, mid arrowsm] (12.5,0) -- (12.5,1.5);
	\draw[-, shorten >= 6, shorten <= 9, shift={(0.07,0.05)}, mid arrowsm] (12.5,1.5) -- (12.5,0);
	\draw[blue] (12.5,2) node {\scriptsize{$Y_2$}};
	
	\wigT (g8) -- (f2); \draw (13.25,-0.3) node {$+$};
	\draw (g1) to[out=-60,in=0] (1.5,-0.6) to[out=180,in=-120] (g1);
	\draw (g2) to[out=60,in=0] (3,0.6) to[out=180,in=120] (g2); \draw[blue] (3,0.75) node {\scriptsize{$\tau$}};
	\draw (g3) to[out=60,in=0] (4.5,0.6) to[out=180,in=120] (g3); \draw[blue] (4.5,0.75) node {\scriptsize{$\tau$}};
	\draw (g4) to[out=60,in=0] (6,0.6) to[out=180,in=120] (g4); \draw[blue] (6,0.75) node {\scriptsize{$\tau$}};
	\draw (g5) to[out=60,in=0] (8,0.6) to[out=180,in=120] (g5); \draw[blue] (8,0.75) node {\scriptsize{$\tau$}};
	\draw (g6) to[out=60,in=0] (9.5,0.6) to[out=180,in=120] (g6); \draw[blue] (9.5,0.75) node {\scriptsize{$\tau$}};
	\draw (g7) to[out=60,in=0] (11,0.6) to[out=180,in=120] (g7); \draw[blue] (11,0.75) node {\scriptsize{$\tau$}};
	\draw (g8) to[out=-60,in=0] (12.5,-0.6) to[out=180,in=-120] (g8);
	
	\draw[blue] (4.5,-0.5) node {\scriptsize{$(X_1)$}}; \draw[blue] (6,-0.5) node {\scriptsize{$(X_2-X_1)$}};
	\draw[blue] (8,-0.5) node {\scriptsize{$(X_F$-$X_{F-1})$}}; \draw[blue] (9.5,-0.5) node {\scriptsize{$(-X_F)$}};
	
\end{tikzpicture}\ee
To avoid cluttering, in the picture we have not included the singlets coming from the dualization of the trivial bifundamentals. This procedure corresponds to using the set of partition function identities \eqref{eq:basic_moves_F} and \eqref{eq:basic_moves_asymm} in the partition function \eqref{eq:step_1} to obtain:\footnote{Notice that the trivial $\CS$-walls on the l.h.s and on the r.h.s have trivial partition functions and we will  drop them in the next pictures.}
\begin{align}\label{eq:step_2}
	Z_{SQCD}=&Z_{\text{step 1}}= \prod_{j=2}^N s_b(\frac{iQ}{2} - j \tau)^2 \int \prod_{a=1}^{F+3} \big( d\vec{Z}_N^{(a)} \D_N(\vec{Z}^{(a)},\tau) \big) 
	 \nn \\
	& Z_{\CS^{-1}}^{(N)}(\{ \frac{N-1}{2}\tau + Y_1, \ldots, \frac{1-N}{2}\tau + Y_1 \},\vec{Z}^{(1)},\tau) Z_{\CS}^{(N)}(\vec{Z}^{(1)},\vec{Z}^{(2)},\tau) \nn \\
	& \prod_{a=1}^{F} Z_{NS}^{(N)}(\vec{Z}^{(a+1)},\vec{Z}^{(a+2)},\tau, B_a, X_a) 
	Z_{\CS^{-1}}^{(N)}(\vec{Z}^{(F+2)},\vec{Z}^{(F+3)},\tau)  \nn \\
	& Z_{\CS}^{(N)}(\vec{Z}^{(F+3)},\{ \frac{N-1}{2}\tau + Y_2, \ldots, \frac{1-N}{2}\tau + Y_2 \},\tau)=Z_{\text{step 2}} \,.
\end{align}
Where we  named as $\vec{Z}^{(a)}$ the Cartans of the $a$-th $U(N)$ gauge group. 
On the l.h.s and on the r.h.s of the quiver, 
the integration over the first and the last  $U(N)$ node  (over $\vec{Z}^{(a)}$  and $\vec{Z}^{(F+3)}$) fuse a  symmetric and an asymmetric $\CS$-wall
to an asymmetric $\mathbb{I}-wall$ \eqref{fig:FTdeltaasymm}. The effect of these asymmetric  $\mathbb{I}$-walls is in turn to 
deform the first and last improved bifundamentals into asymmetric improved bifundamentals defined in \eqref{fig:FM_asymm_symbol}, by
breaking the $U(N)$ symmetries to $U(1)$:
\be\label{fig:step3}
\begin{tikzpicture}[thick,node distance=3cm,gauge/.style={circle,draw,minimum size=5mm},flavor/.style={rectangle,draw,minimum size=5mm}]
	
	\path (0,1.5) node[flavor] (x1) {$\!1\!$} -- (0,0) node[flavor] (g2) {$\!0\!$} -- (2,0) node[gauge] (g3) {$\!\!\!N\!\!\!$} -- 
			(4,0) node[gauge] (g4) {$\!\!\!N\!\!\!$} -- (8,0) node[gauge] (g5) {$\!\!\!N\!\!\!$} -- (10,0) node[gauge] (g6) {$\!\!\!N\!\!\!$} -- 
			(12,0) node[flavor] (g7) {$\!0\!$} -- (12,1.5) node[flavor] (x2) {$\!1\!$};
						
	\wigM (x1) -- (1,0);  \draw[blue] (0,2) node {\scriptsize{$Y_1$}};
	\wigM (g2) -- (g3); \draw[blue] (1.4,0.3) node {\scriptsize{$B_1$}};
	\wigM (g3) -- (g4); \draw[blue] (3,0.3) node {\scriptsize{$B_2$}};
	\wigM (g4) -- (5.5,0);
	\draw (6,0) node {$\cdots$};
	\wigM (g5) -- (6.5,0);
	\wigM (g5) -- (g6); \draw[blue] (9,0.3) node {\scriptsize{$B_{F-1}$}};
	\wigM (g6) -- (g7); \draw[blue] (10.6,0.3) node {\scriptsize{$B_F$}};
	\wigM (x2) -- (11,0); \draw[blue] (12,2) node {\scriptsize{$Y_2$}};
	
	\draw (g3) to[out=60,in=0] (2,0.6) to[out=180,in=120] (g3); \draw[blue] (2,0.75) node {\scriptsize{$\tau$}};
	\draw (g4) to[out=60,in=0] (4,0.6) to[out=180,in=120] (g4); \draw[blue] (4,0.75) node {\scriptsize{$\tau$}};
	\draw (g5) to[out=60,in=0] (8,0.6) to[out=180,in=120] (g5); \draw[blue] (8,0.75) node {\scriptsize{$\tau$}};
	\draw (g6) to[out=60,in=0] (10,0.6) to[out=180,in=120] (g6); \draw[blue] (10,0.75) node {\scriptsize{$\tau$}};
	
	\draw[blue] (2,-0.5) node {\scriptsize{$(X_2-X_1)$}}; \draw[blue] (4,-0.5) node {\scriptsize{$(X_3-X_2)$}};
	\draw[blue] (8,-0.5) node {\scriptsize{$(X_{F-1}$-$X_{F-2})$}}; \draw[blue] (10,-0.5) node {\scriptsize{$(X_F$-$X_{F-1})$}};
	
\end{tikzpicture}\ee
At the level of partition functions the integral over $\vec{Z}^{(1)}$ and $\vec{Z}^{(F+3)}$ in \eqref{eq:step_2}
generating the asymmetric Identity-walls produces a set  of delta functions as explained in \eqref{FTasymm}. Implementing these delta functions  freezes the $\vec{Z}^{(2)}$ and $\vec{Z}^{(F+2)}$ Carans in terms of $\tau,Y_1,Y_2$ and we find:
\begin{align}
	Z_{\text{SQCD}}=&Z_{\text{step 1}}=Z_{\text{step 2}}= \prod_{j=2}^N s_b(\frac{iQ}{2} - j \tau)^2 \int \prod_{a=3}^{F+1} \big( d\vec{Z}_N^{(a)} \D_N(\vec{Z}^{(a)},\tau) \big) \nn \\
	& Z_{NS}^{(N)}(\{ \frac{N-1}{2}\tau + Y_1, \ldots, \frac{1-N}{2}\tau + Y_1 \}, \vec{Z}^{(3)},\tau,B_1,X_1) \nn \\
	& \prod_{a=2}^{F-1} Z_{NS}^{(N)}(\vec{Z}^{(a+1)},\vec{Z}^{(a+2)},\tau, B_a, X_a) \nn \\
	& Z_{NS}^{(N)}(\vec{Z}^{(F+2)},\{ \frac{N-1}{2}\tau + Y_2, \ldots, \frac{1-N}{2}\tau + Y_2 \},\tau,B_F, X_F) = Z_{\text{step 3}} \,.
\end{align}

Finally we can exploit the duality relating  a $U(N)\times U(1)$ asymmetric improved bifundamental  to a flipped flavor
discussed in \eqref{fig:FMtoFlav} to replace the asymmetric improved bifundamentals on the l.h.s and on the r.h.s to 
 land on the mirror dual theory:
\be
\begin{tikzpicture}[thick,node distance=3cm,gauge/.style={circle,draw,minimum size=5mm},flavor/.style={rectangle,draw,minimum size=5mm}] 	 
\begin{scope}
	
	\path (5,0) node[gauge](g1) {$\!\!\!N\!\!\!$} -- (7,0) node[gauge](g2) {$\!\!\!N\!\!\!$} -- (10,0) node[gauge](g3) {$\!\!\!N\!\!\!$} 
		-- (12,0) node[gauge](g4) {$\!\!\!N\!\!\!$} -- (5,1.5) node[flavor](y1) {$\!1\!$} -- (12,1.5) node[flavor](y2) {$\!1\!$};	 
		
	\wigM (g1) -- (g2); \draw[blue] (6,0.4) node {\scriptsize{$B_2$}};
	\wigM (g2) -- (8,0); 
	\wigM (g3) -- (9,0); 
	\wigM (g3) -- (g4); \draw[blue] (11,0.4) node {\scriptsize{$B_{F-1}$}};
	\draw[-, shorten >= 9, shorten <= 6, shift={(-0.05,0.05)}, mid arrowsm] (5,0) -- (5,1.5); 
	\draw[blue] (5,0.75) node[left] {\scriptsize{$\frac{1-N}{2}\tau + B_1$}}; \draw[blue] (5,2) node {\scriptsize{$Y_1$}};
	\draw[-, shorten >= 4, shorten <= 11, shift={(0.05,0.15)}, mid arrowsm] (5,1.5) -- (5,0);
	\draw[-, shorten >= 9, shorten <= 6, shift={(0.05,0.05)}, mid arrowsm] (12,0) -- (12,1.5);
	\draw[blue] (12,0.75) node[right] {\scriptsize{$\frac{1-N}{2}\tau + B_F$}}; \draw[blue] (12,2) node {\scriptsize{$Y_2$}};
	\draw[-, shorten >= 4, shorten <= 11, shift={(-0.05,0.15)}, mid arrowsm] (12,1.5) -- (12,0); 
	\draw (5,0.45) node[cross]{}; \draw (12,0.45) node[cross]{};	
	\draw (8.5,0) node {$\cdots$};
	\draw[-] (g1) to[out=150,in=90] (4.4,0) to[out=-90,in=-150] (g1); \draw[blue] (4.5,0) node[left] {\scriptsize{$\tau$}};
	\draw[-] (g2) to[out=60,in=0] (7,0.6) to[out=180,in=120] (g2); \draw[blue] (7,0.6) node[right] {\scriptsize{$\tau$}};
	\draw[-] (g3) to[out=60,in=0] (10,0.6) to[out=180,in=120] (g3); \draw[blue] (10,0.6) node[right] {\scriptsize{$\tau$}};
	\draw[-] (g4) to[out=30,in=90] (12.6,0) to[out=-90,in=-30] (g4); \draw[blue] (12.5,0) node[right] {\scriptsize{$\tau$}};
	
	\draw[blue] (5,-0.5) node {\scriptsize{$(X_2-X_1)$}}; \draw[blue] (7,-0.5) node {\scriptsize{$(X_3-X_2)$}};
	\draw[blue] (10,-0.5) node {\scriptsize{$(X_{F-1}$-$X_{F-2})$}}; \draw[blue] (12,-0.5) node {\scriptsize{$(X_F$-$X_{F-1})$}};
	
	
\end{scope} \end{tikzpicture}
\label{mthe}
\ee
Which corresponds to the final partition function:\footnote{The Identity in  \eqref{eq:SQCDmirr_parfun} is recovered  by redefining $Y_1 = Y + Y_2$ and then shifting all the gauge parameters as $Z^{(a)} \to Z^{(a)} + Y_2$}
\begin{align}
Z_{\text{SQCD}}=&Z_{\text{step 1}}=Z_{\text{step 2}}=Z_{\text{step 3}}= e^{2 \pi i N (Y_1 X_1 - Y_2 X_F)} \int \prod_{a=1}^{F-1} \big( d\vec{Z}_N^{(a)}  \D_N (\vec{Z}^{(a)},\tau)  \big) \nonumber \\ 
	& \prod_{j=1}^N \big[ s_b\left( \frac{iQ}{2} - \frac{1-N}{2}\tau - B_1 \pm (Z_j^{(1)} - Y_1) \right) s_b(-\frac{iQ}{2} + (j-N)\tau + 2B_1)  \nn \\
	& s_b \left( \frac{iQ}{2} - \frac{1-N}{2}\tau - B_F \pm (Z_j^{(F-1)}-Y_2) \right) s_b(-\frac{iQ}{2} + (j-N)\tau + 2B_F) \big] \nn \\
	& \prod_{a=2}^{F-1} Z_{NS}^{(N)} \left( \vec{Z}^{(a-1)},\vec{Z}^{(a)},\tau,B_{a}, X_{a} \right) = Z_{\widecheck{\text{SQCD}}} \,.
\end{align}

\subsubsection*{Comments on the $F=1$ case}
We now discuss the case $F=1$ which leads to the duality presented in figure \eqref{fig:SQCD_1flav}.
In this case we need to dualize two trivial bifundamental blocks and a single generalized flavor block to get:
\be
\begin{tikzpicture}[thick,node distance=3cm,gauge/.style={circle,draw,minimum size=5mm},flavor/.style={rectangle,draw,minimum size=5mm}]

	\path (1.5,0) node[flavor] (g1) {$\!0\!$} -- (1.5,1.5) node[flavor] (x1) {$\!1\!$} 
		-- (3,0) node[gauge] (g2) {$\!\!\!N\!\!\!$} -- (4.5,0) node[gauge] (g3) {$\!\!\!N\!\!\!$} -- (6,0) node[gauge] (g4) {$\!\!\!N\!\!\!$} 
		-- (7.5,0) node[gauge] (g5) {$\!\!\!N\!\!\!$} -- (9,0) node[flavor] (g6) {$\!0\!$} -- (9,1.5) node[flavor] (x2) {$\!1\!$};

	\wigT (g1) -- (g2); \draw (2.25,-0.3) node {$-$}; 
	\wigT (x1) -- (2.25,0); \draw[blue] (1.5,2) node {\scriptsize{$Y_1$}};
	\wigT (g2) -- (g3); \draw (3.75,-0.3) node {$+$};
	\wigM (g3) -- (g4); \draw[blue] (5.25,0.3) node {\scriptsize{$B$}};
	\wigT (g4) -- (g5); \draw (6.75,-0.3) node {$-$};
	\wigT (g5) -- (g6); \draw (8.25,-0.3) node {$+$};
	\wigT (x2) -- (8.25,0); \draw[blue] (9,2) node {\scriptsize{$Y_2$}};
	
	\draw (g2) to[out=60,in=0] (3,0.6) to[out=180,in=120] (g2); \draw[blue] (3,0.75) node {\scriptsize{$\tau$}};
	\draw (g3) to[out=60,in=0] (4.5,0.6) to[out=180,in=120] (g3); \draw[blue] (4.5,0.75) node {\scriptsize{$\tau$}};
	\draw (g4) to[out=60,in=0] (6,0.6) to[out=180,in=120] (g4); \draw[blue] (6,0.75) node {\scriptsize{$\tau$}};
	\draw (g5) to[out=60,in=0] (7.5,0.6) to[out=180,in=120] (g5); \draw[blue] (7.5,0.75) node {\scriptsize{$\tau$}};
	
	\draw[blue] (4.5,-0.5) node {\scriptsize{$(X)$}}; \draw[blue] (6,-0.5) node {\scriptsize{$(-X)$}};

\end{tikzpicture}\ee
Where we have already remove all the trivial blocks from the picture. We can now implement the  asymmetric $\mathbb{I}$-wall on the left with the effect of Higgsing the second $U(N)$ node down to $U(1)$  rendering improved bifundamental asymmetric.   We then use the duality \eqref{fig:FMtoFlav} relating the asymmetric  improved bifundamental to a flipped flavor to
 land on:
\be
\begin{tikzpicture}[thick,node distance=3cm,gauge/.style={circle,draw,minimum size=5mm},flavor/.style={rectangle,draw,minimum size=5mm}]

	\path  (6,1.5) node[flavor] (x1) {$\!1\!$} -- (6,0) node[gauge] (g4) {$\!\!\!N\!\!\!$} 
		-- (7.5,0) node[gauge] (g5) {$\!\!\!N\!\!\!$} -- (9,0) node[flavor] (g6) {$\!0\!$} 
		-- (9,1.5) node[flavor] (x2) {$\!1\!$};

	\draw[-, shorten >= 9, shorten <= 6, shift={(-0.05,0.05)}, mid arrowsm] (6,0) -- (6,1.5); \draw (6,0.5) node[cross] {};
	\draw[blue] (5.9,0.75) node[left] {\scriptsize{$\frac{1-N}{2}\tau + B$}}; \draw[blue] (6,2) node {\scriptsize{$Y_1$}};
	\draw[-, shorten >= 4, shorten <= 11, shift={(0.05,0.15)}, mid arrowsm] (6,1.5) -- (6,0);
	
	\wigT (g4) -- (g5); \draw (6.75,-0.3) node {$-$};
	\wigT (g5) -- (g6); \draw (8.25,-0.3) node {$+$};
	\wigT (x2) -- (8.25,0); \draw[blue] (9,2) node {\scriptsize{$Y_2$}};

	\draw (g4) to[out=150,in=90] (5.4,0) to[out=-90,in=-150] (g4); \draw[blue] (5.4,0) node[left] {\scriptsize{$\tau$}};
	\draw (g5) to[out=60,in=0] (7.5,0.6) to[out=180,in=120] (g5); \draw[blue] (7.5,0.75) node {\scriptsize{$\tau$}};
	
	\draw[blue] (6,-0.5) node {\scriptsize{$(-X)$}};

\end{tikzpicture}\ee
Where we are not depicting all the singlets produced by the procedure to avoid cluttering.
To this step is associated the following partition function:
\begin{align}
Z^{F=1}_{\text{SQCD}} = Z_{\text{step 3}'} = & \prod_{j=2}^N s_b( \frac{iQ}{2} - j \tau ) \int \prod_{a=1}^2 \big( d\vec{Z}^{(a)}_N \D_N(\vec{Z}^{(a)},\tau) \big)
e^{2\pi i X (NY_1 - \sum_{j=1}^N Z_j^{(1)} )} \nn \\
& \prod_{j=1}^N s_b( \frac{iQ}{2} - \frac{1-N}{2}\tau + B \pm (Z_j^{(1)} - Y_1) ) \prod_{j=1}^N s_b( -\frac{iQ}{2} + (j-N)\tau -2B) \nn \\
& Z_{\CS^{-1}}^{(N)} (\vec{Z}^{(1)}, \vec{Z}^{(2)}, \tau)
Z_{\CS}^{(N)} (\vec{Z}^{(2)}, \{\frac{N-1}{2}\tau + Y_2, \ldots, \frac{1-N}{2}\tau - Y_2 \}, \tau)  \,.
\end{align}
Now we implement  the second asymmetric $\mathbb{I}$-wall, which Higgses the first $U(N)$ gauge group transforming the $U(N)$ flavor into a collection of
$2 N$ chirals.
At the level of the partition function the Higgsing corresponds to specializing the Cartan $\vec{Z}^{(1)}$ in terms of the $\tau$ and $Y_2$ parameters.
Taking into account all the singlets the partition function of the final theory is:
\begin{align}
Z^{F=1}_{\text{SQCD}}=Z_{\text{step 3}'} = & e^{2\pi i X N(Y_1 - Y_2)} \prod_{j=2}^N s_b( \frac{iQ}{2} - j \tau )
\prod_{j=1}^N s_b( \frac{iQ}{2} - (1-j)\tau - B \pm (Y_2 - Y_1) )  \nn \\
& \prod_{j=1}^N s_b( -\frac{iQ}{2} + (j-N)\tau + 2B)
 = Z^{F=1}_{\widecheck{\text{SQCD}}} = Z_{\text{WZ}} \,.
\end{align}
The last set of singlets maps to the traces of the adjoint chiral of the SQCD. Therefore, if we flip them we land precisely on the duality \eqref{fig:SQCD_1flav}.
In particular the charges of the chiral fields are compatible with the cubic superpotential in  \eqref{fig:SQCD_1flav}.

\section{$3d$ $\cN=2$ linear brane setups and improved bifundamentals}\label{sec:branes}
The $\CN=2$ algorithm discussed in the previous sections, allows us to advance our understanding of Hanany-Witten brane setup with $4$ supercharges. In this section we make a proposal for the $3d$ $\cN=2$ gauge theory living on brane setups composed of $D3$, $NS$ and $D5'$ branes. As we will see, our proposal differs from the \emph{naive} quiver gauge theory in that the bifundamentals are improved instead of standard. \\

Let us start by defining the Hanany-Witten brane setup we are interested in. There are $D3$ branes, filling the $0126$ directions, stretching along the $6$ direction between $NS$ branes (filling the $012345$ directions) or  $NS'$ branes (filling the $012389$ directions). Flavors are added inserting $D5$ ($012789$) or so called $D5'$ ($012457$) branes. 
\begin{equation}\label{tab:branes}
\begin{tabular}{|c|cccccccccc|}\hline
  & $0$ & $1$ & $2$ & $3$ & $4$ & $5$ & $6$ & $7$ & $8$ & $9$ \\ \hline
$D3$  & x & x & x &    &   &   & x &   &    &  \\ 
$NS$  & x & x & x & x &x  &x  &   &   &   &  \\ 
$D5$  & x & x & x &    &   &    &   &x  &x  &x \\ 
$NS'$  & x & x & x & x  &  &  &  &   &  x & x \\ 
$D5'$  & x & x & x & &x   &x   &   & x &  & \\ \hline
\end{tabular}
\end{equation}
If all the branes are present the system preseves $3d$ $\cN=2$ supersymmetry. The $U(1)^2$ symmetry rotating the $45$ and $89$ directions becomes the
\be U(1)_R \times U(1)_\tau \ee
symmetry in the IR QFT. $U(1)_R$ is the $\cN=2$ R-symmetry while $U(1)_\tau$ is an additional global symmetry always present in the QFT's associated to brane setups with the branes of \eqref{tab:branes}.\footnote{One can break this $U(1)_\tau$ symmetry rotating some $5$-branes to generic angles along the $45$ and $89$ directions, without breaking the $\cN=2$ supersymmetry.  In this paper we do not study such configurations, but they should be obtained turning on superpotential deformations from the setups we study.} 
If only $D3$, $NS$ and $D5$ (or $D3$, $NS'$ and $D5'$) are present, the system preserves $8$ supercharges and the low energy theory living on it is well known to be a $\cN=4$ quiver with standard bifundamental matter. 
For instance the $3d$ $\CN=4$  $U(N)$ theory with no flavors  is associated to a Type IIB brane setup with $N$ $D3$ branes stretching between $2$ $NS$  branes. We can add $F$ flavors, adding $D5$ or  $D5'$  branes. 
The $D5$ branes preserve the $8$ supersymmetries, while the $D5'$ break half of the superymmetry, from $\CN=4$ to $\CN=2$.

In $\CN=2$ language, $N$ $D3$ branes stretching between $2$ $NS$ branes with $F$ $D5$ branes in the middle provide adjoint $U(N)$ with $F$ flavors and a cubic superpotential coupling the flavor to the adjoint (equivalently, we could use $N$ $D3$ branes stretching between $2$ $NS'$  branes with $F$ $D5'$ branes in the middle).
On the other hand, $N$ $D3$ branes stretching between $2$ $NS$ branes with $F$ $D5'$ branes in the middle,\footnote{Equivalently, we could use $N$ $D3$ branes stretching between $2$ $NS'$ branes  with $F$ $D5$ branes in the middle.} as in the left of the setup in figure \eqref{branesqcd}, give rise to  $U(N)$ with an adjoint and $F$ flavors and a \emph{vanishing} superpotential, $\CW=0$ \cite{Aharony:1997ju,Elitzur:1997fh,Giveon_1999}.
We can exclude a superpotential counting the motion of the $D3$ brane segments as follows. In the $F$ flavors case there are $N(F-1)$ $D5'-D5'$ segments (providing $N(F-1)$ quaternionic directions) and $2N$ $D5'-NS$ segments (providing $2N$ complex directions), so there must be a branch in the moduli space of vacua of the theory of complex dimension $2 N F$. Such a branch exists if $\cW=0$, parameterized by  $N F$ $Q$'s, $N F$ $\tilde{Q}$'s, $N^2$ $A$'s minus $N^2$ gauge symmetries, but a non zero superpotential, e.g. of the form $(Q\tilde{Q})^2$, would lift part of these $2 N F$ complex directions.\footnote{One can also argue for the absence of a cubic superpotential of the type $A Q \tilde{Q}$ noticing that the $D3$ branes, when moving along the $45$ directions (which corresponds to a vev for the adjoint $A$), remain in contact with the $D5'$ branes, hence the $D3-D5$ strings (which correspond to the flavor fields) remain at zero length, so the flavors remain massless.} 

The  $U(N)$ adjoint SQCD with $F$ flavors, $\CW=0$, is precisely the theory we studied in the previous sections, for which we found the mirror dual with $F-1$ gauge groups linked by  improved bifundamentals. Now, as shown in picture \eqref{branesqcd} we apply Type IIB $\CS$-duality to its associated brane setup. Modulo rotating the branes, 
\footnote{For convenience in the pictures we will always present the action of $\CS$-duality combined with the rotation acting by $NS' \to NS$ and $D5 \to D5'$. Clearly the QFT description is invariant under this rotation.} the $\CS$-dual setup is $N$ $D3$ branes stretching between $2$ $D5'$ branes with $F$ $NS$ branes in the middle. 
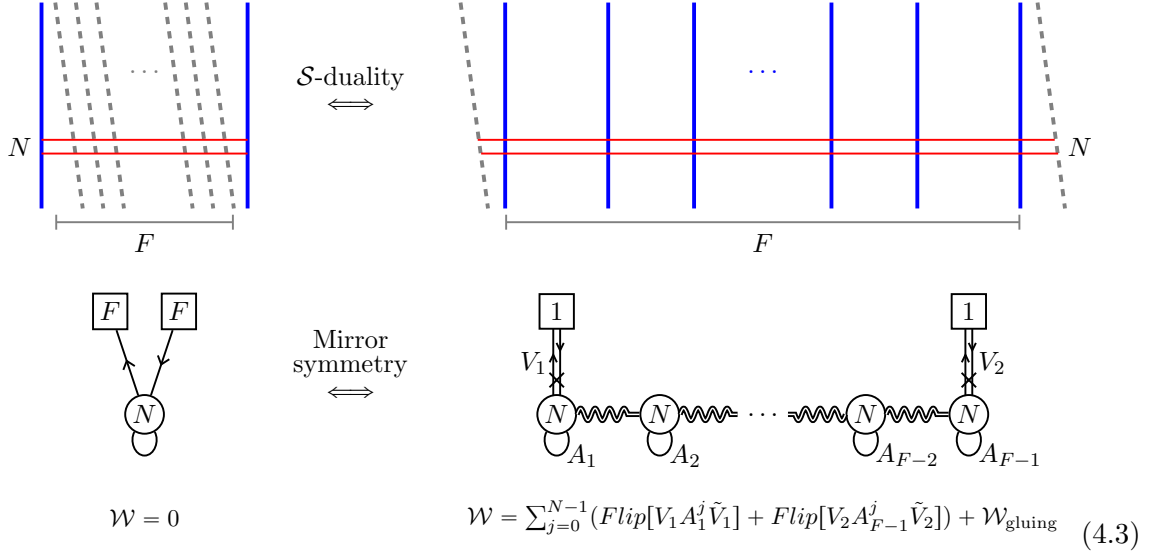
\be
\resizebox{.95\hsize}{!}{
 \bpic[thick,node distance=3cm,gauge/.style={circle,draw,minimum size=5mm},flavor/.style={rectangle,draw,minimum size=5mm}] 
\begin{scope}[shift={(0,0)}]

	\NSbrane (-1.5, 1.5) -- (-1.5, -1.5);
	\Dbrane (-1.3, 1.5) -- (-0.9, -1.5);
	\Dbrane (-1,1.5) -- (-0.6, -1.5);
	\Dbrane (-0.7, 1.5) -- (-0.3, -1.5);
	\draw (0,0.5) node[thick, gray] {$\ldots$};
	\Dbrane (0.3, 1.5) -- (0.7, -1.5);
	\Dbrane (0.6,1.5) -- (1, -1.5);
	\Dbrane (0.9, 1.5) -- (1.3, -1.5);
	\NSbrane (1.5, 1.5) -- (1.5, -1.5);	
	
	\Dthree (-1.5,-0.5) -- (1.5,-0.5);
	\Dthree (-1.5,-0.7) -- (1.5,-0.7);
	 	
	\draw  (-1.5,-0.6) node[left] {$N$};
	\draw[|-|,gray] (-1.3,-1.7) -- (1.3,-1.7);	\draw  (0,-2) node {$F$};	
\end{scope}
	
	\draw (3,0.4) node {$\CS$-duality};
	\draw (3,0) node {$\Longleftrightarrow$};
	
\begin{scope}[shift={(6,0)}]
	
	\Dbrane (-1.4, 1.5) -- (-1, -1.5);
	\NSbrane (-0.75, 1.5) -- (-0.75, -1.5);
	\NSbrane (0.75, 1.5) -- (0.75, -1.5);
	\NSbrane (2, 1.5) -- (2, -1.5);
	\draw (3,0.5) node[thick,blue] {$\ldots$};
	\NSbrane (4, 1.5) -- (4, -1.5);
	\NSbrane (5.25, 1.5) -- (5.25, -1.5);
	\NSbrane (6.75, 1.5) -- (6.75, -1.5);
	\Dbrane (7,1.5) -- (7.4,-1.5);
	
	\Dthree (-1.15,-0.5) -- (7.25,-0.5);
	\Dthree (-1.1,-0.7) -- (7.3,-0.7); 	
	
	\draw  (7.3,-0.6) node[right] {$N$};
	\draw[|-|,gray] (-0.75,-1.7) -- (6.75,-1.7);	\draw  (3,-2) node {$F$};
	
\end{scope}

\begin{scope}[shift={(0,-4.5)}]

    \path (0,0) node[gauge](g1) {$\!\!\!N\!\!\!$}  -- (-0.5,1.5) node[flavor] (x1) {$\!F\!$} -- (0.5,1.5) node[flavor] (x2) {$\!F\!$} 
    	-- (0,-1.5) node (g1p) {\small{$\CW=0$}} ;
    \chir (g1) -- (x1); 
    \chir (x2) -- (g1);
    \draw[-] (g1) to[out=-60,in=0] (0,-0.6) to[out=180,in=-120] (g1);
\end{scope}

	\draw (3,-3.4) node {Mirror};
	\draw (3,-3.8) node {symmetry};
	\draw (3,-4.2) node {$\Longleftrightarrow$};

\begin{scope}[shift={(6,-4.5)}]
	
	\path (0,0) node[gauge](g1) {$\!\!\!N\!\!\!$} -- (1.5,0) node[gauge](g2) {$\!\!\!N\!\!\!$} -- (3,0) node(gi) {$\ldots$}
		-- (4.5,0) node[gauge](g3) {$\!\!\!N\!\!\!$} -- (6,0) node[gauge](g4) {$\!\!\!N\!\!\!$} 
		-- (0,1.5) node[flavor](y1) {$\!1\!$} -- (6,1.5) node[flavor](y2) {$\!1\!$};
		
	\wigM (g1) -- (g2); 
	\wigM (g2) -- (gi); 
	\wigM (g3) -- (gi); 
	\wigM (g3) -- (g4); 
	
	\draw[-, shorten >= 3, shorten <= 12, shift={(-0.05,-0.15)}, middx arrowsm] (0,0) -- (0,1.5);
	\draw[-, shorten >= 8, shorten <= 7, shift={(0.05,0)}, midsx arrowsm] (0,1.5) -- (0,0);
	\draw (0,0.5) node[cross]{}; \draw (0,0.75) node[left] {$V_1$};
	
	\draw[-, shorten >= 3, shorten <= 12, shift={(-0.05,-0.15)}, middx arrowsm] (6,0) -- (6,1.5);
	\draw[-, shorten >= 8, shorten <= 7, shift={(0.05,0)}, midsx arrowsm] (6,1.5) -- (6,0);
	\draw (6,0.5) node[cross]{}; \draw (6,0.75) node[right] {$V_2$};
	
	\draw[-] (g1) to[out=-60,in=0] (0,-0.6) to[out=180,in=-120] (g1); \draw (0,-0.6) node[right] {$A_1$};
	\draw[-] (g2) to[out=-60,in=0] (1.5,-0.6) to[out=180,in=-120] (g2); \draw (1.5,-0.6) node[right] {$A_2$};
	\draw[-] (g3) to[out=-60,in=0] (4.5,-0.6) to[out=180,in=-120] (g3); \draw (4.5,-0.6) node[right] {$A_{F-2}$};
	\draw[-] (g4) to[out=-60,in=0] (6,-0.6) to[out=180,in=-120] (g4); \draw (6,-0.6) node[right] {$A_{F-1}$};
	\draw (3,-1.5) node {\small{$\cW = \sum_{j=0}^{N-1} ( Flip[V_1 A_1^j \tilde{V}_1] + Flip[V_2 A_{F-1}^j \tilde{V}_2] )+\mathcal{W}_{\text{gluing}}$} };
\end{scope}
	 
\epic}
\label{branesqcd}
\ee
Looking at the web of dualities in figure \eqref{branesqcd} it is natural to propose that the IR QFT associated to the brane setup on the right hand side is 
our mirror SQCD quiver obtained via the dualization algorithm, with improved, instead of standard bifundamentals  (in section \ref{oldprop} we will  comment on the relation between our proposal and previous ones).
Building on this observation and on the  $\cN=2$ algorithm perspective, for  an $\cN=2$ brane setup made of a constant  number of $D3$ branes
stretching between an arbitrary sequence of $NS$ and $D5'$ branes, we formulate the following

\vspace{0.4cm}

\begin{proposal}\label{proposalP}
\emph{The IR QFT associated to $N$ $D3$ branes stretching along an arbitrary ordered sequence of $g+1$ $NS$ branes and $F$ $D5'$ branes 
consists of a linear quiver with $g$ $U(N)$ adjoint nodes, $g-1$ improved bifundamentals and a total of $F$ flavors distributed among the $g$ nodes, according to the position of the $D5'$ branes. The superpotential is $\cW_{gluing}$, which couples the adjoint of each $U(N)$ node to the adjoint operators of the nearby improved bifundamentals. The flavors do not enter the superpotential, the only exception is if at the beginning (or at the end) of the sequence of $5$-branes there is a single $D5'$ brane, then the dressed mesons made with the associated flavor are flipped.}
\end{proposal}

\vspace{0.4cm}

We claim that this proposal is consistent with $\CS$-duality, that is two  improved quivers corresponding to $\CS$-dual brane setups are mirror dual, and one can construct the dual using the $\cN=2$ algorithm. We provide a few examples in section \ref{classexamples}.
The improved quiver theories associated to these brane setups have interesting patterns of symmetry enhancement and, as we discuss in \ref{sec:GBU}, we have a notion of {\it balanced nodes} leading to symmetry enhancement, in analogy with the $\mathcal{N}=4$ case \cite{Gaiotto:2008ak}. \\

Let us discuss some special sequences of $5$-branes. \\
If at the beginning of the sequence of $5$-branes there are $h>1$ $NS$ branes, as we show in section \ref{sec:GBU}, the associated theory is {\it ugly} and  we can sequentially confine a string of $h-1$ improved bifundamentals generating $(h-1) N$ free hypers. So the interacting part of the theory is associated to the set up where the first $h-1$ $NS$ branes have been removed. In particular,  our proposal for the theory associated to a sequence of $h$ $NS$ branes, a $U(N)^{h-1}$  improved quiver with no flavors flows in the IR to $h N$ free hypers, exactly as the  {\it bad} $\mathcal{N}=4$ $U(N)^{h-1}$  quiver theory with standard bifundamentals. \\
Analogously, by $\CS$-duality, if at the beginning of the sequence of $5$-branes there are $h>1$ $D5'$ branes, the QFT is given by $(h-1)N$ free hypers plus the QFT associated the brane setup where the first $h-1$ $D5'$ have been removed. \\
The last example is the short sequence $D5'-NS-D5'$, this sequence is not associated to an improved bifundamental (which in our prescription always connects gauge nodes) but  to an improved bifundamental where both the
$U(N)$ symmetries are broken to $U(1)$s. This deformation  reduces the improved bifundamental to  the Wess-Zumino model on the r.h.s. of \eqref{fig:SQCD_1flav} (as shown in section \ref{alsqcd}) which is indeed mirror dual to the SQCD with one flavor associated to the 
$\CS$-dual brane setup $NS-D5'-NS$. \\

Let us also mention that we actually understand some instances of more general situations. \\
We can describe brane systems where an arbitrary number of $D5'$s sit on top of an $NS$, that is the $NS$ and the $D5'$ form a  $(p,q)$-web of rectangular shape. In \ref{sec:pqwebs} by extending the logic of \cite{Benvenuti:2016wet} from the abelian to the non abelian case, we propose the QFT corresponding to the $\CS$-dual  $(p,q)$-web, that is many $NS$'s sitting on top of a $D5'$.
\\
We can turn real mass deformations in our quivers to generate Chern-Simons interactions and/or theories with chiral matter (different number of fundamentals vs anti-fundamentals). The corresponding brane setup might include $(p,q)$ 5-branes and non-rectangular  $(p,q)$-webs. 
We will discuss these theories in \cite{BCP3}, using the chiral improved bifundamental introduced in \cite{BCP1}.\\

Let us conclude saying that the most general $3d$ $\cN=2$ setup would involve all four types of $5$-branes ($NS$, $NS'$, $D5$, $D5'$) and a non-constant number of $D3$ branes  along the brane setup. To describe such setups we need a new object,  an  improved bifundamental, with non-abelian global symmetry $S[U(N_1) \times U(N_2)]$.   We plan to investigate it in the future.\\

\subsection{More examples of $\mathcal{N}=2$ mirror quivers }\label{classexamples}
In this subsection we consider brane setups with $N$ $D3$ branes stretched along the sequence $NS-(D5')^{F_1}-NS^K-(D5')^{F_2}-NS$, which is mirror to $D5'-NS^{F_1}-(D5')^K-NS^{F_2}-D5$. We write down the associated QFT's, discuss the chiral rings and global symmetries, and prove the IR duality between them.

\subsubsection{Electric theory with $2$ nodes}\label{k1}
\begin{figure}
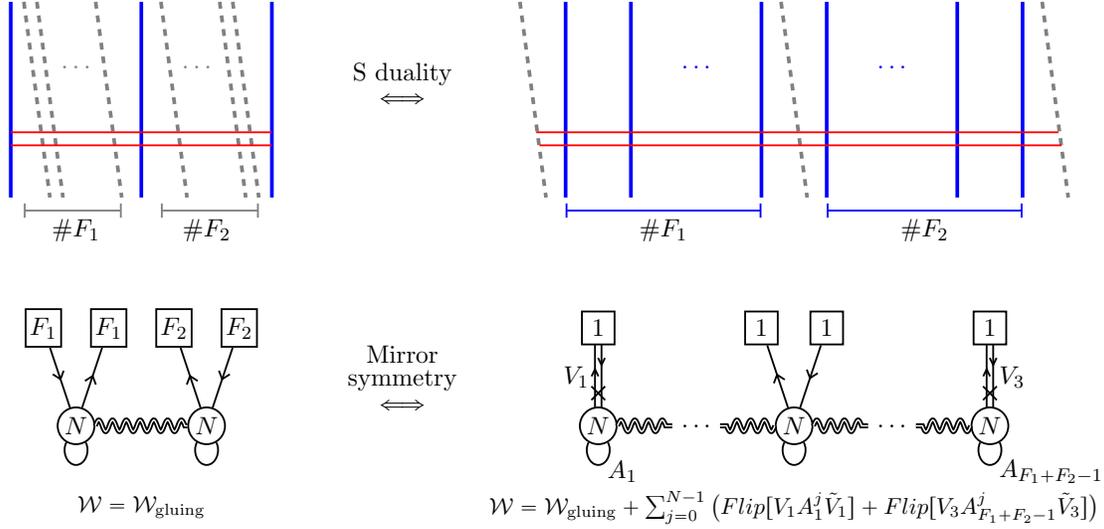

\centering
\resizebox{.95\hsize}{!}{
\bpic[thick,node distance=3cm,gauge/.style={circle,draw,minimum size=5mm},flavor/.style={rectangle,draw,minimum size=5mm}] 

\begin{scope}[shift={(0,0)}]	

	\NSbrane (-1, 1.5) -- (-1, -1.5);
	\Dbrane (-0.8, 1.5) -- (-0.4, -1.5);
	\Dbrane (-0.6, 1.5) -- (-0.2, -1.5);
	\draw (0,0.5) node[thick,gray] {$\ldots$};
	\Dbrane (0.3, 1.5) -- (0.7, -1.5);
	\NSbrane (1, 1.5) -- (1, -1.5);
	\Dbrane (1.3, 1.5) -- (1.7, -1.5);
	\draw (1.85,0.5) node[thick,gray] {$\ldots$};
	\Dbrane (2.2, 1.5) -- (2.6,-1.5);
	\Dbrane (2.4,1.5) -- (2.8, -1.5);
	\NSbrane (3,1.5) -- (3, -1.5);
	
	\Dthree (-1,-0.5) -- (3,-0.5);
	\Dthree (-1,-0.7) -- (3,-0.7);
	
	\draw[|-|,gray] (-0.8,-1.7) -- (0.7,-1.7);	\draw  (0,-2) node {$\# F_1$};
	\draw[|-|,gray] (1.3,-1.7) -- (2.8,-1.7);	\draw  (2,-2) node {$\# F_2$};
\end{scope}	 
	
	\draw (5,0.4) node {S duality};
	\draw (5,0) node {$\Longleftrightarrow$};
	
\begin{scope}[shift={(8,0)}]

	\Dbrane (-1.2,1.5) -- (-0.8,-1.5);
	\NSbrane (-0.5,1.5) -- (-0.5,-1.5);
	\NSbrane (0.5,1.5) -- (0.5,-1.5);
	\draw (1.5,0.5) node[thick,blue] {$\ldots$};
	\NSbrane (2.5,1.5) -- (2.5,-1.5);
	\Dbrane (2.8,1.5) -- (3.2,-1.5);
	\NSbrane (3.5,1.5) -- (3.5,-1.5);
	\draw (4.5,0.5) node[thick,blue] {$\ldots$};
	\NSbrane (5.5,1.5) -- (5.5,-1.5);
	\NSbrane (6.5,1.5) -- (6.5,-1.5);
	\Dbrane (6.8,1.5) -- (7.2,-1.5);
	
	\Dthree (-0.95,-0.5) -- (7.05,-0.5);
	\Dthree (-0.9,-0.7) -- (7.1,-0.7);	
	
	\draw[|-|,blue] (-0.5,-1.7) -- (2.5,-1.7);	\draw  (1,-2) node {$\# F_1$};
	\draw[|-|,blue] (3.5,-1.7) -- (6.5,-1.7);	\draw  (5,-2) node {$\# F_2$};
\end{scope}
\begin{scope}[shift={(0,-5)}]	

	\path (0,0) node[gauge](g1) {$\!\!\!N\!\!\!$} -- (2,0) node[gauge](g2) {$\!\!\!N\!\!\!$} 
		-- (0.5,1.5) node[flavor] (x1) {$\!F_1\!$} -- (-0.5,1.5) node[flavor] (x2) {$\!F_1\!$} 
		-- (1.5,1.5) node[flavor] (y1) {$\!F_2\!$} -- (2.5,1.5) node[flavor] (y2) {$\!F_2\!$}; 
	
	\wigM (g1) -- (g2);
	\chir (g1) -- (x1);
	\chir (x2) -- (g1);
	\chir (g2) -- (y1);
	\chir (y2) -- (g2);
	\draw[-] (g1) to[out=-60,in=0] (0,-0.6) to[out=180,in=-120] (g1);
	\draw[-] (g2) to[out=-60,in=0] (2,-0.6) to[out=180,in=-120] (g2);
	
	\draw (1,-1.25) node {\small{$\CW = \CW_{\text{gluing}}$}};
	
\end{scope}	 
	
	\draw (5,-3.9) node {Mirror};
	\draw (5,-4.3) node {symmetry};
	\draw (5,-4.7) node {$\Longleftrightarrow$};
	
\begin{scope}[shift={(8,-5)}]

	\path (0,0) node[gauge](g1) {$\!\!\!N\!\!\!$} -- (1.5,0) node(gi) {$\ldots$} -- (3,0) node[gauge] (g2) {$\!\!\!N\!\!\!$} 
		-- (4.5,0) node (gii) {$\ldots$} -- (6,0) node[gauge] (g3) {$\!\!\!N\!\!\!$}
		-- (0,1.5) node[flavor] (x1) {$\!1\!$} 
		-- (2.5,1.5) node[flavor] (x2) {$\!1\!$} -- (3.5,1.5) node[flavor] (x22) {$\!1\!$} 
		-- (6,1.5) node[flavor] (x3) {$\!1\!$} ;
		
 	\wigM (g1) -- (gi);
	\wigM (g2) -- (gi); 
	\wigM (g2) -- (gii); 
	\wigM (g3) -- (gii);
	
	\draw[-, shorten >= 3, shorten <= 12, shift={(-0.05,-0.15)}, middx arrowsm] (0,0) -- (0,1.5);
	\draw[-, shorten >= 8, shorten <= 7, shift={(0.05,0)}, midsx arrowsm] (0,1.5) -- (0,0);
	\draw (0,0.5) node[cross] {}; \draw (0,0.75) node[left] {$V_1$};
	
	\chir (g2) -- (x2);
	\chir (x22) -- (g2);
	
	\draw[-, shorten >= 3, shorten <= 12, shift={(-0.05,-0.15)}, middx arrowsm] (6,0) -- (6,1.5);
	\draw[-, shorten >= 8, shorten <= 7, shift={(0.05,0)}, midsx arrowsm] (6,1.5) -- (6,0);
	\draw (6,0.5) node[cross] {}; \draw (6,0.75) node[right] {$V_3$};
	
	\draw[-] (g1) to[out=-60,in=0] (0,-0.6) to[out=180,in=-120] (g1); \draw (0,-0.7) node[right] {$A_1$};
	\draw[-] (g2) to[out=-60,in=0] (3,-0.6) to[out=180,in=-120] (g2);
	\draw[-] (g3) to[out=-60,in=0] (6,-0.6) to[out=180,in=-120] (g3); \draw (6,-0.7) node[right] {$A_{F_1+F_2-1}$};
	
	\draw (3,-1.25) node {\small{$\CW = \CW_{\text{gluing}} + \sum_{j=0}^{N-1} \big( Flip[V_1 A_1^j \tilde{V}_1] + Flip[V_3 A_{F_1+F_2-1}^j \tilde{V}_3] \big)$} };
    
\end{scope}
\epic}
\caption{In  the top left corner we have the electric brane set up with three $NS$ branes,  $F_1$ $D5'$ branes in the first interval and $F_2$ in the second and a  $N$ $D3$ branes stretching from the first $NS$ to the third. The associated  quiver theory in the bottom left corner has two gauge nodes linked by an improved bifundamental, $F_1$ flavor on the first node and $F_2$ on the second. On the top  right corner we have the $S$-dual brane setup
and on  the bottom right corner its associated quiver description. The leftmost and rightmost flavors are associated to the D5' located outside the $SN$ branes hence the corresponding  dressed mesons are flipped. We denoted by $V_a$, with $a=1,2,3$ the   flavors  from left to right and by $A_n$  the adjoint of the $n$-th gauge node. The two quiver theories are mirorr dual.}
\label{fig:TwoNodes_ex}
\end{figure}
Let us start from the simplest example, $K=1$, corresponding to the brane setup in the top left corner of figure \ref{fig:TwoNodes_ex}.
According to our proposal \ref{proposalP} the  associated  theory is the two nodes improved quiver on the bottom left corner. As in the case of the SQCD, it is convenient to reparameterize the electric theory as: 
\be \label{fig:TwoNodes_Elec_manifest}
\bpic[thick,node distance=3cm,gauge/.style={circle,draw,minimum size=5mm},flavor/.style={rectangle,draw,minimum size=5mm}] 
 
	\path (2,0) node[gauge](g1) {$\!\!\!N\!\!\!$} -- (5,0) node[gauge](g2) {$\!\!\!N\!\!\!$} 
		-- (2.75,1.5) node[flavor] (x1) {$\!1\!$} -- (1.25,1.5) node[flavor] (x2) {$\!1\!$} -- (4.25,1.5) node[flavor] (y1) {$\!1\!$} 
		-- (5.75,1.5) node[flavor] (y2) {$\!1\!$}; 
	
	\wigM (g1) -- (g2); \draw[blue] (3.5,-0.4) node {\scriptsize{$D$}};
	
	\draw[-, shorten >= 6, shorten <= 9, shift={(-0.07,0.02)}, mid arrowsm] (1.25,1.5) -- (2,0);
	\draw[-, shorten >= 8, shorten <= 9, shift={(0.1,0)}, mid arrowsm] (2,0) -- (1.25,1.5); 
	\draw[blue] (1.55,0.9) node[left] {\scriptsize{$1-B_1$}}; \draw[blue] (1.25,2) node {\scriptsize$X_1$};
	
	\draw[-, shorten >= 8, shorten <= 8, shift={(-0.1,0.02)}, mid arrowsm] (2,0) -- (2.75,1.5);
	\draw[-, shorten >= 7, shorten <= 8, shift={(0.05,0)}, mid arrowsm] (2.75,1.5) -- (2,0); 
	\draw[blue] (2.55,0.9) node[right] {\scriptsize{$1-B_{F_1}$}}; \draw[blue] (2.75,2) node {\scriptsize$X_{F_1}$};
	
	\draw (2,1.5) node {$\cdots$}; \draw[blue] (2,-0.5) node {\scriptsize{$(W_1 - W_2)$}};

	\draw[-, shorten >= 6, shorten <= 9, shift={(-0.07,0.02)}, mid arrowsm] (4.25,1.5) -- (5,0);
	\draw[-, shorten >= 8, shorten <= 9, shift={(0.1,0)}, mid arrowsm] (5,0) -- (4.25,1.5); 
	\draw[blue] (4.85,0.4) node[left] {\scriptsize{$1-C_1$}}; \draw[blue] (4.25,2) node {\scriptsize$Y_1$};
	
	\draw[-, shorten >= 8, shorten <= 8, shift={(-0.1,0.02)}, mid arrowsm] (5,0) -- (5.75,1.5);
	\draw[-, shorten >= 7, shorten <= 8, shift={(0.05,0)}, mid arrowsm] (5.75,1.5) -- (5,0); 
	\draw[blue] (5.25,0.4) node[right] {\scriptsize{$1-C_{F_2}$}}; \draw[blue] (5.75,2) node {\scriptsize$Y_{F_2}$};
	
	\draw (5,1.5) node {$\cdots$}; \draw[blue] (5,-0.5) node {\scriptsize{$(W_2 - W_3)$}};
	
	\draw[-] (g1) to[out=150,in=90] (1.4,0) to[out=-90,in=-150] (g1); \draw[blue] (1.4,0) node[left] {\scriptsize{$\tau$}};
	\draw[-] (g2) to[out=30,in=90] (5.6,0) to[out=-90,in=-30] (g2);	 \draw[blue] (5.6,0) node[right] {\scriptsize{$\tau$}};	
\epic\ee
The global symmetry group of this theory is given by:\footnote{We can factorise a $U(1)$ vector-like factor from $U(F_1)^2 \times U(F_2)^2$ by a gauge transformation. This consists in imposing the constraint: $\sum_{j=1}^{F_1} X_j + \sum_{j=1}^{F_2} Y_j = 0$.}
\begin{align}\label{eq:TwoNodes_ElecSymm}
	S[ U(F_1)^2 \times U(F_2)^2 ] \times U(1)_D \times U(1)_{W_1-W_2} \times \U(1)_{W_2-W_3} \times U(1)_\tau \,,
\end{align}
Where the parameterization of \eqref{fig:TwoNodes_Elec_manifest} recombines as:
\begin{align}
	\prod_{j=1}^{F_1} \big( U(1)_{B_j} \times U(1)_{X_j} \big) = U(F_1)^2 \,, \nn \\
	\prod_{j=1}^{F_2} \big( U(1)_{C_j} \times U(1)_{Y_j} \big) = U(F_2)^2 \,.
\end{align}
At this point we run the dualization algorithm, as shown in appendix \ref{app:Quiver_algorithm}, and find the mirror dual quiver theory:
\be \label{fig:TwoNodes_Magn}
\resizebox{.95\hsize}{!}{
 \bpic[thick,node distance=3cm,gauge/.style={circle,draw,minimum size=5mm},flavor/.style={rectangle,draw,minimum size=5mm}] 

	\path (0,0) node[gauge](g1) {$\!\!\!N\!\!\!$} -- (2,0) node[gauge](g2) {$\!\!\!N\!\!\!$} -- (3.5,0) node {$\ldots$} 
		-- (5,0) node[gauge] (g3) {$\!\!\!N\!\!\!$} -- (7,0) node[gauge] (g4) {$\!\!\!N\!\!\!$} -- (9,0) node[gauge] (g5) {$\!\!\!N\!\!\!$} 
		-- (10.5,0) node {$\ldots$}-- (12,0) node[gauge] (g6) {$\!\!\!N\!\!\!$} -- (14,0) node[gauge] (g7) {$\!\!\!N\!\!\!$} 
		-- (0,1.5) node[flavor] (x1) {$\!1\!$} -- (7,1.5) node[flavor] (x2) {$\!1\!$} -- (14,1.5) node[flavor] (x3) {$\!1\!$} ;
		
 	\wigM (g1) -- (g2); \draw[blue] (1,0.4) node {\scriptsize{$B_2$}}; 
	\wigM (g2) -- (3,0); 
	\wigM (g3) -- (4,0); 
	\wigM (g3) -- (g4); \draw[blue] (6,0.4) node {\scriptsize{$B_{F_1}$}};
	\wigM (g5) -- (g4); \draw[blue] (8,0.4) node {\scriptsize{$C_1$}};
	\wigM (g5) -- (10,0);
	\wigM (g6) -- (11,0);
	\wigM (g6) -- (g7); \draw[blue] (13,0.4) node {\scriptsize{$C_{F-1}$}};
	
	\draw[-, shorten >= 3, shorten <= 12, shift={(-0.05,-0.15)}, middx arrowsm] (0,0) -- (0,1.5);
	\draw[-, shorten >= 8, shorten <= 7, shift={(0.05,0)}, midsx arrowsm] (0,1.5) -- (0,0);
	\draw (0,0.5) node[cross] {}; \draw[blue] (0,0.75) node[left] {\scriptsize{$\frac{1-N}{2}\tau + B_1$}}; \draw[blue] (0,2) node {\scriptsize{$W_1$}};
	
	\draw[-, shorten >= 3, shorten <= 12, shift={(-0.05,-0.15)}, mid arrowsm] (7,0) -- (7,1.5);
	\draw[-, shorten >= 8, shorten <= 7, shift={(0.05,0)}, mid arrowsm] (7,1.5) -- (7,0);
	\draw[blue] (7,1) node[left] {\scriptsize{$1-D$}}; \draw[blue] (7,2) node {\scriptsize{$W_2$}};
	
	\draw[-, shorten >= 3, shorten <= 12, shift={(-0.05,-0.15)}, middx arrowsm] (14,0) -- (14,1.5);
	\draw[-, shorten >= 8, shorten <= 7, shift={(0.05,0)}, midsx arrowsm] (14,1.5) -- (14,0);
	\draw (14,0.5) node[cross] {}; \draw[blue] (14,0.75) node[right] {\scriptsize{$\frac{1-N}{2}\tau + C_F$}}; \draw[blue] (14,2) node {\scriptsize{$W_3$}};
	
	\draw[blue] (0,-0.5) node {\scriptsize$(X_2-X_1)$}; \draw[blue] (2,-0.5) node {\scriptsize$(X_3-X_2)$}; 
	\draw[blue] (5,-0.5) node {\scriptsize$(X_{F_1}$-$X_{F_1-1})$}; \draw[blue] (7,-0.5) node {\scriptsize$($-$X_{F_1}$+$Y_1)$};
	\draw[blue] (9,-0.5) node {\scriptsize$(Y_2-Y_1)$}; \draw[blue] (12,-0.5) node {\scriptsize$(Y_{F-1}$-$Y_{F-2})$}; 
	\draw[blue] (14,-0.5) node {\scriptsize$(Y_F$ - $Y_{F-1})$};
	\draw[-] (g1) to[out=150,in=90] (-0.6,0) to[out=-90,in=-150] (g1); \draw[blue] (-0.6,0) node[left] {\scriptsize{$\tau$}};
	\draw[-] (g2) to[out=60,in=0] (2,0.6) to[out=180,in=120] (g2); \draw[blue] (2,0.8) node {\scriptsize{$\tau$}};
	\draw[-] (g3) to[out=60,in=0] (5,0.6) to[out=180,in=120] (g3); \draw[blue] (5,0.8) node {\scriptsize{$\tau$}};
	\draw[-] (g4) to[out=30,in=-30] (7.4,0.4) to[out=150,in=60] (g4); \draw[blue] (7.4,0.6) node {\scriptsize{$\tau$}};
	\draw[-] (g5) to[out=60,in=0] (9,0.6) to[out=180,in=120] (g5); \draw[blue] (9,0.8) node {\scriptsize{$\tau$}};
	\draw[-] (g6) to[out=60,in=0] (12,0.6) to[out=180,in=120] (g6); \draw[blue] (12,0.8) node {\scriptsize{$\tau$}};
	\draw[-] (g7) to[out=30,in=90] (14.6,0) to[out=-90,in=-30] (g7); \draw[blue] (14.6,0) node[right] {\scriptsize{$\tau$}};
    

\epic}\ee
Where $F=F_1+F_2$. As expected, the  mirror dual  quiver \eqref{fig:TwoNodes_Magn},  obtained via the algorithm,  coincides with  the quiver in bottom right corner of figure \ref{fig:TwoNodes_ex} (with a different parameterization of the central flavor) which we wrote down starting from the  $\CS$-dual brane configuration and applying our proposal. The manifest global symmetry group is given by:
\begin{align}
	& S \left[ \prod_{j=1}^3 U(1)_{W_j} \right] \times U(1)_D \times \prod_{j=1}^{F_1} U(1)_{B_j} \times \prod_{j=1}^{F_2} U(1)_{C_j} \times \nn \\
	& \times \prod_{j=1}^{F_1-1} U(1)_{X_{j+1}-X_j} \prod_{j=1}^{F_2-1} U(1)_{Y_{j+1}-Y_j} \times U(1)_{Y_1-X_{F_1}} \times U(1)_\tau \,.
\end{align} 
The pattern of symmetry enhancement is similar to the SQCD case, in particular we observe that the topological and axial symmetries enhance as:
\begin{align}
	\prod_{j=1}^{F_1} U(1)_{B_j} \times \prod_{j=1}^{F_1-1} U(1)_{X_{j+1}-X_j} \to S[ U(F_1)^2 ] \,, \nn \\
	\prod_{j=1}^{F_2} U(1)_{C_j} \times \prod_{j=1}^{F_2-1} U(1)_{Y_{j+1}-Y_j} \to S[ U(F_2)^2 ] \,.
\end{align}
The complete IR global symmetry of the mirror theory is then:
\begin{align}
	& S[ U(F_1)^2 ] \times S[ U(F_2)^2 ] \times U(1)_{Y_1-X_{F_1}} \times U(1)_D \times S\left[ \prod_{j=1}^3 U(1)_{W_j} \right] \times U(1)_\tau\,,
\end{align}
which upon a redefinition of some $U(1)$ factors,  matches precisely with the IR global symmetry of the original theory in \eqref{eq:TwoNodes_ElecSymm}. \\
Notice that the pattern of symmetry enhancement in the mirror theory is quite non-trivial but thanks to the parameterization obtained from the dualization algorithm, it is easier to collect together operator with the same R-charge and therefore construct representations of the emergent symmetries.

\subsubsection*{Operator Map}
The operator map works as follows:
\begin{itemize}
	\item In the electric theory we can build mesonic operators in the $\bar{F}_1 \times F_1$ bifundamental. In the magnetic theory these are mapped into a collection of monopoles and singlets. 
	In particular, using the results of appendix \ref{app:FMmonopoles} we can check that monopoles $\mathfrak{M}^{\pm(0,\ldots,0,1,\ldots,1,0,\ldots,0|0|0,\ldots,0)}$, with topological charge
	given by strings of contiguous $1$ (or $-1$) under the topological symmetries $U_{X_{j+1}-X_j}$ of the $F_1-1$ nodes on the l.h.s. of the central node, they all have the same R-charge. We can then collect these
	$F_1(F_1-1)$ monopoles with the $\mathsf{B}^{(j)}_{1,1}$ singlet in the $F_1-1$ improved bifundamentals on the left of the central flavor plus the flipper of the left flavor $\mathcal{F}[V_1 A_1^{N-1} \tilde{V}_1]$ in a matrix transforming in the $\bar{F}_1 \times F_1$ bifundamental of the emergent $U(F_1)^2$ symmetry. 
	
	\item  Similarly have an electric mesonic operators in the $\bar{F}_2 \times F_2$ bifundamental to $F_2(F_2-1)$. This is mapped to a collection of monopoles $\mathfrak{M}^{\pm(0,\ldots,0|0|0,\ldots,0,1,\ldots,1,0,\ldots,0)}$, with topological charge
	given by strings of contiguous $1$ (or $-1$) under the topological symmetries $U_{Y_{j+1}-Y_j}$ of the $F_2-1$ nodes on the r.h.s. of the central  node  , and the $\mathsf{B}^{(j)}_{1,1}$ singlets in the $F_2-1$ improved bifundamentals on the right of the central flavor plus the flipper of the right  flavor $\mathcal{F}[V_3 A_{F_1+F_2-1}^{N-1} \tilde{V}_3]$. Collecting all these operators we can assemble a matrix transforming in the bifundamental $\bar{F}_2 \times F_2$ of the emergent $U(F_2)^2$ symmetry.

	\item Electric long mesons in the $F_1 \times \bar{F}_2$ and $\bar{F}_1 \times F_2$,
	involving the improved bifundamental, are mapped into magnetic
	 monopole operators 	$\mathfrak{M}^{\pm(0,\ldots,0,1,\ldots,1|1|1,\ldots,1,0,\ldots,0)}$
	with topological charge given by a string of $\pm1$ extending  from the central node, to the left and to the right.
	Using the results in \ref{app:FMmonopoles} we can check that all these $2F_1 \times F_2$
	 operators have the same R-charge and can be
	assembled into two matrices. Collecting all the positively charged monopoles we assemble a matrix transforming in the $\bar{F}_1 \times F_2$ which therefore maps to the corresponding electric mesons. Similarly, the negatively charged monopoles form a matrix mapping to the $F_1 \times \bar{F}_2$ mesons.
	
	\item We also have electric monopoles charged under the topological symmetry of the left gauge node $\mathfrak{M}^{\pm(1,0)}$. 
	The positively charged one is mapped into  the long meson in the magnetic theory built by joining
	the chirals $\tilde{V}_1$ and $V_2$ with the string of improved bifundamental operators connecting them.
	The negatively charged monopole is instead mapped in the conjugate long meson built by similarly joining
	$V_1$ and $\tilde{V}_2$.
	
	\item  Similarly, we have electric monopoles charged under the topological symmetry of the right gauge node $\mathfrak{M}^{\pm(0,1)}$ 
	These are mapped respectively into the long meson in the magnetic theory built by joining
	the chirals $\tilde{V}_2$ and $V_3$ with the string of improved bifundamentals connecting them and its conjugate.
	
	\item  Electric  monopoles charged under both the topological symmetries $\mathfrak{M}^{\pm(1,1)}$ are respectively mapped into long mesons built by joining  $\tilde{V}_1$ and $V_3$ with the string of improved bifundamental connecting them and its conjugate.
	
	\item The singlets $\mathsf{B}_{n,m}$ (with R-charge $2n-2D+(m-n)\tau$) contained in the improved bifundamental of the electric theory are mapped into magnetic dressed mesons obtained from the central flavor: $V_2 \mathbf{A}^{n-1} A^{m-1}  \tilde{V}_2$. Where $A$ is the adjoint at the central node while $\mathbf{A}$ is the moment map of the improved bifundamental to the right or to the left of the central node, that are identified due to the F-term relations coming from the field $A$. Notice that in the electric  quiver all the $\mathsf{B}_{n,m}$'s are non trivial in the chiral ring, while in the mirror quiver  the $\mathsf{B}_{n \neq 1, m}$'s of each improved bifundamental are trivial in the chiral ring.\footnote{We will turn on these deformations in section \ref{sec4dquiver} in order to find a $3d$ mirror of a $4d$ quiver coming from a linear Hanany-Witten brane setup with $4$ supercharges.} This is consistent with the fact that operators of the type $V_b \mathbf{A}_b^{n-1}  A_b^{m-1} \tilde{V}_b$ for  $n \neq 1$  (with the flavor $V_b$ living at the boundary of an improved  quiver) are zero in the chiral ring, because the moment map operator $\mathbf{A}_b$ attached to a boundary node is set to zero by the F-terms of $A_b$, the adjoint of the boundary node.
	
\end{itemize}

All the presented operators can be also dressed with powers of the adjoint, unlike in the $\cN=4$ theories where the cubic superpotential sets them to zero. The generalization of the map to dressed operators is straightforward. 
For mesonic operators in the electric theory, their dressed version is mapped into a collection of dressed monopoles plus a set of singlets. 
For any electric mesons dressed with $j<N$ powers of an adjoint, we consider singlets in the magnetic theory that are given by: the $\mathsf{B}^{(j)}_{1,j+1}$ singlets in the improved bifundamental theories and the flip of the flavors dressed $j$ times $Flip[V_a A^j \tilde{V}_a]$. Analogously, dressed monopoles in the electric theory are mapped into dressed mesons in the magnetic theory.

\subsubsection{Electric theory with $K+1$ nodes}\label{generick}
\begin{figure}
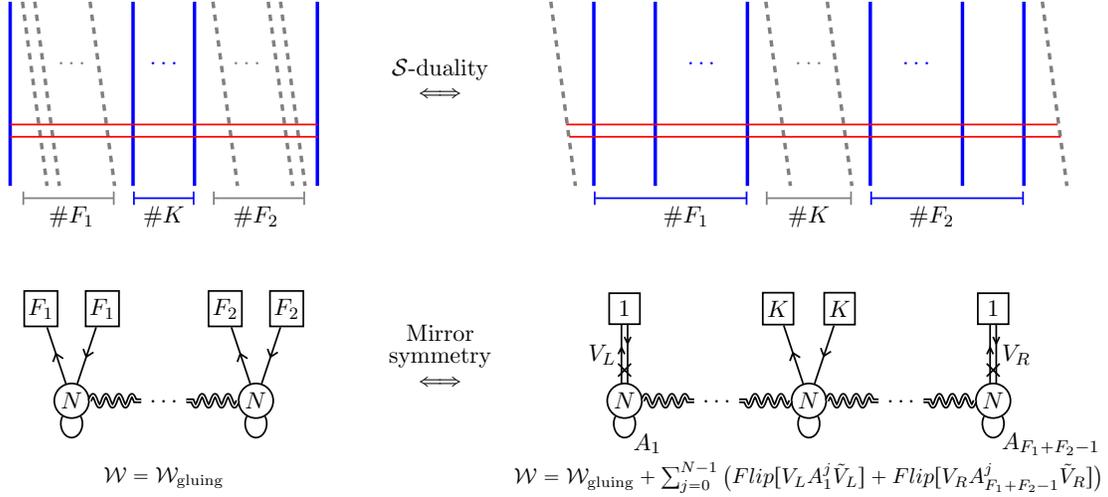

\centering
\resizebox{.95\hsize}{!}{
\bpic[thick,node distance=3cm,gauge/.style={circle,draw,minimum size=5mm},flavor/.style={rectangle,draw,minimum size=5mm}] 

\begin{scope}[shift={(0,0)}]	
	
	\NSbrane (-1, 1.5) -- (-1, -1.5);
	\Dbrane (-0.8, 1.5) -- (-0.4, -1.5);
	\Dbrane (-0.6, 1.5) -- (-0.2, -1.5);
	\draw (0,0.5) node[thick,gray] {$\ldots$};
	\Dbrane (0.3, 1.5) -- (0.7, -1.5);
	\NSbrane (1, 1.5) -- (1, -1.5);
	\draw (1.5,0.5) node[thick,blue] {$\ldots$};
	\NSbrane (2,1.5) -- (2,-1.5);
	\Dbrane (2.3, 1.5) -- (2.7, -1.5);
	\draw (2.85,0.5) node[thick,gray] {$\ldots$};
	\Dbrane (3.2, 1.5) -- (3.6,-1.5);
	\Dbrane (3.4,1.5) -- (3.8, -1.5);
	\NSbrane (4,1.5) -- (4, -1.5);

	\Dthree (-1,-0.5) -- (4,-0.5);
	\Dthree (-1,-0.7) -- (4,-0.7);
	
	\draw[|-|,gray] (-0.8,-1.7) -- (0.7,-1.7);	\draw  (0,-2) node {$\# F_1$};
	\draw[|-|,blue] (1,-1.7) -- (2,-1.7);	\draw  (1.5,-2) node {$\# K$};
	\draw[|-|,gray] (2.3,-1.7) -- (3.8,-1.7);	\draw  (3,-2) node {$\# F_2$};
\end{scope}	

\begin{scope}[shift={(0,-5)}]	
		
	\path (0,0) node[gauge](g1) {$\!\!\!N\!\!\!$} --  (1.5,0) node(gi) {$\ldots$} -- (3,0) node[gauge](g2) {$\!\!\!N\!\!\!$} 
		-- (-0.5,1.5) node[flavor] (x1) {$\!F_1\!$} -- (0.5,1.5) node[flavor] (x2) {$\!F_1\!$} 
		-- (2.5,1.5) node[flavor] (y1) {$\!F_2\!$} -- (3.5,1.5) node[flavor] (y2) {$\!F_2\!$};
 
	\wigM (g1) -- (gi);
	\wigM (g2) -- (gi);
	\chir (g1) -- (x1);
	\chir (x2) -- (g1);
	\chir (g2) -- (y1);
	\chir (y2) -- (g2);
	\draw[-] (g1) to[out=-60,in=0] (0,-0.6) to[out=180,in=-120] (g1);
	\draw[-] (g2) to[out=-60,in=0] (3,-0.6) to[out=180,in=-120] (g2);
	
	\draw (1.5,-1.25) node {\small{$\CW = \CW_{\text{gluing}}$} };	
	
\end{scope}	 

\draw (6,0.4) node {$\CS$-duality};
\draw (6,0) node {$\Longleftrightarrow$};

\begin{scope}[shift={(9,0)}]

	\Dbrane (-1.2,1.5) -- (-0.8,-1.5);
	\NSbrane (-0.5,1.5) -- (-0.5,-1.5);
	\NSbrane (0.5,1.5) -- (0.5,-1.5);
	\draw (1.25,0.5) node[thick,blue] {$\ldots$};
	\NSbrane (2,1.5) -- (2,-1.5);
	\Dbrane (2.3,1.5) -- (2.7,-1.5);
	\draw (3,0.5) node[thick,gray] {$\ldots$};
	\Dbrane (3.3,1.5) -- (3.7,-1.5);
	\NSbrane (4,1.5) -- (4,-1.5);
	\draw (4.75,0.5) node[thick,blue] {$\ldots$};
	\NSbrane (5.5,1.5) -- (5.5,-1.5);
	\NSbrane (6.5,1.5) -- (6.5,-1.5);
	\Dbrane (6.8,1.5) -- (7.2,-1.5);
	
	\Dthree (-0.95,-0.5) -- (7.05,-0.5);
	\Dthree (-0.9,-0.7) -- (7.1,-0.7);	
	
	\draw[|-|,blue] (-0.5,-1.7) -- (2,-1.7);	\draw  (1,-2) node {$\# F_1$};
	\draw[|-|,gray] (2.3,-1.7) -- (3.7,-1.7);	\draw  (3,-2) node {$\# K$};
	\draw[|-|,blue] (4,-1.7) -- (6.5,-1.7);	\draw  (5,-2) node {$\# F_2$};	
	
\end{scope}

\draw (6,-3.9) node {Mirror};
\draw (6,-4.3) node {symmetry};
\draw (6,-4.7) node {$\Longleftrightarrow$};

\begin{scope}[shift={(9,-5)}]
	\path (0,0) node[gauge](g1) {$\!\!\!N\!\!\!$} -- (1.5,0) node(gi) {$\ldots$} -- (3,0) node[gauge] (g2) {$\!\!\!N\!\!\!$} 
	 		-- (4.5,0) node (gii) {$\ldots$} -- (6,0) node[gauge] (g3) {$\!\!\!N\!\!\!$} 
	 		-- (0,1.5) node[flavor] (x1) {$\!1\!$} -- (2.5,1.5) node[flavor] (x2) {$\!K\!$}
	 		-- (3.5,1.5) node[flavor] (x22) {$\!K\!$} -- (6,1.5) node[flavor] (x3) {$\!1\!$};
 
	\wigM (g1) -- (gi);
	\wigM (g2) -- (gi); 
	\wigM (g2) -- (gii);
	\wigM (g3) -- (gii);
	
	\draw[-, shorten >= 3, shorten <= 12, shift={(-0.05,-0.15)}, middx arrowsm] (0,0) -- (0,1.5);
	\draw[-, shorten >= 8, shorten <= 7, shift={(0.05,0)}, midsx arrowsm] (0,1.5) -- (0,0);
	\draw (0,0.5) node[cross] {}; \draw (0,0.75) node[left] {$V_L$};
	
	\chir (g2) -- (x2);
	\chir (x22) -- (g2);
	
	\draw[-, shorten >= 3, shorten <= 12, shift={(-0.05,-0.15)}, middx arrowsm] (6,0) -- (6,1.5);
	\draw[-, shorten >= 8, shorten <= 7, shift={(0.05,0)}, midsx arrowsm] (6,1.5) -- (6,0);
	\draw (6,0.5) node[cross] {}; \draw (6,0.75) node[right] {$V_R$};
	 
	\draw[-] (g1) to[out=-60,in=0] (0,-0.6) to[out=180,in=-120] (g1); \draw (0,-0.7) node[right] {$A_1$};
	\draw[-] (g2) to[out=-60,in=0] (3,-0.6) to[out=180,in=-120] (g2);
	\draw[-] (g3) to[out=-60,in=0] (6,-0.6) to[out=180,in=-120] (g3); \draw (6,-0.7) node[right] {$A_{F_1+F_2-1}$};
	
	\draw (3,-1.25) node {\small{$\CW = \CW_{\text{gluing}} + \sum_{j=0}^{N-1} \big( Flip[V_L A_1^j \tilde{V}_L] + Flip[V_R A_{F_1+F_2-1}^j \tilde{V}_R] \big)$} };
	
\end{scope}

\epic}
\caption{On the top left corned  the electric brane set up with $K+2$ $NS$ branes, $F_1$ $D5'$ branes in the first interval and $F_2$ in the last and $N$ $D3$ branes.
The associated quiver theory, in the bottom left corner, has $K+1$ gauge nodes linked by $K$ improved bifundamental, $F_1$ flavor on the first node and $F_2$ on the last. On the top right corner the $S$-dual brane setup and on the right bottom corner its associated quiver theory.
The leftmost  and rightmost flavors are associated to the $D5'$ located outside the $NS$ branes hence the corresponding  dressed mesons are flipped. The two quiver theories are mirror dual.}
\label{fig:MoreNodes_ex}
\end{figure}

We now consider the  electric  brane setup on the top left corner of  figure \ref{fig:MoreNodes_ex} and its associated quiver theory
with $K+1$ gauge nodes linked by  $K$ improved bifundamental. It is convenient to consider the following parametrization:
\be
 \bpic[thick,node distance=3cm,gauge/.style={circle,draw,minimum size=5mm},flavor/.style={rectangle,draw,minimum size=5mm}] 
		
	\path (2,0) node[gauge](g1) {$\!\!\!N\!\!\!$} --  (4,0) node[gauge](g2) {$\!\!\!N\!\!\!$} --  (5.5,0) node(gi) {$\ldots$} 
		-- (7,0) node[gauge](g3) {$\!\!\!N\!\!\!$} -- (9,0) node[gauge](g4) {$\!\!\!N\!\!\!$} -- (1.3,1.5) node[flavor] (x1) {$\!1\!$} 
		-- (2.7,1.5) node[flavor] (x2) {$\!1\!$} -- (8.3,1.5) node[flavor] (y1) {$\!1\!$} -- (9.7,1.5) node[flavor] (y2) {$\!1\!$};
 
	\wigM (g1) -- (g2); \draw[blue] (3,0.3) node {\scriptsize{$D_1$}};
	\wigM (g2) -- (gi);
	\wigM (g3) -- (gi);
	\wigM (g3) -- (g4); \draw[blue] (8,0.3) node {\scriptsize{$D_K$}};
	
	\draw[-, shorten >= 6, shorten <= 9, shift={(-0.07,0.02)}, mid arrowsm] (1.25,1.5) -- (2,0);
	\draw[-, shorten >= 8, shorten <= 9, shift={(0.1,0)}, mid arrowsm] (2,0) -- (1.25,1.5); 
	\draw[blue] (1.55,0.9) node[left] {\scriptsize{$1-B_1$}}; \draw[blue] (1.25,2) node {\scriptsize$X_1$};
	
	\draw[-, shorten >= 8, shorten <= 8, shift={(-0.1,0.02)}, mid arrowsm] (2,0) -- (2.75,1.5);
	\draw[-, shorten >= 7, shorten <= 8, shift={(0.05,0)}, mid arrowsm] (2.75,1.5) -- (2,0); 
	\draw[blue] (2.55,0.9) node[right] {\scriptsize{$1-B_{F_1}$}}; \draw[blue] (2.75,2) node {\scriptsize$X_{F_1}$};
	
	\draw (2,1.5) node {$\cdots$};

	\draw[-, shorten >= 6, shorten <= 9, shift={(-0.07,0.02)}, mid arrowsm] (8.25,1.5) -- (9,0);
	\draw[-, shorten >= 8, shorten <= 9, shift={(0.1,0)}, mid arrowsm] (9,0) -- (8.25,1.5); 
	\draw[blue] (8.55,0.9) node[left] {\scriptsize{$1-C_1$}}; \draw[blue] (8.3,2) node {\scriptsize{$Y_1$}};
	
	\draw[-, shorten >= 8, shorten <= 8, shift={(-0.1,0.02)}, mid arrowsm] (9,0) -- (9.75,1.5);
	\draw[-, shorten >= 7, shorten <= 8, shift={(0.05,0)}, mid arrowsm] (9.75,1.5) -- (9,0); 
	\draw[blue] (9.55,0.9) node[right] {\scriptsize{$1-C_{F_2}$}}; \draw[blue] (9.7,2) node {\scriptsize{$Y_{F_2}$}};
	
	\draw (9,1.5) node {$\cdots$};

	\draw[blue] (2,-0.5) node {\scriptsize{$(W_L-W_1)$}}; \draw[blue] (4,-0.5) node {\scriptsize{$(W_1-W_2)$}};
	\draw[blue] (7,-0.5) node {\scriptsize{$(W_{K-1}$-$W_{K})$}}; \draw[blue] (9,-0.5) node {\scriptsize{$(W_K-W_R)$}};
	\draw[-] (g1) to[out=150,in=90] (1.4,0) to[out=-90,in=-150] (g1); \draw[blue] (1.4,0) node[left] {\scriptsize{$\tau$}};
	\draw[-] (g2) to[out=60,in=0] (4,0.6) to[out=180,in=120] (g2); \draw[blue] (4,0.7) node[right] {\scriptsize{$\tau$}};
	\draw[-] (g3) to[out=60,in=0] (7,0.6) to[out=180,in=120] (g3); \draw[blue] (7,0.7) node[right] {\scriptsize{$\tau$}};
	\draw[-] (g4) to[out=30,in=90] (9.6,0) to[out=-90,in=-30] (g4); \draw[blue] (9.6,0) node[right] {\scriptsize{$\tau$}};
\epic
 \label{fig:MoreNodes_Elec_manifest}
\ee
The manifest global symmetry group of the theory is given by:
\begin{align}
	S[ U(F_1)^2 \times U(F_2)^2 ] \times \prod_{j=1}^{K} U(1)_{D_j} \times \prod_{j=1}^{K-1} U(1)_{W_j-W_{j+1}} \times U(1)_{W_L-W_1} \times U(1)_{W_K-W_R} \times U(1)_\tau \,,
\end{align}
where  $U(F_1)^2$ and $U(F_2)^2$ are realised as:
\begin{align}
	&\prod_{j=1}^{F_1} \big( U(1)_{X_j} \times U(1)_{B_j} \big) = U(F_1)^2 \,, \nn\\
	&\prod_{j=1}^{F_2} \big( U(1)_{Y_j} \times U(1)_{C_j} \big) = U(F_2)^2 \,.
\end{align}
In  this case also in the electric theory we have a non-trivial symmetry enhancement in the IR where the $K-1$ topological symmetries $U(1)_{W_j-W_{j+1}}$
with $j=1,\ldots K-1$, together with the $K$ symmetries $U(1)_{D_j}$, associated to the improved bifundamentals, enhance as:
\begin{align}
	\prod_{a=1}^K U(1)_{D_a} \times \prod_{j=1}^{K-1} U(1)_{W_j - W_{j+1}} \to S[ U(K)^2 ] \,.
\end{align}
Hence the full IR global symmetry of the theory is:
\begin{align}\label{eq:ManyNodes_ElecSymm}
	S[ U(F_1)^2 \times U(F_2)^2 ] \times S[ U(K)^2 ] \times U(1)_{W_L-W_1} \times U(1)_{W_K-W_R} \times U(1)_\tau \,,
\end{align}

Now we run the dualization algorithm, as shown in appendix \ref{app:Quiver_algorithm}, and find the mirror dual quiver theory with the following parameterization:
\be 
\resizebox{.95\hsize}{!}{
 \bpic[thick,node distance=3cm,gauge/.style={circle,draw,minimum size=5mm},flavor/.style={rectangle,draw,minimum size=5mm}] 
	 
	\path (0,0) node[gauge](g1) {$\!\!\!N\!\!\!$} -- (2,0) node[gauge](g2) {$\!\!\!N\!\!\!$} -- (5,0) node[gauge] (g3) {$\!\!\!N\!\!\!$} 
	 		-- (7,0) node[gauge] (g4) {$\!\!\!N\!\!\!$} -- (9,0) node[gauge] (g5) {$\!\!\!N\!\!\!$} -- (12,0) node[gauge] (g6) {$\!\!\!N\!\!\!$} 
	 		-- (14,0) node[gauge] (g7) {$\!\!\!N\!\!\!$} -- (0,1.5) node[flavor] (x1) {$\!1\!$} -- (6.25,1.5) node[flavor] (x2) {$\!1\!$}
	 		-- (7.75,1.5) node[flavor] (x22) {$\!1\!$} -- (14,1.5) node[flavor] (x3) {$\!1\!$};
 
	\wigM (g1) -- (g2); \draw[blue] (1,0.3) node {\scriptsize{$B_2$}};
	\wigM (g2) -- (3,0); 
	\draw (3.5,0) node {$\cdots$};
	\wigM (g3) -- (4,0);
	\wigM (g3) -- (g4); \draw[blue] (6,0.3) node {\scriptsize{$B_{F_1}$}};
	\wigM (g4) -- (g5); \draw[blue] (8,0.3) node {\scriptsize{$C_1$}};
	\wigM (g5) -- (10,0);
	\draw (10.5,0) node {$\cdots$};
	\wigM (g6) -- (11,0);
	\wigM (g6) -- (g7); \draw[blue] (13,0.3) node {\scriptsize{$C_{F_2-1}$}};
	\hyper (g1) -- (x1); \draw (0,0.75) node[cross] {}; \draw[blue] (0,0.75) node [left] {\scriptsize{$\frac{1-N}{2}\tau + B_1$}}; \draw[blue] (0,2) node {\scriptsize{$W_L$}};
	
	\draw[-, shorten >= 6, shorten <= 9, shift={(-0.07,0.02)}, mid arrowsm] (6.25,1.5) -- (7,0);
	\draw[-, shorten >= 8, shorten <= 9, shift={(0.1,0)}, mid arrowsm] (7,0) -- (6.25,1.5); 
	\draw[blue] (6.625,0.75) node[left] {\scriptsize{$1-D_1$}}; \draw[blue] (6.25,2) node {\scriptsize{$W_1$}};
	
	\draw[-, shorten >= 8, shorten <= 8, shift={(-0.1,0.02)}, mid arrowsm] (7,0) -- (7.75,1.5);
	\draw[-, shorten >= 7, shorten <= 8, shift={(0.05,0)}, mid arrowsm] (7.75,1.5) -- (7,0); 
	\draw[blue] (7.4,0.75) node[right] {\scriptsize{$1-D_K$}}; \draw[blue] (7.75,2) node {\scriptsize{$W_K$}};
	
	\draw (7,1.5) node {$\cdots$};

	\hyper (g7) -- (x3); \draw (14,0.75) node[cross] {}; \draw[blue] (14,0.75) node [right] {\scriptsize{$\frac{1-N}{2}\tau + C_{F_2}$}}; \draw[blue] (14,2) node {\scriptsize{$W_R$}};
	\draw[blue] (0,-0.5) node {\scriptsize{$(X_2 - X_1)$}}; \draw[blue] (2,-0.5) node {\scriptsize{$(X_3 - X_2)$}};
	\draw[blue] (5,-0.5) node {\scriptsize{$(X_{F_1}$ - $X_{F_1-1})$}}; \draw[blue] (7,-1) node {\scriptsize{$(Y_1$ - $X_{F_1})$}};
	\draw[blue] (9,-0.5) node {\scriptsize{$(Y_2 - Y_1)$}}; \draw[blue] (11.9,-0.5) node {\scriptsize{$(Y_{F_2-1}$ - $Y_{F_2-2})$}};
	\draw[blue] (14.1,-0.5) node {\scriptsize{$(Y_{F_2}$ - $Y_{F_2-1})$}};
	\draw[-] (g1) to[out=150,in=90] (-0.6,0) to[out=-90,in=-150] (g1); \draw[blue] (-0.6,0) node[left] {\scriptsize{$\tau$}};
	\draw[-] (g2) to[out=60,in=0] (2,0.6) to[out=180,in=120] (g2); \draw[blue] (2,0.7) node[right] {\scriptsize{$\tau$}};
	\draw[-] (g3) to[out=60,in=0] (5,0.6) to[out=180,in=120] (g3); \draw[blue] (5,0.7) node[right] {\scriptsize{$\tau$}};
	\draw[-] (g4) to[out=-60,in=0] (7,-0.6) to[out=180,in=-120] (g4); \draw[blue] (7,-0.7) node[right] {\scriptsize{$\tau$}};
	\draw[-] (g5) to[out=60,in=0] (9,0.6) to[out=180,in=120] (g5); \draw[blue] (9,0.7) node[right] {\scriptsize{$\tau$}};
	\draw[-] (g6) to[out=60,in=0] (12,0.6) to[out=180,in=120] (g6); \draw[blue] (12,0.7) node[right] {\scriptsize{$\tau$}};
	\draw[-] (g7) to[out=30,in=90] (14.6,0) to[out=-90,in=-30] (g7); \draw[blue] (14.6,0) node[right] {\scriptsize{$\tau$}};

\epic}
\label{fig:MoreNodes_Magn_manifest}
\ee
The set of $K$ flavor on the central node can be reparameterized so that they are rotated by a $U(K)^2$ symmetry obtained as:
\begin{align}
	\prod_{j=1}^K \big( U(1)_{D_j} \times U(1)_{W_j} \big) = U(K)^2 \,,
\end{align}
so that the manifest global symmetry of the theory is given by:
\begin{align}
	& S[ U(K)^2 \times U(1)_{W_L} \times U(1)_{W_R} ] \times \prod_{j=1}^{F_1} U(1)_{B_j} \times \prod_{j=1}^{F_2} U(1)_{C_j} \times \nn \\
	& \times \prod_{j=1}^{F_1-1} U(1)_{X_{j+1}-X_j} \times \prod_{j=1}^{F_2-1} U(1)_{Y_{j+1}-Y_j} \times U(1)_\tau \,.
\end{align}
It is trivial to check that, after the reparameterization, the theory in \eqref{fig:MoreNodes_Magn_manifest} coincides with the quiver
in the bottom right corner in the figure \ref{fig:MoreNodes_ex} which we wrote applying our proposal \ref{proposalP} to the $\CS$-dual brane setup. \\
Similarly to the previous example, the $U(1)_{B_j}$ and $U(1)_{C_j}$ symmetries acting on each improved bifundamental and the topological symmetries
recombine to produce the enhanced IR symmetry:
\begin{align}
	\prod_{j=1}^{F_1} U(1)_{B_j} \times \prod_{j=1}^{F_1-1} U(1)_{X_{j+1}-X_j} \to S[ U(F_1)^2 ] \,, \nn \\
	\prod_{j=1}^{F_2} U(1)_{C_j} \times \prod_{j=1}^{F_2-1} U(1)_{Y_{j+1}-Y_j} \to S[ U(F_2)^2 ] \,.
\end{align}
The complete IR global symmetry of the mirror theory is then:
\begin{align}
	S[ U(F_1)^2 ] \times S[ U(F_2)^2 ] \times U(1)_{Y_1-X_{F_1}} \times S[ U(K)^2 \times U(1)_{W_L} \times U(1)_{W_R} ] \times U(1)_\tau \,,
\end{align}
that upon a redefinition of some $U(1)$ factors, it matches precisely with the IR global symmetry of the original theory in \eqref{eq:ManyNodes_ElecSymm}.

\subsection{$3d$ $\cN=2$ improved quivers: the good, the bad and  the ugly}\label{sec:GBU}
In this section we extend the $\cN=4$ quivers notion of balanced ($N_F=2N_C$), good ($N_F \geq 2N_C$), ugly ($N_F=2N_C-1$), and bad ($N_F<2N_C-1$) nodes
of \cite{Gaiotto:2008ak}, to the $\cN=2$ improved quivers with constant ranks. We expect that a similar story holds for $\cN=2$ improved quivers with non-constant ranks.

\subsubsection*{Comments on symmetry enhancement: balancing condition}
Looking back at all the theories presented up to this point, namely the $1$, $2$ and $K$ node examples together with their mirror duals, a recurring pattern of global symmetry enhancement can be seen, we now want to collect all these hints to formulate a general rule to recognize the enhancement of symmetries. \\
Let us start from the $K$ nodes example which is given in  \eqref{fig:MoreNodes_Elec_manifest}. 
Let us isolate from the theory the sequence of improved bifundamentals:
\be
 \bpic[thick,node distance=3cm,gauge/.style={circle,draw,minimum size=5mm},flavor/.style={rectangle,draw,minimum size=5mm}] 
		
	\path (2,0) node[gauge](g1) {$\!\!\!N\!\!\!$} --  (4,0) node[gauge](g2) {$\!\!\!N\!\!\!$} --  (5.5,0) node(gi) {$\ldots$} 
		-- (7,0) node[gauge](g3) {$\!\!\!N\!\!\!$} -- (9,0) node[gauge](g4) {$\!\!\!N\!\!\!$};
 
 	\draw (0.5,0) node {$\cdots$};
 	\wigM (g1) -- (1,0); \draw[blue] (1.5,0.3) node {\scriptsize{$D_1$}};
	\wigM (g1) -- (g2); \draw[blue] (3,0.3) node {\scriptsize{$D_2$}};
	\wigM (g2) -- (gi);
	\wigM (g3) -- (gi);
	\wigM (g3) -- (g4); \draw[blue] (8,0.3) node {\scriptsize{$D_{K-1}$}};
	\wigM (g4) -- (10,0); \draw[blue] (9.5,0.3) node {\scriptsize{$D_K$}};
	\draw (10.5,0) node {$\cdots$};
	\draw[blue] (2,-0.5) node {\scriptsize{$(W_1-W_2)$}}; \draw[blue] (4,-0.5) node {\scriptsize{$(W_2-W_3)$}};
	\draw[blue] (7,-0.5) node {\scriptsize{$(W_{K-2}$-$W_{K-1})$}}; \draw[blue] (9,-0.5) node {\scriptsize{$(W_{K-1}$-$W_{K})$}}; 
	\draw[-] (g1) to[out=60,in=0] (2,0.6) to[out=180,in=120] (g1); \draw[blue] (2,0.7) node[right] {\scriptsize{$\tau$}};
	\draw[-] (g2) to[out=60,in=0] (4,0.6) to[out=180,in=120] (g2); \draw[blue] (4,0.7) node[right] {\scriptsize{$\tau$}};
	\draw[-] (g3) to[out=60,in=0] (7,0.6) to[out=180,in=120] (g3); \draw[blue] (7,0.7) node[right] {\scriptsize{$\tau$}};
	\draw[-] (g4) to[out=60,in=0] (9,0.6) to[out=180,in=120] (g4); \draw[blue] (9,0.7) node[right] {\scriptsize{$\tau$}};
 
\epic\ee
This structure of $K$ improved bifundamentals gives a global symmetry enhancement obtained from the $K$ $U(1)_{D_a}$ axial symmetries associated to the improved bifundamentals and the $K-1$ $U(1)_{W_j-W_{j+1}}$ topological symmetries. Together these symmetries enhance to a $U(K)^2/U(1)$ non-abelian global symmetry group. Let's take the Cartan's of $U(K)^2$ to be $\vec{M}$ and $\vec{N}$, we have the following relations:
\begin{align}\label{eq:UK_parametrization}
	& M_a = W_a - D_a \,, \nn \\
	& N_a = W_a + D_a \,,
\end{align}
from which we notice that the $U(1)_{D_a}$ symmetries parameterize the axial-like and the $U(1)_{W_a}$ the vector-like subgroup of $U(K)^2$. \\
We then look at the mirror theory given in \eqref{fig:MoreNodes_Magn_manifest}, both on the left and on the right of the quiver we observe a string of improved bifundamental ending with a single flipped flavor. Let's focus on the left part only and isolate the following structure:
\be
 \bpic[thick,node distance=3cm,gauge/.style={circle,draw,minimum size=5mm},flavor/.style={rectangle,draw,minimum size=5mm}] 
		
	\path (2,0) node[gauge](g1) {$\!\!\!N\!\!\!$} --  (4,0) node[gauge](g2) {$\!\!\!N\!\!\!$} --  (5.5,0) node(gi) {$\ldots$} 
		-- (7,0) node[gauge](g3) {$\!\!\!N\!\!\!$} -- (9,0) node[gauge](g4) {$\!\!\!N\!\!\!$} -- (2,1.5) node[flavor](x) {$\!1\!$};
 
 	\draw[-, shorten >= 3, shorten <= 12, shift={(-0.05,-0.15)}, middx arrowsm] (2,0) -- (2,1.5);
	\draw[-, shorten >= 8, shorten <= 7, shift={(0.05,0)}, midsx arrowsm] (2,1.5) -- (2,0);
 	\draw (2,0.5) node[cross] {}; \draw[blue] (2,0.75) node[left] {\scriptsize{$\frac{1-N}{2}\tau + B_1$}};
 	
	\wigM (g1) -- (g2); \draw[blue] (3,0.3) node {\scriptsize{$B_2$}};
	\wigM (g2) -- (gi);
	\wigM (g3) -- (gi);
	\wigM (g3) -- (g4); \draw[blue] (8,0.3) node {\scriptsize{$B_{K-1}$}};
	\wigM (g4) -- (10,0); \draw[blue] (9.5,0.3) node {\scriptsize{$B_K$}};
	\draw (10.5,0) node {$\cdots$};
	\draw[blue] (2,-0.5) node {\scriptsize{$(X_2-X_1)$}}; \draw[blue] (4,-0.5) node {\scriptsize{$(X_3-X_2)$}};
	\draw[blue] (7,-0.5) node {\scriptsize{$(X_{K-1}$-$X_{K-2})$}}; \draw[blue] (9,-0.5) node {\scriptsize{$(X_{K}$-$X_{K-1})$}}; 
	\draw[-] (g1) to[out=150,in=90] (1.4,0) to[out=-90,in=-150] (g1); \draw[blue] (1.4,0) node[left] {\scriptsize{$\tau$}};
	\draw[-] (g2) to[out=60,in=0] (4,0.6) to[out=180,in=120] (g2); \draw[blue] (4,0.7) node[right] {\scriptsize{$\tau$}};
	\draw[-] (g3) to[out=60,in=0] (7,0.6) to[out=180,in=120] (g3); \draw[blue] (7,0.7) node[right] {\scriptsize{$\tau$}};
	\draw[-] (g4) to[out=60,in=0] (9,0.6) to[out=180,in=120] (g4); \draw[blue] (9,0.7) node[right] {\scriptsize{$\tau$}};
 
\epic\ee
We observe that the $U(1)_{B_a}$ and $U(1)_{X_{j+1}-X_j}$ symmetries enhance to a $U(F_1)^2/U(1)$, where the parameterization is given analogously to that in \eqref{eq:UK_parametrization}. The same enhancement happens also in the mirror of the two node theory in \eqref{fig:TwoNodes_Magn}. \\
Let us now look finally to the mirror of the SQCD in figure \ref{fig:SQCD_Dual}, we have a string of improved bifundamentals ending on both sides with a flipped flavor:
\be
 \bpic[thick,node distance=3cm,gauge/.style={circle,draw,minimum size=5mm},flavor/.style={rectangle,draw,minimum size=5mm}] 
		
	\path (2,0) node[gauge](g1) {$\!\!\!N\!\!\!$} --  (4,0) node[gauge](g2) {$\!\!\!N\!\!\!$} --  (5.5,0) node(gi) {$\ldots$} 
		-- (7,0) node[gauge](g3) {$\!\!\!N\!\!\!$} -- (9,0) node[gauge](g4) {$\!\!\!N\!\!\!$} -- (2,1.5) node[flavor](x1) {$\!1\!$}
		-- (9,1.5) node[flavor](x2) {$\!1\!$};
 
 	\draw[-, shorten >= 3, shorten <= 12, shift={(-0.05,-0.15)}, middx arrowsm] (2,0) -- (2,1.5);
	\draw[-, shorten >= 8, shorten <= 7, shift={(0.05,0)}, midsx arrowsm] (2,1.5) -- (2,0);
	\draw (2,0.5) node[cross] {}; \draw[blue] (2,0.75) node[left] {\scriptsize{$\frac{1-N}{2}\tau + B_1$}};
	
	\wigM (g1) -- (g2); \draw[blue] (3,0.3) node {\scriptsize{$B_2$}};
	\wigM (g2) -- (gi);
	\wigM (g3) -- (gi);
	\wigM (g3) -- (g4); \draw[blue] (8,0.3) node {\scriptsize{$B_{F-1}$}};
	
	\draw[-, shorten >= 3, shorten <= 12, shift={(-0.05,-0.15)}, middx arrowsm] (9,0) -- (9,1.5);
	\draw[-, shorten >= 8, shorten <= 7, shift={(0.05,0)}, midsx arrowsm] (9,1.5) -- (9,0);
	\draw (9,0.5) node[cross] {}; \draw[blue] (9,0.75) node[right] {\scriptsize{$\frac{1-N}{2}\tau + B_F$}};
	
	\draw[blue] (2,-0.5) node {\scriptsize{$(X_2-X_1)$}}; \draw[blue] (4,-0.5) node {\scriptsize{$(X_3-X_2)$}};
	\draw[blue] (7,-0.5) node {\scriptsize{$(X_{F-1}$-$X_{F-2})$}}; \draw[blue] (9,-0.5) node {\scriptsize{$(X_{F}$-$X_{F-1})$}}; 
	\draw[-] (g1) to[out=150,in=90] (1.4,0) to[out=-90,in=-150] (g1); \draw[blue] (1.4,0) node[left] {\scriptsize{$\tau$}};
	\draw[-] (g2) to[out=60,in=0] (4,0.6) to[out=180,in=120] (g2); \draw[blue] (4,0.7) node[right] {\scriptsize{$\tau$}};
	\draw[-] (g3) to[out=60,in=0] (7,0.6) to[out=180,in=120] (g3); \draw[blue] (7,0.7) node[right] {\scriptsize{$\tau$}};
	\draw[-] (g4) to[out=30,in=90] (9.6,0) to[out=-90,in=-30] (g4); \draw[blue] (9.6,0) node[right] {\scriptsize{$\tau$}};
 
\epic\ee
As discussed in section \ref{sqcdmirror} we have  symmetry enhancement involving all the $U(1)_{B_i}$ and $U(1)_{X_i}$ symmetries.\\
Collecting all these observations we can give a definition for a balanced node:

\vspace{0.4cm}

\begin{quote}\emph{
A node is balanced if it joins two improved bifundamentals or if it joins an improved bifundamental to a flipped flavor.
}\end{quote}


\subsubsection*{Ugly and Bad quivers}
We now study the following brane setup and its QFT description:
\be
 \bpic[thick,node distance=3cm,gauge/.style={circle,draw,minimum size=5mm},flavor/.style={rectangle,draw,minimum size=5mm}] 
	
	\NSbrane (-8,1.5) -- (-8,-1.5);
	\NSbrane (-7.5,1.5) -- (-7.5,-1.5);
	\draw (-7,0) node {$\cdots$};
	\NSbrane (-6.5,1.5) -- (-6.5,-1.5);
	\Dbrane (-6.2,1.5) -- (-5.6,-1.5);
	\draw (-5.5,0) node {$\cdots$};
	\Dbrane (-5.2,1.5) -- (-4.6,-1.5);
	\NSbrane (-4.5,1.5) -- (-4.5,-1.5);
	\Dthree (-8,-0.5) -- (-4.5,-0.5);
	\Dthree (-8,-0.4) -- (-4.5,-0.4);
	\draw[|-|,blue] (-8,-1.7) -- (-6.5,-1.7);	\draw  (-7.25,-2) node {$\# K$};
	\draw[|-|,gray] (-6.2,-1.7) -- (-4.6,-1.7);	\draw  (-5.4,-2) node {$\# F$};
	
	\draw (-3.5,0) node {$\Longleftrightarrow$};

\begin{scope}[shift={(0,-0.75)}]
	\path (-2,0) node[gauge](g1) {$\!\!\!N\!\!\!$} -- (0,0)  node[gauge](g2) {$\!\!\!N\!\!\!$}
		-- (3,0)  node[gauge](g3) {$\!\!\!N\!\!\!$} -- (5,0)  node[gauge](g4) {$\!\!\!N\!\!\!$} 
		-- (4.5,1.5) node[flavor](x1) {$\!F\!$} -- (5.5,1.5)  node[flavor](x2) {$\!F\!$};
		
	\wigM (g1) -- (g2);
	\wigM (g2) -- (1,0);
	\draw (1.5,0) node {$\cdots$};
	\wigM (g3) -- (2,0);
	\wigM (g3) -- (g4);
	\chir (g4) -- (x1);
	\chir (x2) -- (g4);
	\draw[-] (g1) to[out=-60,in=0] (-2,-0.6) to[out=180,in=-120] (g1); 
	\draw[-] (g2) to[out=-60,in=0] (0,-0.6) to[out=180,in=-120] (g2);
	\draw[-] (g3) to[out=-60,in=0] (3,-0.6) to[out=180,in=-120] (g3);
	\draw[-] (g4) to[out=-60,in=0] (5,-0.6) to[out=180,in=-120] (g4);
\end{scope}

\epic
\label{fig:ugly_example}
\ee
We focus on the QFT. We have a sequence of $K-1$ improved bifundamentals ending with $F$ flavors on the last node.
Using  the fact that an improved bifundamental gauged on one side  {\it confines} to $N$ free hypers  as
explained in appendix \ref{app:3d_braid}, we can  sequentially confine all the  improved bifundamentals.
So the QFT associated to this brane setup is given by  $(K-1)\times N$  free hypers and a $U(N)$ adjoint SQCD with $F$ flavors:
\be\label{fig:ugly_exaple_elec}
\resizebox{.9\hsize}{!}{
 \bpic[thick,node distance=3cm,gauge/.style={circle,draw,minimum size=5mm},flavor/.style={rectangle,draw,minimum size=5mm}] 
	 
	\path (-2,0) node[gauge](g1) {$\!\!\!N\!\!\!$} -- (0,0)  node[gauge](g2) {$\!\!\!N\!\!\!$}
		-- (3,0)  node[gauge](g3) {$\!\!\!N\!\!\!$} -- (5,0)  node[gauge](g4) {$\!\!\!N\!\!\!$} 
		-- (4.5,1.5) node[flavor](x1) {$\!F\!$} -- (5.5,1.5)  node[flavor](x2) {$\!F\!$};
		
	\wigM (g1) -- (g2);
	\wigM (g2) -- (1,0);
	\draw (1.5,0) node {$\cdots$};
	\wigM (g3) -- (2,0);
	\wigM (g3) -- (g4);
	\chir (g4) -- (x1);
	\chir (x2) -- (g4);
	\draw[-] (g1) to[out=-60,in=0] (-2,-0.6) to[out=180,in=-120] (g1);
	\draw[-] (g2) to[out=-60,in=0] (0,-0.6) to[out=180,in=-120] (g2);
	\draw[-] (g3) to[out=-60,in=0] (3,-0.6) to[out=180,in=-120] (g3);
	\draw[-] (g4) to[out=-60,in=0] (5,-0.6) to[out=180,in=-120] (g4);
	
	\draw (6.5,0) node {$=$};
	
\begin{scope}[shift={(-0.5,0)}]
	\path (9.5,0) node {$(\text{Free Hyper})^{N\times(K-1)}$} -- (12,0) node[gauge](g) {$\!\!\!N\!\!\!$} 
		-- (11.5,1.5) node[flavor](x1) {$\!F\!$} -- (12.5,1.5) node[flavor](x2) {$\!F\!$};
		
	\chir (g) -- (x1);
	\chir (x2) -- (g);
	\draw[-] (g) to[out=-60,in=0] (12,-0.6) to[out=180,in=-120] (g);
\end{scope}

\epic}\ee

We thus call a node attached only to an improved bifundamental an \emph{ugly} node, in analogy with $\cN=4$ $U(N)$ with $2N-1$ flavors whose monopole has $\Delta=\frac{1}{2}$ and is a free field.

A $U(N)$ node attached to two improved bifundamentals  with $F\geq 0$ flavors, or a $U(N)$ node attached to one improved bifundamental  with $F\geq 1$ flavors is \emph{good} and does not provide free decoupled fields.

Let's now consider the $\CS$-dual configuration of \eqref{fig:ugly_example}, which is given by:
\be
 \bpic[thick,node distance=3cm,gauge/.style={circle,draw,minimum size=5mm},flavor/.style={rectangle,draw,minimum size=5mm}] 
	
	\Dbrane (-8.2,1.5) -- (-7.7,-1.5);
	\Dbrane (-7.7,1.5) -- (-7.2,-1.5);
	\draw (-7,0) node {$\cdots$};
	\Dbrane (-6.7,1.5) -- (-6.2,-1.5);
	\NSbrane (-6,1.5) -- (-6,-1.5);
	\NSbrane (-5,1.5) -- (-5,-1.5);
	\draw (-5.5,0) node {$\cdots$};
	\Dbrane (-4.7,1.5) -- (-4.2,-1.5);
	\Dthree (-7.89,-0.5) -- (-4.37,-0.5);
	\Dthree (-7.91,-0.4) -- (-4.35,-0.4);
	\draw[|-|,gray] (-8.2,-1.7) -- (-6.2,-1.7);	\draw  (-7.25,-2) node {$\# K$};
	\draw[|-|,blue] (-6,-1.7) -- (-5,-1.7);	\draw  (-5.5,-2) node {$\# F$};
	 
%
	
\epic\ee
We can find the QFT description of this setup by applying the dualization algorithm to the electric theory.  After we have dualized each block composing the theory, we find the following intermediate step:
\be\label{fig:ugly_example_step}
 \bpic[thick,node distance=3cm,gauge/.style={circle,draw,minimum size=5mm},flavor/.style={rectangle,draw,minimum size=5mm}] 
	
	\path (0,0) node[flavor](x1) {$\!0\!$} -- (2,0) node[gauge](g1) {$\!\!\!N\!\!\!$} -- (4,0) node[gauge](g2) {$\!\!\!N\!\!\!$} 
		-- (6,0) node[gauge](g3) {$\!\!\!N\!\!\!$} -- (9,0) node[gauge](g4) {$\!\!\!N\!\!\!$} -- (11,0) node[gauge](g5) {$\!\!\!N\!\!\!$}
		-- (11,1.5) node[flavor](x2) {$\!1\!$} -- (0,1.5) node[flavor](v1) {$\!1\!$}
		-- (3.5,1.5) node[flavor](y1) {\!\tiny{$K$-$1$}\!} -- (4.5,1.5) node[flavor](y2) {\!\tiny{$K$-$1$}\!};
	
	\wigT (x1) -- (g1); \draw (1,-0.3) node {$-$};
	\wigT (v1) -- (1,0);
	\wigT (g1) -- (g2); \draw (3,-0.3) node {$+$};
	\wigM (g2) -- (g3);
	\wigM (g3) -- (7,0);
	\wigM (g4) -- (8,0);
	\wigM (g4) -- (g5);
	
	\draw[-, shorten >= 3, shorten <= 12, shift={(-0.05,-0.15)}, middx arrowsm] (11,0) -- (11,1.5);
	\draw[-, shorten >= 8, shorten <= 7, shift={(0.05,0)}, midsx arrowsm] (11,1.5) -- (11,0);
	\draw (11,0.5) node[cross] {};
	
	\chir (g2) -- (y1);
	\chir (y2) -- (g2);
	\draw[-] (g1) to[out=-60,in=0] (2,-0.6) to[out=180,in=-120] (g1);
	\draw[-] (g2) to[out=-60,in=0] (4,-0.6) to[out=180,in=-120] (g2);
	\draw[-] (g3) to[out=-60,in=0] (6,-0.6) to[out=180,in=-120] (g3);
	\draw[-] (g4) to[out=-60,in=0] (9,-0.6) to[out=180,in=-120] (g4);
	\draw[-] (g5) to[out=-60,in=0] (11,-0.6) to[out=180,in=-120] (g5);
	
\epic\ee
We now implement the asymmetric $\mathbb{I}$-wall on the left given by the two $\CS$-walls glued together, which,  repeating the steps in  section \ref{algsec}, leads to the following theory:
\be
 \bpic[thick,node distance=3cm,gauge/.style={circle,draw,minimum size=5mm},flavor/.style={rectangle,draw,minimum size=5mm}] 
	
	\path (1,0) node {$(\text{Free Hyper})^{N\times(K-1)}$} -- (4,0) node[gauge](g2) {$\!\!\!N\!\!\!$} 
		-- (6,0) node[gauge](g3) {$\!\!\!N\!\!\!$} -- (9,0) node[gauge](g4) {$\!\!\!N\!\!\!$} -- (11,0) node[gauge](g5) {$\!\!\!N\!\!\!$}
		-- (4,1.5) node[flavor](x1) {$\!1\!$} -- (11,1.5) node[flavor](x2) {$\!1\!$};
	
	\draw[-, shorten >= 3, shorten <= 12, shift={(-0.05,-0.15)}, middx arrowsm] (4,0) -- (4,1.5);
	\draw[-, shorten >= 8, shorten <= 7, shift={(0.05,0)}, midsx arrowsm] (4,1.5) -- (4,0);
	\draw (4,0.5) node[cross] {};
	
	\wigM (g2) -- (g3);
	\wigM (g3) -- (7,0);
	\wigM (g4) -- (8,0);
	\wigM (g4) -- (g5);
	
	\draw[-, shorten >= 3, shorten <= 12, shift={(-0.05,-0.15)}, middx arrowsm] (11,0) -- (11,1.5);
	\draw[-, shorten >= 8, shorten <= 7, shift={(0.05,0)}, midsx arrowsm] (11,1.5) -- (11,0);
	\draw (11,0.5) node[cross] {};
	
	\draw[-] (g2) to[out=-60,in=0] (4,-0.6) to[out=180,in=-120] (g2);
	\draw[-] (g3) to[out=-60,in=0] (6,-0.6) to[out=180,in=-120] (g3);
	\draw[-] (g4) to[out=-60,in=0] (9,-0.6) to[out=180,in=-120] (g4);
	\draw[-] (g5) to[out=-60,in=0] (11,-0.6) to[out=180,in=-120] (g5);

\epic
\label{fig:ugly_example_magn}
\ee
More precisely,  implementing the asymmetric $\mathbb{I}$-wall  has the effect of Higgsing completely the second $U(N)$ gauge node
 in \eqref{fig:ugly_example_step} down to a flavor $U(1)$. Let us now analyze separately the effect of this Higgsing on the first improved bifundamental and on the $K-1$ flavors to show how the result \eqref{fig:ugly_example_magn} is obtained. The first improved bifundamental becomes asymmetric (see appendix \ref{app:FM}), using the duality \eqref{fig:FMtoFlav} it is dual to a flipped flavor for the third $U(N)$ gauge node. By taking this effect into account, we see that we have a string of $F-2$ improved bifundamentals with a flipped flavor on the two sides, which is the theory depicted on the right side of  \eqref{fig:ugly_example_magn}. After the Higgsing, the $K-1$ flavors for the second $U(N)$ gauge node in \eqref{fig:ugly_example_step} become just a set of $2N(K-1)$ chirals that do not interact, therefore these are $N(K-1)$ free hypermultiplets, as written on the left of  \eqref{fig:ugly_example_magn}.\\
We can compare the result for the mirror theory in \eqref{fig:ugly_example_magn} with that of the electric theory in \eqref{fig:ugly_exaple_elec} to see that in both frames we have $K-1$ set of $N$ free hypers times an interacting theory. We can notice that the two interacting theories are mirror dual to each other by means of the duality proposed in \ref{fig:SQCD_Dual}. This analysis suggest the following recipe: given a $\CN=2$ brane setup (with constant number of $N$ D3 branes) starting with a sequence of $K$ $NS$ or $D5'$ branes, the first $K-1$ branes will decouple from the theory giving a set of $N$ free hypers. 

\vspace{0.3cm}

Let us now consider the following brane setup and its QFT description:
\be\label{fig:bad_example}
 \bpic[thick,node distance=3cm,gauge/.style={circle,draw,minimum size=5mm},flavor/.style={rectangle,draw,minimum size=5mm}] 
	
	\NSbrane (-8,1.5) -- (-8,-1.5);
	\NSbrane (-7.5,1.5) -- (-7.5,-1.5);
	\draw (-7,0) node {$\cdots$};
	\NSbrane (-6.5,1.5) -- (-6.5,-1.5);
	\draw[|-|,blue] (-8,-1.7) -- (-6.5,-1.7);	\draw  (-7.25,-2) node {$\# K$};

\begin{scope}[shift={(-2,0)}]
	\path (-3.5,0) node {$\Longrightarrow$} -- (-2,0) node[gauge](g1) {$\!\!\!N\!\!\!$} -- (0,0)  node[gauge](g2) {$\!\!\!N\!\!\!$}
		-- (3,0)  node[gauge](g3) {$\!\!\!N\!\!\!$} -- (5,0)  node[gauge](g4) {$\!\!\!N\!\!\!$};
		
	\wigM (g1) -- (g2);
	\wigM (g2) -- (1,0);
	\draw (1.5,0) node {$\cdots$};
	\wigM (g3) -- (2,0);
	\wigM (g3) -- (g4);
	\draw[-] (g1) to[out=-60,in=0] (-2,-0.6) to[out=180,in=-120] (g1);
	\draw[-] (g2) to[out=-60,in=0] (0,-0.6) to[out=180,in=-120] (g2);
	\draw[-] (g3) to[out=-60,in=0] (3,-0.6) to[out=180,in=-120] (g3);
	\draw[-] (g4) to[out=-60,in=0] (5,-0.6) to[out=180,in=-120] (g4);
\end{scope}

\epic\ee
We notice that this setup preserves eight supercharges and not just four. Therefore, we expect that our prescription gives the same result as the $\CN=4$ case. 
As in the previous case we can sequentailly confine all the improved bifundamentals  producing  $(K-1)\times N $  free hypers:
\be\label{fig:bad_exaple_elec}
 \bpic[thick,node distance=3cm,gauge/.style={circle,draw,minimum size=5mm},flavor/.style={rectangle,draw,minimum size=5mm}] 
	 
	\path (-2,0) node[gauge](g1) {$\!\!\!N\!\!\!$} -- (0,0)  node[gauge](g2) {$\!\!\!N\!\!\!$}
		-- (3,0)  node[gauge](g3) {$\!\!\!N\!\!\!$} -- (5,0)  node[gauge](g4) {$\!\!\!N\!\!\!$};
		
	\wigM (g1) -- (g2);
	\wigM (g2) -- (1,0);
	\draw (1.5,0) node {$\cdots$};
	\wigM (g3) -- (2,0);
	\wigM (g3) -- (g4);
	\draw[-] (g1) to[out=-60,in=0] (-2,-0.6) to[out=180,in=-120] (g1);
	\draw[-] (g2) to[out=-60,in=0] (0,-0.6) to[out=180,in=-120] (g2);
	\draw[-] (g3) to[out=-60,in=0] (3,-0.6) to[out=180,in=-120] (g3);
	\draw[-] (g4) to[out=-60,in=0] (5,-0.6) to[out=180,in=-120] (g4);
	
	\path (6,0) node {$=$} -- (8.5,0) node {$(\text{Free Hyper})^{N\times(K-1)}$} -- (11,0) node[gauge](g) {$\!\!\!N\!\!\!$};
	
	\draw[-] (g) to[out=-60,in=0] (11,-0.6) to[out=180,in=-120] (g);
	
\epic\ee
In addition to the free hypers we are left with a $\CN=4$ $U(N)$ pure SYM theory which,  as shown in \cite{Giacomelli:2023zkk,Assel:2018exy}, is a bad theory  described by $N$ free hypers. All in all, the theory \eqref{fig:bad_example} is just given as $K\times N$ free hypers and is a \emph{bad} theory.

\subsection{Comments on previous proposals of mirror symmetry with $4$ supercharges}\label{oldprop}
Non-abelian $3d$  mirror symmetry with $4$ supercharges has been discussed in \cite{deBoer:1997ka, Aharony:1997ju, Giacomelli:2017vgk, Benvenuti:2018bav, Giacomelli:2019blm}.

In this subsection we focus on the case of $U(N)$ SQCD, and compare our proposal for the mirror dual  with the \emph{naive} proposal 
by \cite{Benvenuti:2018bav} depicted in Figure  \ref{fig:SQCD_mirror_old}.
 The naive mirror dual  is a quiver  theory with $F-1$ $U(N)$ gauge nodes linked by standard flipped bifundamental fields that come coupled to the adjoint fields via cubic superpotential terms. Moreover we have two towers of singlets flipping the meson built from the two vertical flavors dressed with the adjoint fields, exactly as in the theories described in this paper. The proposal in figure  \ref{fig:SQCD_mirror_old} was based on a \emph{naive} reading on the magnetic brane setup of \eqref{branesqcd}, analogous to \cite{deBoer:1997ka, Aharony:1997ju, Giacomelli:2017vgk, Benvenuti:2018bav, Giacomelli:2019blm}. A very similar proposal (with standard bifundamentals) for the mirror of $3d$ $\cN=2$ $U(N)$ SQCD without adjoint appeared before in \cite{Giacomelli:2017vgk}. Both for $SU(N)$ and $U(N)$, as was already noticed, the naive proposals suffer from a mismatch in the number of UV global symmetries, namely the mirror quiver has much fewer global symmetries than the SQCD, which, having zero superpotential enjoies a chiral $U(F)^2/U(1)$ symmetry.  For instance, the naive  mirror dual of figure \ref{fig:SQCD_mirror_old}  has UV global symmetry $U(1)^{F-1}\times U(1)_\tau\times U(1) $, only half of the Cartans of the electric theory.

\begin{figure}[h!]
\centering
\begin{tikzpicture}[thick,node distance=3cm,gauge/.style={circle,draw,minimum size=5mm},flavor/.style={rectangle,draw,minimum size=5mm}] 
 
\begin{scope}[shift={(-0.5,0)}]
	\path (0,0) node[gauge](g) {$\!\!\!N\!\!\!$} -- (-0.5,1.5) node[flavor] (x1) {$\!F\!$} -- (0.5,1.5) node[flavor] (x2) {$\!F\!$} 
		-- (2,0) node{$\Longleftrightarrow$};
	
	\chir (g) -- (x1); \draw (-0.3,0.6) node[left] {$Q$};
	\chir (x2) -- (g); \draw (0.3,0.6) node[right] {$\tilde{Q}$};
	\draw[-] (g) to[out=-60,in=0] (0,-0.6) to[out=180,in=-120] (g); \draw (0,-0.6) node[right] {$A$};
	
	\draw (0,-1.5) node{$\cW = 0$};
\end{scope}

	\path (5,0) node[gauge](g1) {$\!\!\!N\!\!\!$} -- (6.5,0) node[gauge](g2) {$\!\!\!N\!\!\!$} -- (8.5,0) node[gauge](g3) {$\!\!\!N\!\!\!$} -- (10,0) node[gauge](g4) {$\!\!\!N\!\!\!$} -- (5,1.5) node[flavor](y1) {$\!1\!$} -- (10,1.5) node[flavor](y2) {$\!1\!$};
	 
	\draw[-, shorten >= 5, shorten <= 11, shift={(-0.1,0.05)}, middx arrowsm] (5,0) -- (6.5,0); 
	\draw[-, shorten >= 5, shorten <= 11, shift={(0.1,-0.05)}, midsx arrowsm] (6.5,0) -- (5,0);
	\draw (5.5,0) node[cross] {}; \draw (5.75,0.35) node {$\Pi_2$};
	
	\draw[-, shorten >= 0, shorten <= 7, shift={(0.05,0.05)}, middx arrowsm] (6.5,0) -- (7.1,0); 
	\draw[-, shorten >= 5, shorten <= 2, shift={(0.1,-0.05)}, mid arrowsm] (7.1,0) -- (6.5,0);
	
	\draw[-, shorten >= 2, shorten <= 3, shift={(-0.2,0.05)}, middx arrowsm] (7.9,0) -- (8.5,0); 
	\draw[-, shorten >= 2, shorten <= 4, shift={(-0.15,-0.05)}, mid arrowsm] (8.5,0) -- (7.9,0);
	
	\draw[-, shorten >= 5, shorten <= 11, shift={(-0.1,0.05)}, middx arrowsm] (8.5,0) -- (10,0); 
	\draw[-, shorten >= 5, shorten <= 11, shift={(0.1,-0.05)}, midsx arrowsm] (10,0) -- (8.5,0);
	\draw (9,0) node[cross] {}; \draw (9.25,0.35) node {$\Pi_{F-1}$};
	
	\draw[-, shorten >= 3, shorten <= 12, shift={(-0.05,-0.15)}, middx arrowsm] (5,0) -- (5,1.5);
	\draw[-, shorten >= 8, shorten <= 7, shift={(0.05,0)}, midsx arrowsm] (5,1.5) -- (5,0);
	\draw (5,0.5) node[cross] {}; \draw (5,0.75) node[left] {$V_1$};
	
	\draw[-, shorten >= 3, shorten <= 12, shift={(-0.05,-0.15)}, middx arrowsm] (10,0) -- (10,1.5);
	\draw[-, shorten >= 8, shorten <= 7, shift={(0.05,0)}, midsx arrowsm] (10,1.5) -- (10,0);
	\draw (10,0.5) node[cross] {}; \draw (10,0.75) node[right] {$V_2$};
	
	\draw (7.5,0) node {$\cdots$};
	
	\draw[-] (g1) to[out=-60,in=0] (5,-0.6) to[out=180,in=-120] (g1); \draw (5,-0.6) node[right] {$A_1$};
	\draw[-] (g2) to[out=-60,in=0] (6.5,-0.6) to[out=180,in=-120] (g2); \draw (6.5,-0.6) node[right] {$A_2$};
	\draw[-] (g3) to[out=-60,in=0] (8.5,-0.6) to[out=180,in=-120] (g3); \draw (8.5,-0.6) node[right] {$A_{F-2}$};
	\draw[-] (g4) to[out=-60,in=0] (10,-0.6) to[out=180,in=-120] (g4); \draw (10,-0.6) node[right] {$A_{F-1}$};
	
	\draw (7.5,-1.5) node {$\cW = \sum_{j=2}^{F-2} A_j (\Pi_j \tilde{\Pi}_j + \Pi_{j+1} \tilde{\Pi}_{j+1}) + \sum_{j=2}^{F-1} Flip[\Pi_j \tilde{\Pi}_j] +$};
	\draw (7.5,-2.2) node {$\sum_{j=0}^{N-1} ( Flip[V_1 A_1^j \tilde{V}_1] + Flip[V_2 A_{F-1}^j \tilde{V}_2] ) $  };
	 
\end{tikzpicture}
\caption{Naive proposal for the mirror pair of the $\CN=2$ $U(N)$ adjoint SQCD. The superpotential $\CW_{\CN=4}$ in the mirror theory contains all the superpotential terms coupling each adjoint field to the bifundamentals besides it.}
\label{fig:SQCD_mirror_old}
\end{figure}

One argument in support of the proposals in \cite{Giacomelli:2017vgk,Benvenuti:2018bav} was provided in \cite{Giacomelli:2017vgk, Giacomelli:2019blm} showing that  the naive mirror pairs can be obtained  by starting from a well established $\CN=4$ mirror  pair by turning on  a  superpotential deformations to land on the $\CN=2$ dualities of  \cite{Giacomelli:2017vgk, Benvenuti:2018bav}. Because of this strategy, the resulting mirror theory inherits from the $\CN=4$ superpotential the cubic couplings between bifundamentals and the adjoint fields. However this strategy also produces additional superpotential terms. In the case of $U(N)$ with adjoint there is an additional $\cW=V_1 A_1^{N} \tilde{V}_1+V_2 A_{F-1}^{N} \tilde{V}_2$. These terms are zero in the chiral ring of the theory on the right of figure  \ref{fig:SQCD_mirror_old}, hence they violate the chiral stability condition \cite{Benvenuti:2017lle}. Simply removing these two terms, if $F>2$, breaks the degeneracy between the monopole operators that are supposed to map to the electric mesons, hence rendering the mapping of the operators problematic.

As already mentioned, our mirror dual can not be deformed to the naive one of figure \ref{fig:SQCD_mirror_old}. In order to do such a deformation,  we would need to iron all our improved bifundamentals to standard ones. To do so we have two options. 

The first one corresponds to  adding linearly to the superpotential the singlet $\mathsf{B}^{(k)}_{1,2}$ as we did in section \ref{ironing} when discussing the $\mathcal{N}=4$ limit.
 Keeping track of the adjoints appearing in the ironing duality \eqref{ironduality} we would then find that all  nodes, apart from the first and the last one, will have an adjoint of charge $2-\tau$ which couples to the bifundamentals to its right and to its left. So we reach a theory different from the  mirror dual of \ref{fig:SQCD_mirror_old}.  

To reach a mirror theory where also the first and last gauge node have an adjoint we could consider the second option  to iron the improved bifundamentals
by adding linearly to the superpotential the singlet $\mathsf{B}^{(k)}_{2,1}$, which have the effect of ironing the improved bifundamentals  to  bifundamentals without any extra adjoint field as shown in \eqref{fig:FM_c=1-t/2}. Therefore, if we use this deformation on all the improved bifundamentals we reach exactly the theory in the r.h.s. of figure \ref{fig:SQCD_mirror_old}. 
 However, as we noticed when discussing the operator map in our SQCD mirror pair in section \ref{sec:3d_SQCD_map},  the $\mathsf{B}^{(k)}_{2,1}$ singlets
(unlike $\mathsf{B}^{(k)}_{1,2}$  which map to dressed mesons)  are trivial in the chiral ring, they can not map to any operator in the $U(N)$ SQCD chiral ring.

The general lesson is that our mirrors with improved bifundamentals and the naive mirrors with standard bifundamentals (or the mirrors obtained deforming $\cN=4$ dualities) differ by turning on or off in the superpotential holomorphic operators which are zero in the chiral ring, hence they provide different UV completions of the same IR SCFT. 

While both the naive and our mirrors are \emph{correct},  the mirrors with improved bifundamentals discussed in this paper are  more \emph{useful}, since they encode a full rank  UV global symmetry and allow us to study the IR SCFT's in a transparent way. In particular the  present technology allows us to compute the superconformal index, the $S_b^3$ partition functions, the chiral ring and the moduli space of vacua of the IR SCFT's using our UV quivers with improved bifundamentals. Moreover, the $\cN=2$ algorithm proves that the UV improved quivers associated to $\CS$-dual brane setup flow to the same IR SCFT.

\section{$3d$ $\cN=2$ brane setups with $(p,q)$-webs}\label{sec:pqwebs}
\begin{figure}
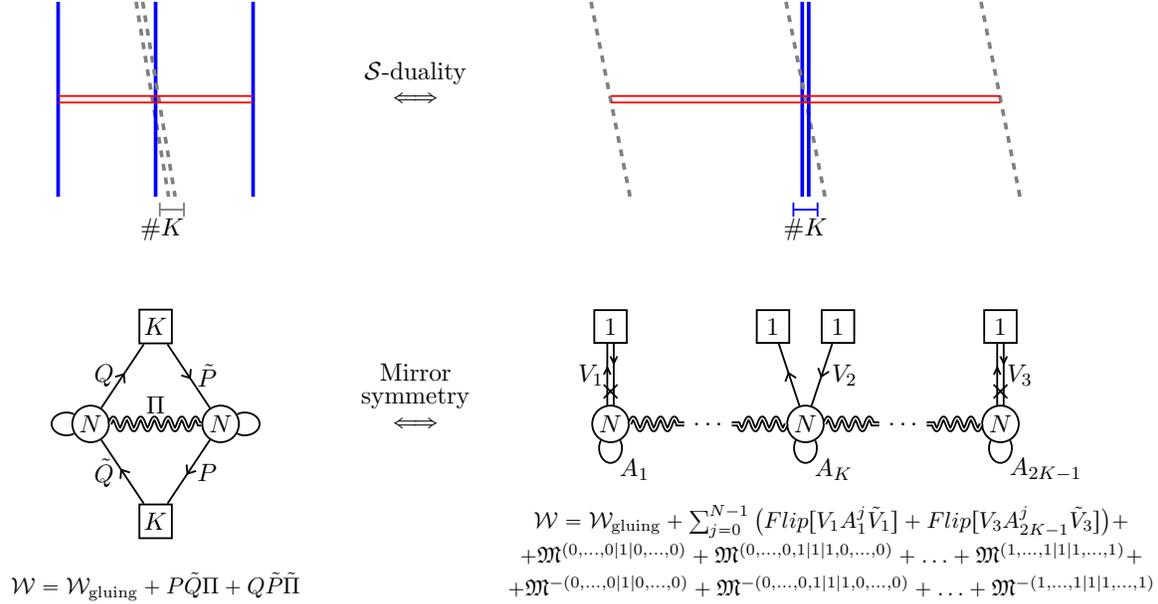

\centering
\resizebox{\hsize}{!}{
\bpic[thick,node distance=3cm,gauge/.style={circle,draw,minimum size=5mm},flavor/.style={rectangle,draw,minimum size=5mm}] 

\begin{scope}[shift={(0,0)}]	

	\NSbrane (-0.5, 1.5) -- (-0.5, -1.5);
	\Dbrane (0.7,1.5) -- (1.2,-1.5);
	\NSbrane (1, 1.5) -- (1, -1.5);
	\Dbrane (0.8,1.5) -- (1.3,-1.5);
	\NSbrane (2.5,1.5) -- (2.5, -1.5);
	
	\Dthree (-0.5,0.05) -- (2.5,0.05);
	\Dthree (-0.5,-0.05) -- (2.5,-0.05);
	
	\draw[|-|,gray] (1.05,-1.7) -- (1.45,-1.7);	\draw  (1.1,-2) node {$\# K$};
\end{scope}	 
	
	\draw (5,0.4) node {$\CS$-duality};
	\draw (5,0) node {$\Longleftrightarrow$};
	
\begin{scope}[shift={(8,0)}]

	\Dbrane (-0.3,1.5) -- (0.3,-1.5);
	\NSbrane (2.95,1.5) -- (2.95,-1.5);
	\Dbrane (2.7,1.5) -- (3.3,-1.5);
	\NSbrane (3.05,1.5) -- (3.05,-1.5);
	\Dbrane (5.7,1.5) -- (6.3,-1.5);
	
	\Dthree (0,0.05) -- (6,0.05);
	\Dthree (0,-0.05) -- (6,-0.05);	
	
	\draw[|-|,blue] (2.8,-1.7) -- (3.2,-1.7);	\draw  (3,-2) node {$\# K$};
\end{scope}
\begin{scope}[shift={(0,-5)}]	

	\path (0,0) node[gauge](g1) {$\!\!\!N\!\!\!$} -- (2,0) node[gauge](g2) {$\!\!\!N\!\!\!$} 
		-- (1,1.5) node[flavor] (x1) {$\!K\!$} -- (1,-1.5) node[flavor] (x2) {$\!K\!$};
	
	\wigM (g1) -- (g2); \draw (1,0.3) node {$\Pi$};
	\chir (g1) -- (x1); \draw (0.5,0.75) node[left] {$Q$};
	\chir (x2) -- (g1); \draw (0.5,-0.75) node[left] {$\tilde{Q}$};
	\chir (x1) -- (g2); \draw (1.5,0.75) node[right] {$\tilde{P}$};
	\chir (g2) -- (x2); \draw (1.5,-0.75) node[right] {$P$};
	\draw[-] (g1) to[out=150,in=90] (-0.6,0) to[out=-90,in=-150] (g1);
	\draw[-] (g2) to[out=30,in=90] (2.6,0) to[out=-90,in=-30] (g2);
	
	\draw (1,-2.5) node {\small{$\CW = \CW_{\text{gluing}} + P \tilde{Q} \Pi + Q \tilde{P} \tilde{\Pi} $}};
	
\end{scope}	 
	
	\draw (5,-4.2) node {Mirror};
	\draw (5,-4.6) node {symmetry};
	\draw (5,-5) node {$\Longleftrightarrow$};
	
\begin{scope}[shift={(8,-5)}]
	\path (0,0) node[gauge](g1) {$\!\!\!N\!\!\!$} -- (1.5,0) node(gi) {$\ldots$} -- (3,0) node[gauge] (g2) {$\!\!\!N\!\!\!$} 
		-- (4.5,0) node (gii) {$\ldots$} -- (6,0) node[gauge] (g3) {$\!\!\!N\!\!\!$}
		-- (0,1.5) node[flavor] (x1) {$\!1\!$} 
		-- (2.5,1.5) node[flavor] (x2) {$\!1\!$} -- (3.5,1.5) node[flavor] (x22) {$\!1\!$} 
		-- (6,1.5) node[flavor] (x3) {$\!1\!$} ;
		
 	\wigM (g1) -- (gi);
	\wigM (g2) -- (gi); 
	\wigM (g2) -- (gii); 
	\wigM (g3) -- (gii);
	
	\draw[-, shorten >= 3, shorten <= 12, shift={(-0.05,-0.15)}, middx arrowsm] (0,0) -- (0,1.5);
	\draw[-, shorten >= 8, shorten <= 7, shift={(0.05,0)}, midsx arrowsm] (0,1.5) -- (0,0);
	\draw (0,0.5) node[cross] {}; \draw (-0.3,0.75) node {$V_1$};
	
	\chir (g2) -- (x2);
	\chir (x22) -- (g2); \draw (3.6,0.75) node {$V_2$};
	
	\draw[-, shorten >= 3, shorten <= 12, shift={(-0.05,-0.15)}, middx arrowsm] (6,0) -- (6,1.5);
	\draw[-, shorten >= 8, shorten <= 7, shift={(0.05,0)}, midsx arrowsm] (6,1.5) -- (6,0);
	\draw (6,0.5) node[cross] {}; \draw (6.3,0.75) node {$V_3$};
	
	\draw[-] (g1) to[out=-60,in=0] (0,-0.6) to[out=180,in=-120] (g1); \draw (0,-0.7) node[right] {$A_1$};
	\draw[-] (g2) to[out=-60,in=0] (3,-0.6) to[out=180,in=-120] (g2); \draw (3,-0.7) node[right] {$A_K$};
	\draw[-] (g3) to[out=-60,in=0] (6,-0.6) to[out=180,in=-120] (g3);\draw (6,-0.7) node[right] {$A_{2K-1}$};
\end{scope}
	
\begin{scope}[shift={(8,-5.5)}]
\draw (3.4,-1) node {\small{$\CW = \CW_{\text{gluing}} + \sum_{j=0}^{N-1} \big( Flip[V_1 A_1^j \tilde{V}_1] + Flip[V_3 A_{2K-1}^j \tilde{V}_3] \big)+ $} };
\draw (3.4,-1.5) node {\small{$ + \mathfrak{M}^{(0,\ldots,0|1|0,\ldots,0)} + \mathfrak{M}^{(0,\ldots,0,1|1|1,0,\ldots,0)} + \ldots + \mathfrak{M}^{(1,\ldots,1|1|1,\ldots,1)} +$}};
\draw (3.4,-2) node {\small{$ + \mathfrak{M}^{-(0,\ldots,0|1|0,\ldots,0)} + \mathfrak{M}^{-(0,\ldots,0,1|1|1,0,\ldots,0)} + \ldots + \mathfrak{M}^{-(1,\ldots,1|1|1,\ldots,1)} $}};
\end{scope}
    
\epic}
\caption{$\CS$-duality and mirror symmetry for a brane setup containing  $(1_{NS}, K_{D5'}) \leftrightarrow (1_{D5'}, K_{NS})$ $(p,q)$-webs. The $K$ $D5'$ branes in the $(p,q)$-web on the left provide $K+K$ flavors, coupled through a cubic superpotential. On the mirror side, the $K$ $NS$ branes in the $(p,q)$-web provide $2K-1$ gauge groups, whose topological symmetries are broken by superpotential terms linear in the monopoles.}
\label{fig:pqweb_ex}
\end{figure}

$(p,q)$-webs, introduced in \cite{Aharony:1997ju,Aharony:1997bh}, are Hanany-Witten brane setups for $5$ dimensional $\cN=1$ theories. In our $3d$ setups, if we move some $D5'$ branes ($012457$) on top of some $NS$ branes ($012345$), we produce a rectangular $(p,q)$-web extending along $37$, which, if isolated would provide a $5d$ QFT living on the $01245$ space-time directions. 

We are interested in putting such $(p,q)$-webs in our sequence of $NS$ and $D5'$ branes, at fixed $x_6$ position and with $N$ $D3$ branes stretching along the sequence. We focus on the case of $(p,q)$-webs made by $K$ $D5'$ on top of a single $NS$ (the $(1_{NS}, K_{D5'})$-web), or its $\CS$-dual $(p,q)$-web made by $K$ $NS$ on top of a single $D5'$ (the $(K_{NS}, 1_{D5'})$-web), extending the results of \cite{Benvenuti:2016wet} for a single $D3$ to the situation with $N$ $D3$'s.

For definiteness, we study the QFT associated to a brane setup given by the sequence $NS - (1_{NS},K_{D5'}) - NS$, with $N$ constant $D3$ branes stretching, depicted in the top left corner of figure \ref{fig:pqweb_ex}. We propose that the corresponding QFT is the one in the bottom-left corner of figure \ref{fig:pqweb_ex}. This is a theory of two $U(N)$ gauge nodes linked by an improved bifundamental, associated to the 3 $NS$ branes. The strings stretching between the $N$ left (right) $D3$ branes and the $K$ stacked $D5'$ branes (which are broken in two halfes by the $NS$) provide $K$ massless flavors for  the left (right)  gauge node. The flavors are  coupled in a cubic fashion to the bifundamental operator
in the improved bifundamental. 

$\CS$-duality sends the  $NS - (1_{NS},K_{D5'}) - NS$ sequence into the $D5'-(K_{NS},1_{D5'})-D5'$ sequence, as in the top-right corner of figure \ref{fig:pqweb_ex}. We now wish to understand the QFT associated to the $\CS$-dual brane setup, and doing this is non trivial. In order to make progress we follow the strategy of \cite{Benvenuti:2016wet}. We first consider an auxiliary sequence, $NS - (D5')^K - NS- (D5')^K - NS$, S-dual of $D5' - NS^K - D5'- NS^K - D5$. This example and the associated $3d$ mirror QFT's are studied in section \ref{k1}, setting $F_1 = F_2 = K$. 

Now we deform the duality of section \ref{k1}, interpreting the action of stacking the $D5'$ branes on top of the $NS$ as the introduction of a cubic superpotential coupling the flavors with the improved bifudamental:
\begin{align}\label{cubic}
	\d \CW = \sum_{i=1}^K (P^i \tilde{Q}_i \Pi + Q^i \tilde{P}_i \tilde{\Pi} ) \,. 
\end{align}
This superpotential breaks the $U(K)^4/U(1)$ flavor symmetry rotating independently $Q,\tilde{Q},P,\tilde{P}$ down to $U(K)^2/U(1)$. \\
The operator map discussed in section \ref{k1} tells us that the deformation \eqref{cubic} is mapped to a monopole superpotential on the mirror dual:
\begin{align}
	\d \CW &= \mathfrak{M}^{(0,\ldots,0|1|0,\ldots,0)} + \mathfrak{M}^{(0,\ldots,0,1|1|1,0,\ldots,0)} + \ldots + \mathfrak{M}^{(1,\ldots,1|1|1,\ldots,1)}+\\
	& + \mathfrak{M}^{-(0,\ldots,0|1|0,\ldots,0)} + \mathfrak{M}^{-(0,\ldots,0,1|1|1,0,\ldots,0)} + \ldots + \mathfrak{M}^{-(1,\ldots,1|1|1,\ldots,1)} \,.
\end{align}
Hence we propose the QFT for the mirror dual as the $U(N)^{2K-1}$ QFT with monopole superpotential appearing in the bottom-right corner of figure \ref{fig:pqweb_ex}.

The monopole superpotential breaks $2K$ $U(1)$ topological symmetries of the UV theory. We then have an enhanced IR symmetry group given by $U(K)^2/U(1)$, matching with that of the electric theory. \\
The operator map for the duality in \ref{fig:pqweb_ex} can be easily inferred from the operator map of mirror pair in figure \ref{fig:TwoNodes_ex} discussed in section \ref{k1}, by taking into account the extra constraints provided by the extra superpotential terms. 

Let us perform a simple consistency check. If we turn on a mass term for  the $j$-th $Q, \Qt$ flavor on the l.h.s. of the duality in  \ref{fig:pqweb_ex}
\be \d \cW=  Q^j \Qt_j \, \ee
and integrate out the two massive $Q^j, \Qt_j$ fields, we are left with a two node quiver with superpotential
\be \cW = \cW_{gluing} + \sum_{i \neq j}( P_i \tilde{Q}^i \Pi + Q^i \tilde{P}_i \tilde{\Pi} )+ \xcancel{\Pi^a_A \tilde{\Pi}^A_b\, P_a^j \, \Pt^b_j}\,, \ee
where \emph{chiral ring stability} \cite{Benvenuti:2017lle} removes the term  $\Pi^a_A \tilde{\Pi}^A_b\, P_a^j \, \Pt^b_j$, because the operator 
$\Pi^a_A \tilde{\Pi}^A_b$ is zero in the chiral ring of the improved bifundamental  theories, as explained in \cite{BCP1}. In other words, turning on a mass term for one of the left flavors corresponds to moving a $D5'$ brane out of the $(p,q)$-web, to the right, so that it becomes an ordinary flavor brane for the right gauge node.\\
On the dual side, a mass term $\d \cW= Q^j \Qt_j$ is mapped to the $\mathsf{B}^{(j)}_{1,1}$ singlet in an improved bifundamental on the  the left side of the quiver. Such a deformation turns the improved bifundamental theory to an $\mathbb{I}$-wall, hence the left sequence of $U(N)$ nodes is shortened by one unit. So this deformation corresponds to moving a single $NS$ brane out of the  $(K_{NS},1_{D5'})$-web, to the right, as expected.

The trick above, of starting from a sequence with a doubled number of $D5'$ or $NS$ branes and then adding a cubic or monopole superpotential, can be easily generalized to any situation where $(1_{NS},K_{D5'})$-webs or $(K_{NS},1_{D5'})$-webs appear in a linear sequence together with $NS$ and $D5'$ branes. 

Let us close this section recalling that understanding the $3d$ QFT associated to $N$ $D3$ branes ending on more general $(p,q)$-webs, with internal faces, remains an open problem, both for $N=1$ and $N>1$.

\section{$3d$ mirror (a.k.a. magnetic quiver) of $4d$ $\CN=1$ $SU(N)$ quivers}\label{3dmirror}

In this section we show in simple examples how the results of this paper can provide the $3d$ mirror of $4d$ theories with $4$ supercharges. 

In the case of $4d$, $5d$ and $6d$ theories with $8$ supercharges, the $3d$ mirror, or magnetic quiver, played in the recent years a very important role in uncovering the strong coupling properties of many models, see for instance \cite{Benini:2010uu, Xie:2012hs, DelZotto:2014kka, Cremonesi:2015lsa, Ferlito:2017xdq, Hanany:2018uhm, Hanany:2018vph, Cabrera:2018jxt, Cabrera:2019izd, Bourget:2019rtl, Cabrera:2019dob}. This is especially true for QFT's defined by higher dimensional constructions like class S, or from string theory constructions, like F-theory, string/M theories on Calabi-Yau cones or $5d/6d$ brane setups, which often lack a Lagrangian description, and even when some Lagrangian description is available, it typically does not see the full global symmetry.

In this section we show how our techniques easily allow us to handle the $\cN=2$ $3d$ mirrors of a simple class of $4d$ $\cN=1$ Lagrangian theories, namely standard non chiral quivers with $SU(N)$ gauge nodes, adjoint matter for each node, standard bifundamentals.

\subsection{$3d$ mirror  of $4d$ $\CN=1$ $SU(N)$ adjoint-SQCD}\label{sec:SUN_SQCD}
We start  from  $4d$ $\cN=1$ adjoint $SU(N)$ SQCD. Reducing this theory to $3d$, no monopole superpotential is generated, so the only difference with respect to the $U(N)$ adjoint SQCD we already studied is the gauge group, $SU(N)$ vs $U(N)$.

Starting from the duality for the adjoint $U(N)$ SQCD in figure \ref{fig:SQCD_Dual}, we can obtain the mirror dual of the $SU(N)$ SQCD with an adjoint field. In order to go from unitary to special unitary gauge group we can gauge the topological symmetry of the $U(N)$ theory. 

At level of  the partition function with start with the  $U(N)$ SQCD partition function  given in \eqref{eq:SQCD_parfun}:
\begin{align}
	Z_{SQCD} \left(\tau,\vec{B},\vec{X},Y\right) = & \int d\vec{Z}_n \D_n (\vec{Z},\tau) e^{2\pi i Y \sum_{j=1}^N Z_j} \prod_{j=1}^N \prod_{a=1}^F s_b\left( B_a \pm (Z_j - X_a) \right) \,.
\end{align}
We gauge the topological symmetry $U(1)_Y$ obtaining the partition function of the $SU(N)$ adjoint SQCD as:
\begin{align}
	Z_{SU(N)} \left(\tau,\vec{B},\vec{X}\right) = & \int dY e^{-2\pi i N b Y} Z_{SQCD} \left(\tau,\vec{B},\vec{X},Y\right) = \nn \\
	= & \int dY d\vec{Z}_N \D_N (\vec{Z},\tau) s_b(-\frac{iQ}{2} + \tau) e^{2\pi i Y (\sum_{j=1}^N Z_j - Nb)} \nn \\
	& \prod_{j=1}^N \prod_{a=1}^F s_b\left( B_a \pm (Z_j - X_a) \right) \,,
\end{align} 
where we have also added an FI parameter $(-N b)$ for the topological symmetry associated to the new $U(1)$ gauge symmetry. We can now redefine the $\vec{Z}$ parameters as: $Z_i \to \tilde{Z}_i + Z$, with the constraint $\sum_{j=1}^N \tilde{Z}_i = 0$. With the new parameterization we get:
\begin{align}
	Z_{SU(N)} \left(\tau,\vec{B},\vec{X}\right) = \int dY dZ d\vec{\tilde{Z}}_{SU(N)} \D_N (\vec{\tilde{Z}},\tau) e^{2\pi i Y N (Z - b)} \prod_{j=1}^N \prod_{a=1}^F s_b\left( B_a \pm (\tilde{Z}_j +Z- X_a) \right) \,.
\end{align}
Where now we have defined for $SU(N)$ a short notation for the integration measure:
\begin{align}
	d\vec{\tilde{Z}}_{SU(N)} = d\vec{\tilde{Z}}_N  \,\, \d\left(\sum_{j=1}^N \tilde{Z}_j\right) \,.
\end{align}
The $Y$ and $Z$ integrals now only involve the exponential term, therefore the $Y$ integration gives a $\d(Z-b)$ which we implement by performing the $Z$ integration  setting $Z=b$. We then obtain the following result:
\begin{align}
	Z_{SU(N)} \left(\tau,\vec{B},\vec{X}\right) = \int d\vec{\tilde{Z}}_{SU(N)} \D_N (\vec{\tilde{Z}},\tau) \prod_{j=1}^N \prod_{a=1}^F s_b\left( B_a \pm (\tilde{Z}_j +b- X_a) \right) \,,
\end{align}
which we recognize to be the partition function of the $\CN=2$ adjoint $SU(N)$ SQCD, with $b$ the real mass for the baryonic $U(1)_b$ symmetry
assigning charge $\pm1$ to the fundamental/anti-fundamental chirals.\\

\begin{figure}
\centering
\begin{tikzpicture}[thick,node distance=3cm,gauge/.style={circle,draw,minimum size=5mm},flavor/.style={rectangle,draw,minimum size=5mm}] 

\begin{scope}[shift={(1,0)}]
	\path (0,0) node[gauge,double](g) {$\!\!\!N\!\!\!$} -- (-0.5,1.5) node[flavor] (x1) {$\!F\!$} -- (0.5,1.5) node[flavor] (x2) {$\!F\!$} 
		-- (2,0) node{$\Longleftrightarrow$};
	
	\chir (g) -- (x1); \draw (-0.3,0.6) node[left] {$Q$};
	\chir (x2) -- (g); \draw (0.3,0.6) node[right] {$\tilde{Q}$};
	\draw[-] (g) to[out=-60,in=0] (0,-0.6) to[out=180,in=-120] (g); \draw (0,-0.6) node[right] {$A$};
	
	\draw (0,-1.5) node{$\cW = 0$};
\end{scope}
	 
	\path (5,0) node[gauge](g1) {$\!\!\!N\!\!\!$} -- (7,0) node[gauge](g2) {$\!\!\!N\!\!\!$} -- (10,0) node[gauge](g3) {$\!\!\!N\!\!\!$} -- (12,0) node[gauge](g4) {$\!\!\!N\!\!\!$} -- (5,1.5) node[gauge](y1) {$\!\!\!1\!\!\!$} -- (12,1.5) node[flavor](y2) {$\!1\!$};
	 
	\wigM (g1) -- (g2); \draw (6,0.4) node {$\Pi_2$};
	\wigM (g2) -- (8,0); 
	\wigM (g3) -- (9,0); 
	\wigM (g3) -- (g4); \draw (11,0.4) node {$\Pi_{F-1}$};
	
	\draw[-, shorten >= 9, shorten <= 6, shift={(-0.05,0.05)}, mid arrowsm] (5,0) -- (5,1.5); \draw (5,0.75) node[left] {$V_1$};
	\draw[-, shorten >= 4, shorten <= 11, shift={(0.05,0.15)}, mid arrowsm] (5,1.5) -- (5,0);
	\draw[-, shorten >= 9, shorten <= 6, shift={(0.05,0.05)}, mid arrowsm] (12,0) -- (12,1.5); \draw (12,0.75) node[right] {$V_2$};
	\draw[-, shorten >= 4, shorten <= 11, shift={(-0.05,0.15)}, mid arrowsm] (12,1.5) -- (12,0); 
	\draw (5,0.45) node[cross]{}; \draw (12,0.45) node[cross]{};	
	\draw (8.5,0) node {$\cdots$};
	
	\draw[-] (g1) to[out=-60,in=0] (5,-0.6) to[out=180,in=-120] (g1); \draw (5,-0.7) node[right] {$A_1$};
	\draw[-] (g2) to[out=-60,in=0] (7,-0.6) to[out=180,in=-120] (g2); \draw (7,-0.7) node[right] {$A_2$};
	\draw[-] (g3) to[out=-60,in=0] (10,-0.6) to[out=180,in=-120] (g3); \draw (10,-0.7) node[right] {$A_{F-2}$};
	\draw[-] (g4) to[out=-60,in=0] (12,-0.6) to[out=180,in=-120] (g4); \draw (12,-0.7) node[right] {$A_{F-1}$};
	
	\draw (4,-1.5) node[right]{$\cW = \sum_{j=0}^{N-1} ( Flip[V_1 A_1^j \tilde{V}_1] + Flip[V_2 A_{F-1}^j \tilde{V}_2] )+$  };
	\draw (4.7,-2.1) node[right]{$ + Flip[\Tr A_1] + \mathcal{W}_{\text{gluing}}$};
	 
\end{tikzpicture}
\caption{Mirror duality for the $\CN=2$ adjoint $SU(N)$ SQCD. The double circle node denotes the $SU(N)$ gauge group.}
\label{fig:SU_Dual} 
\end{figure}
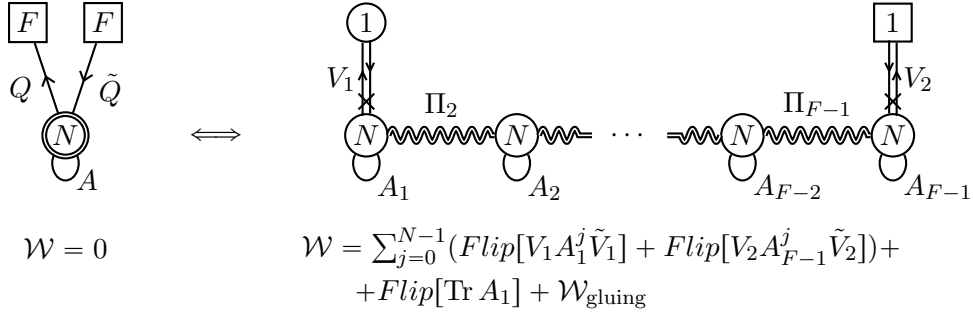

Now we can perform the same steps in the mirror partition function \eqref{eq:SQCDmirr_parfun} obtaining:
\begin{align}
	Z_{\widecheck{SU(N)}} \left(\tau, \vec{B}, \vec{X}\right) & = 
	\int dY  e^{2 \pi i N (X_1-b) Y} \prod_{a=1}^{F-1} \big( d\vec{Z}_n^{(a)}  \D_n (\vec{Z}^{(a)},\tau) e^{2\pi i (X_{a+1}-X_{a}) \sum_{j=1}^N Z_j }  \big) \nn \\ 
	& \prod_{j=1}^N \big[ s_b\left( \frac{iQ}{2} - \frac{1-N}{2}\tau - B_1 \pm (Z_j^{(1)} - Y) \right) s_b( -\frac{iQ}{2} + (j-N)\tau + 2B_1 ) \nn\\
	& s_b \left( \frac{iQ}{2} - \frac{1-N}{2}\tau - B_F \pm Z_j^{(F-1)} \right) s_b( -\frac{iQ}{2} + (j-N)\tau + 2B_F ) \big] \nn \\
	& \prod_{a=1}^{F-2} Z_{FM}^{(N)} \left( \vec{Z}^{(a)},\vec{Z}^{(a+1)},\tau,B_{a+1} \right) \,.
\end{align}
The mirror pair read from the partition function identity is depicted in figure \ref{fig:SU_Dual}. 
In the electric theory we have the following global symmetry:
\begin{align}
	\prod_{j=1}^F  U(1)_{B_a} \times S\left[ \prod_{j=1}^F  U(1)_{X_a} ) \right] \times U(1)_\tau\times U(1)_b & = \nn \\
	& = SU(F)_U \times SU(F)_V  \times U(1)_m \times U(1)_\tau \times U(1)_b\,,
\end{align}
where the two $SU(F)$ and the $U(1)_m$ global symmetries are obtained from the $U(1)_{B_j} \times U(1)_{X_j}$ as usual with the  redefinitions in eqs. \eqref{axdef}, \eqref{uvdef}.\\

In the mirror theory  the global symmetry enhances in the IR as:
\begin{align}
	\prod_{j=1}^F U(1)_{B_j} \times \prod_{j=1}^{F-1} U(1)_{X_{j+1}-X_j} \times U(1)_{N (X_1-b)} =  SU(F)_U \times SU(F)_V \times U(1)_b \times U(1)_m \,,
\end{align}
where as usual $U(1)_{B_j}$ symmetries are the axial-like symmetries rotating the two vertical flavors and the improved bifundamentals, while $X_{j+1}-X_j$ is the FI parameters for the topological symmetry associated to the $j$-th gauge node. Finally ${N (X_1-b)}$ is the FI parameter of the new $U(1)$ gauge node, which as expected by mirror duality, is related to the electric baryonic symmetry $U(1)_b$.
\\

In conclusion, let us mention that the trick of gauging a topological symmetry to transform a $U$ node into a $SU$ can be played also in more generic theories to make quiver with $U/SU$ nodes. 

\subsection*{Operator map}
Let us now discuss how the operator map works for the case of the adjoint $SU(N)$ SQCD. There are three gauge invariant operators that we want to map:
\begin{itemize}
	\item The meson matrix $Q \tilde{Q}$ in the $\bar{F} \times F$ of $SU(F)_U \times SU(F)_V$, with $R$-charge 2, $m$-charge $-2$ and zero charge under the remaining $U(1)$ symmetries. These operators are mapped exactly as in the SQCD case, as explained in detail in section \ref{sec:3d_SQCD_map}, meaning that we collect $F$ singlets and $F(F-1)$ monopoles to form a matrix transforming in the $\bar{F} \times F$ of the emergent $U(F) \times U(F)$ symmetry.
	
	\item The lowest dimensional $SU(N)$ monopole is characterised by the magnetic flux $(1,0,\ldots,0,-1)$. This operator is a singlet under all the non-abelian global symmetries, it has R-charge 0, $\tau$-charge $(1-N)$, $m$-charge $2$ and $b$-charge zero. This operator is mapped to the simplest mesonic operator that we can construct in the mirror theory obtained as:
	\begin{align} \label{longmesdual2M}
		\tilde{V}_2 \tilde{\Pi}_{F-1} \ldots \tilde{\Pi}_2 V_1 \tilde{V}_1 \Pi_2 \ldots \Pi_{F-1} V_2 \,,
	\end{align}
	which indeed is charged only under the abelian global symmetries with the correct charges in order to be mapped to the $SU$ monopole.
	
	\item In the $SU(N)$ SQCD we also have baryons and antibaryons, differently from the $U(N)$ case. These are constructed by taking the antisymmetrized product of $N$ fundamentals $Q$, to obtain baryons, or of $N$ antifundamentals $\tilde{Q}$, to obtain antibaryons. We then have two sets of $\binom{F}{N}$ operators, the baryons are in the conjugate $N$-antisymmetric representation of $SU(F)_U$ while antibaryons are in the $N$-antisymmetric of $SU(F)_V$, and they have R-charge $2$, $m$-charge $-N$ and baryonic $b$-charge $\pm N$. Notice that only if $F \geq N$ we can have baryons in the theory. \\
	These operators are mapped to a collection of suitably charged monopoles that have all R-charge $2$, $m$-charge $-N$ and $\pm N$ $b$-charge (see appendix \ref{app:FMmonopoles}). Let us focus on the antibaryons, these are mapped to the collection of all the monopoles with a topological charge satisfying the following set of rules:\footnote{We adopt the following notation: by having a topological charge $k \geq 0$ we mean that the magnetic flux is given by $k$ entries equal to 1 while the remaining are zero canonically ordered as $(1,\ldots,1,0,\ldots,0)$. With a negative $-k$ charge we take a similar vector with $k$ entries equal to $-1$ while the remaining are zero as $(0,\ldots,0,-1,\ldots,-1)$.}
	\begin{itemize}
		\item It must have charge $+1$ under the topological symmetry of the $U(1)$ gauge node.
				\item It must have a topological charge of $N$ or $(N-1)$ under the topological symmetry of the first $U(N)$ gauge node. 
		\item The remaining charges are fixed by requiring that 
		 the charge under the $j$-th topological symmetry is either equal or one unit less than  the charge under the $(j-1)$-th topological symmetry.
		\item  The last non-zero charge must be $1$. 
	\end{itemize}

Collecting all the monopoles satisfying this set of rules we can form a $N$-antisymmetric representation under one $U(F)$ emergent global symmetry. 
For example, for $N=3$ and $F=4$  we have the following collection of monopoles mapping to the $\binom{4}{3}=4$ antibaryons:
	\begin{align}
		\mathfrak{M}^{(1|2,1,0)} \,,\, \mathfrak{M}^{(1|2,1,1)} \,,\, \mathfrak{M}^{(1|2,2,1)} \,,\, \mathfrak{M}^{(1|3,2,1)} \,.
	\end{align}
\end{itemize}
There are also dressed operators. Dressed mesons are mapped to singlets or dressed monopoles, exactly as in the $U(N)$ case. The dressed $SU(N)$ monopole maps to the long meson with the same level of dressing. Dressed baryons are less trivial to map, the reason being that the number of dressed baryons increases rapidly with the level of dressing. Also, in the mirror theory it is not easy to compute the R-charge of all the monopoles due to the presence improved bifundamental. However, we suspect that dressed baryons maps either to dressed monopoles or to the monopoles that are not included in the set of rules presented above to map undressed baryons. 

\subsection{$3d$ mirror of $4d$ $SU(N)$ SQCD without adjoint}\label{sec4dQCDnoadj}
We now consider $4d$ $\CN=1$ $SU(N)$ SQCD with $F$ flavors, no adjoint and $\CW_{4d}=0$. When compactified to $3d$, the theory develops a monopole superpotential $\CW_{3d}=\M$, where $\M$ is the lowest dimensional $SU(N)$ monopole with magnetic flux $(1,0,\ldots,0,-1)$ and we obtain the theory on the l.h.s of \eqref{3dMnoadjsqcd}.
\be
\resizebox{.9\hsize}{!}{
 \bpic[thick,node distance=3cm,gauge/.style={circle,draw,minimum size=5mm},flavor/.style={rectangle,draw,minimum size=5mm}] 
\begin{scope}[shift={(1,-4.5)}]
    \path (0,0) node[gauge,double](g1) {$\!\!\!N\!\!\!$}  -- (-0.5,1.5) node[flavor] (x1) {$\!F\!$} -- (0.5,1.5) node[flavor] (x2) {$\!F\!$} 
    	-- (0,-1) node (g1p) {\small{$\CW = \M^{(1,0,\ldots,0,-1)}$}} ;
    \chir (g1) -- (x1); \draw (0.4,0.8) node[right] {$\tilde{Q}$}; 
    \chir (x2) -- (g1); \draw (-0.4,0.8) node[left] {$Q$}; 
\end{scope}
	\draw (3.5,-4.2) node {$\Longleftrightarrow$};
\begin{scope}[shift={(5.75,-4.5)}]	
	\path (0,0) node[gauge](g1) {$\!\!\!N\!\!\!$} -- (1.5,0) node[gauge](g2) {$\!\!\!N\!\!\!$} -- (3,0) node(gi) {$\ldots$}
		-- (4.5,0) node[gauge](g3) {$\!\!\!N\!\!\!$} -- (6,0) node[gauge](g4) {$\!\!\!N\!\!\!$} 
		-- (0,1.5) node[gauge](y1) {$\!\!1\!\!$} -- (6,1.5) node[flavor](y2) {$\!1\!$};
	\wigM (g1) -- (g2); \draw (0.7,0.4) node {$\Pi_2$};
	\wigM (g2) -- (gi);
	\wigM (g3) -- (gi);
	\wigM (g3) -- (g4);  \draw (5.3,0.4) node {$\Pi_{F\!-\!1}$};
	\draw[-, shorten >= 3, shorten <= 12, shift={(-0.05,-0.15)}, middx arrowsm] (0,0) -- (0,1.5);
	\draw[-, shorten >= 8, shorten <= 7, shift={(0.05,0)}, midsx arrowsm] (0,1.5) -- (0,0);
	\draw (0,0.5) node[cross]{}; \draw (-0.1,0.75) node[left] {$V_1$};
	\draw[-, shorten >= 3, shorten <= 12, shift={(-0.05,-0.15)}, middx arrowsm] (6,0) -- (6,1.5);
	\draw[-, shorten >= 8, shorten <= 7, shift={(0.05,0)}, midsx arrowsm] (6,1.5) -- (6,0);
	\draw (6,0.5) node[cross]{}; \draw (6.1,0.75) node[right] {$V_2$};
	\draw (3,-1) node {\small{$\cW = \sum_{j=0}^{N-1} ( Flip[V_1 (\mathsf{A}_{L}^{(2)})^j \tilde{V}_1] + Flip[V_2 (\mathsf{A}_{R}^{(F-1)})^j \tilde{V}_2] )+$} };
	\draw (3,-1.6) node {\small{$+\sum_{I=2}^{F-2} \mathsf{A}_{R}^{(I)}\mathsf{A}_{L}^{(I+1)} 
	+ \tilde{V}_2 \tilde{\Pi}_{F-1} \ldots \tilde{\Pi_2} V_1 \tilde{V}_1 \Pi_2 \ldots \Pi_{F-1} V_2 $} };
\end{scope}
\epic}\label{3dMnoadjsqcd}
\ee
The duality in \eqref{3dMnoadjsqcd} can easily be obtained as a deformation of the $SU(N)$ adjoint SQCD mirror pair in figure \ref{fig:SU_Dual}.
On the electric side we turn on a mass term for the adjoint, as in Section \ref{sec:noadjQCDduality} and then the monopole superpotential
which has the effect of breaking the $U(1)_m$ axial symmetry. Therefore there are no $U(1)$ global symmetries that can mix with the trial R-charge and  the R-charge of the  fundamental/antifundamental chirals in the electric theory on the l.h.s of \eqref{3dMnoadjsqcd}
  is completely fixed by the superpotential as 
  \begin{align}
  R[Q]=R[\tilde Q]=1-\frac{N}{F}
  \end{align}
   and the global symmetry is:
 \begin{align}\label{eq:3dsunglobalsymmetry}
	SU(F)_U \times SU(F)_V \times U(1)_b \,.
\end{align}
Similarly the theory on the r.h.s of figure in \eqref{3dMnoadjsqcd},  the $3d$ mirror of $4d$ $\CN=1$ $SU(N)$ SQCD  and $\CW=0$,
is reached from the magnetic theory in \ref{fig:SU_Dual} by adding a mass term for the adjoint and turning on the operator dual to $\M$ which has been identified in  \eqref{longmesdual2M} as the long meson in the quiver, hence we simply need to add this operator to the quiver superpotential.

Again on the mirror side the R-charges are completely fixed to  
 \begin{align}
R[V_{1,2}]=R[\tilde V_{1,2}]=\frac{N}{F}-\frac{N-1}{2}\,, \qquad R[\Pi_j]=R[\tilde \Pi_j]=\frac{N}{F}
\end{align}
 and
the remaining   global symmetries are
\begin{align}
	S[ \prod_{j=1}^F U(1)_{B_j} ] \times \prod_{j=1}^{F-1} U(1)_{X_{j+1}-X_j} \times U(1)_{N(X_1-b)}\,,
\end{align}
where as usual $U(1)_{B_j}$ are the axial-like symmetries rotating the two vertical flavors and the improved bifundamentals, while  $X_{j+1}-X_j$ is the FI parameter associated to the $j$-th $U(N)$ and $N(X_1-b)$ the FI parameter of the extra $U(1)$ node and
enhance in the IR to the group in \eqref{eq:3dsunglobalsymmetry}. 



The global symmetries, the chiral ring and the moduli space of vacua of the $3d$ SQCD on the l.h.s.~of \eqref{3dMnoadjsqcd} and those of the $4d$ $SU(N)$ SQCD should be exactly the same. In this sense the $3d$ mirror of $4d$ $\cN=1$ SQCD with no adjoint behaves in the same way to the $3d$ mirror of theories with $8$ supercharges.

The map of the chiral ring generators is similar to the map discussed for adjoint SQCD in the previous subsection. The difference is that now there are no dressed operators.

Notice that if we simply deform the duality of the previous subsection by the mass term of the adjoint, we obtain a different result, missing the monopole superpotential and its dual. The operations of deforming by a superpotential and compactifying to $3d$ do not commute.

\subsection{$3d$ mirror of $4d$ $SU(N)$ quivers}\label{sec4dquiver}
We start from the duality in figure \ref{fig:MoreNodes_ex}, where on the l.h.s. there is a $U(N)^{K+1}$ quiver with improved bifundamentals:
\be\label{3ddualquiv}  \bpic[thick,node distance=3cm,gauge/.style={circle,draw,minimum size=5mm},flavor/.style={rectangle,draw,minimum size=5mm}] 
\begin{scope}[shift={(0,-5)}]			
	\path (0,0) node[gauge](g1) {$\!\!\!N\!\!\!$} --  (1.5,0) node(gi) {$\ldots$} -- (3,0) node[gauge](g2) {$\!\!\!N\!\!\!$} 
		-- (-0.5,1.5) node[flavor] (x1) {$\!F_1\!$} -- (0.5,1.5) node[flavor] (x2) {$\!F_1\!$} 
		-- (2.5,1.5) node[flavor] (y1) {$\!F_2\!$} -- (3.5,1.5) node[flavor] (y2) {$\!F_2\!$};
	\wigM (g1) -- (gi);
	\wigM (g2) -- (gi);
	\chir (g1) -- (x1);
	\chir (x2) -- (g1);
	\chir (g2) -- (y1);
	\chir (y2) -- (g2);
	\draw[-] (g1) to[out=-60,in=0] (0,-0.6) to[out=180,in=-120] (g1);
	\draw[-] (g2) to[out=-60,in=0] (3,-0.6) to[out=180,in=-120] (g2);	
	\draw (1.5,-1.5) node {\small{$\CW = \CW_{\text{gluing}}$} };		
\end{scope}	 

\draw (5,-4.6) node {Mirror};
\draw (5,-5) node {$\Longleftrightarrow$};

\begin{scope}[shift={(7,-5)}]
	\path (0,0) node[gauge](g1) {$\!\!\!N\!\!\!$} -- (1.5,0) node(gi) {$\ldots$} -- (3,0) node[gauge] (g2) {$\!\!\!N\!\!\!$} 
	 		-- (4.5,0) node (gii) {$\ldots$} -- (6,0) node[gauge] (g3) {$\!\!\!N\!\!\!$} 
	 		-- (0,1.5) node[flavor] (x1) {$\!1\!$} -- (2.5,1.5) node[flavor] (x2) {$\!K\!$}
	 		-- (3.5,1.5) node[flavor] (x22) {$\!K\!$} -- (6,1.5) node[flavor] (x3) {$\!1\!$};
 
	\wigM (g1) -- (gi);
	\wigM (g2) -- (gi); 
	\wigM (g2) -- (gii);
	\wigM (g3) -- (gii);
	
	\draw[-, shorten >= 3, shorten <= 12, shift={(-0.05,-0.15)}, middx arrowsm] (0,0) -- (0,1.5);
	\draw[-, shorten >= 8, shorten <= 7, shift={(0.05,0)}, midsx arrowsm] (0,1.5) -- (0,0);
	\draw (0,0.5) node[cross] {}; \draw (-0.4,0.75) node {$V_L$};
	
	\chir (g2) -- (x2);
	\chir (x22) -- (g2);
	
	\draw[-, shorten >= 3, shorten <= 12, shift={(-0.05,-0.15)}, middx arrowsm] (6,0) -- (6,1.5);
	\draw[-, shorten >= 8, shorten <= 7, shift={(0.05,0)}, midsx arrowsm] (6,1.5) -- (6,0);
	\draw (6,0.5) node[cross] {}; \draw (6.4,0.75) node {$V_R$};
	
	\draw[-] (g1) to[out=-60,in=0] (0,-0.6) to[out=180,in=-120] (g1); \draw (0,-0.6) node[right] {$A_1$};
	\draw[-] (g2) to[out=-60,in=0] (3,-0.6) to[out=180,in=-120] (g2);
	\draw[-] (g3) to[out=-60,in=0] (6,-0.6) to[out=180,in=-120] (g3); \draw (6,-0.6) node[right] {$A_{F_1+F_2-1}$};
	
	\draw (3,-1.5) node {\small{$\CW = \CW_{\text{gluing}} + \sum_{j=0}^{N-1} \big( Flip[V_L A_1^j \tilde{V}_L] +$}};
	\draw (3,-2.2) node {\small{$ + Flip[V_R A_{F_1+F_2-1}^j \tilde{V}_R] \big)$} };
\end{scope}
\epic\ee
The global symmetry is $U(F_1)^2 \times U(F_2)^2 \times U(K)^2 \times U(1)_{\tau}$, see section \ref{generick} for more details.\\
\eqref{3ddualquiv} is a $3d$ $\cN=2$ duality such that neither side can be the circle reduction of a simple $4d$ Lagrangian theory. \\
We modify duality \eqref{3ddualquiv} as follows:
\begin{itemize}
\item on the l.h.s. we gauge all the $K+1$ topological symmetries, making all the gauge nodes $SU(N)$ instead of $U(N)$. This maps on the r.h.s. to gauging $K$ $U(1)$ flavor symmetries associated to the $K$ central flavors and one $U(1)$ symmetry associated to a boundary flavor.
\item on the l.h.s.  we turn on the $K$ operators $\mathsf{B}_{2,1}^{(i)} $, $i=1,\ldots,K$ in the superpotential, ironing the $K$ improved bifundamentals into flipped bifundamentals $b_i, \tilde{b}_i$, without producing new adjoints. This maps on the r.h.s. to turning on $K$ cubic superpotential terms for the $K$ flavors in the middle, $\delta \cW = Tr ( A_{F_1} \sum_{i=1}^K  V_i \tilde{V}_i )$.  
\item on the l.h.s. we flip the gauge singlets which are flipping the squares of the bifundamentals $Tr( b_i \tilde{b}_i)$.  This maps on the r.h.s. to flipping the $K$ mesons made with the $K$ flavors in the middle, $Tr(V_i \tilde{V}_i)$.
\end{itemize}
These modifications break the $U(K)^2$ global symmetry to $U(1)^{K}$.
\be \label{4ddualquiv} 
\resizebox{.95\hsize}{!}{ 
\bpic[thick,node distance=3cm,gauge/.style={circle,draw,minimum size=5mm},flavor/.style={rectangle,draw,minimum size=5mm}] 

\begin{scope}[shift={(0,-5)}]	
		
	\path (0,0) node[gauge,double](g1) {$\!\!\!N\!\!\!$} --  (1.5,0) node(gi) {$\ldots$} -- (3,0) node[gauge,double](g2) {$\!\!\!N\!\!\!$} 
		-- (-0.5,1.5) node[flavor] (x1) {$\!F_1\!$} -- (0.5,1.5) node[flavor] (x2) {$\!F_1\!$} 
		-- (2.5,1.5) node[flavor] (y1) {$\!F_2\!$} -- (3.5,1.5) node[flavor] (y2) {$\!F_2\!$};
 
	\draw[-, shorten >= 0, shorten <= 9, shift={(0,0.05)}, middx arrowsm] (0,0) -- (1,0);
	\draw[-, shorten >= 9, shorten <= 0, shift={(0,-0.05)}, midsx arrowsm] (1,0) -- (0,0);
	\draw (0.5,0.3) node {$b_1$};

	\draw[-, shorten >= 0, shorten <= 9, shift={(0,0.05)}, middx arrowsm] (1.7,0) -- (2.7,0);
	\draw[-, shorten >= 9, shorten <= 0, shift={(0,-0.05)}, midsx arrowsm] (2.7,0) -- (1.7,0);
	\draw (2.3,0.3) node {$b_{K+1}$};

	\chir (g1) -- (x1);
	\chir (x2) -- (g1);
	\chir (g2) -- (y1);
	\chir (y2) -- (g2);
	\draw[-] (g1) to[out=-60,in=0] (0,-0.6) to[out=180,in=-120] (g1); 
	\draw[-] (g2) to[out=-60,in=0] (3,-0.6) to[out=180,in=-120] (g2); 
	
	\draw (1.5,-1.5) node {\small{$\CW = \sum_{i=1}^{K} b_i (a_i - a_{i+1}) \tilde{b}_i  $} };	
	
\end{scope}	 

\draw (5,-4.6) node {Mirror};\draw (5,-5) node {$\Longleftrightarrow$};

\begin{scope}[shift={(7,-5)}]
	\path (0,0) node[gauge](g1) {$\!\!\!N\!\!\!$} -- (1.5,0) node(gi) {$\ldots$} -- (3,0) node[gauge] (g2) {$\!\!\!N\!\!\!$} 
	 		-- (4.5,0) node (gii) {$\ldots$} -- (6,0) node[gauge] (g3) {$\!\!\!N\!\!\!$} 
	 		-- (0,1.5) node[gauge] (x1) {$\!1\!$} -- (6,1.5) node[flavor] (x3) {$\!1\!$}
	 		-- (2,1.5) node[gauge] (y1) {$\!1\!$} -- (3,1.5) node (y2) {$\cdots$} -- (4,1.5) node[gauge] (y3) {$\!1\!$};
	 		
	\wigM (g1) -- (gi);
	\wigM (g2) -- (gi); 
	\wigM (g2) -- (gii);
	\wigM (g3) -- (gii);
	
	\draw[-, shorten >= 4, shorten <= 12, shift={(-0.05,-0.15)}, middx arrowsm] (0,0) -- (0,1.5);
	\draw[-, shorten >= 8, shorten <= 9, shift={(0.05,0)}, midsx arrowsm] (0,1.5) -- (0,0);
	\draw (0,0.5) node[cross] {}; \draw (-0.4,0.75) node {$V_L$};
	
	\draw[-, shorten >= 6, shorten <= 10, shift={(-0.07,0.02)}, mid arrowsm] (2,1.5) -- (3,0);
	\draw[-, shorten >= 6, shorten <= 9, shift={(0.1,0)}, mid arrowsm] (3,0) -- (2,1.5); 
	\draw (2,0.75) node {$V_1$};
	
	\draw[-, shorten >= 7, shorten <= 9, shift={(-0.1,0.02)}, mid arrowsm] (3,0) -- (4,1.5);
	\draw[-, shorten >= 7, shorten <= 9.5, shift={(0.05,0)}, mid arrowsm] (4,1.5) -- (3,0); 
	\draw (4,0.75) node {$V_{K}$};
	
	\draw[-, shorten >= 3, shorten <= 12, shift={(-0.05,-0.15)}, middx arrowsm] (6,0) -- (6,1.5);
	\draw[-, shorten >= 8, shorten <= 7, shift={(0.05,0)}, midsx arrowsm] (6,1.5) -- (6,0);
	\draw (6,0.5) node[cross] {}; \draw (6.4,0.75) node {$V_R$};
	
	\draw[-] (g1) to[out=-60,in=0] (0,-0.6) to[out=180,in=-120] (g1); \draw (0,-0.7) node[right] {$A_1$};
	\draw[-] (g2) to[out=-60,in=0] (3,-0.6) to[out=180,in=-120] (g2); \draw (3,-0.7) node[right] {$A_{F_1}$};
	\draw[-] (g3) to[out=-60,in=0] (6,-0.6) to[out=180,in=-120] (g3); \draw (6,-0.7) node[right] {$A_{F_1+F_2-1}$};
	
	\draw (3,-1.5) node {\small{$\CW = \CW_{\text{gluing}} + A_{F_1}  \sum_{i=1}^K V_i \tilde{V}_i  + \sum_{i=1}^{K}Flip[V_i \tilde{V}_i] + $}};
	\draw (3.4,-2.2) node {\small{$ +\sum_{j=0}^{N-1} \big( Flip[V_L A_1^j \tilde{V}_L] + Flip[V_R A_{F_1+F_2-1}^j \tilde{V}_R] \big)$} };
\end{scope}
\epic}
\ee
The global symmetry is $U(F_1)^2 \times U(F_2)^2 \times U(1)^K \times U(1)_{\tau}$.

Now, the $SU(N)^{K+1}$ quiver on the l.h.s. of \eqref{4ddualquiv} is precisely the circle reduction of a $4d$ quiver, with the same matter content and the same superpotential. 
This is the $4d$ quiver associated to a brane setup with $N$ $D4$ branes stretching along the sequence 
\be \label{setup4d} NS' - D6^{F_1} - (NS')^K -  D6^{F_2} - NS' \,.\ee

We claim that the r.h.s. of  \eqref{3ddualquiv} is the $3d$ mirror of such $4d$  quiver. 

Notice that since no monopole superpotential is generated in the circle reduction, the global symmetry of the reduced theory contains two additional $U(1)$ factors with respect to the $4d$ quiver, this is because the two axial $U(1)$'s inside $U(F_1)^2$ and $U(F_2)^2$ are anomalous in $4d$.

On the r.h.s. of \eqref{4ddualquiv} there is an $S_K$ discrete global symmetry permuting the bouquet of $K$ $U(1)$ nodes. As found in $6d$ in \cite{Hanany:2018vph,Cabrera:2019izd}, it is possible to gauge this symmetry, with the effect of replacing the $K$ $U(1)$ nodes with a single $U(K)$ node. Such a move corresponds to taking the $K$ $NS'$ branes in the setup \eqref{setup4d} to be coincident, an infinite coupling situation from the point of view of the electric $4d$ quiver. Indeed the central piece of the quiver on the r.h.s. of \eqref{4ddualquiv} is the $3d$ mirror of a string of $K$ $NS'$ branes and is $\cN=4$ supersymmetric (the gauge singlets $\mathcal{F}[V_i \tilde{V}_i]$ can be seen as supersymmetric partners of the $U(1)$ gauge nodes), hence the results of  \cite{Hanany:2018vph,Cabrera:2019izd}, obtained for theories with $8$ supercharges, carry over to the central piece of our $3d$ mirror.


\section{$4d$ mirror dualities}\label{4dsec}

It was shown in \cite{Hwang:2020wpd} that $3d$ $\CN=4$ mirror dualities can be uplifted to a class of $4d$ $\CN=1$ theories with symplectic gauge groups, that enjoy mirror-like dualities. This strategy   can be extended also to the $3d$ $\CN=2$ theories considered in this work. \\
In this section we present the mirror-like dual of the $4d$ $\CN=1$ $USp(2N)$ antisymmetric SQCD, discussing the operator map and various deformations. This duality exhibits many similarities with that of $3d$ $\CN=2$ adjoint SQCD, described in section \ref{sqcdmirror}. Indeed the two mirror pairs are related by a dimensional reduction limit. We then explain how to uplift all the $3d$ $\CN=2$ mirror dualities described in section \ref{sec:branes}. Finally, we also describe how $4d$ mirror-like dualities can be proven using the dualization algorithm.

\subsection{$4d$ $\CN=1$ antisymmetric $USp(2N)$ SQCD and its mirror pair}
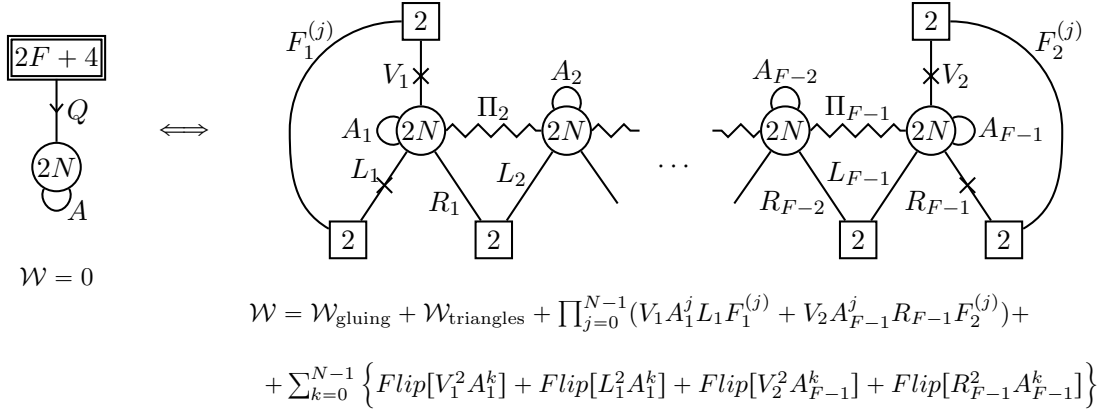
\begin{figure}
\centering 
\resizebox{.95\hsize}{!}{
\begin{tikzpicture}[thick,node distance=3cm,gauge/.style={circle,draw,minimum size=5mm},flavor/.style={rectangle,draw,minimum size=5mm}] 
\begin{scope}[shift={(0,-0.5)}]
	\path (0,0) node[gauge](g) {$\!\!\!2N\!\!\!$} -- (0,1.5) node[flavor,double] (x) {$\!2F+4\!$};
	
	\chir (x) -- (g); \draw (0,0.75) node[right]{$Q$};
	\draw[-] (g) to[out=-60,in=0] (0,-0.6) to[out=180,in=-120] (g); \draw (0,-0.6) node[right] {$A$};
	
	\draw (0,-1.5) node{\small{$\cW = 0$}};
\end{scope}

	\draw (1.75,0) node {$\Longleftrightarrow$}; 

	\path (5,0) node[gauge](g1) {$\!\!\!2N\!\!\!$} -- (7,0) node[gauge](g2) {$\!\!\!2N\!\!\!$} -- (10,0) node[gauge](g3) {$\!\!\!2N\!\!\!$} -- (12,0) node[gauge](g4) {$\!\!\!2N\!\!\!$} -- (4,-1.5) node[flavor](x1) {$\!2\!$} -- (6,-1.5) node[flavor](x2) {$\!2\!$} -- (11,-1.5) node[flavor](x3) {$\!2\!$} -- (13,-1.5) node[flavor](x4) {$\!2\!$} -- (5,1.5) node[flavor](y1) {$\!2\!$} -- (12,1.5) node[flavor](y2) {$\!2\!$};
	 
	\wigE (g1) -- (g2); \draw (6,0.3) node {$\Pi_2$};
	\wigE (g2) -- (8,0);
	\wigE (g3) -- (9,0);
	\wigE (g3) -- (g4);	 \draw (11,0.3) node {$\Pi_{F-1}$};
	\draw[-] (g1) -- (x1); \draw (4.6,-0.55) node[left]{$L_1$}; 
	\draw[-] (g1) -- (x2); \draw (5.7,-1) node[left]{$R_1$};
	\draw[-] (g1) -- (y1); \draw (5,0.75) node[left]{$V_1$};
	\draw[-] (g2) -- (x2); \draw (6.6,-0.6) node[left]{$L_2$};
	\draw[-] (g2) -- (7.7,-1);
	\draw (8.5,-0.5) node {$\cdots$};
	\draw[-] (g3) -- (9.3,-1);
	\draw[-] (g3) -- (x3); \draw (10.7,-1) node[left]{$R_{F-2}$};
	\draw[-] (g4) -- (x3); \draw (11.6,-0.6) node[left]{$L_{F-1}$}; 
	\draw[-] (g4) -- (x4); \draw (12.7,-1) node[left]{$R_{F-1}$};
	\draw[-] (g4) -- (y2); \draw (12,0.75) node[right]{$V_2$};
	\draw (4.5,-0.75) node[cross]{};
	\draw (5,0.75) node[cross]{};
	\draw (12.5,-0.75) node[cross]{};
	\draw (12,0.75) node[cross]{};
	\draw[-] (y1) to[out=180,in=90] (3.2,0) to[out=-90,in=150] (x1); \draw (3,1.25) node[right]{$F_1^{(j)}$};
	\draw[-] (y2) to[out=0,in=90] (13.8,0) to[out=-90,in=30] (x4); \draw (14.25,1.25) node[left]{$F_2^{(j)}$};
	\draw[-] (g1) to[out=150,in=90] (4.4,0) to[out=-90,in=-150] (g1); \draw (4.5,0) node[left]{$A_1$};
	\draw[-] (g2) to[out=60,in=0] (7,0.6) to[out=180,in=120] (g2); \draw (7,0.8) node {$A_2$};
	\draw[-] (g3) to[out=60,in=0] (10,0.6) to[out=180,in=120] (g3); \draw (10,0.8) node {$A_{F-2}$};
	\draw[-] (g4) to[out=30,in=90] (12.6,0) to[out=-90,in=-30] (g4); \draw (12.5,0) node[right]{$A_{F-1}$};
	
	\draw (2.5,-2.5) node[right]{\small{$\cW = \cW_{\text{gluing}} + \cW_{\text{triangles}} + \prod_{j=0}^{N-1} (V_1 A_1^j L_1 F_1^{(j)} + V_2 A_{F-1}^j R_{F-1} F_2^{(j)})  + $}};
	\draw (2.7,-3.5) node[right] {\small{$ + \sum_{k=0}^{N-1} \Big\{Flip[V_1^2A_1^k] +Flip[ L_1^2 A_1^{k}]+ Flip[V_2^2A_{F-1}^{k}]+Flip[ R_{F-1}^2 A_{F-1}^{k}]\Big\}$ }};
	
\end{tikzpicture}}
\caption{Miror pair of the $4d$ $\CN=1$ $USp(2N)$ antisymmetric SQCD. Throughout this section all the nodes, square or round, are gauge or flavor $USp(2N)$ groups. Nodes depicted with a double line are instead $SU$ groups. Lines are fields in the fundamentals of the groups to whom they are linked, arches are traceless antisymmetric fields. In the mirror theory we have also zig-zag lines representing improved bifundamentals, that are $FE[USp(2N)]$ theories, and crosses denote flipping singlets. To avoid cluttering, we will not indicate the name of the flipping singlets, however their presence can be read from the superpotential given below the theory. By $\CW_{\text{gluing}}$ we denote all the superpotential terms coupling the traceless antisymmetric chirals $A_i$ to the traceless antisymmetric operators inside the improved bifundamentals. Also, in $\CW_{\text{triangles}}$ we collect all the terms associated to triangles in the theory as: $\CW = \Pi_{j+1} R_j L_{j+1}$.} 
\label{fig:SQCD_4d_Dual}
\end{figure}

In this section we present the mirror dual of the $\cN=1$ $USp(2N)$ SQCD with one antisymmetric and $2F+4$ fundamental chirals, which is depicted in figure \ref{fig:SQCD_4d_Dual}.
The global symmetry of the electric theory is: 
\begin{align}
	SU(2F+4)\times U(1)_\tau\,.
\end{align}
We assign a trial R-charge $0$ and $\tau$-charge $1$ to the antisymmetric field $A$, then the R-charge of the fundamentals $Q$ are fixed by requiring the vanishing of the NSVZ $\b$-function. We have:
\begin{align}
	& R[A] = \tau \,, \nn \\
	& R[Q] = r_Q = \frac{F}{2+F} + \frac{1-N}{2+F}\tau \,.
\end{align}

\begin{figure}
\centering
\begin{tikzpicture}[thick,node distance=3cm,gauge/.style={circle,draw,minimum size=5mm},flavor/.style={rectangle,draw,minimum size=5mm}] 

	\path (1,0) node[gauge](g) {$\!\!\!2N\!\!\!$} -- (1,1.5) node[flavor,double] (x) {$\!2F+4\!$} -- (2.75,0) node{$=$};
	
	\chir (x) -- (g); \draw (1,0.75) node[right]{$Q$};
	\draw[-] (g) to[out=-60,in=0] (1,-0.6) to[out=180,in=-120] (g); \draw (1,-0.6) node[right] {$A$};
	
	\path (5,0) node[gauge](g) {$\!\!\!2N\!\!\!$} -- (4,1.5) node[flavor](x1) {$\!2\!$} -- (6,1.5) node[flavor] (x2) {$\!2\!$} 
		-- (4.25,-1.5) node[flavor](y1) {$\!2\!$} -- (5.75,-1.5) node[flavor](y2) {$\!2\!$};
	
	\draw[-] (g) -- (x1); \draw (4.35,0.9) node[left]{$Q_1$};
	\draw[-] (g) -- (x2); \draw (5.65,0.9) node[right]{$Q_F$};
	\draw[-] (g) -- (y1); \draw (4.5,-0.9) node[left]{$P_1$};
	\draw[-] (g) -- (y2); \draw (5.5,-0.9) node[right]{$P_2$};
	\draw[-] (g) to[out=30,in=90] (5.6,0) to[out=-90,in=-30] (g); \draw (5.55,0) node[right]{$A$};
	\draw (5,1.5) node{$\cdots$};
	
	\draw[blue] (3.5,1.5) node {$x_1$}; \draw[blue] (6.5,1.5) node {$x_F$};
	\draw[blue] (3.75,-1.5) node {$y_1$}; \draw[blue] (6.25,-1.5) node {$y_2$};
	
	\draw (11,0) node { \begin{tabular}{|c|c|}
		\hline
				& R-charge \\
		\hline
		$Q_k$ & $r_Q + B_k$ \\
		$P_1$ & $r_Q + C$ \\
		$P_2$ & $r_Q - \sum_{k=1}^{F} B_k - C$ \\
		$A$ & $\tau$ \\
		\hline
	\end{tabular}};
	
\end{tikzpicture}
\caption{Reparameterization of the electric theory along with the list of the R-charges of the fields in the reparameterized theory. The $USp(2)$ symmetries are labeled in blue.}
\label{fig:SQCD_4d_manifest}
\end{figure}
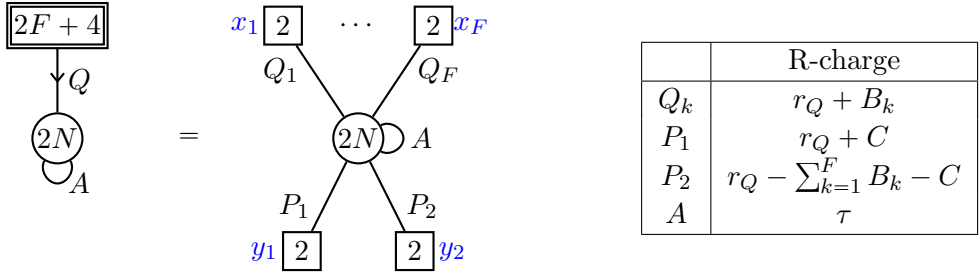

It will be also convenient to consider the theory in the different parameterization where we split the fundamental flavor $Q$ into $F+2$ flavors  as depicted in figure \ref{fig:SQCD_4d_manifest}, where the chirals $Q_j$ are $USp(2N) \times USp(2)_{x_j}$ bifundamentals while $P_{1,2}$ are $USp(2N) \times USp(2)_{y_{1,2}}$ bifundamentals.
Each doublet is then rotated by the $U(1)$ charges reported in table \ref{fig:SQCD_4d_manifest} and the global symmetry recombines as:
\begin{align}\label{eq:4d_symm_decomposition}
\prod_{j=1}^F USp(2)_{x_j}\times  USp(2)_{y_1} \times  USp(2)_{y_2}\times \prod_{j=1}^F U(1)_{B_j} \times U(1)_C = 
SU(2F+4) \,, \nn\\ 
\end{align}
with the fundamental decomposing with the branching rule:
\begin{align}\label{eq:4d_fund_br}
\textbf{2N+4} \to (\textbf{2},\textbf{1},\ldots,\textbf{1})_{(1,0,\ldots,0)} \oplus \ldots (\textbf{1},\ldots,\textbf{1},\textbf{2},\textbf{1})_{(0,\ldots,0,1)} \oplus (\textbf{1},\ldots,\textbf{1},\textbf{2})_{(-1,\ldots,-1)} \,.
\end{align}

\paragraph{Dual quiver}
Let us now discuss the mirror theory: 
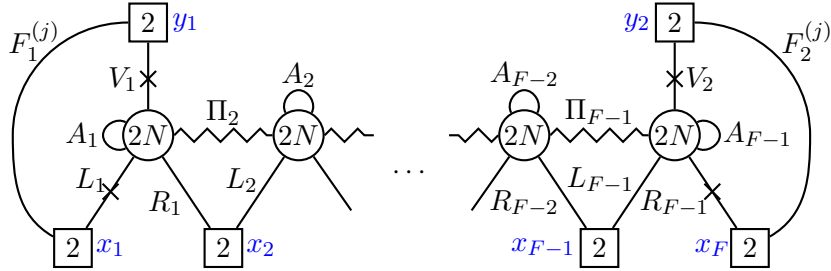
\be
\begin{tikzpicture}[thick,node distance=3cm,gauge/.style={circle,draw,minimum size=5mm},flavor/.style={rectangle,draw,minimum size=5mm}]
	
	\path (5,0) node[gauge](g1) {$\!\!\!2N\!\!\!$} -- (7,0) node[gauge](g2) {$\!\!\!2N\!\!\!$} -- (10,0) node[gauge](g3) {$\!\!\!2N\!\!\!$} -- (12,0) node[gauge](g4) {$\!\!\!2N\!\!\!$} -- (4,-1.5) node[flavor](x1) {$\!2\!$} -- (6,-1.5) node[flavor](x2) {$\!2\!$} -- (11,-1.5) node[flavor](x3) {$\!2\!$} -- (13,-1.5) node[flavor](x4) {$\!2\!$} -- (5,1.5) node[flavor](y1) {$\!2\!$} -- (12,1.5) node[flavor](y2) {$\!2\!$};
	 
	\wigE (g1) -- (g2); \draw (6,0.3) node {$\Pi_2$};
	\wigE (g2) -- (8,0);
	\wigE (g3) -- (9,0);
	\wigE (g3) -- (g4);	 \draw (11,0.3) node {$\Pi_{F-1}$};
	\draw[-] (g1) -- (x1); \draw (4.6,-0.6) node[left]{$L_1$}; 
	\draw[-] (g1) -- (x2); \draw (5.6,-0.9) node[left]{$R_1$};
	\draw[-] (g1) -- (y1); \draw (5,0.75) node[left]{$V_1$};
	\draw[-] (g2) -- (x2); \draw (6.6,-0.6) node[left]{$L_2$};
	\draw[-] (g2) -- (7.7,-1);
	\draw (8.5,-0.5) node {$\cdots$};
	\draw[-] (g3) -- (9.3,-1);
	\draw[-] (g3) -- (x3); \draw (10.6,-0.9) node[left]{$R_{F-2}$};
	\draw[-] (g4) -- (x3); \draw (11.6,-0.6) node[left]{$L_{F-1}$}; 
	\draw[-] (g4) -- (x4); \draw (12.6,-0.9) node[left]{$R_{F-1}$};
	\draw[-] (g4) -- (y2); \draw (12,0.75) node[right]{$V_2$};
	\draw (4.5,-0.75) node[cross]{};
	\draw (5,0.75) node[cross]{};
	\draw (12.5,-0.75) node[cross]{};
	\draw (12,0.75) node[cross]{};
	\draw[-] (y1) to[out=180,in=90] (3.2,0) to[out=-90,in=150] (x1); \draw (3,1.25) node[right]{$F_1^{(j)}$};
	\draw[-] (y2) to[out=0,in=90] (13.8,0) to[out=-90,in=30] (x4); \draw (14.25,1.25) node[left]{$F_2^{(j)}$};
	\draw[-] (g1) to[out=150,in=90] (4.4,0) to[out=-90,in=-150] (g1); \draw (4.5,0) node[left]{$A_1$};
	\draw[-] (g2) to[out=60,in=0] (7,0.6) to[out=180,in=120] (g2); \draw (7,0.8) node{$A_2$};
	\draw[-] (g3) to[out=60,in=0] (10,0.6) to[out=180,in=120] (g3); \draw (10,0.8) node{$A_{F-2}$};
	\draw[-] (g4) to[out=30,in=90] (12.6,0) to[out=-90,in=-30] (g4); \draw (12.5,0) node[right]{$A_{F-1}$};
	
	\draw[blue] (5.5,1.5) node {$y_1$}; \draw[blue] (11.5,1.5) node {$y_2$};
	\draw[blue] (4.5,-1.5) node {$x_1$}; \draw[blue] (6.5,-1.5) node {$x_2$};
	\draw[blue] (10.25,-1.5) node {$x_{F-1}$}; \draw[blue] (12.5,-1.5) node {$x_{F}$};
	
	\draw (2.5,-2.5) node[right]{\small{$\cW =\cW_{\text{gluing}} + \cW_{\text{triangles}} + \prod_{j=0}^{N-1} (V_1 A_1^j L_1 F_1^{(j)} + V_2 A_{F-1}^j R_{F-1} F_2^{(j)})  + $}};
	\draw (2.7,-3.5) node[right] {\small{$ + \sum_{k=0}^{N-1} \Big\{Flip[V_1^2A_1^k] +Flip[ L_1^2 A_1^{k}]+ Flip[V_2^2A_{F-1}^{k}]+Flip[ R_{F-1}^2 A_{F-1}^{k}]\Big\}$ }};

\end{tikzpicture}
\label{mirr4d}
\ee
It is the linear quiver of $F-1$ $USp(2N)$ gauge nodes linked by $F-2$ improved bifundamentals which  are identified with the $FE[USp(2N)]$ theories introduced in \cite{Pasquetti:2019hxf}, which we describe in appendix \ref{app:FE}. 
We denote this theory in short by a zig-zag line connecting the two non-abelian $USp(2N)$ IR symmetries. In addition to them the improved bifundamental has a $U(1)_\tau \times U(1)_B$ global abelian symmetry. The spectrum of this theory includes two traceless antisymmetric operators, one for each $USp(2N)$ symmetry, a bifundamental $\Pi$ and a matrix of singlets under the two $USp(2N)$ symmetries $\mathsf{B}_{n,m}$. The two antisymmetric operators carry the same R-charge and are both rotated by the $U(1)_\tau$ symmetry, while the $\Pi$ operator is charged only under the $U(1)_B$ symmetry. \\
On the two sides of the quiver we have flavors $V_{1,2}$ charged under the $USp(2)_{y_{1,2}}$ symmetries. Each gauge node is attached to two teeth of the {\it saw} by the chirals $L_j$ and $R_j$ that are respectively charged under the $USp(2)_{x_j}$ and $USp(2)_{x_{j+1}}$ symmetries. \\
\begin{table}[h!]
\centering
\renewcommand{\arraystretch}{1.2}
\begin{tabular}{|c|c|}
	\hline
		& R-charge \\
	\hline
	$V_1$ & $1 - r_Q + \frac{1-N}{2}\tau - B_1$ \\
	$V_2$ & $1 - r_Q + \frac{1-N}{2}\tau - B_F$ \\
	$L_k$ &	$k(1 - r_Q) + \frac{1-N}{2}\tau - \sum_{j=1}^{k-1} B_j - C$ \\
	$R_k$ & $2 - (k+2)(1 - r_Q) - \frac{1-N}{2}\tau + \sum_{j=1}^{k+1} B_j + C$ \\
	$\Pi_k$ & $1 - r_Q - B_k$ \\
	$F_1^{(k)}$ & $2r_Q + (N+k-2)\tau + B_1 + C$ \\
	$F_2^{(k)}$ & $2r_Q + (k-1)\tau - \sum_{j=1}^{F-1} B_j - C$ \\
	$A_j$ & $\tau$ \\
	\hline
\end{tabular}
\caption{List of the R-charges of the fields and operators in the mirror theory given in figure \ref{mirr4d}. Recall that to specify completely the parameterization of the two $U(1)$ symmetries of an improved bifundamental, it is sufficient to specify the R-charge of the anti-symmetric and of the bifundamental operator.}
\label{tab:SQCD_4d_Charges_mirror}
\end{table}
The improved bifundamentals are {\it glued} together by gauging a diagonal $USp(2N)$ symmetry and adding
a traceless antisymmetric chiral $A_j$ at each node, which couples to the traceless antisymmetric operators inside the improved bifundamentals as: $A_j(\mathsf{A}^{(j)}_L + \mathsf{A}^{(j+1)}_R)$, where $\mathsf{A}^{(j)}_L$ is the antisymmetric operator of the improved bifundamental on the left of the gauge node, while $\mathsf{A}^{(j+1)}_R$ is that of the improved bifundamental on the right.
We collect all the gluing superpotentials inside $\CW_{\text{gluing}}$. Notice that when we glue a string of improved bifundamentals,  all the $U(1)_\tau$ symmetries are identified while the $U(1)_{B_j}$ symmetries acting on each improved bifundamental are all preserved.
The $R_j$ and $L_j$ chirals are coupled to the bifundamental operators $\Pi_j$ as: $\Pi_j R_{j+1} L_j$. Each term corresponds to a triangle composing the saw, we then collect them in $\CW_{\text{triangles}}$. 
$F^{(j)}_1$ enter in the superpotential flipping all the meson constructed from $V_1$ and $L_1$ in the bifundamental of $USp(2)_{y_{1}} \times USp(2)_{x_{1}}$ dressed with powers of the first antisymmetric $A_1$, similarly $F^{(j)}_2$ flips the dressed mesons built from $V_2$ and $R_{F-1}$. Finally, we flip all the square mesons built from $V_1, V_2, L_1, R_{F-1}$ dressed with powers of the antisymmetric. \\

The manifest global symmetry is 
\be
\prod_{j=1}^F USp(2)_{x_j}\times  USp(2)_{y_1} \times  USp(2)_{y_2}\times \prod_{j=1}^F U(1)_{B_j} \times U(1)_c \times U(1)_\tau \,,
\ee
which enhances in the IR to $SU(2F+4)\times U(1)_\tau$, we will provide many evidences of this enhancement throughout the section.

\paragraph{Anomaly matching}
As a first check of the proposed duality we can show how the anomalies of the two theories match. \\
In the electric theory we can compute two anomalies for the flavor group $SU(2F+4)$ that are:
\begin{align}
	\Tr SU(2F+4)^2U(1)_\tau = \frac{N(1-N)}{2+F}  \quad, \quad \Tr SU(2F+4)^2U(1)_R = -\frac{2N}{2+F} \,.
\end{align}
In the magnetic theory the $SU(2F+4)$ symmetry is only emergent and we can't  directly calculate its anomaly. Nevertheless we can
can calculate  the following anomalies involving  $USp(2)_{x_i}$, $USp(2)_{y_i}$, $U(1)_{b_i}$ and $U(1)_c$   with either $U(1)_\tau$ and $U(1)_R$:
\begin{align}
	\Tr USp(2)^2_{x_i,y_{j}}U(1)_\tau = \frac{N(1-N)}{2+F} \quad, \quad \Tr USp(2)^2_{x_i,y_{j}}U(1)_R = -\frac{2N}{2+F} \,, \nn \\
	\Tr U(1)^2_{b_i,c} U(1)_\tau = 8\frac{N(1-N)}{2+F} \quad , \quad \Tr U(1)^2_{b_i,c} U(1)_R = -8\frac{2N}{2+F} \,,
\end{align}
and check they are compatible with the enhancement. Indeed given a decomposition of a group $G$ into $H$, for which we have a branching rule:
\begin{align}
	r \rightarrow \oplus_{j=1}^{K} \tilde{r}_j \,,
\end{align}
where $r$ is some representation of $G$ and $\tilde{r}_j$ are representations of $H$, the embedding index is defined as:
\begin{align}\label{eq:4d_embind_def}
	I ( H \hookrightarrow G ) = \frac{\sum_{j=1}^{K}T(\tilde{r}_j)}{T(r)} \,,
\end{align}
where we have denoted as $T(r)$ the Dynkin index of a representation $r$. The result is independent on the choice of the branching rule. 
Once we have computed the embedding index, the anomalies of the manifest symmetries are constrained by the anomalies of the emergent symmetries to satisfy:
\begin{align}\label{eq:EmbIndAnom}
	I(H \hookrightarrow G) \Tr G^2 U(1) = \Tr H^2U(1) \,.
\end{align}
Using the branching rule in \eqref{eq:4d_fund_br} and the definition in \eqref{eq:4d_embind_def}, we get:
\begin{align}\label{eq:embind_SU2F+4}
	& I(USp(2)_{x_j} \hookrightarrow SU(2F+4)) = I(USp(2)_{y_j} \hookrightarrow SU(2F+4)) = 1 \nonumber \,,\\
	& I(U(1)_{b_j} \hookrightarrow SU(2F+4)) = I(U(1)_c \hookrightarrow SU(2F+4)) = 8  \,.
\end{align}
The results found are are exactly the results expected: all the anomalies for the $USp(2)$s factors coincide with the anomalies of the electric theory (since the embedding index is 1); while the anomalies involving factors of $U(1)$s differ by a factor 8 (since the embedding index is 8).

As a final remark, we checked that also the anomalies involving only the $U(1)$ groups match, as for example $\Tr U(1)_\tau, \Tr U(1)_R, \Tr U(1)_\tau^2 U(1)_R, \ldots$

\paragraph{Superconformal indexes}
We now give the superconformal index identity for the SQCD mirror pair in figure \ref{fig:SQCD_4d_Dual}. To write the superconformal index we first define the fugacities related to the $U(1)$ symmetries as:
\begin{align}
	t = (pq)^{\tau/2} \quad, \quad b_j = (pq)^{B_j/2} \quad, \quad c = (pq)^{C/2} \,,
\end{align}
and also the vector $\vec{y}$ and $\vec{x}$ as fugacities for the $USp(2)_{y_j}$ and $USp(2)_{x_j}$ symmetries respectively. 
The duality \ref{fig:SQCD_4d_Dual} consist in the following superconformal index identity:
\begin{align}\label{eq:4d_ind_id}
	\CI_{SQCD}(\vec{x},\vec{y},\vec{b},c,t) = \CI_{\widecheck{SQCD}}(\vec{x},\vec{y},\vec{b},c,t) \,.
\end{align}
Where we define the superconformal index of the SQCD, parameterized as  in \ref{fig:SQCD_4d_manifest}, as:
\begin{align}\label{eq:4d_ind_elec}
	\CI_{SQCD}(\vec{x},\vec{y},\vec{b},c,t) = \oint & \, d\vec{z}_N \D_N(\vec{z},t) \prod_{j=1}^N \big( \prod_{a=1}^F \Ge( pq^{r_Q/2} b_a z_j^\pm x_a^\pm) \nn \\
	& \Ge( pq^{r_Q/2} c z_j^\pm y_1^\pm) \Ge( pq^{r_Q/2} \prod_{a=1}^{F} b_a^{-1} c^{-1} z_j^\pm y_2^\pm ) \big) \,.
\end{align}
The index of the mirror theory is instead given as:
\begin{align}\label{eq:4d_ind_magn}
	\CI_{\widecheck{SQCD}}(\vec{x},\vec{y},\vec{b},c,t) = & \prod_{j=1}^N \big[ \Ge( pq^{r_Q} t^{N+j-2} b_1 c x_1^\pm y_1^\pm ) 
	\Ge( pq^{r_Q} t^{j-1} (b_1 \ldots b_{F-1} c)^{-1} x_F^\pm y_2^\pm ) \nn \\
	& \Ge( pq^{r_Q} t^{N-1-j} b_1^{-2} ) \Ge( pq^{r_Q} t^{N-1-j} b_F^{-2} ) \Ge( pq^{r_Q} t^{N-1-j} c^2 ) \nn \\
	&  \Ge( pq^{(F-1)(1-r_Q)} t^{N-1-j} (b_1 \ldots b_{F} c)^{-2} )	\big] \nn \\
	& \oint \prod_{a=1}^{F-1} \big( d\vec{z}^{(a)}_N \D_N(\vec{z}^{(a)},t) \big) \prod_{a=2}^{F-1} \CI_{FE}^{(N)} (\vec{z}^{(a-1)},\vec{z}^{(a)},\tau,b_a) \nn \\
	& \prod_{j=1}^N \big[ \Ge( pq^{\frac{1-r_Q}{2}} t^{\frac{1-N}{2}} b_1^{-1} z_j^{(1)\pm} y_1^\pm ) 
	\Ge( pq^{\frac{1-r_Q}{2}} t^{\frac{1-N}{2}} b_F^{-1} z_j^{(F-1)\pm} y_2^\pm ) \big] \nn \\
	& \prod_{l=1}^{F-1} \prod_{j=1}^N \big[ \Ge( pq^{l\frac{1-r_Q}{2}} t^{\frac{1-N}{2}} (b_1 \ldots b_{l-1}c)^{-1} z_j^{(a)\pm} x_l^\pm ) \nn \\
	& \Ge( pq^{1-(l+2)\frac{1-r_Q}{2}} t^{\frac{N-1}{2}} b_1 \ldots b_{l+1} c z_j^{(a)\pm} x_{l+1}^\pm ) \big] \,.
\end{align}
The convention used to write the superconformal indexes can be found in appendix \ref{inpaconv}.

\subsubsection{Comments on $F=1,2$ cases}
The cases $F=1,2$ are already discussed in literature, in this section we wish to comment on how our result reconciles with these known results.

Let us start from the case $F=2$. The mirror dual proposed in \ref{fig:SQCD_4d_Dual} reduces to a theory of a single $USp(2N)$ gauge group with no improved bifundamentals. The duality is then a self-duality modulo flips:
\be
\resizebox{.95\hsize}{!}{
\begin{tikzpicture}[thick,node distance=3cm,gauge/.style={circle,draw,minimum size=5mm},flavor/.style={rectangle,draw,minimum size=5mm}] 

	\path (1,0) node[gauge](g) {$\!\!\!2N\!\!\!$} -- (0,1.5) node[flavor] (x1) {$\!2\!$} -- (2,1.5) node[flavor] (x2) {$\!2\!$} 
		-- (0,-1.5) node[flavor] (y1) {$\!2\!$} -- (2,-1.5) node[flavor] (y2) {$\!2\!$}
		-- (3.5,0) node{$=$};
	
	\draw[-] (g) -- (x1); \draw (0.5,0.75) node[left]{$Q_1$};
	\draw[-] (g) -- (x2); \draw (1.5,0.75) node[right]{$Q_2$};
	\draw[-] (g) -- (y1); \draw (0.5,-0.75) node[left]{$P_1$};
	\draw[-] (g) -- (y2); \draw (1.5,-0.75) node[right]{$P_1$};
	\draw[-] (g) to[out=30,in=90] (1.6,0) to[out=-90,in=-30] (g); \draw (1.6,0) node[right] {$A$};
	
	\draw[blue] (0.5,1.5) node {$x_1$}; \draw[blue] (1.5,1.5) node {$x_2$};
	\draw[blue] (0.5,-1.5) node {$y_1$}; \draw[blue] (1.5,-1.5) node {$y_2$};
	
	\draw (1,-2.5) node {$\CW = 0$};

\begin{scope}[shift={(2,0)}]
	\path (5,0) node[gauge](g) {$\!\!\!2N\!\!\!$} -- (4,1.5) node[flavor](x1) {$\!2\!$} -- (6,1.5) node[flavor] (x2) {$\!2\!$} 
		-- (4,-1.5) node[flavor](y1) {$\!2\!$} -- (6,-1.5) node[flavor](y2) {$\!2\!$};
	
	\draw[-] (g) -- (x1); \draw (4.5,0.75) node[cross] {}; \draw (4.5,0.75) node[left]{$Q'_1$};
	\draw[-] (g) -- (x2); \draw (5.5,0.75) node[cross] {}; \draw (5.5,0.75) node[right]{$Q'_2$};
	\draw[-] (g) -- (y1); \draw (4.5,-0.75) node[cross] {}; \draw (4.5,-0.75) node[left]{$P'_1$};
	\draw[-] (g) -- (y2); \draw (5.5,-0.75) node[cross] {}; \draw (5.5,-0.75) node[right]{$P'_2$};
	\draw[-] (g) to[out=30,in=90] (5.6,0) to[out=-90,in=-30] (g); \draw (5.55,0) node[right]{$A'$};
	
	\draw[-] (x1) to[out=-150,in=90] (3,0) to[out=-90,in=150] (y1); \draw (3.4,1) node[left] {$F_1^{(j)}$};
	\draw[-] (x2) to[out=-30,in=90] (7,0) to[out=-90,in=30] (y2); \draw (6.6,1) node[right] {$F_2^{(j)}$};
	
	\draw[blue] (4.5,1.5) node {$x_1$}; \draw[blue] (5.5,1.5) node {$x_2$};
	\draw[blue] (4.5,-1.5) node {$y_1$}; \draw[blue] (5.5,-1.5) node {$y_2$};
	
	\draw (5,-2.5) node {$\CW = \sum_{a=1}^2 \sum_{j=1}^N \big( F_a^{(j)} Q'_a A^{j-1} P'_a +$};
	\draw (5,-3.2) node {$ + Flip[{Q'_a}^2 A^{j-1}] + Flip[{P'_a}^2 A^{j-1}] \big) $};
	
	\draw (13,0) node { 
	\renewcommand{\arraystretch}{1.5}
	\begin{tabular}{|c|c|}
		\hline
				& R-charge \\
		\hline 
		$Q_{1,2}$ & $r_Q + B_{1,2}$ \\
		$P_1$ & $r_Q + C$ \\
		$P_2$ & $r_Q - \sum_{k=1}^{F} B_k - C$ \\
		\hline 
		$Q'_1$ & $1-r_Q + \frac{1-N}{2}\tau - C$ \\
		$Q'_2$ & $-2 - 4r_Q + \frac{N-1}{2}\tau + B_1 + B_2 + C$\\
		$P'_{1,2}$ & $1 - r_Q + \frac{1-N}{2}\tau - B_{1,2}$ \\
		\hline 
		$A,A'$ & $\tau$ \\
		\hline
	\end{tabular}};
\end{scope}
	
\end{tikzpicture}}
\ee
This result coincides with the CSST self-duality derived in \cite{Csaki:1997cu}. A similar self-duality, with different number of flippers, can be also obtained using the sequential deconfinement technique, as shown in \cite{Bajeot:2022lah}. 

For the  $F=1$ case, which can't be  read directly from the duality in figure \ref{fig:SQCD_4d_Dual}, we can  run the dualization algorithm, as in the $3d$ $F=1$ case  described in section \ref{alsqcd},   
 the result produced is consistent with the earlier  duality proposed in \cite{Csaki:1996sm} shown in figure
\eqref{conf4d} which  was discussed in  \cite{Benvenuti:2018bav} and  derived via sequential deconfinement in \cite{Bottini:2022vpy,Bajeot:2022lah}.
\be
\resizebox{.9\hsize}{!}{
\begin{tikzpicture}[thick,node distance=3cm,gauge/.style={circle,draw,minimum size=5mm},flavor/.style={rectangle,draw,minimum size=5mm}] 

	\path (1,0) node[gauge](g) {$\!\!\!2N\!\!\!$} -- (1,1.5) node[flavor] (x) {$\!2\!$}
		-- (0,-1.5) node[flavor] (y1) {$\!2\!$} -- (2,-1.5) node[flavor] (y2) {$\!2\!$}
		-- (3.5,0) node{$=$};
	
	\draw[-] (g) -- (x); \draw (1,0.75) node[left]{$Q$};
	\draw[-] (g) -- (y1); \draw (0.5,-0.75) node[left]{$P_1$};
	\draw[-] (g) -- (y2); \draw (1.5,-0.75) node[right]{$P_2$};
	\draw[-] (g) to[out=30,in=90] (1.6,0) to[out=-90,in=-30] (g); \draw (1.6,0) node[right] {$A$};
	
	\draw[blue] (0.5,1.5) node {$x$};
	\draw[blue] (-0.5,-1.5) node {$y_1$}; \draw[blue] (2.5,-1.5) node {$y_2$};
	
	\draw (1,-2.5) node {\small{$\CW = 0$}};

\begin{scope}[shift={(2,0)}]
	\path (4,-0.75) node[flavor](y1) {$\!2\!$} -- (6,-0.75) node[flavor](y2) {$\!2\!$} -- (5,0.75) node[flavor](x) {$\!2\!$};
	
	\draw[-] (y1) -- (y2); \draw (5,-0.75) node[cross] {}; \draw (5,-1.15) node {$R_j$};
	\draw[-] (y2) -- (x); \draw (5.5,0) node[cross] {}; \draw (5.5,0) node[right] {$S_j$};
	\draw[-] (x) -- (y1); \draw (4.5,0) node[cross] {}; \draw (4.5,0) node[left] {$T_j$};
	
	\draw[blue] (5,1.25) node {$x$};
	\draw[blue] (3.5,-0.75) node {$y_1$}; \draw[blue] (6.5,-0.75) node {$y_2$};
	
	\draw (5,-2.5) node {\small{$\CW = \sum_{j,k,l=1}^N R_jS_kT_l \d_{j+k+l,N+2}$}};
	
	\draw (11.5,0) node { 
	\renewcommand{\arraystretch}{1.15}
	\begin{tabular}{|c|c|}
		\hline
				& R-charge \\
		\hline 
		$Q$ & $r_Q + B$ \\
		$P_1$ & $r_Q + C$ \\
		$P_2$ & $r_Q - \sum_{k=1}^{F} B_k - C$ \\
		\hline 
		$R_j$ & $1-r_Q + (1-j)\tau - B$ \\
		$S_j$ & $2r_Q + (N+j-2)\tau + B + C$\\
		$T_j$ & $2r_Q + (j-1)\tau - C$ \\
		\hline 
		$A$ & $\tau$ \\
		\hline
	\end{tabular}};
\end{scope}
	
\end{tikzpicture}}
\label{conf4d}
\ee
This duality relates the $USp(2N)$ SQCD with one antisymmetric  and 6 fundamental chirals where we flipped the  tower of powers of the antisymmetric  (which would  all be below the unitarity bound), to a Wess-Zumino model with $15N$ chirals.  The superpotential was proposed in \cite{Benvenuti:2018bav} and  tested  using sequential deconfinement in \cite{Bajeot:2022lah}. 
Starting from the the SQCD on the l.h.s of figure \eqref{conf4d}, the algorithm yields on the dual side a collection of $15N$ chiral fields with a charge assignment  compatible with the superpotential given \eqref{conf4d}.

\subsubsection{Operator map}
In this section we discuss how the operator map works in the duality presented in \ref{fig:SQCD_4d_Dual}.
\begin{itemize}
	\item In the electric theory we have dressed mesonic operators with R-charge:
	\begin{align}
		R[Q^2A^k] = 2r_Q + j\tau \qquad \text{for} ~k=0,\cdots,N-1 \,.
	\end{align}
	For each value of the dressing we have an operator in the antisymmetric representation of $SU(2F+4)$, of dimension $(2F+4)(2F+3)/2$, which we map to a collection of $(F+2)$ singlets and $(F+2)(F+1)/2 \times 4$ mesonic operators in the bifundamental of a pair of two $USp(2)$ global symmetries. It is actually easier to write the explicit map considering the SQCD in the parameterization \ref{fig:SQCD_4d_manifest}: 
\begin{align}
\renewcommand{\arraystretch}{1.2}
\begin{tabular}{c|c}
	$SQCD$ & $\widecheck{SQCD}$ \\
	\hline 
	$Q_1^2 A^j$ & $\CF[ V_1^2 A_1^{N-1-j} ]$ \\
	$Q_F^2 A^j$ & $\CF[ V_2^2 A_{F-1}^{N-1-j} ]$ \\
	$Q_a^2 A^j \, \text{for} \, a \neq 1,F$ & $\mathsf{B}^{(a)}_{1,j}$ \\
	$P_1^2 A^j$ & $\CF[ L_1^2 A_1^{N-1-j} ]$ \\
	$P_2^2 A^j$ & $\CF[ R_{F-1}^2 A_{F-1}^{N-1-j} ]$ \\
	\hline
	$Q_a A^j Q_b \, \text{for} \, a \neq b$ & $L_a A^j \Pi_{a+1} \ldots \Pi_{b-1} R_{b-1} $ \\ 
	$Q_1 A^j P_1 $ & $F^{(N-j)}_1$ \\
	$Q_a A^j P_1 \, \text{for} \, a \neq 1$ & $V_1 A^j \Pi_2 \ldots \Pi_{a-1} R_{a-1} $ \\
	$Q_F A^j P_2 $ & $F^{(N-j)}_2$ \\
	$Q_a A^j P_2 \, \text{for} \, a \neq F$ & $L_a A^j \Pi_{a+1} \ldots \Pi_{F-1} V_2 $ 
\end{tabular}
\end{align}
For the magnetic mesonic operators the dressing is performed using any antisymmetric chiral $A_k$, which we denote simply by $A$, all the possible choices of dressing are identified by quantum relations.

\item In the electric theory we then have the traces of the antisymmetric $A$ with charge:
	\begin{align}
		R[\Tr A^l] = j \tau \qquad \text{for} ~l=2,\ldots,N \,.
 	\end{align}
	In the magnetic theory they are simply mapped into traces of any antisymmetric chiral $A_j$, that are all identified due to quantum relations.
\end{itemize}
We also point out that under the duality it seems that all the $\mathsf{B}^{(a)}_{n,m}$ operators in the magnetic theory with $n>1$ are not mapped. We suspect that these operators are trivial in the chiral ring.

\subsubsection{Deformations and consistency checks}
In this section we study the effect of some interesting deformations, providing also nontrivial consistency checks of the duality in figure \ref{fig:SQCD_4d_Dual}. \\

Before discussing the deformations let us mention that, as in the $3d$ case, we have a freedom of rearranging flavors and improved bifundamentals. There are three {\it swapping} dualities that allow us to perform this reshuffling. \\

The first one is the  duality  \ref{fig:4d_starstar} that allows us to exchange any pair of consequent improved bifundamentals. The effect of this duality is to swap the two $U(1)_{B_j}$ symmetries rotating the improved bifundamentals and also the $USp(2)_{x_j}$ symmetries associated to the saw structure:
\be\label{fig:MagSwap_B2B3} 
\resizebox{.95\hsize}{!}{
 	 \bpic[thick,node distance=3cm,gauge/.style={circle,draw,minimum size=5mm},flavor/.style={rectangle,draw,minimum size=5mm}] 
 
		\path (-2,0) node[gauge](g1) {$\!\!\!2N\!\!\!$} -- (0,0) node[gauge](g2) {$\!\!\!2N\!\!\!$} -- (2,0) node[gauge](g3) {$\!\!\!2N\!\!\!$}  
			-- (-1,-1.5) node[flavor](x2) {$\!2\!$} -- (1,-1.5) node[flavor](x3) {$\!2\!$}
			-- (4,0) node{$\Longleftrightarrow$};
	 
		\wigE (g1) -- (-3,0);		
		\wigE (g1) -- (g2); \draw (-1,0.3) node {$\Pi_j$};
		\wigE (g2) -- (g3); \draw (1,0.3) node {$\Pi_{j+1}$};
		\wigE (g3) -- (3,0);
		\draw[-] (g1) -- (-2.7,-1);
		\draw[-] (g1) -- (x2); 
		\draw[-] (g2) -- (x2);
		\draw[-] (g2) -- (x3);
		\draw[-] (g3) -- (x3);
		\draw[-] (g3) -- (2.7,-1);
		\draw[-] (g1) to[out=60,in=0] (-2,0.6) to[out=180,in=120] (g1);
		\draw[-] (g2) to[out=60,in=0] (0,0.6) to[out=180,in=120] (g2);
		\draw[-] (g3) to[out=60,in=0] (2,0.6) to[out=180,in=120] (g3);
		\draw[blue] (-0.8,-1.5) node[right] {\scriptsize{$x_j$}};
		\draw[blue] (1.2,-1.5) node[right] {\scriptsize{$x_{j+1}$}};
		
		\draw (-2,-3) node[right] { \begin{tabular}{c|c}
			$\Pi_j$ & $1 - r_Q - B_j$ \\
			$\Pi_{j+1}$ & $1 - r_Q - B_{j+1}$
		\end{tabular}};
	 
		\path (6,0) node[gauge](g1) {$\!\!\!2N\!\!\!$} -- (8,0) node[gauge](g2) {$\!\!\!2N\!\!\!$} -- (10,0) node[gauge](g3) {$\!\!\!2N\!\!\!$}
			-- (7,-1.5) node[flavor](x2) {$\!2\!$} -- (9,-1.5) node[flavor](x3) {$\!2\!$};
	 
		\wigE (g1) -- (5,0);		
		\wigE (g1) -- (g2); \draw (7,0.35) node {$\Pi'_j$};
		\wigE (g2) -- (g3); \draw (9,0.35) node {$\Pi'_{j+1}$};
		\wigE (g3) -- (11,0);
		\draw[-] (g1) -- (5.3,-1);
		\draw[-] (g1) -- (x2);
		\draw[-] (g2) -- (x2);
		\draw[-] (g2) -- (x3);
		\draw[-] (g3) -- (x3);
		\draw[-] (g3) -- (10.7,-1);
		\draw[-] (g1) to[out=60,in=0] (6,0.6) to[out=180,in=120] (g1);
		\draw[-] (g2) to[out=60,in=0] (8,0.6) to[out=180,in=120] (g2);
		\draw[-] (g3) to[out=60,in=0] (10,0.6) to[out=180,in=120] (g3);
		\draw[blue] (7.2,-1.5) node[right] {\scriptsize{$x_{j+1}$}};
		\draw[blue] (9.2,-1.5) node[right] {\scriptsize{$x_j$}};
		
		\draw (6,-3) node[right] { \begin{tabular}{c|c}
			$\Pi'_j$ & $1 - r_Q - B_{j+1}$ \\
			$\Pi'_{j+1}$ & $1 - r_Q - B_j$
		\end{tabular}};
		 
\epic}\ee	
Notice that under this duality the matrix $\mathsf{B}^{(j)}_{n,m}$ is mapped to ${¶\mathsf{B}}'^{(j+1)}_{n,m}$ and vice-versa.
Also the charges of the chirals composing the saw are non-trivially mapped under the duality above, for all the details we refer the reader 
to the discussion in  appendix \ref{fig:4d_starstar}. \\

The second duality given in \eqref{fig:4d_starstar_nilmass},  allows us to swap the left vertical flavor $V_1$ with the first improved bifundamental $\Pi_2$, meaning that we swap the $U(1)_{B_1} \times USp(2)_{x_1}$ and $U(1)_{B_2} \times USp(2)_{x_2}$ symmetries.
\be\label{fig:MagSwap_V1B2} 
 	 \bpic[thick,node distance=3cm,gauge/.style={circle,draw,minimum size=5mm},flavor/.style={rectangle,draw,minimum size=5mm}] 
 
		\path (-2,0) node[gauge](g1) {$\!\!\!2N\!\!\!$} -- (0,0) node[gauge](g2) {$\!\!\!2N\!\!\!$} 
			-- (-3,-1.5) node[flavor](x1) {$\!2\!$} -- (-1,-1.5) node[flavor](x2) {$\!2\!$} -- (-2,1.5) node[flavor](y1) {$\!2\!$} 
			-- (2,0) node{$\Longleftrightarrow$};
	 
		\wigE (g1) -- (g2); \draw (-1,0.3) node {$\Pi_2$};
		\wigE (g2) -- (1,0);
		\draw[-] (g1) -- (x1); 
		\draw[-] (g1) -- (x2);
		\draw[-] (g1) -- (y1); \draw (-2,0.75) node[right] {$V_1$};
		\draw[-] (g2) -- (x2);
		\draw[-] (g2) -- (0.7,-1);
		\draw (-2.5,-0.75) node[cross]{};
		\draw (-2,0.75) node[cross]{};
		\draw[-] (y1) to[out=180,in=90] (-3.8,0) to[out=-90,in=150] (x1);
		\draw[-] (g1) to[out=150,in=90] (-2.6,0) to[out=-90,in=-150] (g1);
		\draw[-] (g2) to[out=60,in=0] (0,0.6) to[out=180,in=120] (g2);
		\draw[blue] (-2.8,-1.5) node[right] {\scriptsize{$x_1$}};
		\draw[blue] (-0.8,-1.5) node[right] {\scriptsize{$x_2$}};
		
		\draw (-3.5,-3) node[right] { \begin{tabular}{c|c}
			$V_1$ & $1 - r_Q - \frac{1-N}{2}\tau + B_1$ \\
			$\Pi_2$ & $1 - r_Q - B_2$
		\end{tabular}};
	 
		\path (5,0) node[gauge](g1) {$\!\!\!2N\!\!\!$} -- (7,0) node[gauge](g2) {$\!\!\!2N\!\!\!$} 
			-- (4,-1.5) node[flavor](x1) {$\!2\!$} -- (6,-1.5) node[flavor](x2) {$\!2\!$} -- (5,1.5) node[flavor](y1) {$\!2\!$};
	 
		\wigE (g1) -- (g2); \draw (6,0.3) node {$\Pi'_2$};
		\wigE (g2) -- (8,0);
		\draw[-] (g1) -- (x1);
		\draw[-] (g1) -- (x2);
		\draw[-] (g1) -- (y1); \draw (5,0.75) node[right] {$V'_1$};
		\draw[-] (g2) -- (x2);
		\draw[-] (g2) -- (7.7,-1);
		\draw (4.5,-0.75) node[cross]{};
		\draw (5,0.75) node[cross]{};
		\draw[-] (y1) to[out=180,in=90] (3.2,0) to[out=-90,in=150] (x1);
		\draw[-] (g1) to[out=150,in=90] (4.4,0) to[out=-90,in=-150] (g1);
		\draw[-] (g2) to[out=60,in=0] (7,0.6) to[out=180,in=120] (g2);
		\draw[blue] (4.2,-1.5) node[right] {\scriptsize{$x_2$}};
		\draw[blue] (6.2,-1.5) node[right] {\scriptsize{$x_1$}};
		
		\draw (3.5,-3) node[right] { \begin{tabular}{c|c}
			$V'_1$ & $1 - r_Q - \frac{1-N}{2}\tau + B_2$ \\
			$\Pi'_2$ & $1 - r_Q - B_1$
		\end{tabular}};
		 
\epic\ee
Under this duality the matrix of singlets $\mathsf{B}^{(2)}_{n,m}$ is partially mapped to the tower of singlets of the vertical flavor as: $\mathsf{B}^{(2)}_{1,m} \leftrightarrow \CF[{V'}_1^2 A^{(N-m)}]$ and viceversa. Notice that the rest of the matrix of singlets of the improved bifundamental doesn not map under this duality since  these operators are zero in the chiral ring. The same strategy can be used to swap the last improved bifundamental $\Pi_{F-1}$ with the right vertical flavor $V_2$. \\

The last swapping move consist in exchanging the left vertical flavor $V_1$ with the first diagonal leg $L_1$ or, analogously, the right vertical flavor $V_2$ with the last diagonal leg $R_{F-1}$. This is a trivial move since it just amounts to a redefinition of the fields:
\be\label{fig:MagSwap_V1L1} 
 	 \bpic[thick,node distance=3cm,gauge/.style={circle,draw,minimum size=5mm},flavor/.style={rectangle,draw,minimum size=5mm}] 
 
		\path (-2,0) node[gauge](g1) {$\!\!\!2N\!\!\!$} -- (0,0) node[gauge](g2) {$\!\!\!2N\!\!\!$} 
			-- (-3,-1.5) node[flavor](x1) {$\!2\!$} -- (-1,-1.5) node[flavor](x2) {$\!2\!$} -- (-2,1.5) node[flavor](y1) {$\!2\!$} 
			-- (2,0) node{$\Longleftrightarrow$};
	 
		\wigE (g1) -- (g2); 
		\wigE (g2) -- (1,0);
		\draw[-] (g1) -- (x1); \draw (-2.5,-0.75) node[left] {$L_1$};
		\draw[-] (g1) -- (x2); 
		\draw[-] (g1) -- (y1); \draw (-2,0.75) node[right] {$V_1$};
		\draw[-] (g2) -- (x2);
		\draw[-] (g2) -- (0.7,-1);
		\draw (-2.5,-0.75) node[cross]{};
		\draw (-2,0.75) node[cross]{};
		\draw[-] (y1) to[out=180,in=90] (-3.8,0) to[out=-90,in=150] (x1);
		\draw[-] (g1) to[out=150,in=90] (-2.6,0) to[out=-90,in=-150] (g1);
		\draw[-] (g2) to[out=60,in=0] (0,0.6) to[out=180,in=120] (g2);
		\draw[blue] (-1.8,1.5) node[right] {\scriptsize{$y_1$}};
		\draw[blue] (-2.8,-1.5) node[right] {\scriptsize{$x_1$}};
		
		\draw (-3.5,-3) node[right] { \begin{tabular}{c|c}
			$V_1$ & $1 - r_Q - \frac{1-N}{2}\tau + B_1$ \\
			$L_1$ & $1 - r_Q - \frac{1-N}{2}\tau + C$
		\end{tabular}};
	 
		\path (5,0) node[gauge](g1) {$\!\!\!2N\!\!\!$} -- (7,0) node[gauge](g2) {$\!\!\!2N\!\!\!$} 
			-- (4,-1.5) node[flavor](x1) {$\!2\!$} -- (6,-1.5) node[flavor](x2) {$\!2\!$} -- (5,1.5) node[flavor](y1) {$\!2\!$};
	 
		\wigE (g1) -- (g2);
		\wigE (g2) -- (8,0);
		\draw[-] (g1) -- (x1); \draw (4.5,-0.75) node[left] {$L'_1$};
		\draw[-] (g1) -- (x2);
		\draw[-] (g1) -- (y1); \draw (5,0.75) node[right] {$V'_1$};
		\draw[-] (g2) -- (x2);
		\draw[-] (g2) -- (7.7,-1);
		\draw (4.5,-0.75) node[cross]{};
		\draw (5,0.75) node[cross]{};
		\draw[-] (y1) to[out=180,in=90] (3.2,0) to[out=-90,in=150] (x1);
		\draw[-] (g1) to[out=150,in=90] (4.4,0) to[out=-90,in=-150] (g1);
		\draw[-] (g2) to[out=60,in=0] (7,0.6) to[out=180,in=120] (g2);
		\draw[blue] (5.2,1.5) node[right] {\scriptsize{$x_1$}};
		\draw[blue] (4.2,-1.5) node[right] {\scriptsize{$y_1$}};
		
		\draw (3.5,-3) node[right] { \begin{tabular}{c|c}
			$V'_1$ & $1 - r_Q - \frac{1-N}{2}\tau + C$ \\
			$L'_1$ & $1 - r_Q - \frac{1-N}{2}\tau + B_1$
		\end{tabular}};
		 
\epic\ee
It is clear that those three actions, combined and iterated appropriately, are sufficient to realize any possible rearranging of improved bifundamentals or flavors. \\

Let us also mention, in conclusion, that these dualities realize a subgroup of the Weyl symmetry of the $SU(2F+4)$ global symmetry group. It consists, in fact, in swapping together pairs of $SU(2)_{x_j} \times U(1)_{b_j}$ or $SU(2)_{y_1} \times U(1)_c$ symmetries, which indeed is a symmetry of the branching rule \eqref{eq:4d_fund_br} used to decompose the fundamental representation of $SU(2F+4)$.

\paragraph{Shortening} \mbox{}\\
\newline
The first type of deformations that we consider consists in  a mass term for one flavor in the electric theory: $\d \CW = Q_j^2$ 
or $\d \CW = P_{1,2}^2$.  By means of the {\it swapping} dualities \eqref{fig:MagSwap_B2B3}, \eqref{fig:MagSwap_V1B2} and \eqref{fig:MagSwap_V1L1}
we can restrict the analysis to the case $\d \CW = Q_j^2$ with $j=2,\ldots,F-1$ which maps to the
linear superpotential term $\d \CW = \mathsf{B}^{(j)}_{1,1}$ in the mirror dual side.

As explained in appendix \ref{app:FE}, the effect of such superpotential is to transform an improved bifundamental into  an Identity-wall, which identifies the two $USp(2N)$ groups which is connecting, shortening the string of improved bifundamental by one unit, as shown below:
\be 
\resizebox{.95\hsize}{!}{
 	 \bpic[thick,node distance=3cm,gauge/.style={circle,draw,minimum size=5mm},flavor/.style={rectangle,draw,minimum size=5mm}] 
 
		\path (-4,0) node[gauge] (g0) {$\!\!\!2N\!\!\!$} -- (-2,0) node[gauge](g1) {$\!\!\!2N\!\!\!$} -- (0,0) node[gauge](g2) {$\!\!\!2N\!\!\!$} 
			-- (2,0) node[gauge](g3) {$\!\!\!2N\!\!\!$} -- (-3,-1.5) node[flavor](x1) {$\!2\!$} -- (-1,-1.5) node[flavor](x2) {$\!2\!$} 
			-- (1,-1.5) node[flavor](x3) {$\!2\!$} -- (4,0) node{$\Longrightarrow$};
			
		\wigE (g0) -- (-5,0);
		\wigE (g0) -- (g1);	\draw (-3,0.3) node {$\Pi_{j-1}$};
		\wigE (g1) -- (g2); \draw (-1,0.3) node {$\Pi_j$};
		\wigE (g2) -- (g3); \draw (1,0.3) node {$\Pi_{j+1}$};
		\wigE (g3) -- (3,0);
		\draw[-] (g0) -- (-4.7,-1);
		\draw[-] (g0) -- (x1);
		\draw[-] (g1) -- (x1);
		\draw[-] (g1) -- (x2); \draw (-1.3,-0.95) node[left] {$R_{j-1}$};
		\draw[-] (g2) -- (x2); \draw (-0.4,-0.55) node[left] {$L_j$};
		\draw[-] (g2) -- (x3);
		\draw[-] (g3) -- (x3);
		\draw[-] (g3) -- (2.7,-1);
		\draw[-] (g0) to[out=60,in=0] (-4,0.6) to[out=180,in=120] (g0);
		\draw[-] (g1) to[out=60,in=0] (-2,0.6) to[out=180,in=120] (g1);
		\draw[-] (g2) to[out=60,in=0] (0,0.6) to[out=180,in=120] (g2);
		\draw[-] (g3) to[out=60,in=0] (2,0.6) to[out=180,in=120] (g3);
		
		\draw (-1,-2.5) node {$\CW = \CW_{\text{gluing}} + \CW_{\text{triangles}} + \mathsf{B}^{(j)}_{1,1}$};
	 
		\path (6,0) node[gauge](g0) {$\!\!\!2N\!\!\!$} -- (8,0) node[gauge](g1) {$\!\!\!2N\!\!\!$} -- (10,0) node[gauge](g2) {$\!\!\!2N\!\!\!$} 
		-- (7,-1.5) node[flavor](x1) {$\!2\!$} -- (9,-1.5) node[flavor](x2) {$\!2\!$};
	 
	 	\wigE (g0) -- (5,0);
		\wigE (g0) -- (g1); \draw (7,0.3) node {$\Pi_{j-1}$};		
		\wigE (g1) -- (g2); \draw (9,0.3) node {$\Pi_{j+1}$};
		\wigE (g2) -- (11,0);
		\draw[-] (g0) -- (5.3,-1);
		\draw[-] (g0) -- (x1);
		\draw[-] (g1) -- (x1);
		\draw[-] (g1) -- (x2);
		\draw[-] (g2) -- (x2);
		\draw[-] (g2) -- (10.7,-1);
		\draw[-] (g0) to[out=60,in=0] (6,0.6) to[out=180,in=120] (g0);
		\draw[-] (g1) to[out=60,in=0] (8,0.6) to[out=180,in=120] (g1);
		\draw[-] (g2) to[out=60,in=0] (10,0.6) to[out=180,in=120] (g2);
		
		\draw (8,-2.5) node {$\CW = \CW_{\text{gluing}} + \CW_{\text{triangles}}$};
		 
\epic}\ee

More intuitively, one can think that the linear term $\d \CW = \mathsf{B}^{(j)}_{1,1}$ has the effect of giving a VEV to the $\Pi_j$ operator, which after the deformation has R-charge 0. This VEV Higgs the $USp(2N) \times USp(2N)$ down to the diagonal $USp(2N)$. In addition, after $\Pi_j$ acquires a VEV the triangle superpotential $\Pi_j L_j R_{j-1}$ becomes a mass term for $L_j$ and $R_{j-1}$.
Therefore, under this deformation, we can see that the mirror theory reduces correctly to the mirror dual of the SQCD with $F-1$ flavors. \\

There is a second type of mass term that we can consider which is $\d \CW= Q_jQ_i$, with $j\neq i$ 
or $\d \CW= P_{1,2}Q_j$ or $\d \CW= P_1 P_2$. These deformations have the effect of giving a mass to two flavors in the electric theory.
Again by means of the {\it swapping} dualities \eqref{fig:MagSwap_B2B3}, \eqref{fig:MagSwap_V1B2} and \eqref{fig:MagSwap_V1L1}
we can restrict the analysis to the case $\d \CW= Q_jQ_{j+1}$.
This superpotential term maps in  the magnetic theory to the term $\d \CW = L_j R_j$, which is a mass terms for both $L_j$ and $R_j$. We are then left with two consecutive improved bifundamental  theories glued together which fuse to  an $\mathbb{I}$-wall as in \eqref{FEdelta}, having the effect of shortening the sequence of improved bifundamentals by two unit. After the shortening, it is generated a new superpotential term: $\d \CW = R_{j-1}L_{j+1}$, which has the effect of giving a mass also to both the $R_{j-1}$ and $L_{j+1}$ legs. All in all we have the following schematic situation:
\be 
 	 \bpic[thick,node distance=3cm,gauge/.style={circle,draw,minimum size=5mm},flavor/.style={rectangle,draw,minimum size=5mm}] 
 
		\path (-2,0) node[gauge](g1) {$\!\!\!2N\!\!\!$} -- (0,0) node[gauge](g2) {$\!\!\!2N\!\!\!$} -- (2,0) node[gauge](g3) {$\!\!\!2N\!\!\!$}  
			-- (-1,-1.5) node[flavor](x2) {$\!2\!$} -- (1,-1.5) node[flavor](x3) {$\!2\!$} 
			-- (4,0) node{$\Longrightarrow$};
	 
	 	\wigE (g1) -- (-3,0); \draw (-3,0.3) node {$\Pi_{j-1}$};
		\wigE (g1) -- (g2); \draw (-1,0.3) node {$\Pi_j$};
		\wigE (g2) -- (g3); \draw (1,0.3) node {$\Pi_{j+1}$};
		\wigE (g3) -- (3,0); \draw (3,0.3) node {$\Pi_{j+2}$};
		\draw[-] (g1) -- (-2.7,-1); 
		\draw[-] (g1) -- (x2); \draw (-1.3,-0.95) node[left]{$R_{j-1}$};
		\draw[-] (g2) -- (x2); \draw (-0.4,-0.55) node[left]{$L_j$};
		\draw[-] (g2) -- (x3); \draw (0.7,-0.95) node[left]{$R_j$};
		\draw[-] (g3) -- (x3); \draw (1.6,-0.55) node[left]{$L_{j+1}$};
		\draw[-] (g3) -- (2.7,-1);
		\draw[-] (g1) to[out=60,in=0] (-2,0.6) to[out=180,in=120] (g1);
		\draw[-] (g2) to[out=60,in=0] (0,0.6) to[out=180,in=120] (g2);
		\draw[-] (g3) to[out=60,in=0] (2,0.6) to[out=180,in=120] (g3);
		
		\draw (0,-2.5) node {$\CW = \CW_{\text{gluing}} + \CW_{\text{triangles}} + R_{j} L_j $};
	 
		\path (7,0) node[gauge](g1) {$\!\!\!2N\!\!\!$};
	 	
	 	\wigE (g1) -- (6,0); \draw (6,0.3) node {$\Pi_{j-1}$};
		\wigE (g1) -- (8,0); \draw (8,0.3) node {$\Pi_{j+2}$};
		\draw[-] (g1) -- (6.3,-1);
		\draw[-] (g1) -- (7.7,-1);
		\draw[-] (g1) to[out=60,in=0] (7,0.6) to[out=180,in=120] (g1);
		
		\draw (7,-2.5) node {$\CW = \CW_{\text{gluing}} + \CW_{\text{triangles}}$};
		 
\epic\ee
We can see that the net effect of the deformation in the mirror theory is to shorten the sequence of improved bifundamental by two, leading to the correct mirror dual of the SQCD with $F-2$ flavors.

\paragraph{Ironing} \mbox{}\\
\newline
Another type of deformation that we want to consider consist in turning on cubic superpotential terms for the flavors and the antisymmetric field as: $\d \CW = Q_j^2 A$ or $\d \CW = P_{1,2}^2 A$.
Again by means of the {\it swapping} dualities \eqref{fig:MagSwap_B2B3}, \eqref{fig:MagSwap_V1B2} and \eqref{fig:MagSwap_V1L1}
we can restrict the analysis to the case $\d \CW = Q_j^2 A$ with  $j=2,\ldots F-1$ which maps in the mirror dual to $\d \CW = \mathsf{B}^{(j)}_{1,2}$.
The effect of this deformation  in the magnetic theory, is to iron an improved bifundamental into a standard one along with two antisymmetric fields to which it is coupled, see \eqref{fig:FE_c=t/2}. Graphically this deformation consist in:
\be\label{fig:N=4second}
 \bpic[thick,node distance=3cm,gauge/.style={circle,draw,minimum size=5mm},flavor/.style={rectangle,draw,minimum size=5mm}] 
 
	\path (-2,0) node[gauge](g1) {$\!\!\!2N\!\!\!$} -- (0,0) node[gauge](g2) {$\!\!\!2N\!\!\!$} -- (-1,-1.5) node[flavor](x1) {$\!2\!$} 
		-- (3,0) node{$\Longrightarrow$};
	 
	\wigE (g1) -- (-3,0);	
	\wigE (g1) -- (g2); \draw (-1,0.4) node {$\Pi_j$};
	\wigE (g2) -- (1,0);
	\draw[-] (g1) -- (-2.7,-1);
	\draw[-] (g1) -- (x1); 
	\draw[-] (g2) -- (x1); 
	\draw[-] (g2) -- (0.7,-1);
	\draw[-] (g1) to[out=60,in=0] (-2,0.6) to[out=180,in=120] (g1);
	\draw[-] (g2) to[out=60,in=0] (0,0.6) to[out=180,in=120] (g2);
	
	\draw (-1,-2.5) node {$\CW = \CW_{\text{gluing}} + \CW_{\text{triangles}} + \mathsf{B}^{(j)}_{1,2} $};
	 
	\path (6,0) node[gauge](g1) {$\!\!\!2N\!\!\!$} -- (8,0) node[gauge](g2) {$\!\!\!2N\!\!\!$} -- (7,-1.5) node[flavor](x1) {$\!2\!$} ;
	 
	\wigE (g1) -- (5,0);	
	\draw[-] (g1) -- (g2); \draw (7,0) node[cross] {}; \draw (7,0.4) node {$\Pi'_j$};
	\wigE (g2) -- (9,0);
	\draw[-] (g1) -- (5.3,-1);
	\draw[-] (g1) -- (x1); 
	\draw[-] (g2) -- (x1); 
	\draw[-] (g2) -- (8.7,-1);
	
	\draw (7,-2.5) node {$\CW =Flip[{\Pi'}_j^2]+ {\Pi'}^2_j (\mathsf{A}_L + \mathsf{A}_R) + $};
	\draw (7,-3.1) node {$ + \CW_{\text{triangles}} $};
		 
\epic\ee
On the r.h.s. the standard bifundamental $\Pi'_j$ is coupled to $\mathsf{A}_L$ and $\mathsf{A}_R$, that are the antisymmetric operators inside the improved bifundamentals  on its  left and on its right.

It is interesting to study the result when we introduce all the $\sum_{j=1}^F Q_j^2 A$ terms since it leads to the $4d$ uplift of the 3d $\cN=4$ $U(N)$ SQCD proposed in \cite{Hwang:2020wpd} for $F \geq 2N$. The effect of the deformation is to iron all the improved bifundamentals. Also, keeping track of all the antisymmetric fields produced from this deformation, we see that each gauge node except the first and last one have an antisymmetric field coupled to the standard bifundamental on its right and left, also all the bifundamentals are flipped. We collect these terms in a superpotential called $\CW_{\CN={4}\text{-like}}^{\text{partial}}$. On the first and last gauge node we do not have any antisymmetric fields, however the flavors are now coupled to an antisymmetric operator obtained from the bifundamental on its side. In addition,
on the mirror side  we turn on the linear terms in the flipping singlets  $\CF[V_1^2 (\Pi_2^2)^{N-2}]$ and $\CF[V_2^2 (\Pi_{F-1}^2)^{N-2}]$.
All in all we have the following duality:
\be
\resizebox{.95\hsize}{!}{
\begin{tikzpicture}[thick,node distance=3cm,gauge/.style={circle,draw,minimum size=5mm},flavor/.style={rectangle,draw,minimum size=5mm}] 
 
	\path (0,0) node[gauge](g) {$\!\!\!2N\!\!\!$} -- (0,1.5) node[flavor] (x) {$\!2F\!$} 
		-- (-0.75,-1.5) node[flavor] (y1) {$\!2\!$} -- (0.75,-1.5) node[flavor] (y2) {$\!2\!$}
		-- (2,0) node{$\Longleftrightarrow$};
	
	\draw[-] (g) -- (x); \draw (0,0.75) node[right] {$Q$};
	\draw[-] (g) -- (y1); \draw (-0.35,-0.75) node[left] {$P_1$};
	\draw[-] (g) -- (y2); \draw (0.35,-0.75) node[right] {$P_2$};
	\draw[-] (g) to[out=30,in=90] (0.6,0) to[out=-90,in=-30] (g); \draw (0.55,0) node[right] {$A$};
	
	\draw (0,-2.5) node{\small{$\cW = Q^2 A$}};	
	 
	\path (5,0) node[gauge](g1) {$\!\!\!2N\!\!\!$} -- (7,0) node[gauge](g2) {$\!\!\!2N\!\!\!$} -- (10,0) node[gauge](g3) {$\!\!\!2N\!\!\!$} 
		-- (12,0) node[gauge](g4) {$\!\!\!2N\!\!\!$} -- (4,-1.5) node[flavor](x1) {$\!2\!$} -- (6,-1.5) node[flavor](x2) {$\!2\!$} 
		-- (11,-1.5) node[flavor](x3) {$\!2\!$} -- (13,-1.5) node[flavor](x4) {$\!2\!$} -- (5,1.5) node[flavor](y1) {$\!2\!$} 
		-- (12,1.5) node[flavor](y2) {$\!2\!$};
	 
	\draw[-] (g1) -- (g2); \draw (6,0) node[cross] {}; \draw (6,0.35) node {$\Pi_2$};
	\draw[-] (g2) -- (8,0);
	\draw[-] (g3) -- (9,0);
	\draw[-] (g3) -- (g4); \draw (11,0) node[cross] {}; \draw (11,0.35) node {$\Pi_{F-1}$};
	\draw[-] (g1) -- (x1); \draw (4.55,-0.6) node[left]{$L_1$}; 
	\draw[-] (g1) -- (x2); \draw (5.7,-1) node[left]{$R_1$};
	\draw[-] (g1) -- (y1); \draw (5,0.75) node[left]{$V_1$};
	\draw[-] (g2) -- (x2); \draw (6.6,-0.6) node[left]{$L_2$};
	\draw[-] (g2) -- (7.7,-1);
	\draw (8.5,-0.5) node {$\cdots$};
	\draw[-] (g3) -- (9.3,-1);
	\draw[-] (g3) -- (x3); \draw (10.7,-1) node[left]{$R_{F-2}$};
	\draw[-] (g4) -- (x3); \draw (11.6,-0.6) node[left]{$L_{F-1}$}; 
	\draw[-] (g4) -- (x4); \draw (12.7,-1) node[left]{$R_{F-1}$};
	\draw[-] (g4) -- (y2); \draw (12,0.75) node[right]{$V_2$};
	\draw (4.5,-0.75) node[cross]{};
	\draw (5,0.75) node[cross]{};
	\draw (12.5,-0.75) node[cross]{};
	\draw (12,0.75) node[cross]{};
	\draw[-] (y1) to[out=180,in=90] (3.2,0) to[out=-90,in=150] (x1); \draw (3,1.25) node[right]{$F_1^{(j)}$};
	\draw[-] (y2) to[out=0,in=90] (13.8,0) to[out=-90,in=30] (x4); \draw (14.15,1.25) node[left]{$F_2^{(j)}$};
	\draw[-] (g2) to[out=60,in=0] (7,0.6) to[out=180,in=120] (g2); \draw (7,0.8) node {$A_2$};
	\draw[-] (g3) to[out=60,in=0] (10,0.6) to[out=180,in=120] (g3); \draw (10,0.8) node {$A_{F-2}$};
	
	\draw (2.5,-2.5) node[right]{\small{$\cW = \CW_{\CN={4}\text{-like}}^{\text{partial}} + \cW_{\text{triangles}} + \prod_{j=0}^{N-1} (V_1 A_1^j L_1 F_1^{(j)} + V_2 A_{F-1}^j R_{F-1} F_2^{(j)})  + $}};
	\draw (3,-3.5) node[right] {\small{$ + \sum_{k=0}^{N-1} \Big\{Flip[V_1^2 (\Pi_2^2)^k] +Flip[ L_1^2 (\Pi_2^2)^{k}] + Flip[V_2^2 (\Pi_{F-1}^2)^{k}] + $}};
	\draw (3,-4.5) node[right] {\small{$ + Flip[ R_{F-1}^2 (\Pi_{F-1}^2)^{k}]\Big\} + \CF[V_1^2 (\Pi_2^2)^{N-2}] + \CF[V_2^2 (\Pi_{F-1}^2)^{N-2}] $}};

\end{tikzpicture}}\ee
Notice that introducing in the superpotential the singlets $\CF[V_1^2 (\Pi_2^2)^{N-2}]$ and $\CF[V_2^2 (\Pi_{F-1}^2)^{N-2}]$ causes the operators $V_1^2 (\Pi_2^2)^{N-2}$ and $V_2^2 (\Pi_{F-1}^2)^{N-2}$ to acquire a VEV.  As shown in \cite{Comi:2022aqo} the two VEVs have the effect to propagate reconstructing a tail of increasing ranks from 1 to $N$. Also we have a plateau of gauge nodes with rank $N$ with a flavor on the two sides. Taking into account also the singlets, we obtain the known $\mathcal{N}=4$-like mirror dual for the 4d $\CN=1$ $USp(2N)$ SQCD:
\be
\resizebox{.85\hsize}{!}{
\begin{tikzpicture}[thick,node distance=3cm,gauge/.style={circle,draw,minimum size=5mm},flavor/.style={rectangle,draw,minimum size=5mm}] 
 
	\path (0,0) node[gauge](g) {$\!\!\!2N\!\!\!$} -- (0,1.5) node[flavor] (x) {$\!2F\!$} 
		-- (-0.75,-1.5) node[flavor] (y1) {$\!2\!$} -- (0.75,-1.5) node[flavor] (y2) {$\!2\!$}
		-- (2,0) node{$\Longleftrightarrow$};
	
	\draw[-] (g) -- (x); 
	\draw[-] (g) -- (y1); 
	\draw[-] (g) -- (y2); 
	\draw[-] (g) to[out=30,in=90] (0.6,0) to[out=-90,in=-30] (g); 
	
	\draw (0,-2.5) node {$\CW= \CW_{\CN=4\text{-like}}$};
	

\begin{scope}[shift={(-1,0)}]
	\path (5,0) node[gauge](g1) {$\!\!\!2\!\!\!$} -- (6,0) node (gi) {$\ldots$} -- (7,0) node[gauge](g2) {\!\!\tiny{$2N$-$2$}\!\!\!} 
		-- (8.5,0) node[gauge](g3) {$\!\!\!2N\!\!\!$} -- (9.5,0) node (gii) {$\ldots$} -- (10.5,0) node[gauge] (g4) {$\!\!\!2N\!\!\!$} 
		-- (12,0) node[gauge] (g5) {\!\!\tiny{$2N$-$2$}\!\!\!} -- (13,0) node (giii) {$\ldots$} -- (14,0) node[gauge] (g6) {$\!\!\!2\!\!\!$}
		-- (4.25,-1.5) node[flavor] (x1) {$\!2\!$} -- (7.75,-1.5) node[flavor] (x2) {$\!2\!$} -- (11.25,-1.5) node[flavor] (x3) {$\!2\!$}
		-- (14.75,-1.5) node[flavor] (x4) {$\!2\!$}
		-- (8.5,1.5) node[flavor] (y1) {$\!2\!$} -- (10.5,1.5) node[flavor] (y2) {$\!2\!$};
		
	\draw[-] (g1) -- (gi);
	\draw[-] (g2) -- (gi);
	\draw[-] (g2) -- (g3);
	\draw[-] (g3) -- (gii);
	\draw[-] (g4) -- (gii);
	\draw[-] (g4) -- (g5);
	\draw[-] (g5) -- (giii);
	\draw[-] (g6) -- (giii);
	\draw[-] (g1) -- (x1); \draw (4.65,-0.7) node[cross] {};
	\draw[-] (g1) -- (5.35,-0.7);
	\draw[-] (g2) -- (6.65,-0.7);
	\draw[-] (g2) -- (x2);
	\draw[-] (g3) -- (x2); \draw (8.15,-0.7) node[cross] {};
	\draw[-] (g3) -- (8.85,-0.7);
	\draw[-] (g4) -- (10.15,-0.7);
	\draw[-] (g4) -- (x3); \draw (10.85,-0.7) node[cross] {};
	\draw[-] (g5) -- (x3);
	\draw[-] (g5) -- (12.35,-0.7);
	\draw[-] (g6) -- (13.65,-0.7);
	\draw[-] (g6) -- (x4); \draw (14.35,-0.7) node[cross] {};
	\draw[-] (g3) -- (y1);
	\draw[-] (g4) -- (y2);
	
	\draw (y1) to[out=180,in=120] (x1); \draw (y1) to[out=-120,in=90] (x2);
	\draw (y2) to[out=-60,in=90] (x3); \draw (y2) to[out=0,in=60] (x4);
	
	\draw[-] (g1) to[out=60,in=0] (5,0.5) to[out=180,in=120] (g1);
	\draw[-] (g2) to[out=60,in=0] (7,0.6) to[out=180,in=120] (g2);
	\draw[-] (g3) to[out=20,in=-30] (9,0.5) to[out=150,in=70] (g3);
	\draw[-] (g4) to[out=160,in=-150] (10,0.5) to[out=30,in=110] (g4);
	\draw[-] (g5) to[out=60,in=0] (12,0.6) to[out=180,in=120] (g5);
	\draw[-] (g6) to[out=60,in=0] (14,0.5) to[out=180,in=120] (g6);
	
	\draw (9.5,-2.5) node {$\CW= \CW_{\CN=4\text{-like}}$};

\end{scope}

\end{tikzpicture}}\ee
Where in the picture above all the antisymmetrics are taken to be tracefull.

\subsection{Reduction to $3d$ and {\it uplifts}}\label{3dred}
It is interesting to observe how the $4d$ SQCD mirror pair reduces to our $3d$ result ins section \ref{sqcdmirror}.
The first step of the $3d$ reduction limit consists in compactifying the mirror pair in figure \ref{fig:SQCD_4d_Dual} on a circle. This limit can be performed by redefining the set of fugacities appearing in the SCI identity in \eqref{eq:4d_ind_id} as:
\begin{align}
	& x_j = e^{2\pi i r X_j} \quad, \quad y_j = e^{2\pi i r Y_j} \quad, \quad z_j = e^{2\pi i r Z_j} \,, \nn \\
	& t = e^{2\pi i r \tau} \quad, \quad b_j = e^{2\pi i r B_j} \quad, \quad c = e^{2\pi i r \D} \,, \nn \\
	& p = e^{-2 \pi rb} \quad, \quad q = e^{- 2 \pi rb^{-1}} \,,
\end{align}
where the capital letter variables are real variables taking values in $[-\frac{1}{2r},\frac{1}{2r}]$, with $r$ being the radius of the $S^1$ circle of the $S^3 \times S^1$ space. We then perform the limit $r \to 0$ to land on a $3d$ theory. The superconformal index reduces to the $S_b^3$ partition function of the resulting $3d$ theory and can be obtained using the relation between elliptic-gamma and double-sine functions:
\begin{align}
	\lim_{r \to 0} \Ge( e^{2\pi i x}; p=e^{-2 \pi rb}, q=e^{-2 \pi rb^{-1}}) = e^{-\frac{i\pi}{6}(\frac{iQ}{2}-x)}s_b(\frac{iQ}{2}-x) \,,
\end{align} 
with $Q=b+b^{-1}$. Performing this limit in the $4d$ SQCD pair we obtain a $\CN=2$ $3d$ duality that is identical to the $4d$ one, the only difference is that a superpotential linear in the KK monopole $\mathcal{W}=\mathfrak{M}$ is generated, as argued in \cite{Aharony:2013dha}.
This monopole superpotential ensures that the $3d$ and $4d$ theories (where we have the anomaly cancellation condition) have the same global symmetry.
Notice that on the mirror side we have $3d$ $FE[USp(2N)]$ theories whose UV description is given as a $3d$ $\CN=2$ quiver theory identical to \ref{fig:FE_quiver}, where each node has $\mathcal{W}=\mathfrak{M}$ turned on.

We can now perform some deformations.
For example we can proceed as in \cite{Benini:2017dud} and perform a combination of real mass deformation for the non-abelian flavor symmetries and Coulomb branch VEVs breaking the gauge groups from symplectic to unitary obtaining the following duality:
\be
\resizebox{.95\hsize}{!}{
\begin{tikzpicture}[thick,node distance=3cm,gauge/.style={circle,draw,minimum size=5mm},flavor/.style={rectangle,draw,minimum size=5mm}] 
\begin{scope}[shift={(-6,0)}]
	\path (5,0) node[gauge](g) {$\!\!\!N\!\!\!$} -- (4,1.5) node[flavor](x1) {$\!1\!$} -- (6,1.5) node[flavor] (x2) {$\!1\!$} 
		-- (4.25,-1.5) node[flavor](y1) {$\!1\!$} -- (5.75,-1.5) node[flavor](y2) {$\!1\!$};
	
	\draw[-, shorten >= 6.5, shorten <= 9, shift={(-0.07,0.02)}, mid arrowsm] (4,1.5) -- (5,0);
	\draw[-, shorten >= 7.5, shorten <= 9, shift={(0.1,0)}, mid arrowsm] (5,0) -- (4,1.5);
	\draw (4.5,0.75) node[left]{$Q_1$};
	
	\draw[-, shorten >= 7.5, shorten <= 9, shift={(-0.1,0.02)}, mid arrowsm] (5,0) -- (6,1.5);
	\draw[-, shorten >= 7, shorten <= 9, shift={(0.05,0)}, mid arrowsm] (6,1.5) -- (5,0); 
	\draw (5.5,0.75) node[right]{$Q_F$};
	
	\draw[-, shorten >= 7, shorten <= 7.5, shift={(-0.1,0.02)}, mid arrowsm] (4.25,-1.5) -- (5,0);
	\draw[-, shorten >= 8, shorten <= 8, shift={(0.05,0)}, mid arrowsm] (5,0) -- (4.25,-1.5); 
	\draw (4.55,-0.75) node[left]{$P_1$};
	
	\draw[-, shorten >= 7, shorten <= 9, shift={(-0.07,0.02)}, mid arrowsm] (5,0) -- (5.75,-1.5);
	\draw[-, shorten >= 6.5, shorten <= 8, shift={(0.1,0)}, mid arrowsm] (5.75,-1.5) -- (5,0); 
	\draw (5.45,-0.75) node[right]{$P_2$};
	
	\draw[-] (g) to[out=30,in=90] (5.6,0) to[out=-90,in=-30] (g); \draw (5.55,0) node[right]{$A$};
	\draw (5,1.5) node{$\cdots$};
	
	\draw (5,-2.5) node {$\cW = \M^+ + \M^-$};
	
\end{scope}

	\draw (1.75,0) node {$\Longleftrightarrow$}; 

	\path (5,0) node[gauge](g1) {$\!\!\!N\!\!\!$} -- (7,0) node[gauge](g2) {$\!\!\!N\!\!\!$} -- (10,0) node[gauge](g3) {$\!\!\!N\!\!\!$} 
		-- (12,0) node[gauge](g4) {$\!\!\!N\!\!\!$} 
		-- (4,-1.5) node[flavor](x1) {$\!1\!$} -- (6,-1.5) node[flavor](x2) {$\!1\!$} -- (11,-1.5) node[flavor](x3) {$\!1\!$} 
		-- (13,-1.5) node[flavor](x4) {$\!1\!$} -- (5,1.5) node[flavor](y1) {$\!1\!$} -- (12,1.5) node[flavor](y2) {$\!1\!$};
	 
	\wigM (g1) -- (g2); \draw (6,0.3) node {$\Pi_2$};
	\wigM (g2) -- (8,0);
	\wigM (g3) -- (9,0);
	\wigM (g3) -- (g4);	 \draw (11,0.3) node {$\Pi_{F-1}$};
	
	\draw[-, shorten >= 6.5, shorten <= 8, shift={(-0.1,0.02)}, middx arrowsm] (4,-1.5) -- (5,0);
	\draw[-, shorten >= 9, shorten <= 9, shift={(0.05,0)}, midsx arrowsm] (5,0) -- (4,-1.5); 
	\draw (4.3,-1) node[rotate={-30}] {\LARGE{$\times$}}; \draw (4.55,-0.5) node[left]{$L_1$};
	
	\draw[-, shorten >= 8, shorten <= 10, shift={(-0.07,0.02)}, mid arrowsm] (5,0) -- (6,-1.5);
	\draw[-, shorten >= 6.5, shorten <= 9, shift={(0.1,0)}, mid arrowsm] (6,-1.5) -- (5,0); 
	\draw (5.7,-1) node[left]{$R_1$};
	
	\draw[-, shorten >= 6, shorten <= 9.5, shift={(-0.07,-0.05)}, middx arrowsm] (5,0) -- (5,1.5);
	\draw[-, shorten >= 6.5, shorten <= 9, shift={(0.07,0.05)}, midsx arrowsm] (5,1.5) -- (5,0); 
	\draw (5,0.5) node[cross]{}; \draw (5,0.75) node[left]{$V_1$};
	
	\draw[-, shorten >= 6.5, shorten <= 8, shift={(-0.1,0.02)}, mid arrowsm] (6,-1.5) -- (7,0);
	\draw[-, shorten >= 9, shorten <= 9, shift={(0.05,0)}, mid arrowsm] (7,0) -- (6,-1.5);  
	\draw (6.5,-0.5) node[left]{$L_2$};
	
	\draw[-, shorten >= -5, shorten <= 10, shift={(-0.07,0.02)}, middx arrowsm] (7,0) -- (7.75,-0.9);
	\draw[-, shorten >= 5.5, shorten <= 0, shift={(0.1,0)}, midsx arrowsm] (7.75,-0.9) -- (7,0);
	\draw (8.5,-0.5) node {$\cdots$};
	\draw[-, shorten >= 6.5, shorten <= 0, shift={(-0.1,0.02)}, midsx arrowsm] (9.25,-0.9) -- (10,0);
	\draw[-, shorten >= -3, shorten <= 9, shift={(0.05,0)}, middx arrowsm] (10,0) -- (9.25,-0.9);
	
	\draw[-, shorten >= 8, shorten <= 10, shift={(-0.07,0.02)}, mid arrowsm] (10,0) -- (11,-1.5);
	\draw[-, shorten >= 6.5, shorten <= 9, shift={(0.1,0)}, mid arrowsm] (11,-1.5) -- (10,0); 
	\draw (10.7,-1) node[left]{$R_{F-2}$};
	
	\draw[-, shorten >= 6.5, shorten <= 8, shift={(-0.1,0.02)}, mid arrowsm] (11,-1.5) -- (12,0);
	\draw[-, shorten >= 9, shorten <= 9, shift={(0.05,0)}, mid arrowsm] (12,0) -- (11,-1.5);
	\draw (11.6,-0.5) node[left]{$L_{F-1}$}; 
	
	\draw[-, shorten >= 8, shorten <= 10, shift={(-0.07,0.02)}, midsx arrowsm] (12,0) -- (13,-1.5);
	\draw[-, shorten >= 6.5, shorten <= 9, shift={(0.1,0)}, middx arrowsm] (13,-1.5) -- (12,0);  
	\draw (12.7,-1) node[rotate={30}] {\LARGE{$\times$}}; \draw (12.7,-1) node[left]{$R_{F-1}$};
	
	\draw[-, shorten >= 6.5, shorten <= 9.5, shift={(-0.07,-0.05)}, middx arrowsm] (12,0) -- (12,1.5);
	\draw[-, shorten >= 6.5, shorten <= 9, shift={(0.07,0.05)}, midsx arrowsm] (12,1.5) -- (12,0); 
	\draw (12,0.5) node[cross]{}; \draw (12,0.75) node[right]{$V_2$};
	
	\draw[-, shorten >= 7, shorten <= 6, shift={(-0.05,0.05)}, mid arrowsm] (5,1.5) to[out=180,in=90] (3.55,0) to[out=-90,in=120] (4,-1.5); 
	\draw[-, shorten >= 10, shorten <= 9, shift={(0.1,-0.03)}, mid arrowsm] (4,-1.5) to[out=120,in=-90] (3.5,0) to[out=90,in=180] (5,1.5); 
	\draw (3,1.25) node[right]{$F_1^{(j)}$};
	
	\draw[-, shorten >= 7, shorten <= 6, shift={(0.05,0.05)}, mid arrowsm] (12,1.5) to[out=0,in=90] (13.55,0) to[out=-90,in=60] (13,-1.5); 
	\draw[-, shorten >= 10, shorten <= 9, shift={(-0.1,-0.03)}, mid arrowsm] (13,-1.5) to[out=60,in=-90] (13.6,0) to[out=90,in=0] (12,1.5);  
	\draw (14.15,1.25) node[left]{$F_2^{(j)}$};
	
	\draw[-] (g1) to[out=150,in=90] (4.4,0) to[out=-90,in=-150] (g1); \draw (4.5,0) node[left]{$A_1$};
	\draw[-] (g2) to[out=60,in=0] (7,0.6) to[out=180,in=120] (g2); \draw (7,0.8) node {$A_2$};
	\draw[-] (g3) to[out=60,in=0] (10,0.6) to[out=180,in=120] (g3); \draw (10,0.8) node {$A_{F-2}$};
	\draw[-] (g4) to[out=30,in=90] (12.6,0) to[out=-90,in=-30] (g4); \draw (12.5,0) node[right]{$A_{\text{$F$-$1$}}$};
	
	\draw (3,-2.5) node[right] {\small{$\cW = \cW_{\text{gluing}} + \cW_{\text{triangles}} + \cW_{\text{monopoles}} + \prod_{j=0}^{N-1} (V_1 A_1^j L_1 F_1^{(j)} +
	\tilde{V}_1 A_1^j \tilde{L}_1 \tilde{F}_1^{(j)}  $}};
	\draw (3.5,-3.1) node[right] {\small{$ + V_2 A_{F-1}^j R_{F-1} F_2^{(j)}+ \tilde{V}_2 A_{F-1}^j \tilde{R}_{F-1} \tilde{F}_2^{(j)}) + \sum_{k=0}^{N-1} \Big\{Flip[V_1A_1^k \tilde{V}_1] + $}};
	\draw (3.5,-3.7) node[right] {\small{$ + Flip[ L_1 A_1^{k}\tilde{L}_1] + Flip[V_2A_{F-1}^{k}\tilde{V}_2]+Flip[ R_{F-1} A_{F-1}^{k}\tilde{R}_{F-1} ]\Big\}$ }};
	
\end{tikzpicture}}
\ee
Notice all the antisymmetric chirals become adjoints. 
On the electric side this limit yields an adjoint SCQD with $F+2$ flavors.
This flow has the effect of generating non-perturbative contributions due to the $USp(2N) \to U(N)$ breaking
of the gauge group. These contributions together with the original KK monopoles combine
in a contribution to superpotential consisting in the sum of the two fundamental monopole $\mathcal{W}=\mathfrak{M}^+ +  \mathfrak{M}^- $.
Also in the mirror theory at each node we have $\mathcal{W}=\mathfrak{M}^+ +  \mathfrak{M}^- $, all these terms are collected in short into $\cW_{\text{monopoles}}$.
This RG flow also   reduces the $3d$ $FE[USp(2N)]$ theories to $3d$ $FM[U(N)]$ theories as shown in appendix \ref{app:FM}.
The charges of the fields are again the same as the $4d$ ones in given in table \ref{tab:SQCD_4d_Charges_mirror}.

Finally we turn on a real mass deformation for the $U(1)_c$ symmetry.
On the electric side the  $P_1,\tilde P_1$ and $P_2,\tilde P_2$ flavors become massive and when integrated out they generate mixed Chern-Simons couplings and restore the topological symmetry at each node lifting the monopole superpotential. 
In this way we obtain the $3d$ adjoint SQCD with $F$ flavors.
Similarly on the mirror side the flavors $R_j,\tilde R_j$ and $L_j,\tilde L_j$ forming the saw all become massive and when integrated out they generate mixed Chern-Simons couplings and restore the topological symmetry at each node lifting the monopole superpotential. 
At the level of the partition function this consists in taking the limit $\D \to +\infty$ and using the limit behavior of the double-sine function:
\begin{align}
	\lim_{x_\to \pm \infty} s_b(x) = e^{\pm\frac{i\pi}{2}} \,.
\end{align}
Performing this limit leads to the partition function identity \eqref{eq:SQCD_Zidentity}.

One can generalize the strategy above to construct improved $4d$ $\mathcal{N}=1$ quivers which {\it uplifts} the $3d$ quivers associated to brane setups preserving four supercharges described in section \ref{sec:branes}. 
Where by {\it uplifts} we mean that under the $3d$ reduction described above, any mirror-like pair of $4d$ $\CN=1$ improved quivers reduces to a $3d$ $\CN=2$ mirror pair.
Intuitively the strategy to uplift a $3d$ $\CN=2$ mirror pair is the following:
\begin{itemize}
	\item We replace each $U(N)$ gauge node with a $USp(2N)$ gauge node. 
	
	\item We replace pairs of $3d$ chirals/antichirals in the fundamental of $U(N)$ with $4d$ chiral doublets in the fundamental of $USp(2N)$. 
		
	\item We replace  $3d$ improved bifundamentals, that are $FM[U(N)]$ theories, by $FE[USp(2N)]$ theories, the  $4d$ improved bifundamentals.
	
	\item We add the saw-like structure.
	
	\item Finally, if in the $3d$ theory there is a single vertical flipped flavor on the leftmost or rightmost gauge node of the theory, we add flipping fields as in the mirror dual of the SQCD in \ref{fig:SQCD_4d_Dual}.
\end{itemize}
For example, following this strategy, we find that  the uplift of the mirror pair discussed in section \ref{k1} is given by:
\be
\resizebox{.95\hsize}{!}{
\bpic[thick,node distance=3cm,gauge/.style={circle,draw,minimum size=5mm},flavor/.style={rectangle,draw,minimum size=5mm}] 
\begin{scope}[shift={(0,0)}]	

	\path (0,0) node[gauge](g1) {$\!\!\!N\!\!\!$} -- (2,0) node[gauge](g2) {$\!\!\!N\!\!\!$} 
		-- (0.5,1.5) node[flavor] (x1) {$\!F_1\!$} -- (-0.5,1.5) node[flavor] (x2) {$\!F_1\!$} 
		-- (1.5,1.5) node[flavor] (y1) {$\!F_2\!$} -- (2.5,1.5) node[flavor] (y2) {$\!F_2\!$}; 
	
	\wigM (g1) -- (g2);
	\chir (g1) -- (x1);
	\chir (x2) -- (g1);
	\chir (g2) -- (y1);
	\chir (y2) -- (g2);
	\draw[-] (g1) to[out=-60,in=0] (0,-0.6) to[out=180,in=-120] (g1);
	\draw[-] (g2) to[out=-60,in=0] (2,-0.6) to[out=180,in=-120] (g2);
	
	\draw (1,-1.25) node {\small{$\CW = \CW_{\text{gluing}}$}};
	
\end{scope}	 
	
	\draw (5,1.3) node {$3d$ Mirror};
	\draw (5,0.8) node {symmetry};
	\draw (5,0.4) node {$\Longleftrightarrow$};
	
\begin{scope}[shift={(9,0)}]

	\path (0,0) node[gauge](g1) {$\!\!\!N\!\!\!$} -- (1.5,0) node(gi) {$\ldots$} -- (3,0) node[gauge] (g2) {$\!\!\!N\!\!\!$} 
		-- (4.5,0) node (gii) {$\ldots$} -- (6,0) node[gauge] (g3) {$\!\!\!N\!\!\!$}
		-- (0,1.5) node[flavor] (x1) {$\!1\!$} 
		-- (2.5,1.5) node[flavor] (x2) {$\!1\!$} -- (3.5,1.5) node[flavor] (x22) {$\!1\!$} 
		-- (6,1.5) node[flavor] (x3) {$\!1\!$} ;
		
 	\wigM (g1) -- (gi);
	\wigM (g2) -- (gi); 
	\wigM (g2) -- (gii); 
	\wigM (g3) -- (gii);
	
	\draw[-, shorten >= 6.5, shorten <= 9.5, shift={(-0.07,-0.05)}, middx arrowsm] (0,0) -- (0,1.5);
	\draw[-, shorten >= 6.5, shorten <= 9, shift={(0.07,0.05)}, midsx arrowsm] (0,1.5) -- (0,0);
	\draw (0,0.5) node[cross] {}; \draw (0,0.75) node[left] {$V_1$};
	
	\chir (g2) -- (x2); \draw (2.75,0.75) node[left] {$V_2$};
	\chir (x22) -- (g2); 
	
	\draw[-, shorten >= 6.5, shorten <= 9.5, shift={(-0.07,-0.05)}, middx arrowsm] (6,0) -- (6,1.5);
	\draw[-, shorten >= 6.5, shorten <= 9, shift={(0.07,0.05)}, midsx arrowsm] (6,1.5) -- (6,0);
	\draw (6,0.5) node[cross] {}; \draw (6,0.75) node[right] {$V_3$};
	
	\draw[-] (g1) to[out=-60,in=0] (0,-0.6) to[out=180,in=-120] (g1); \draw (0,-0.7) node[right] {$A_1$};
	\draw[-] (g2) to[out=-60,in=0] (3,-0.6) to[out=180,in=-120] (g2); \draw (3,-0.7) node[right] {$A_{F_1}$};
	\draw[-] (g3) to[out=-60,in=0] (6,-0.6) to[out=180,in=-120] (g3); \draw (6,-0.7) node[right] {$A_{F_1+F_2-1}$};
	
	\draw (3,-1.25) node {\small{$\CW = \CW_{\text{gluing}} + \sum_{j=0}^{N-1} \big( Flip[V_1 A_1^j \tilde{V}_1] + Flip[V_3 A_{F_1+F_2-1}^j \tilde{V}_3] \big)$} };
    
\end{scope}

\begin{scope}[shift={(0,-4.5)}]	

	\path (0,0) node[gauge](g1) {$\!\!\!2N\!\!\!$} -- (2,0) node[gauge](g2) {$\!\!\!2N\!\!\!$} 
		-- (0,1.5) node[flavor,double] (x1) {$\!2F_1\!$} -- (2,1.5) node[flavor,double] (x2) {$\!2F_2\!$}
		-- (-1,-1.5) node[flavor] (y1) {$\!2\!$} -- (1,-1.5) node[flavor] (y2) {$\!2\!$} -- (3,-1.5) node[flavor] (y3) {$\!2\!$};
	
	\wigE (g1) -- (g2);
	\chir (g1) -- (x1);
	\chir (g2) -- (x2);
	\draw[-] (g1) -- (y1);
	\draw[-] (g1) -- (y2);
	\draw[-] (g2) -- (y2);
	\draw[-] (g2) -- (y3);
	\draw[-] (g1) to[out=150,in=90] (-0.6,0) to[out=-90,in=-150] (g1);
	\draw[-] (g2) to[out=30,in=90] (2.6,0) to[out=-90,in=-30] (g2);
	
	\draw (1,-2.5) node {\small{$\CW = \CW_{\text{gluing}} + \CW_{\text{triangles}}$}};
	
\end{scope}	 
	
	\draw (5,-3.6) node {$4d$ Mirror};
	\draw (5,-4.1) node {symmetry};
	\draw (5,-4.4) node {$\Longleftrightarrow$};

\begin{scope}[shift={(9,-4.5)}]

	\path (0,0) node[gauge](g1) {$\!\!\!2N\!\!\!$} -- (1.5,0) node(gi) {$\ldots$} -- (3,0) node[gauge] (g2) {$\!\!\!2N\!\!\!$} 
		-- (4.5,0) node (gii) {$\ldots$} -- (6,0) node[gauge] (g3) {$\!\!\!2N\!\!\!$}
		-- (0,1.5) node[flavor] (x1) {$\!2\!$} -- (3,1.5) node[flavor] (x2) {$\!2\!$} -- (6,1.5) node[flavor] (x3) {$\!2\!$}
		-- (-1,-1.5) node[flavor] (y1) {$\!2\!$} -- (7,-1.5) node[flavor] (y2) {$\!2\!$}; 
		
 	\wigE (g1) -- (gi);
	\wigE (g2) -- (gi); 
	\wigE (g2) -- (gii); 
	\wigE (g3) -- (gii);
	\draw[-] (g1) -- (x1); \draw (0,0.75) node[cross] {}; \draw (0,0.75) node[right] {$V_1$};
	\draw[-] (g2) -- (x2); \draw (3,0.75) node[left] {$V_2$};
	\draw[-] (g3) -- (x3); \draw (6,0.75) node[cross] {}; \draw (6,0.75) node[left] {$V_3$};
	
	\draw[-] (g1) -- (y1); \draw (-0.5,-0.75) node[cross] {}; \draw (-0.5,-1) node[right] {$L_1$};
	\draw[-] (g1) -- (0.75,-0.75);
	\draw[-] (g2) -- (2.25,-0.75);
	\draw[-] (g2) -- (3.75,-0.75);
	\draw[-] (g3) -- (5.25,-0.75);
	\draw[-] (g3) -- (y2); \draw (6.5,-0.75) node[cross] {}; \draw (6.5,-1) node[left] {$R_{F_1+F_2-1}$};
	
	\draw[-] (x1) to[out=180,in=90] (-1.5,0) to[out=-90,in=120] (y1); \draw (-1.1,1) node[left] {$F_1^{(j)}$};
	\draw[-] (x3) to[out=0,in=90] (7.6,0) to[out=-90,in=60] (y2); \draw (7.25,1) node[right] {$F_2^{(j)}$};
	
	\draw[-] (g1) to[out=150,in=90] (-0.6,0) to[out=-90,in=-150] (g1); \draw (0,0.45) node[left] {$A_1$};
	\draw[-] (g2) to[out=-60,in=0] (3,-0.6) to[out=180,in=-120] (g2); \draw (3,-0.85) node {$A_{F_1}$};
	\draw[-] (g3) to[out=30,in=90] (6.6,0) to[out=-90,in=-30] (g3); \draw (6,0.45) node[right] {\small{$A_{\text{$F_1$+$F_2$-$1$}}$}};
	
	\draw (-1.5,-2.5) node[right] {\small{$\CW = \CW_{\text{gluing}} + \CW_{\text{triangles}} + \sum_{j=0}^{N-1} \big( Flip[V_1^2 A_1^j] +  $} };
	\draw (-1,-3.1) node[right] {\small{$ + Flip[V_3^2 A_{F_1+F_2-1}^j] + Flip[L_1^2 A_1^j] + $} };
	\draw (-1,-3.6) node[right] {\small{$ + Flip[R_{F_1+F_2-1}^2 A_{F_1+F_2-1}^j] + F^{(j+1)}_1 V_1 A_1^j L_1 + $} };
	\draw (-1,-4.2) node[right] {\small{$F^{(j+1)}_2 V_3 A_{F_1+F_2-1}^j R_{F_1+F_2-1} \big) $}};
    
\end{scope}

\epic}
\ee
As we  explain the next sub-section, we can rigorously construct these $4d$ mirror dualities by running the $4d$ dualization algorithm.

\subsection{$4d$ local dualization algorithm}

We now want to show how the $4d$ SQCD mirror dual can be obtained using the local dualization algorithm.
 The $4d$ algorithm consists of the same steps as the $3d$ one: we chop the theory into basic QFT blocks; we dualize each block using the basic duality moves; we glue back the dualized blocks.  
In \cite{Hwang:2021ulb,Comi:2022aqo} a $4d$ mirror dualization algorithm was formulated to study the special family of $4d$ $\mathcal{N}=1$ theories which are uplifts  of $3d$ theories with eight supercharges constructed in  \cite{Hwang:2020wpd}.
Here we need a generalized version of the algorithm as our $4d$ $\mathcal{N}=1$ theories are uplifts of $3d$ theories with four supercharges.
We will  need  new, generalized, QFT blocks and  new basic duality moves.

\subsection*{Generalized QFT blocks}

\begin{figure}
\centering
\begin{tikzpicture}[thick,node distance=3cm,gauge/.style={circle,draw,minimum size=5mm},flavor/.style={rectangle,draw,minimum size=5mm}]
	
	\path (0,0) node[left] {Generalized flavor:} -- (2,0) node[flavor](g) {$\!2N\!$} -- (2,1.5) node[flavor](y) {$\!2\!$};
	
	\draw[-] (g) -- (y); \draw[blue] (2,0.75) node[right] {$\sqrt{pq} b^{-1}$}; \draw[blue] (2,2) node {$v$};
	\draw (2.5,0) node[right] {\LARGE{${}_{\vec{x}}\mathbb{I}_{\vec{y}}(t)$}};
	\draw (5,0.75) node[right] {$\CW = 0$};
	
\begin{scope}[shift={(0,-1.5)}]
	\path (0,-0.75) node[left] {Improved bifundamental:} -- (2,0) node[flavor](x1) {$\!2N\!$} -- (4,0) node[flavor](x2) {$\!2N\!$}
		-- (3,-1.5) node[flavor](y) {$\!2\!$};
		
	\wigE (x1) -- (x2); \draw[blue] (3,0.3) node {$b$};
	\draw[-] (x1) -- (y); \draw[blue] (2.5,-0.75) node[left] {$\sqrt{pq} c b^{-1}$};
	\draw[-] (x2) -- (y); \draw[blue] (3.5,-0.75) node[right] {$\sqrt{pq} c^{-1} $};
	\draw[blue] (2,0.5) node {$\vec{x}$}; \draw[blue] (4,0.5) node[right] {$\vec{y}$}; \draw[blue] (3.2,-1.5) node[right] {$v$};
	\draw (5,-0.75) node[right] {$\CW = \CW_{\text{triangle}}$};
	
\end{scope}
\end{tikzpicture}
\caption{Definition of the generalized blocks. In the picture we write in blue the parameterization of the two theories. To the generalized flavor we assign a trial R-charge $1$ and charge $-1$ under a $U(1)_b$ symmetry. $v$ denotes the fugacity of the $SU(2)$ symmetry while $\vec{x}$ and $\vec{y}$ are the fugacities of two $USp(2N)$ symmetries. The improved bifundamental block is given by a $FE[USp(2N)]$ theory, for which we assign to the bifundamental operator a trial R-charge of $0$ and $b$-charge $1$. The improved bifundamental block is also equipped with a pair of $USp(2N)\times SU(2)_v$ chirals that are coupled cubically to the improved bifundamental.}
\label{fig:4d_genblock}
\end{figure}
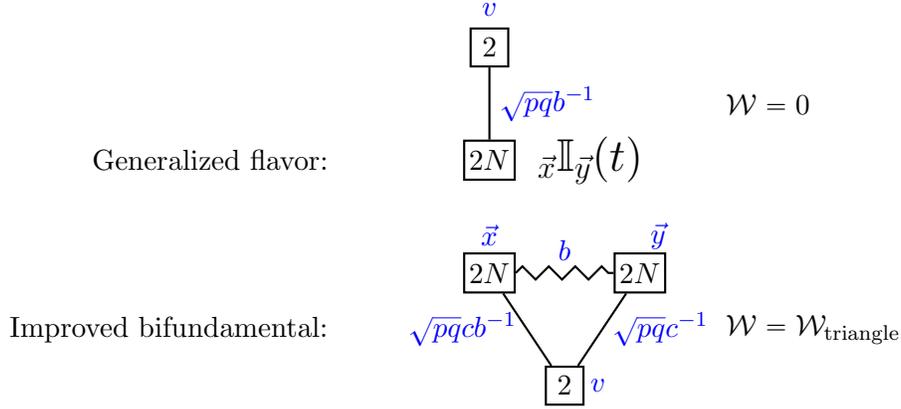

In the first line of in figure \ref{fig:4d_genblock} 
we have the flavor block, parameterized so that it has R-charge $1$, $U(1)_b$ charge $-1$ and it is rotated by a $USp(2N)_x \times USp(2)_v$ symmetry. The flavor comes together with an identity operator whose action consists in identifying the set of fugacities associated to two $USp(2N)$ symmetries. \\
On the second line we have the definition of an improved bifundamental block which is given by a $FE[USp(2N)]$ theory together with two  chirals, $USp(2N)_{x,y} \times USp(2)_v$ bifundamentals. We also introduce a superpotential $\CW_{\text{triangle}}$ coupling cubically the improved bifundamental and the two chirals.
The superconformal index of the  generalized block is given:
\begin{align}
	\CI_{GF}^{(N)}(\vec{x},\vec{y},t,b) = & \prod_{j=1}^N \Ge( \sqrt{pq} b^{-1} x_j^\pm v^\pm ) {}_{\vec{x}}\mathbb{I}_{\vec{y}}(t) \,, \nn \\
	\CI_{GB}^{(N)}(\vec{x},\vec{y},t,b,c) = & \CI_{FE}^{(N)}(\vec{x},\vec{y},t,b) \prod_{j=1}^N \big( \Ge( \sqrt{pq} b^{-1}c x_j^\pm v^\pm) \Ge( \sqrt{pq} c^{-1} y_j^\pm v^\pm) \big) \,.
\end{align}
Where the superconformal index $\CI_{FE}^{(N)}$ is defined in appendix \ref{app:FE}, equation \eqref{eq:FE_SCI}. The identity operator is instead defined as:
\begin{align}\label{eq:4d_idope}
	{}_{\vec{x}}\mathbb{I}_{\vec{y}}(t) = \frac{\prod_{j=1}^N 2\pi i x_j}{\D_N (\vec{x},t) } \sum_{\s \in S_N} \prod_{j=1}^N \d(x_j - y_{\s(j)}^\pm) \,.
\end{align}
Our convention for the $4d$ superconformal index can be found in appendix \ref{inpaconv}.

\subsection*{The $4d$ $\mathsf{S}$-wall theory}
It has been argued in \cite{Bottini:2021vms,Comi:2022aqo} that the $4d$ $\mathsf{S}$-wall theory is given by the $FE[USp(2N)]$. It was also shown that this operator satisfies $PSL(2,\mathbb{Z})$ relations: $(\mathsf{S} \mathsf{T})^3 = 1$ and $\mathsf{S}=\mathsf{S}^{-1}$. The SCI of the $\mathsf{S}$ generator is then defined as:
\begin{align}
	\CI_{\mathsf{S}}^{(N)}(\vec{x},\vec{y},t,c) = \CI_{FE}^{(N)}(\vec{x},\vec{y},t,c) \,.
\end{align}
The $\mathsf{S}=\mathsf{S}^{-1}$ identity corresponds to the Identity-wall property of the $FE[USp(2N)]$ theory:
\begin{align}
	\oint d\vec{z}_N \D_N(\vec{z},t) \CI_{S}^{(N)}(\vec{x},\vec{z},t,c) \CI_{S}^{(N)}(\vec{z},\vec{y},t,c^{-1}) =
	{}_{\vec{x}}\mathbb{I}_{\vec{y}}(t) \,,
\end{align}
where the identity operator is defined as in \eqref{eq:4d_idope} (see also appendix \ref{app:FE}. Graphically this property can be depicted as:
\be\label{FEdelta}
 \bpic[thick,node distance=3cm,gauge/.style={circle,draw,minimum size=5mm},flavor/.style={rectangle,draw,minimum size=5mm}]  
 
	\path (-3,0) node[flavor](x) {$\!2N\!$} -- (-1,0) node[gauge,black](y) {$\!\!\!2N\!\!\!$} -- (1,0) node[flavor](z) {$\!2N\!$} 
		--  (2.5,0) node{$\Longleftrightarrow$};
	
	\wigE (x) -- (y); \draw[blue] (-2,0.4) node {\scriptsize{$c$}};
	\wigE (z) -- (y); \draw[blue] (0,0.4) node {\scriptsize{$c^{-1}$}};
	\draw (4,0) node {\LARGE{${}_{\vec{x}}\mathbb{I}_{\vec{y}}(t)$}};
	\draw[-] (y) to[out=60,in=0] (-1,0.6) to[out=180,in=120] (y); \draw[blue] (-1,0.8) node {\scriptsize{$t$}};
	\draw[blue] (-3,0.5) node {\scriptsize{$\vec{x}$}}; \draw[blue] (1,0.5) node {\scriptsize{$\vec{y}$}};
	
	\draw (-1,-1) node {$\cW = \CW_{\text{gluing}}$};
	
\epic\ee
In \cite{Bottini:2021vms} it was shown that this property can be proven by  iterating the Intriligator-Pouliot duality \cite{Intriligator:1995ne}. 
The $FE[USp(2N)]$ theory therefore plays a double role in improved $4d$ mirror dualities, being both the $\mathsf{S}$-wall theory and the improved bifundamental. It is possible to distinguish the two by the presence of chirals forming a triangular structure in the latter, as in figure \ref{fig:4d_genblock}.

\subsection*{Basic duality moves}
\begin{figure}
\centering
\begin{tikzpicture}[thick,node distance=3cm,gauge/.style={circle,draw,minimum size=5mm},flavor/.style={rectangle,draw,minimum size=5mm}]
	
\begin{scope}
	\path (0,0) node[flavor](x1) {$\!2N\!$} -- (2,0) node[flavor](x2) {$\!2N\!$} -- (1,-1.5) node[flavor](y1) {$\!2\!$};
	
	\wigE (x1) -- (x2); \draw[blue] (1,0.3) node {\scriptsize{$b$}};
	\draw[-] (x1) -- (y1); \draw[blue] (0.5,-0.75) node[left] {\scriptsize{$\sqrt{pq} c b^{-1}$}};
	\draw[-] (x2) -- (y1); \draw[blue] (1.5,-0.75) node[right] {\scriptsize{$\sqrt{pq} c^{-1}$}};
	\draw[blue] (0,0.5) node {\scriptsize{$\vec{x}$}}; \draw[blue] (2,0.5) node {\scriptsize{$\vec{y}$}};
	\draw[blue] (1.2,-1.5) node[right] {\scriptsize{$v$}};
	
	\draw (1,-2.5) node {$\CW = \CW_{\text{triangle}}$};
\end{scope}

	\draw (4,-0.75) node {$\Longleftrightarrow$};	
	
\begin{scope}[shift={(1,-1.5)}]
	\path (5,0) node[flavor](g1) {$\!2N\!$} -- (7,0) node[gauge](g2) {$\!\!\!2N\!\!\!$} -- (9,0) node[flavor](g3) {$\!2N\!$} 
		-- (7,1.5) node[flavor](y2) {$\!2\!$};	
	
	\wigE (g1) -- (g2); \draw[blue] (6,0.3) node {\scriptsize{$c$}};
	\wigE (g2) -- (g3); \draw[blue] (8,0.3) node {\scriptsize{$b c^{-1}$}};
	\draw[-] (g2) -- (y2); \draw[blue] (7,0.75) node[right] {\scriptsize{$\sqrt{pq} b^{-1}$}};
	\draw[-] (g2) to[out=-60,in=0] (7,-0.6) to[out=180,in=-120] (g2); \draw[blue] (7.1,-0.6) node[right] {\scriptsize{$\tau$}};
	
	\draw (7,-1) node {$\CW = W_{\text{gluing}}$};
\end{scope}

\begin{scope}[shift={(0.5,-6)}]
	\path (0,0) node[flavor](f) {$\!2N\!$} -- (0,1.5) node[flavor](y1) {$\!2\!$};
	
	\draw[-] (f) -- (y1); \draw[blue] (0,0.75) node[right] {\scriptsize{$\sqrt{pq} b^{-1} $}}; \draw[blue] (0,2) node {\scriptsize{$v$}};
	\draw (0.3,0) node[right] {\LARGE{${}_{\vec{x}}\mathbb{I}_{\vec{y}}(t)$}};
	
	\draw (0,-1) node[right] {$\CW = 0$};
\end{scope}
	
	\draw (4,-5.25) node {$\Longleftrightarrow$};	
	
\begin{scope}[shift={(2,-4.5)}]
	\path (4,0) node[flavor](g1) {$\!2N\!$} -- (6,0) node[gauge](g2) {$\!\!\!2N\!\!\!$} -- (8,0) node[gauge](g3) {$\!\!\!2N\!\!\!$} 
		-- (10,0) node[flavor](g4) {$\!2N\!$} 
		-- (7,-1.5) node[flavor](y2) {$\!2\!$};
			
	\wigE (g1) -- (g2); \draw[blue] (5,0.3) node {\scriptsize{$c^{-1}$}};
	\wigE (g2) -- (g3); \draw[blue] (7,0.3) node {\scriptsize{$b$}};
	\wigE (g3) -- (g4); \draw[blue] (9,0.3) node {\scriptsize{$c b^{-1}$}};
	\draw[-] (g2) -- (y2); \draw[blue] (6.5,-0.75) node[left] {\scriptsize{$\sqrt{pq} c b^{-1}$}};
	\draw[-] (g3) -- (y2); \draw[blue] (7.5,-0.75) node[right] {\scriptsize{$\sqrt{pq} c^{-1}$}};
	\draw[-] (g2) to[out=60,in=0] (6,0.6) to[out=180,in=120] (g2); \draw[blue] (6,0.7) node[right] {\scriptsize{$\tau$}};
	\draw[-] (g3) to[out=60,in=0] (8,0.6) to[out=180,in=120] (g3); \draw[blue] (8,0.7) node[right] {\scriptsize{$\tau$}};
	\draw[blue] (4,0.5) node {\scriptsize{$\vec{x}$}}; \draw[blue] (10,0.5) node {\scriptsize{$\vec{y}$}};
	\draw[blue] (7.2,-1.5) node[right] {\scriptsize{$v$}};

	\draw (5,-2.5) node[right] {$\CW = W_{\text{gluing}} + \CW_{\text{triangle}} $};
\end{scope}
	
\end{tikzpicture}
\caption{Basic $\mathsf{S}$-duality moves for the $4d$ generalized blocks. On top we have the $\mathsf{S}$-dualization of the flavor block into an improved bifundamental. On the bottom we have the $\mathsf{S}$-dualization of the improved bifundamental into a flavor. The $\mathsf{S}$ operator is identified with the $FE[USp(2N)]$ theory that in these dualities plays a double-role. 
$\CW_{\text{gluing}}$ encodes the superpotential terms coupling the antisymmetric chirals to the antisymmetric operators inside the improved bifundamental and $\mathsf{S}$-wall theories. Also, $\CW_{\text{triangle}}$ means that we couple cubically the improved bifundamental and the chirals in each triangle.}
\label{fig:4d_basicmoves}
\end{figure}
The two basic duality moves encode the mirror dualization of the two blocks and are depicted in figure \ref{fig:4d_basicmoves}. \\

In the first duality move we relate an improved bifundamental block with a generalized flavor on which are acting two $\mathsf{S}$-walls. 
On the r.h.s. $\CW_{\text{gluing}}$ implies that the antisymmetric operator is coupled to the antisymmetric operators inside the two $\mathsf{S}$-walls, the flavor does not enter in the superpotential and therefore is rotated by an independent $USp(2)_v \times U(1)_b$ symmetry. \\

In the second duality we are acting with two $\mathsf{S}$-walls on an improved bifundamental block to obtain a generalized flavor block. On the r.h.s. we have $\CW_{\text{gluing}}$ and $\CW_{\text{triangle}}$ to imply that the antisymmetric chirals are coupled to the antisymmetric operators inside the improved bifundamental and $\mathsf{S}$-wall theories. \\

The first basic duality move in \ref{fig:4d_basicmoves} is also called \emph{braid duality}, while the second duality can be obtained starting from the first one and gluing on the left and on the right an $FE$ theory and using that $\mathsf{S}^2 = 1$.
In \cite{BCP1}, it is shown that braid duality can be proved by induction assuming only the  Intriligator-Pouliot duality, hence also all the dualities obtained from the $4d$ dualization algorithm can be seen as consequences of basic Seiberg-like dualities. \\

As superconformal index identities the basic duality moves can be written as:
\begin{align}
	\CI_{GB}^{(N)}(\vec{x},\vec{y},t,b,c) = & \oint \prod_{a=1}^2 \big( d\vec{z}^{(a)}_N \D_N(\vec{z}^{(a)},t) \big) \prod_{j=1}^N \CI_{\mathsf{S}}^{(N)}(\vec{x},\vec{z}^{(1)},t,c) \nn \\
	& \CI_{GF}^{(N)}(\vec{z}^{(1)},\vec{z}^{(2)},t,b) \CI_{\mathsf{S}}^{(N)}(\vec{z}^{(2)},\vec{y},t,b/c) \,, \nn \\
	\CI_{GF}^{(N)}(\vec{x},\vec{y},t,b) = & \oint \prod_{a=1}^2 \big( d\vec{z}^{(a)}_N \D_N(\vec{z}^{(a)},t) \big) \CI_{\mathsf{S}}^{(N)}(\vec{x},\vec{z}^{(1)},t,c^{-1})  \nn \\
	& \CI_{GB}^{(N)}(\vec{z}^{(1)},\vec{z}^{(2)},t,b,c) \CI_{\mathsf{S}}^{(N)}(\vec{z}^{(2)},\vec{y},t,c/b) \,.
\end{align}
Notice that the superconformal indexes of the $\mathsf{S}$-walls and of the generalized QFT blocks can be seen as matrices carring two $USp(2N)$ fugacities. Multiplying two of them consist in identifying two $USp(2N)$ symmetries and gauging its diagonal subgroup with the integration measure $\D_N(\vec{z},t)$, which contains both a $\CN=1$ vector multiplet and an antisymmetric chiral with charge $+1$ under a $U(1)_t$ symmetry. Notice that the $U(1)_t$ symmetries of all the blocks multiplied are identified, due to the $\CW_{\text{gluing}}$ superpotentials. \\

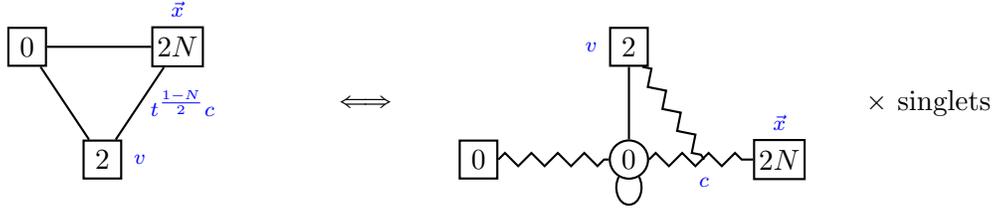
\begin{figure}
\centering
\begin{tikzpicture}[thick,node distance=3cm,gauge/.style={circle,draw,minimum size=5mm},flavor/.style={rectangle,draw,minimum size=5mm}]
	
	\path (0,0) node[flavor](x1) {$\!0\!$} -- (2,0) node[flavor](x2) {$\!2N\!$} -- (1,-1.5) node[flavor](y1) {$\!2\!$}
		-- (4.5,-0.75) node {$\Longleftrightarrow$};
	
	\draw[-] (x1) -- (x2); 
	\draw[-] (x1) -- (y1);
	\draw[-] (x2) -- (y1); \draw[blue] (1.5,-0.75) node[right] {\scriptsize{$t^{\frac{1-N}{2}}c$}};
	\draw[blue] (2,0.5) node {\scriptsize{$\vec{x}$}};
	\draw[blue] (1.5,-1.5) node {\scriptsize{$v$}};
	
\begin{scope}[shift={(2,-1.5)}]
	\path (4,0) node[flavor](g1) {$\!0\!$} -- (6,0) node[gauge](g2) {$\!\!\!0\!\!\!$} -- (8,0) node[flavor](g3) {$\!2N\!$} 
		-- (6,1.5) node[flavor](y1) {$\!2\!$};
		
	\wigE (g1) -- (g2);
	\wigE (g2) -- (g3); \draw[blue] (7,-0.3) node {\scriptsize{$c$}};
	\wigE (y1) -- (7,0); 
	\draw[-] (g2) -- (y1); \draw[blue] (5.5,1.5) node {\scriptsize{$v$}};
	\draw[-] (g2) to[out=-60,in=0] (6,-0.6) to[out=180,in=-120] (g2);
	\draw[blue] (8,0.5) node {\scriptsize{$\vec{x}$}};
	
	\draw (9,0.75) node[right] {$\times$ singlets};
	
\end{scope}

\end{tikzpicture}
\caption{Asymmetric basic duality move relating a $USp(2N)\times USp(0)$ improved bifundamental block with an asymmetric generalized flavor with $S$-walls on the sides.}
\label{fig:4d_asymm_basicmove}
\end{figure}

As already discussed in the $3d$ case, we do not know an ``asymmetric" version of the braid duality, which would relate a generalized $USp(2N)\times USp(2M)$ bifundamental to a flavor. However, in order to run the algorithm we still need the $M=0$ case. The $USp(2N) \times USp(0)$ bifundamental is just given by a single bifundamental chiral of $USp(2N)_x\times SU(2)_v$, its dualization is given in figure \ref{fig:4d_asymm_basicmove}. This duality consist in the following partition function identity:
\begin{align}\label{eq:4d_asymm_basicmove}
	\prod_{j=1}^N \Ge( t^{\frac{1-N}{2}} c x_j^\pm v^\pm ) = \prod_{j=1}^N \big( \Ge(t^j) \Ge(t^{1-j} c^2) \big) 
	\CI_{\mathsf{S}}^{(N)}(\vec{x},\{ t^{\frac{N-1}{2}}v , \ldots, t^{\frac{N-1}{2}}v \}, t, c) \,.
\end{align}
The definition of the asymmetric $S$-wall theory is given in appendix \ref{app:FE}.

\subsubsection*{Useful combined moves}
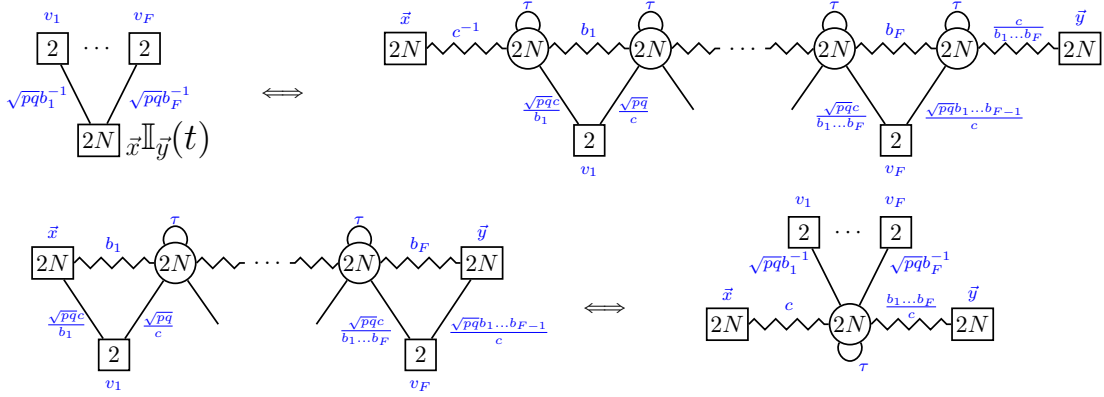
\begin{figure}
\centering
\resizebox{.95\hsize}{!}{
\begin{tikzpicture}[thick,node distance=3cm,gauge/.style={circle,draw,minimum size=5mm},flavor/.style={rectangle,draw,minimum size=5mm}] 
	
\begin{scope}
	\path (0,0) node[flavor](g) {$\!2N\!$} -- (-0.75,1.5) node[flavor](x1) {$\!2\!$} -- (0.75,1.5) node[flavor](x2) {$\!2\!$};
	
	\draw[-] (g) -- (x1); \draw[blue] (-0.35,0.7) node[left] {\scriptsize{$\sqrt{pq} b_1^{-1}$}}; \draw[blue] (-0.75,2) node {\scriptsize{$v_1$}};
	\draw (0,1.5) node {$\cdots$};
	\draw[-] (g) -- (x2); \draw[blue] (0.35,0.7) node[right] {\scriptsize{$\sqrt{pq} b_F^{-1}$}}; \draw[blue] (0.75,2) node {\scriptsize{$v_F$}};
	\draw (0.3,0) node[right] {\LARGE{${}_{\vec{x}}\mathbb{I}_{\vec{y}}(t)$}};

\end{scope}

	\draw (3,0.75) node {$\Longleftrightarrow$};

\begin{scope}[shift={(5,1.5)}]
	\path (0,0) node[flavor](y1) {$\!2N\!$} -- (2,0) node[gauge](g1) {$\!\!\!2N\!\!\!$} -- (4,0) node[gauge](g2) {$\!\!\!2N\!\!\!$}
		-- (5.5,0) node(gi) {$\ldots$} -- (7,0) node[gauge](g3) {$\!\!\!2N\!\!\!$} -- (9,0) node[gauge](g4) {$\!\!\!2N\!\!\!$} 
		-- (11,0) node[flavor](y2) {$\!2N\!$}
		-- (3,-1.5) node[flavor](x1) {$\!2\!$} -- (8,-1.5) node[flavor](x2) {$\!2\!$};
		
	\wigE (y1) -- (g1); \draw[blue] (1,0.3) node {\scriptsize{$c^{-1}$}};
	\wigE (g1) -- (g2); \draw[blue] (3,0.3) node {\scriptsize{$b_1$}};
	\wigE (g2) -- (gi); 
	\wigE (g3) -- (gi);
	\wigE (g3) -- (g4); \draw[blue] (8,0.3) node {\scriptsize{$b_F$}};
	\wigE (g4) -- (y2); \draw[blue] (10,0.3) node {\scriptsize{$\frac{c}{b_1\ldots b_F}$}};
	\draw[-] (g1) -- (x1); \draw[blue] (2.7,-1) node[left] {\scriptsize{$\frac{\sqrt{pq} c}{b_1}$}};
	\draw[-] (g2) -- (x1); \draw[blue] (3.3,-1) node[right] {\scriptsize{$\frac{\sqrt{pq}}{c}$}};
	\draw[-] (g2) -- (4.7,-1);
	\draw[-] (g3) -- (6.3,-1);
	\draw[-] (g3) -- (x2); \draw[blue] (7.7,-1.1) node[left] {\scriptsize{$\frac{\sqrt{pq} c}{b_1 \ldots b_F}$}};
	\draw[-] (g4) -- (x2); \draw[blue] (8.3,-1.1) node[right] {\scriptsize{$\frac{\sqrt{pq} b_1 \ldots b_{F-1}}{c} $}};
	\draw[-] (g1) to[out=60,in=0] (2,0.6) to[out=180,in=120] (g1); \draw[blue] (2,0.7) node {\scriptsize{$\tau$}};
	\draw[-] (g2) to[out=60,in=0] (4,0.6) to[out=180,in=120] (g2); \draw[blue] (4,0.7) node {\scriptsize{$\tau$}};
	\draw[-] (g3) to[out=60,in=0] (7,0.6) to[out=180,in=120] (g3); \draw[blue] (7,0.7) node {\scriptsize{$\tau$}};
	\draw[-] (g4) to[out=60,in=0] (9,0.6) to[out=180,in=120] (g4); \draw[blue] (9,0.7) node {\scriptsize{$\tau$}};
	
	\draw[blue] (0,0.5) node {\scriptsize{$\vec{x}$}}; \draw[blue] (11,0.5) node {\scriptsize{$\vec{y}$}};
	\draw[blue] (3,-2) node {\scriptsize{$v_1$}}; \draw[blue] (8,-2) node {\scriptsize{$v_F$}};
	
\end{scope}

\begin{scope}[shift={(-0.75,-2)}]
	\path (0,0) node[flavor](y1) {$\!2N\!$} -- (2,0) node[gauge](g1) {$\!\!\!2N\!\!\!$} -- (3.5,0) node(gi) {$\ldots$}
		-- (5,0) node[gauge](g2) {$\!\!\!2N\!\!\!$} -- (7,0) node[flavor](y2) {$\!2N\!$}
		-- (1,-1.5) node[flavor](x1) {$\!2\!$} -- (6,-1.5) node[flavor](x2) {$\!2\!$};
		
	\wigE (y1) -- (g1); \draw[blue] (1,0.3) node {\scriptsize{$b_1$}};
	\wigE (g1) -- (gi);
	\wigE (g2) -- (gi);
	\wigE (g2) -- (y2); \draw[blue] (6,0.3) node {\scriptsize{$b_F$}};
	\draw[-] (y1) -- (x1); \draw[blue] (0.7,-1) node[left] {\scriptsize{$\frac{\sqrt{pq} c}{b_1}$}};
	\draw[-] (g1) -- (x1); \draw[blue] (1.3,-1) node[right] {\scriptsize{$\frac{\sqrt{pq}}{c}$}};
	\draw[-] (g1) -- (2.7,-1);
	\draw[-] (g2) -- (4.3,-1);
	\draw[-] (g2) -- (x2); \draw[blue] (5.7,-1.1) node[left] {\scriptsize{$\frac{\sqrt{pq} c}{b_1 \ldots b_F}$}};
	\draw[-] (y2) -- (x2); \draw[blue] (6.3,-1.1) node[right] {\scriptsize{$\frac{\sqrt{pq} b_1 \ldots b_{F-1}}{c} $}};
	\draw[-] (g1) to[out=60,in=0] (2,0.6) to[out=180,in=120] (g1); \draw[blue] (2,0.7) node {\scriptsize{$\tau$}};
	\draw[-] (g2) to[out=60,in=0] (5,0.6) to[out=180,in=120] (g2); \draw[blue] (5,0.7) node {\scriptsize{$\tau$}};
	
	\draw[blue] (0,0.5) node {\scriptsize{$\vec{x}$}}; \draw[blue] (7,0.5) node {\scriptsize{$\vec{y}$}};
	\draw[blue] (1,-2) node {\scriptsize{$v_1$}}; \draw[blue] (6,-2) node {\scriptsize{$v_F$}};
	
\end{scope}

	\draw (8.25,-2.75) node {$\Longleftrightarrow$};

\begin{scope}[shift={(10.25,-3)}]
	\path (0,0) node[flavor](g1) {$\!2N\!$} -- (2,0) node[gauge](g2) {$\!\!\!2N\!\!\!$} -- (4,0) node[flavor](g3) {$\!2N\!$}
		-- (1.25,1.5) node[flavor](x1) {$\!2\!$} -- (2.75,1.5) node[flavor](x2) {$\!2\!$};
	
	\wigE (g1) -- (g2); \draw[blue] (1,0.3) node{\scriptsize{$c$}};
	\wigE (g2) -- (g3); \draw[blue] (3,0.3) node{\scriptsize{$\frac{b_1 \ldots b_F}{c}$}};
	\draw[-] (g2) -- (x1); \draw[blue] (1.5,1) node[left] {\scriptsize{$\sqrt{pq} b_1^{-1}$}}; \draw[blue] (1.25,2) node {\scriptsize{$v_1$}};
	\draw (2,1.5) node {$\cdots$};
	\draw[-] (g2) -- (x2); \draw[blue] (2.5,1) node[right] {\scriptsize{$\sqrt{pq} b_F^{-1}$}}; \draw[blue] (2.75,2) node {\scriptsize{$v_F$}};
	\draw[-] (g2) to[out=-60,in=0] (2,-0.6) to[out=180,in=-120] (g2); \draw[blue] (2,-0.7) node[right] {\scriptsize{$\tau$}};
	
	\draw[blue] (0,0.5) node {\scriptsize{$\vec{x}$}}; \draw[blue] (4,0.5) node {\scriptsize{$\vec{y}$}};

\end{scope}
	
\end{tikzpicture}}
\caption{Duality moves relating a block of $F$ flavors to $F$ improved bifundamental blocks.}
\label{fig:4d_flavors_basicmove}
\end{figure}

 It is convenient to consider the dualization of $F$ flavors, which can be inferred from the dualization of a single flavor block. This is dual to a set of $F$ improved bifundamental blocks with an $\mathsf{S}$-wall on each side. The corresponding duality is depicted on the top  of figure \ref{fig:4d_flavors_basicmove}. On the bottom we have the inverse move, relating $F$ improved bifundamentals to $F$ flavors on which an $\mathsf{S}$-wall acts on each side. We first define the SCI of a set of $F$ flavors as:
\begin{align}
	\CI_{F-GF}^{(N)}(\vec{x},\vec{y},t,\vec{b},\vec{v}) = \prod_{j=1}^N \prod_{a=1}^F \Ge(\sqrt{pq}b_a^{-1} x_j^\pm v_a^\pm) {}_{\vec{x}}\mathbb{I}_{\vec{y}}(t) \,.
\end{align}
The dualities consist in the following SCI identities:
\begin{align}\label{eq:4d_flavors_basicmove}
	& \CI_{F-GF}^{(N)}(\vec{x},\vec{y},t,\vec{b},\vec{v}) = 
	\oint \prod_{a=1}^{F+1} \big( d\vec{z}^{(a)}_N \D_N(\vec{z}^{(a)},t) \big) \CI_{\mathsf{S}}^{(N)}(\vec{x},\vec{z}^{(1)},t,c^{-1}) \nn \\
	& \prod_{a=1}^F \CI_{GB}^{(N)} (\vec{z}^{(a)},\vec{z}^{(a+1)},t,b_a,c(b_{1} \ldots b_{a-1})^{-1} ) 
	\CI_{\mathsf{S}}^{(N)}(\vec{z}^{(F+1)},\vec{y},t,c(b_1 \ldots b_F)^{-1} ) \,, \nn \\
	& \oint \prod_{a=1}^{F-1} \big( d\vec{z}^{(a)}_N \D_N(\vec{z}^{(a)},t) \big) \CI_{GB}(\vec{x},\vec{z}^{(1)},t,b_1,c) \nn \\
	& \prod_{a=2}^{F-1} \CI_{GB}^{(N)} (\vec{z}^{(a-1)},\vec{z}^{(a)},t,b_a,c(b_{1} \ldots b_{a-1})^{-1} ) 
	\CI_{GB}(\vec{z}^{(F-1)},\vec{y},t,b_F,c (b_{1} \ldots b_{F})^{-1} ) = \nn \\
	& = \oint \prod_{a=1}^{2} \big( d\vec{z}^{(a)}_N \D_N(\vec{z}^{(a)},t) \big) 
	\CI_{\mathsf{S}}^{(N)}(\vec{x},\vec{z}^{(1)},t,c) \CI_{F-GF}^{(N)}(\vec{z}^{(1)},\vec{z}^{(2)},t,\vec{b},\vec{v}) \nn \\
	& \CI_{\mathsf{S}}^{(N)}(\vec{z}^{(2)},\vec{y},t, b_1 \ldots b_F c^{-1} )\,.
\end{align}

\subsubsection*{Proving the $\CN=1$ antisymmetric SQCD mirror pair via the dualization algorithm}

We are now ready to derive  the SQCD mirror dual via the algorithm. \\
We start from the antisymmetric SQCD parameterized as in \ref{fig:SQCD_4d_manifest}, we decompose the  theory into two  trivial bifundamental blocks
and a block of $F$ flavors in the center. Notice that two of the original  flavors are used to reconstruct  the trivial bifundamental blocks.
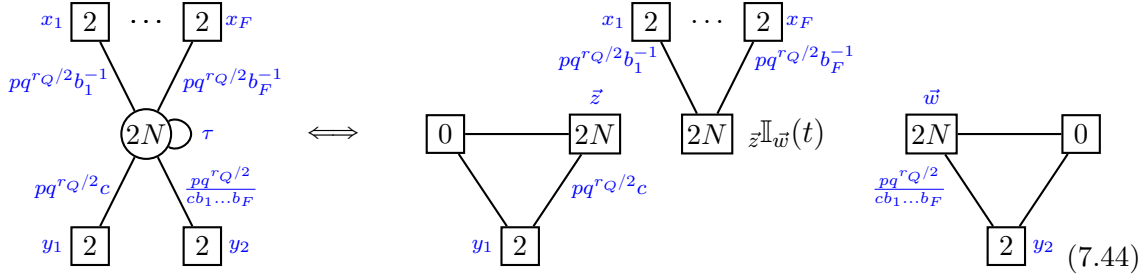
\be
\resizebox{.95\hsize}{!}{
 \bpic[thick,node distance=3cm,gauge/.style={circle,draw,minimum size=5mm},flavor/.style={rectangle,draw,minimum size=5mm}] 
 
	\path (-2,0) node[gauge](g) {$\!\!\!2N\!\!\!$} -- (-2.75,1.5) node[flavor](x1) {$\!2\!$} -- (-1.25,1.5) node[flavor](x2) {$\!2\!$}
		-- (-2.75,-1.5) node[flavor](y1) {$\!2\!$} -- (-1.25,-1.5) node[flavor](y2) {$\!2\!$} -- (0.5,0) node{$\Longleftrightarrow$};
	
	\draw[-] (g) -- (x1); \draw[blue] (-2.35,0.7) node[left] {\scriptsize{$pq^{r_Q/2} b_1^{-1}$}}; \draw[blue] (-3.25,1.5) node {\scriptsize{$x_1$}};
	\draw (-2,1.5) node {$\ldots$};
	\draw[-] (g) -- (x2); \draw[blue] (-1.65,0.7) node[right] {\scriptsize{$pq^{r_Q/2} b_F^{-1}$}}; \draw[blue] (-0.75,1.5) node {\scriptsize{$x_F$}};
	\draw[-] (g) -- (y1); \draw[blue] (-2.35,-0.7) node[left] {\scriptsize{$pq^{r_Q/2} c$}}; \draw[blue] (-3.25,-1.5) node {\scriptsize{$y_1$}};
	\draw[-] (g) -- (y2); \draw[blue] (-1.65,-0.7) node[right] {\scriptsize{$\frac{pq^{r_Q/2}}{c b_1 \ldots b_F} $}}; \draw[blue] (-0.75,-1.5) node {\scriptsize{$y_2$}};
	\draw[-] (g) to[out=30,in=90] (-1.4,0) to[out=-90,in=-30] (g); \draw[blue] (-1.4,0) node[right] {\scriptsize{$\tau$}};
	
	\path (2,0) node[flavor](z1) {$\!0\!$} -- (4,0) node[flavor](g1) {$\!2N\!$} -- (3,-1.5) node[flavor](y1) {$\!2\!$}
		-- (5.5,0) node[flavor](g2) {$\!2N\!$} -- (4.75,1.5) node[flavor](x1) {$\!2\!$} -- (6.25,1.5) node[flavor](x2) {$\!2\!$}
		-- (8.5,0) node[flavor] (g3) {$\!2N\!$} -- (10.5,0) node[flavor] (z2) {$\!0\!$} -- (9.5,-1.5) node[flavor](y2) {$\!2\!$}; 
	
	\draw[-] (z1) -- (g1); \draw[blue] (4,0.5) node {\scriptsize{$\vec{z}$}};
	\draw[-] (g1) -- (y1); \draw[blue] (3.55,-0.7) node[right] {\scriptsize{$pq^{r_Q/2} c$}}; \draw[blue] (2.5,-1.5) node {\scriptsize{$y_1$}};
	\draw[-] (z1) -- (y1); \draw[blue] (8.85,-0.7) node[left] {\scriptsize{$\frac{pq^{r_Q/2}}{c b_1 \ldots b_F} $}}; \draw[blue] (10,-1.5) node {\scriptsize{$y_2$}};
	
	\draw[-] (g2) -- (x1); \draw[blue] (5,1) node[left] {\scriptsize{$pq^{r_Q/2} b_1^{-1}$}}; \draw[blue] (4.25,1.5) node {\scriptsize{$x_1$}};
	\draw (-2,1.5) node {$\ldots$};
	\draw (5.5,1.5) node {$\ldots$};
	\draw[-] (g2) -- (x2); \draw[blue] (6,1) node[right] {\scriptsize{$pq^{r_Q/2} b_F^{-1}$}}; \draw[blue] (6.75,1.5) node {\scriptsize{$x_F$}};
	\draw (5.9,0) node[right] {\large{${}_{\vec{z}}\mathbb{I}_{\vec{w}}(t)$}};	
	
	\draw[-] (g3) -- (z2); \draw[blue] (8.5,0.5) node {\scriptsize{$\vec{w}$}};
	\draw[-] (g3) -- (y2);
	\draw[-] (z2) -- (y2);
	
\epic}\ee
At the level of the superconformal index this step consists in starting from the index of the SQCD, defined in \eqref{eq:4d_ind_elec}, and rewriting it as:
\begin{align}\label{eq:4d_step_1}
	\CI_{SQCD}(\vec{x},\vec{y},\vec{b},c,t) = & \oint d\vec{z}_N \D_N(\vec{z},t) \prod_{j=1}^N \big( \Ge( pq^{r_Q/2} c z_j^\pm y_1^\pm) \nn \\
	& \prod_{a=1}^F \Ge( pq^{r_Q/2} b_a z_j^\pm x_a^\pm) \Ge( pq^{r_Q/2} \prod_{a=1}^{F} b_a^{-1} c^{-1} z_j^\pm y_2^\pm ) \big] = \, \nn \\
	= & \oint d\vec{z}_N d\vec{w}_N \D_N(\vec{z},t) \D_N(\vec{w},t) \prod_{j=1}^N \big[ \Ge( pq^{r_Q/2} c z_j^\pm y_1^\pm) \nn \\
	& \prod_{a=1}^F \Ge( pq^{r_Q/2} b_a z_j^\pm x_a^\pm) {}_{\vec{z}}\mathbb{I}_{\vec{w}}(t) \Ge( pq^{r_Q/2} \prod_{a=1}^{F} b_a^{-1} c^{-1} w_j^\pm y_2^\pm ) \big] = \CI_{\text{Step I}} \,.
\end{align}
The matching between the first and second expression is trivial after using the fact that the ${}_{\vec{z}}\mathbb{I}_{\vec{w}}(t)$ operator behaves as a delta-function identifying $\vec{z}$ and $\vec{w}$, with the normalization:
\begin{align}
	\oint d\vec{z}_N \D_N(\vec{z},t) {}_{\vec{z}}\mathbb{I}_{\vec{w}}(t) =1\,.
\end{align}

In the second step we dualize each block using the basic moves in figure \ref{fig:4d_flavors_basicmove} and \ref{fig:4d_asymm_basicmove}. Gluing 
back the dualized blocks we obtain:
\be
\resizebox{.95\hsize}{!}{
 \bpic[thick,node distance=3cm,gauge/.style={circle,draw,minimum size=5mm},flavor/.style={rectangle,draw,minimum size=5mm}] 
	
	\path (0,0) node[flavor] (z1) {$\!0\!$} -- (2,0) node[gauge] (g1) {$\!\!\!0\!\!\!$} 
			-- (4,0) node[gauge](g2) {$\!\!\!2N\!\!\!$} -- (6,0) node[gauge](g3) {$\!\!\!2N\!\!\!$} -- (8,0) node[gauge](g4) {$\!\!\!2N\!\!\!$} 
			-- (11,0) node[gauge](g5) {$\!\!\!2N\!\!\!$} -- (13,0) node[gauge] (g6) {$\!\!\!2N\!\!\!$} -- (15,0) node[gauge] (g7) {$\!\!\!2N\!\!\!$}
			-- (17,0) node[gauge] (g8) {$\!\!\!0\!\!\!$} -- (19,0) node[flavor] (z2) {$\!0\!$} 
			-- (2,1.5) node[flavor] (y1) {$\!2\!$} -- (17,1.5) node[flavor] (y2) {$\!2\!$} 
			-- (7,-1.5) node[flavor] (x2) {$\!2\!$} -- (12,-1.5) node[flavor] (x3) {$\!2\!$};
	
	\wigE (z1) -- (g1);
	\wigE (g1) -- (g2); \draw[blue] (3,-0.3) node {\scriptsize{$ c_0^{-1} $}};
	\wigE (y1) -- (3,0);
	\wigE (g2) -- (g3); \draw[blue] (5,0.3) node {\scriptsize{$ c_0 $}};
	\wigE (g3) -- (g4); \draw[blue] (7,0.3) node {\scriptsize{$\sqrt{pq} \tilde{b}_1$}};
	\wigE (g4) -- (9,0);
	\wigE (g5) -- (10,0);
	\wigE (g5) -- (g6); \draw[blue] (12,0.3) node {\scriptsize{$\sqrt{pq} \tilde{b}_F$}};
	\wigE (g6) -- (g7); \draw[blue] (14,0.3) node {\scriptsize{$c_F^{-1}$}};
	\wigE (g7) -- (g8); \draw[blue] (16,-0.3) node {\scriptsize{$c_F$}};
	\wigE (y2) -- (16,0);
	\wigE (g8) -- (z2);

	\draw (9.5,-0.5) node {$\ldots$};
	
	\draw[-] (g3) -- (x2); \draw[blue] (6.5,-0.75) node[left] {\scriptsize{$ \sqrt{pq} c_1^{-1}$}};
	\draw[-] (g4) -- (x2); \draw[blue] (7.5,-0.75) node[right] {\scriptsize{$ \sqrt{pq} c_0$}};
	\draw[-] (g5) -- (x3); \draw[blue] (11.5,-0.75) node[left] {\scriptsize{$ \sqrt{pq} c_{F}^{-1}$}};
	\draw[-] (g6) -- (x3); \draw[blue] (12.5,-0.75) node[right] {\scriptsize{$ \sqrt{pq} c_{F-1}^{-1}$}};
	\draw[-] (g4) -- (8.75,-0.75);
	\draw[-] (g5) -- (10.25,-0.75);
	\draw[-] (g1) -- (y1); \draw[blue] (2,2) node {\scriptsize{$y_1$}};
	\draw[-] (g8) -- (y2); \draw[blue] (17,2) node {\scriptsize{$y_1$}};
	
	\draw[blue] (7.5,-1.5) node {\scriptsize{$x_1$}}; \draw[blue] (12.5,-1.5) node {\scriptsize{$x_F$}};	
	
	\draw[-] (g1) to[out=-60,in=0] (2,-0.6) to[out=180,in=-120] (g1);
	\draw[-] (g2) to[out=60,in=0] (4,0.6) to[out=180,in=120] (g2); \draw[blue] (4,0.7) node {\scriptsize{$\tau$}};
	\draw[-] (g3) to[out=60,in=0] (6,0.6) to[out=180,in=120] (g3); \draw[blue] (6,0.7) node {\scriptsize{$\tau$}};
	\draw[-] (g4) to[out=60,in=0] (8,0.6) to[out=180,in=120] (g4); \draw[blue] (8,0.7) node {\scriptsize{$\tau$}};
	\draw[-] (g5) to[out=60,in=0] (11,0.6) to[out=180,in=120] (g5); \draw[blue] (11,0.7) node {\scriptsize{$\tau$}};
	\draw[-] (g6) to[out=60,in=0] (13,0.6) to[out=180,in=120] (g6); \draw[blue] (13,0.7) node {\scriptsize{$\tau$}};
	\draw[-] (g7) to[out=60,in=0] (15,0.6) to[out=180,in=120] (g7); \draw[blue] (15,0.7) node {\scriptsize{$\tau$}};
	\draw[-] (g8) to[out=-60,in=0] (17,-0.6) to[out=180,in=-120] (g8);
	
\epic}\ee
To avoid cluttering we will not write all the singlets coming from the dualization in the figures, we will restore them in the end. For convenience we have also defined the following:
\begin{align}
	& \tilde{b}_a = pq^{r_Q/2} b_a \qquad , \qquad \tilde{c} = pq^{r_Q/2} c \,, \nn \\
	& c_a = pq^{\frac{a}{2}} t^{\frac{1-N}{2}} \tilde{b}_1 \ldots \tilde{b}_a \tilde{c}^{-1} \,.
\end{align}
At the level of the superconformal index this step consists in using the identities \eqref{eq:4d_flavors_basicmove} and \eqref{eq:4d_asymm_basicmove}, corresponding to the duality moves, inside the expression \eqref{eq:4d_step_1} to obtain:
\begin{align}\label{eq:4d_step2}
	\CI_{SQCD}(\vec{x},\vec{y},\vec{b},c,t) = & \CI_{\text{Step I}} = \oint \prod_{a=1}^{F+3} \big( d\vec{z}^{(a)}_N \D_N(\vec{z}^{(a)},t) \big) \prod_{j=1}^N \big( \Ge( pq t^{-j} )^2 \Ge(t^{1-j} c_0^{-2}) \Ge( t^{1-j} c_F^2 ) \big) \nn \\
	& \CI_{\mathsf{S}}^{(N)}(\{ t^{\frac{N-1}{2}} y_1, \ldots , t^{\frac{1-N}{2}} y_1 \},\vec{z}^{(1)},t, c_0^{-1} ) \CI_{\mathsf{S}}^{(N)}( \vec{z}^{(1)}, \vec{z}^{(2)}, t, c_0 ) \nn \\
	& \prod_{a=1}^F \CI_{GB}^{(N)} (\vec{z}^{(a+2)},\vec{z}^{(a+3)}, t, \sqrt{pq} \tilde{b}_1 , \sqrt{pq} c_a^{-1} \tilde{b}_a^{-1} )
	\CI_{\mathsf{S}}^{(N)}( \vec{z}^{(F+2)}, \vec{z}^{(F+3)}, t,  c_F^{-1} ) \nn \\
	& \CI_{\mathsf{S}}^{(N)}( \vec{z}^{(F+3)}, \{ t^{\frac{N-1}{2}} y_2, \ldots, t^{\frac{1-N}{2}} y_2 \}, t, c_F ) = \CI_{\text{Step II}} \,.
\end{align}

We then recognize two asymmetric $\mathbb{I}$-walls given by an asymmetric $\mathsf{S}$-wall theory glued to a standard one. Using the result \ref{FEdeltaA}, we see that the effect of the asymmetric $\mathbb{I}$-wall is to Higgs the second  and second last $USp(2N)$ gauge groups down to a flavor $USp(2)$. The Higgsing also causes the first and last diagonal leg to become $N$ chirals in the bifundamental of $USp(2)_{x_1} \times USp(2)_{y_1}$ and $USp(2)_{x_F} \times USp(2)_{y_2}$. All in all we have:
\be
\resizebox{.95\hsize}{!}{
 \bpic[thick,node distance=3cm,gauge/.style={circle,draw,minimum size=5mm},flavor/.style={rectangle,draw,minimum size=5mm}] 
	
	\path (0,0) node[flavor](g2) {$\!0\!$} -- (2,0) node[gauge](g3) {$\!\!\!2N\!\!\!$} -- (4,0) node[gauge](g4) {$\!\!\!2N\!\!\!$} 
			-- (7,0) node[gauge](g5) {$\!\!\!2N\!\!\!$} -- (9,0) node[gauge] (g6) {$\!\!\!2N\!\!\!$} -- (11,0) node[flavor] (g7) {$\!0\!$} 
			-- (0,1.5) node[flavor] (y1) {$\!2\!$} -- (11,1.5) node[flavor] (y2) {$\!2\!$} 
			-- (1,-1.5) node[flavor] (x1) {$\!2\!$} -- (3,-1.5) node[flavor] (x2) {$\!2\!$} -- (8,-1.5) node[flavor] (x3) {$\!2\!$}
			-- (10,-1.5) node[flavor] (x4) {$\!2\!$};
	
	\wigE (y1) -- (1,0); 
	\wigE (g2) -- (g3); \draw[blue] (1,-0.3) node {\scriptsize{$\sqrt{pq}\tilde{b}_1$}};
	\wigE (g3) -- (g4); \draw[blue] (3,0.3) node {\scriptsize{$\sqrt{pq}\tilde{b}_2$}};
	\wigE (g4) -- (5,0);
	\wigE (g5) -- (6,0);
	\wigE (g5) -- (g6); \draw[blue] (8,0.3) node {\scriptsize{$\sqrt{pq}\tilde{b}_{F-1}$}};
	\wigE (g6) -- (g7); \draw[blue] (10,-0.3) node {\scriptsize{$\sqrt{pq}\tilde{b}_F$}};
	\wigE (y2) -- (10,0);
	
	\draw[blue] (0,2) node {\scriptsize{$y_1$}}; \draw[blue] (11,2) node {\scriptsize{$y_2$}};
	\draw[blue] (1.5,-1.5) node {\scriptsize{$x_1$}}; \draw[blue] (3.5,-1.5) node {\scriptsize{$x_2$}};
	\draw[blue] (7.3,-1.5) node {\scriptsize{$x_{F-1}$}}; \draw[blue] (9.5,-1.5) node {\scriptsize{$x_F$}};

	\draw (5.5,-0.25) node {$\ldots$};
	
	\draw[-] (g3) -- (x1); \draw[blue] (1.5,-0.75) node[left] {\scriptsize{$\sqrt{pq} c_0 $}};
	\draw[-] (g3) -- (x2); \draw[blue] (2.7,-1) node[left] {\scriptsize{$\sqrt{pq} c_2^{-1} $}};
	\draw[-] (g4) -- (x2); \draw[blue] (3.75,-0.5) node[left] {\scriptsize{$\sqrt{pq} c_1 $}};
	\draw[-] (g4) -- (4.75,-0.75);
	\draw[-] (g5) -- (x3); \draw[blue] (7.7,-1) node[left] {\scriptsize{$\sqrt{pq} c_{F \text{-} 1}^{-1} $}};
	\draw[-] (g5) -- (6.25,-0.75);
	\draw[-] (g6) -- (x3); \draw[blue] (8.75,-0.5) node[left] {\scriptsize{$\sqrt{pq} c_{F \text{-} 2} $}};
	\draw[-] (g6) -- (x4); \draw[blue] (9.75,-1) node[left] {\scriptsize{$\sqrt{pq} c_F^{-1} $}};
	\draw[-] (y1) to[out=210,in=90] (-1,0) to[out=-90,in=180] (x1); \draw[blue] (-1,0) node[left] {\scriptsize{$\sqrt{pq} t^{\frac{N+1-2j}{2}} c_1^{-1} $}};
	\draw[-] (y2) to[out=-30,in=90] (12,0) to[out=-90,in=0] (x4); \draw[blue] (12,0) node[right] {\scriptsize{$\sqrt{pq} t^{\frac{N+1-2j}{2}} c_{F-1} $}};
	
	\draw[-] (g3) to[out=60,in=0] (2,0.6) to[out=180,in=120] (g3); \draw[blue] (2,0.7) node {\scriptsize{$\tau$}};
	\draw[-] (g4) to[out=60,in=0] (4,0.6) to[out=180,in=120] (g4); \draw[blue] (4,0.7) node {\scriptsize{$\tau$}};
	\draw[-] (g5) to[out=60,in=0] (7,0.6) to[out=180,in=120] (g5); \draw[blue] (7,0.7) node {\scriptsize{$\tau$}};
	\draw[-] (g6) to[out=60,in=0] (9,0.6) to[out=180,in=120] (g6); \draw[blue] (9,0.7) node {\scriptsize{$\tau$}};
	
\epic}\ee
We can now use the duality  in \eqref{fig:FEtoFlav} to replace the  two asymmetric improved bifundamentals with $N$ chirals  plus flippers. Using this duality and also collecting together all the singlets produced at each step, we obtain the final result:
\be
\resizebox{.95\hsize}{!}{
 \bpic[thick,node distance=3cm,gauge/.style={circle,draw,minimum size=5mm},flavor/.style={rectangle,draw,minimum size=5mm}] 	
	
	\path (5,0) node[gauge](g1) {$\!\!\!2N\!\!\!$} -- (7,0) node[gauge](g2) {$\!\!\!2N\!\!\!$} -- (10,0) node[gauge](g3) {$\!\!\!2N\!\!\!$} 
		-- (12,0) node[gauge](g4) {$\!\!\!2N\!\!\!$} -- (4,-1.5) node[flavor](x1) {$\!2\!$} -- (6,-1.5) node[flavor](x2) {$\!2\!$} 
		-- (11,-1.5) node[flavor](x3) {$\!2\!$} -- (13,-1.5) node[flavor](x4) {$\!2\!$} -- (5,1.5) node[flavor](y1) {$\!2\!$} 
		-- (12,1.5) node[flavor](y2) {$\!2\!$};
	 
	\wigE (g1) -- (g2); \draw[blue] (6,0.3) node {\scriptsize{$\sqrt{pq}\tilde{b}_1$}};
	\wigE (g2) -- (8,0);
	\wigE (g3) -- (9,0);
	\wigE (g3) -- (g4); \draw[blue] (11,0.3) node {\scriptsize{$\sqrt{pq}\tilde{b}_1$}};
	\draw[-] (g1) -- (x1); \draw[blue] (4.75,-0.5) node[left] {\scriptsize{$\sqrt{pq} c_0 $}};
	\draw[-] (g1) -- (x2); \draw[blue] (5.7,-1) node[left] {\scriptsize{$\sqrt{pq} c_2^{-1} $}};
	\draw[-] (g1) -- (y1);
	\draw[-] (g2) -- (x2); \draw[blue] (6.75,-0.5) node[left] {\scriptsize{$\sqrt{pq} c_1 $}};
	\draw[-] (g2) -- (7.7,-1);
	\draw (8.5,-0.5) node {$\cdots$};
	\draw[-] (g3) -- (9.3,-1);
	\draw[-] (g3) -- (x3); \draw[blue] (10.7,-1) node[left] {\scriptsize{$\sqrt{pq} c_{F \text{-} 1}^{-1} $}};
	\draw[-] (g4) -- (x3); \draw[blue] (11.75,-0.5) node[left] {\scriptsize{$\sqrt{pq} c_{F \text{-} 2} $}};
	\draw[-] (g4) -- (x4); \draw[blue] (12.7,-1) node[left] {\scriptsize{$\sqrt{pq} c_F^{-1} $}};
	\draw[-] (g4) -- (y2);
	\draw (4.5,-0.75) node[cross]{};
	\draw (5,0.75) node[cross]{};
	\draw (12.5,-0.75) node[cross]{};
	\draw (12,0.75) node[cross]{};
	\draw[-] (y1) to[out=180,in=90] (3.2,0) to[out=-90,in=150] (x1); \draw[blue] (3.2,0) node[left] {\scriptsize{$\sqrt{pq} t^{\frac{N+1-2j}{2}} c_1^{-1} $}};
	\draw[-] (y2) to[out=0,in=90] (13.8,0) to[out=-90,in=30] (x4); \draw[blue] (13.8,0) node[right] {\scriptsize{$\sqrt{pq} t^{\frac{N+1-2j}{2}} c_{F-1} $}};
	
	\draw[-] (g1) to[out=150,in=90] (4.4,0) to[out=-90,in=-150] (g1); \draw[blue] (4.4,0) node[left] {\scriptsize{$\tau$}};
	\draw[-] (g2) to[out=60,in=0] (7,0.6) to[out=180,in=120] (g2); \draw[blue] (7,0.7) node {\scriptsize{$\tau$}};
	\draw[-] (g3) to[out=60,in=0] (10,0.6) to[out=180,in=120] (g3); \draw[blue] (10,0.7) node {\scriptsize{$\tau$}};
	\draw[-] (g4) to[out=30,in=90] (12.6,0) to[out=-90,in=-30] (g4); \draw[blue] (12.6,0) node[right] {\scriptsize{$\tau$}};
	
	\draw[blue] (5,2) node {\scriptsize{$y_1$}}; \draw[blue] (12,2) node {\scriptsize{$y_2$}};
	\draw[blue] (4.5,-1.5) node {\scriptsize{$x_1$}}; \draw[blue] (6.5,-1.5) node {\scriptsize{$x_2$}};
	\draw[blue] (10.3,-1.5) node {\scriptsize{$x_{F-1}$}}; \draw[blue] (12.5,-1.5) node {\scriptsize{$x_F$}};

\epic}\ee
which is precisely the mirror dual presented in \ref{fig:SQCD_4d_Dual}.

At the level of the index, using the identity \eqref{mtoid} inside \eqref{eq:4d_step2} we obtain:
\begin{align}
	\CI_{SQCD}(\vec{x},\vec{y},\vec{b},c,t) = & \CI_{\text{Step I}} = \CI_{\text{Step II}} = \prod_{j=1}^N \big[ \Ge( \sqrt{pq} t^{\frac{N+1-2j}{2}} c_1^{-1} x_1^\pm y_1^\pm ) 
	\Ge( \sqrt{pq} t^{\frac{N+1-2j}{2}} c_{F-1} x_F^\pm y_2^\pm )  \nn \\
	& \Ge(t^{1-j}c_0^{-2} ) \Ge( t^{j-1} \tilde{b}_1^{-2} ) \Ge( t^{1-j} c_F^2 ) 
	\Ge( t^{j-1} \tilde{b}_F^{-2} )	\big] \nn \\
	& \oint \prod_{a=3}^{F+1} \big( d\vec{z}^{(a)}_N \D_N(\vec{z}^{(a)},t) \big)
	\prod_{a=2}^{F-1} \CI_{GB}^{(N)} (\vec{z}^{(a+2)},\vec{z}^{(a+3)},t, \sqrt{pq} \tilde{b}_a , \sqrt{pq} c_a^{-1} \tilde{b}_a^{-1} ) \nn \\
	& \prod_{j=1}^N \big[ \Ge( \sqrt{pq}t^{\frac{1-N}{2}} \tilde{b}_1 z_j^{(3)\pm} y_1^\pm ) 
	\Ge( \sqrt{pq}t^{\frac{1-N}{2}} \tilde{b}_F z_j^{(F+1)\pm} y_2^\pm ) \big] \nn \\
	& \prod_{j=1}^N \big[ \Ge( \sqrt{pq}c_0 x_l^\pm ) \Ge(\sqrt{pq} c_F^{-1} z_j^{(F+1)\pm} x_{F}^\pm ) \big] = \CI_{\widecheck{SQCD}}(\vec{x},\vec{y},\vec{b},c,t) \,.
\end{align}
which reproduces the  exactly \eqref{eq:4d_ind_id}.

\acknowledgments
We would like to thank Amihay Hanany, Simone Giacomelli and Matteo Sacchi for useful discussions.  RC and SP would also like to thank Chiung Hwang and Fabio Marino  for discussions and collaborations on related topics. SB is partially supported by the INFN “Iniziativa Specifica GAST”.
SB and SP are partially supported by the MUR-PRIN grant No. 2022NY2MXY.


\clearpage
\newpage
\appendix

\section{Notations for $4d$ superconformal index and $3d$ partition function}\label{inpaconv}

\subsection*{$4d$ superconformal index} 
In this section we introduce the notation for the $4d$ $\mathcal{N}=1$ superconformal index \cite{Romelsberger:2005eg,Kinney:2005ej,Dolan:2008qi}.
Let us consider a $4d$ $\CN=1$ gauge theory with gauge group $G$ and matter given by a set of $\CN=1$ chiral multiplets of R-charge $r$, in the representation $R_G$ of $G$ and $R_F$ of some flavor symmetry group $F$. To write the SCI we turn on a set of $(\dim G)$ fugacities $\vec{z}$ for the gauge group $G$ and $(\dim F)$ fugacities $\vec{x}$ for the flavor symmetry $F$. We then write:
\begin{align}
	\CI_G(\vec{x}) = \frac{1}{|W_G|} \oint \prod_{j=1}^{\dim G} \frac{dz_j}{2 \pi i z_j} 
	\frac{[(p;p)_\infty (q;q)_\infty]^{\dim G} }{ \prod_{\vec{\r} \in G} \Ge(\vec{z}^{\, \vec{\r}} \, ) } 
	\prod_{\s_G \in R_G} \prod_{\s_F \in R_F} \Ge \big( (pq)^{r/2} \vec{z}^{\, \vec{s}_G} \vec{x}^{\, \vec{s}_F} \big) \,.
\end{align}
Where $\vec{\r}$ are the roots of $G$, $\vec{\s}_G$ and $\vec{s}_F$ are the weights of the representations $R_G$ and $R_F$. $|W_G|$ is the dimension of the Weyl group of $G$.  We adopted the following notation:
\begin{align}
	\vec{z}^{\, \vec{\r}} = \prod_{j=1}^{\dim G} z_j^{\r_j} \qquad , \qquad
	\vec{z}^{\, \vec{\s}_G} = \prod_{j=1}^{\dim G} z_j^{{\s_G}_j} \qquad , \qquad
	\vec{x}^{\, \vec{\s}_F} = \prod_{j=1}^{\dim F} z_j^{{\s_F}_j} \,.
\end{align}
We define a short notation for the integration measure:
\begin{align}
	d\vec{z}_N = \frac{1}{|W_G|} \prod_{j=1}^{N} \frac{dz_j}{2 \pi i z_j} \,.
\end{align}
In this work we deal mostly with $USp$ gauge groups for which we define the contribution of the vector multiplet as:
\begin{align}
	\D_N(\vec{z}) = \frac{[(p;p)_\infty (q;q)_\infty]^{N} }{ \prod_{j=1}^N \Ge(z_j^{\pm2}) \prod_{j<k}^N \Ge(z_j^\pm z_k^\pm) } \,.
\end{align}
It is convenient to also define the contribution of both a vector and a chiral in the traceless antisymmetric representation:
\begin{align}
	\D_N(\vec{z},t) = \D_N(\vec{z}) \Ge(t)^{N-1} \prod_{j<k}^N \Ge(t z_j^\pm z_k^\pm ) \,.
	\label{adjme}
\end{align}
For a chiral of R-charge $r$ in the bifundamental of $USp(2N)\times USp(2M)$ we have:
\begin{align}
	\CI_{bif} = \prod_{j=1}^N \prod_{a=1}^M \Ge \big( (pq)^{r/2} z_j^\pm x_a^\pm \big) \,.
\end{align}
Suppose that a theory also possesses a $U(1)$ symmetry for which we turn on a fugacity $c$. Along the RG flow this symmetry can mix with the R-symmetry as $r + q_c C$, where $q_C$ is the $U(1)$ charge and $C$ is the mixing coefficient, which is related to the fugacity as:
\begin{align}
	c = (pq)^{C/2} \,.
\end{align}

\subsection*{$3d$ partition function}

In this section we introduce the notation for the $3d$ $\mathcal{N}=2$ $S^3_b$ partition function \cite{Kapustin:2009kz,Jafferis:2010un,Hama:2011ea}.
Let us consider a $3d$ $\CN=2$ gauge theory with gauge group $G$ and matter given by a set of $\CN=2$ chiral multiplets of R-charge $r$, in the representation $R_G$ of $G$ and $R_F$ of some flavor symmetry group $F$. To write the $S^3_b$ partition function we turn of a set of $(\dim G)$ parameters $\vec{Z}$ for the gauge group $G$ and $(\dim F)$ parameters $\vec{X}$ for the flavor symmetry $F$. We then write:
\begin{align}
	Z (Y,k,\vec{X}) = \frac{1}{|W_G|} \int \prod_{j=1}^{\dim G} d Z_j  & Z_{\text{cl}}(Y,k) 
	\frac{1}{\prod_{\vec{\r} \in G} s_b \big( \frac{iQ}{2} - \vec{\r}(\vec{Z}) \big) } \nn \\
	& \prod_{\vec{\s}_G \in R_G} \prod_{\vec{\s}_F \in R_F} s_b \big( \frac{iQ}{2}(1-r) - \vec{\s}_G(\vec{Z}) - \vec{\s}_F(\vec{X}) \big) \,.
\end{align}
Where $\vec{\r}$ are the roots of $G$, $\vec{\s}_G$ and $\vec{s}_F$ are the weights of the representations $R_G$ and $R_F$. $|W_G|$ is the dimension of the Weyl group of $G$. We also adopted the following notation:
\begin{align}
	\vec{\r}(\vec{Z}) = \sum_{j=1}^{\dim G} \r_j Z_j \qquad , \qquad
	\vec{\s}_G(\vec{Z}) = \sum_{j=1}^{\dim G} {\s_G}_j Z_j \qquad , \qquad
	\vec{\s}_F(\vec{X}) = \sum_{j=1}^{\dim F} {\s_F}_j X_j \,.
\end{align}
We also have $Z_{\text{cl}}(Y,k)$ that encodes the contribution of the FI parameter $Y$ associated to a topological symmetry and that of CS term of level $k$:
\begin{align}
	Z_{\text{cl}}(Y,k) = \exp \big[ 2\pi i Y \sum_{j=1}^{\dim G} Z_j + \pi i k \sum_{j=1}^{\dim G} Z_j^2 \big] \,.
\end{align}
We define a short notation for the integration measure:
\begin{align}
	d\vec{Z}_N = \frac{1}{|W_G|} \prod_{j=1}^{N} dZ_j \,.
\end{align}
In this work we deal mostly with $U$ gauge groups. In this case we define the contribution of the vector multiplet as:
\begin{align}
	\D_N(\vec{Z}) = \frac{1}{\prod_{j<k}^N s_b \big( \frac{iQ}{2} \pm (Z_j - Z_k) \big) } \,.
\end{align}
It is convenient to also define the contribution of both a vector and a chiral in the traceless adjoint representation:
\begin{align}
	\D_N(\vec{Z},\tau) = \D_N(\vec{Z}) \,\, s_b( \frac{iQ}{2} - \tau )^{N-1} \prod_{j<k}^N s_b \big( \frac{iQ}{2} - \tau \pm ( Z_j - Z_k ) \big) \,.
\end{align}
For a chiral of R-charge $r$ in the bifundamental $N\times \bar{M}$ of a $U(N)\times U(M)$ gauge group we have:
\begin{align}
	Z_{bif} = \prod_{j=1}^N \prod_{a=1}^M s_b \big( \frac{iQ}{2}(1-r) + Z_j - X_a \big) \,.
\end{align}

\section{Improved bifundamentals and duality walls}\label{app:gen_blocks}

In this appendix we briefly review the three important theories used throughout the work, that are the $FE[USp(2N)]$, $FM[U(N)]$ and $FT[U(N)]$ theories. \\
The $FE[USp(2N)]$ theory was first introduced in \cite{Pasquetti:2019hxf}. Its properties and deformations were later studied in \cite{Hwang:2020wpd,Bottini:2021vms,Comi:2022aqo,BCP1}. \\
The $FM[U(N)]$ theory was first introduced in \cite{Pasquetti:2019tix}. Its properties and deformations were later studied in \cite{Bottini:2021vms,BCP1}. \\
The $FT[U(N)]$ theory is simply the $T[SU(N)]$ theory of \cite{Gaiotto:2008ak}, with the addition of an extra singlet flipping the moment map of the Higgs branch.

\subsection{$4d$ improved bifundamental: the $FE[USp(N)]$ theory}\label{app:FE}
\begin{figure}
\centering
\begin{tikzpicture}[thick,node distance=3cm,gauge/.style={circle,draw,minimum size=5mm},flavor/.style={rectangle,draw,minimum size=5mm}]
	
	\path (0,0) node[gauge](g1) {$\!\!2\!\!$} -- (2,0) node[gauge](g2) {$\!\!4\!\!$} -- (5,0) node[gauge](g3) {\!\!\tiny{$2N$-$2$}\!\!} 
		-- (7,0) node[flavor,red](x) {$\!2N\!$} -- (-1,-1.5) node[flavor,blue](y1) {$\!2\!$} -- (1,-1.5) node[flavor,blue](y2) {$\!2\!$} 
		-- (6,-1.5) node[flavor,blue](y3) {$\!2\!$};
		
	\draw[-] (g1) -- (g2); \draw (1,0) node[cross]{}; \draw (1,0.3) node {$b_1$};
	\draw[-] (g2) -- (3,0);
	\draw[-] (4,0) -- (g3);
	\draw[-] (g3) -- (x); \draw (6,0) node[cross]{}; \draw (6,0.3) node {$b_{N-1}$};
	\draw[-] (g1) -- (y1); \draw (-0.5,-0.75) node[cross] {}; \draw (-0.5,-0.75) node[left] {$d_1$};
	\draw[-] (g1) -- (y2); \draw (0.7,-1) node[left] {$v_1$};
	\draw[-] (g2) -- (y2); \draw (1.5,-0.75) node[cross] {}; \draw (1.5,-0.75) node[left] {$d_2$};
	\draw[-] (g2) -- (2.5,-0.75);
	\draw[-] (3.5,-0.5) node {$\ldots$};
	\draw[-] (g3) -- (4.5,-0.75);
	\draw[-] (g3) -- (y3); \draw (5.7,-1) node[left] {$v_{N-1}$};
	\draw[-] (x) -- (y3); \draw (6.5,-0.75) node[cross] {}; \draw (6.5,-0.75) node[left] {$d_N$};
	\draw[-] (g2) to[out=60,in=0] (2,0.6) to[out=180,in=120] (g2); \draw (2,0.8) node {$a_2$};
	\draw[-] (g3) to[out=60,in=0] (5,0.7) to[out=180,in=120] (g3); \draw (5,0.85) node {$a_{N-1}$};
	\draw[-] (x) to[out=60,in=0] (7,0.6) to[out=180,in=120] (x); \draw (7,0.8) node {$a_N$};
	
	\draw (-1,-2.5) node[right] {$\CW = \sum_{j=1}^{N-1} v_j b_j d_{j+1} + \sum_{j=2}^N a_j( b_{j-1}^2 - b^2_j ) + \sum_{j=1}^{N-1} Flip[b_j^2] + \sum_{j=1}^N Flip[d_j^2] $};
	
	\draw (11,-0.5) node {\begin{tabular}{c|c}
							$b_i$ & $\tau/2$ \\
							$a_i$ & $2-\tau$ \\
							$v_i$ & $2 + \frac{N-i-2}{2}\tau - C$ \\
							$d_i$ & $\frac{i-N}{2}\tau + C$ 
						\end{tabular}};
	
\end{tikzpicture}
\caption{Quiver representation of the UV completion of the $FE[USp(2N)]$ SCFT. Each node, square or round, labeled with a number $2n$, represents respectively a gauge or flavor $USp(2n)$ group. Each line is a field in the fundamental representation of the nodes to whom is attached, except for arch lines that are fields in the traceless antisymmetric representation. Crosses denote the presence of flipping fields. Lastly, all the superpotential terms are written in short by omitting the traces which also include the antisymmetric  of the $USp$ group. Also, on the right, we give the table with the R-charge of all the fields in the theory.
The R-charge is given as a trial value mixed with the other two abelian symmetries of the theory, $U(1)_\tau$ and $U(1)_C$, whose mixing values are given by the two real variables $\tau$ and $C$.}
\label{fig:FE_quiver}
\end{figure}
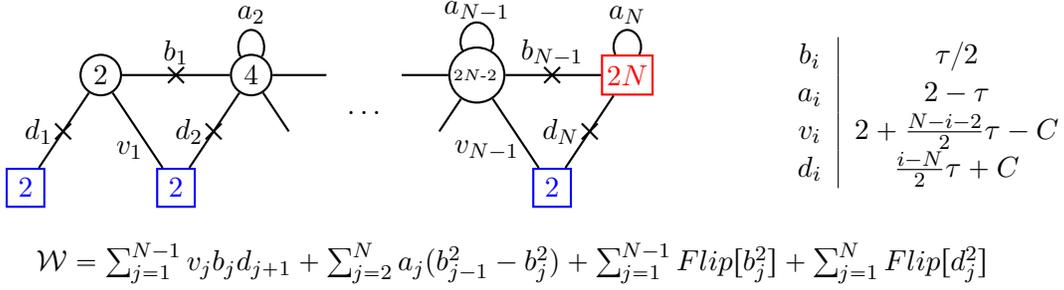
The $FE[USp(2N)]$ theory is a $4d$ $\cN=1$ SCFT denoted by the following symbol:
\be\label{fig:FE_symbol}
\begin{tikzpicture}[thick,node distance=3cm,gauge/.style={circle,draw,minimum size=5mm},flavor/.style={rectangle,draw,minimum size=5mm}]
	
	\path (-1.5,0) node[flavor,blue](y1) {$\!2N\!$} -- (1.5,0) node[flavor,red](y3) {$\!2N\!$};
    \wigE (y1) -- (y3); 
	
\end{tikzpicture}\ee
This theory admits a UV Lagrangian description as a quiver of $N-1$ symplectic gauge nodes as given in figure \ref{fig:FE_quiver}.
The $FE[USp(2N)]$ theory has the UV global symmetry group: 
\begin{align}
	{\color{red}USp(2N)} \times {\color{blue}USp(2)}^N \times U(1)_\tau \times U(1)_C\,,
\end{align}
in addition to the $U(1)_R$ symmetry. At the IR fixed point, the SCFT is characterized by the enhanced global symmetry:
\begin{align}
	{\color{red}USp(2N)} \times {\color{blue}USp(2N)} \times U(1)_\tau \times U(1)_C\,.
\end{align}
The gauge invariant operators indeed reorganize into representations of the IR symmetry group. The list of the chiral ring generators of the $FE[USp(2N)]$ SCFT, along with their charges and representations, is given in table \ref{tab:FE_operators}.
\begin{table}[h!]
\centering
\renewcommand{\arraystretch}{1.15}
\begin{tabular}{|c|cc|c|}\hline
{} & ${\color{red}USp(2N)}$ & ${\color{blue}USp(2N)}$ & R charge\\ \hline
${\color{red}\mathsf{A}}$ & ${\bf N(2N-1)-1}$ & $\bf 1$  & $2-\tau$ \\
${\color{blue}\mathsf{A}}$ & $\bf1$ & ${\bf N(2N-1)-1}$  & $2-\tau$ \\
$\Pi$ & $\bf N$ & $\bf N$  & $C$ \\
$\mathsf{B}_{n, m}$ & $\bf1$ & $\bf1$  & $2n-2C+(m-n)\tau$ \\
 \hline
\end{tabular}
\caption{List of all the gauge invariant operator that compose the spectrum of the $FE[USp(2N)]$ SCFT. The R-charge is given as a trial value mixed with the other two abelian symmetries of the theory, $U(1)_\tau$ and $U(1)_C$, whose mixing values are given by the two real variables $\tau$ and $C$.
The $\mathsf{B}_{n,m}$ matrix is a collection of ${\color{red}USp(2N)}\times{\color{blue}USp(2N)}$ singlets for $n=1,\ldots,N$ and $m=1,\ldots,N+1-n$.}
\label{tab:FE_operators}
\end{table}
The SCI of the $FE[USp(2N)]$ theory can be defined recursively as:\footnote{Notice that in this work we take all the antisymmetric fields to be traceless instead of tracefull, differently from the original definition.} 
\begin{align}\label{eq:FE_SCI}
	\CI_{FE}^{(N)} (\vec{x},\vec{y},t,c) = & \Ge(pq c^{-2}) \prod_{a=1}^{N} \Ge(c y_N^\pm x_a^\pm) 
	\Ge(pqt^{-1})^{N-1} \prod_{a<b}^N \Ge(pq t^{-1} x_a^\pm x_b^\pm) \nn \\
	& \times\oint d\vec{z}_{N-1} \D_{N-1}(\vec{z}) \Ge(pqt^{-1})
	\prod_{j=1}^{N-1} \prod_{a=1}^{N} \Ge(t^{\frac{1}{2}}z_j^\pm x_a^\pm) \nonumber \\
	& \times \prod_{j=1}^{N-1} \Ge(pq t^{-\frac{1}{2}}c^{-1} y_N^\pm z_j^\pm) \CI_{FE}^{(N-1)}(\vec{z},\{ y_1,\ldots,y_{N-1}\},t,t^{-\frac{1}{2}}c) \,,
\end{align}
with the base for the recursion:
\begin{align}
	\CI_{FE}^{(1)} (x,y,t,c) = \Ge(pq c^{-2}) \Ge(c x^\pm y^\pm)  \,.
\end{align}
The fugacities for the $U(1)$ symmetries are related to the R-charge mixing as:
\begin{align}
	c = (pq)^{C/2} , \qquad t = (pq)^{\tau/2} \,,
\end{align}
Also, the vectors $\vec{x}$ and $\vec{y}$ are the fugacities for the manifest and emergent $USp(2N)$ symmetries respectively, and $\vec{z}$ is the fugacity for the gauge group $USp(2N-2)$. The notation for the superconformal index can be found in appendix \ref{inpaconv}.

\paragraph{Self-mirror property}
The $FE[USp(2N)]$ theory enjoys an exact self-duality that acts by exchanging the manifest and emergent $USp(2N)$ symmetries. As a SCI identity we have:
\begin{align}\label{eq:FE_mirror}
	\CI_{FE}^{(N)}(\vec{x},\vec{y},t,c) = \CI_{FE}^{(N)}(\vec{y},\vec{x},t,c) \,.
\end{align}
This property can be thought as the freedom of choosing which of the two $USp(2N)$ symmetries is the manifest one when we consider the UV completion of the $FE$ theory.
The self-mirror property can be demonstrated inductively using the mirror dualization algorithm.

\paragraph{Interesting deformations}
In this section we review  two interesting types of deformation that can be turn on in an $FE[USp(2N)]$ theory: those that are $USp(2N)^2$ preserving and those that are not. 
All the details can be found in \cite{Comi:2022aqo}.\\
Let us start from the former, in this work we will interested just in a small subset of them. The first possibility that we consider is the linear superpotential deformation $\d \CW = \mathsf{B}_{1,1}$, which breaks completely the $U(1)_C$ symmetry while it preserves $U(1)_\tau$. Under this deformation the $FE[USp(2N)]$ theory behave as an identity operator as:
\begin{align}\label{eq:FE_c=1}
	\mathcal{I}_{FE}^{(N)} (\vec{x}, \vec{y}, t, c=1) = {}_{\vec{x}}\mathbb{I}_{\vec{y}}(t) \,,
\end{align}
where the identity operator is defined as:
\begin{align}\label{eq:4d_identity}
	{}_{\vec{x}}\mathbb{I}_{\vec{y}} (t)
	= \frac{\prod_{j=1}^{N} 2\pi i y_i}{\Delta_N(\vec{y},t)} \sum_{\sigma \in S_N} \prod_{j=1}^{N} \delta (x_j - y_{\sigma(j)}^\pm) \,.
\end{align}
The second possibility is given by the linear deformation $\d \CW = \mathsf{B}_{1,2}$, which breaks the $U(1)_C\times U(1)_\tau$ symmetry down to a $U(1)$ diagonal subgroup defined by the constraint $C= \tau/2$, or analogously $c=t^{1/2}$ in terms of the fugacities. This deformation has the effect of deforming the  $FE[USp(2N)]$ theory to a bifundamental coupled to antisymmetric chirals as:
\begin{align}
	\CI_{FE}^{(N)} (\vec{x},\vec{y},t,c = t^{\frac{1}{2}}) = & \Ge(pqt^{-1})^{2N-1} \prod_{j<k}^N \big(\Ge( pqt^{-1} x_j^\pm x_k^\pm) \Ge( pqt^{-1} y_j^\pm y_k^\pm) \big) \nn \\
	& \prod_{j,k=1}^N \Ge( t^{\frac{1}{2}} x_j^\pm y_k^\pm ) \,.
\end{align}
Graphically this deformation can be depicted as:
\be\label{fig:FE_c=t/2}
 \bpic[thick,node distance=3cm,gauge/.style={circle,draw,minimum size=5mm},flavor/.style={rectangle,draw,minimum size=5mm}]  
	\path (-4,0) node[flavor,blue](x) {$\!2N\!$} -- (-1,0) node[flavor,red](y) {$\!2N\!$};
	\wigE (x) -- (y); \draw (-2.5,0.4) node {$\Pi$};
	\draw (-2.5,-1) node {$\CW = \mathsf{B}_{1,2}$};
	
	\draw (0.5,0) node {$\Longleftrightarrow$};
	
	\path (2,0) node[flavor,blue](x1) {$\!2N\!$} -- (5,0) node[flavor,red](y1) {$\!2N\!$};
	\draw[-] (x1) -- (y1); 
	\draw[-] (x1) to[out=120,in=180] (2,0.6) to[out=0,in=60] (x1);
	\draw[-] (y1) to[out=120,in=180] (5,0.6) to[out=0,in=60] (y1);
	\draw (3.5,0) node[cross] {};
	\draw (3.5,0.4) node {$b$}; \draw (2,0.7) node[right] {$\color{blue}a$}; \draw (5,0.7) node[right] {$\color{red}a$};
	\draw (3.5,-1) node {$\cW = ({\color{blue}a} + {\color{red}a})b^2 + Flip[b^2]$};
	
	\path (8,0) node {\begin{tabular}{c|c}
							$\Pi$ & $\tau/2$ \\
							${\color{blue}a}$ & $2-\tau$ \\
							${\color{red}a}$ & $2-\tau$ \\
							$b$ & $\tau/2$ 
						\end{tabular}};	
	
\epic\ee
One can also {\it iron} a $FE[USp(2N)]$ theory to a standard bifundamental by using the deformation $\d \CW = \mathsf{B}_{2,1}$. This has the effect of breaking $U(1)_C \times U(1)_\tau$ down to a $U(1)$ subgroup defined by the constraint $C = 1 - \tau/2$, or analogously $c = \sqrt{pq/t}$ in terms of the fugacities. We have the following property:
\begin{align}
	\CI_{FE}^{(N)}(\vec{x},\vec{y},t,c=\sqrt{pq/t}) = \Ge(t) \prod_{j,k=1}^N \Ge( \sqrt{pq/t} x_j^\pm y_k^\pm ) \,.
\end{align}
Graphically we have:
\be\label{fig:FE_c=1-t/2}
 \bpic[thick,node distance=3cm,gauge/.style={circle,draw,minimum size=5mm},flavor/.style={rectangle,draw,minimum size=5mm}]  
	\path (-4,0) node[flavor,blue](x) {$\!2N\!$} -- (-1,0) node[flavor,red](y) {$\!2N\!$};
	\wigE (x) -- (y); \draw (-2.5,0.4) node {$\Pi$};
	\draw (-2.5,-1) node {$\CW = \mathsf{B}_{2,1}$};
	
	\draw (0.5,0) node {$\Longleftrightarrow$};
	
	\path (2,0) node[flavor,blue](x1) {$\!2N\!$} -- (5,0) node[flavor,red](y1) {$\!2N\!$};
	\draw[-] (x1) -- (y1); \draw (3.5,0) node[cross] {}; \draw (3.5,0.4) node {$b$};
	\draw (3.5,-1) node {$\cW = Flip[b^2]$};
	
	\path (8,0) node {\begin{tabular}{c|c}
							$\Pi$ & $1 - \tau/2$ \\
							$b$ & $1 - \tau/2$ 
						\end{tabular}};	
	
\epic\ee
The second category of deformations is given by $USp(2N)$ breaking superpotential terms. A class of such deformations consist in giving VEVs (or masses) to the antisymmetric operators in the $FE$ theory in the form of Jordan matrices. The  VEVs  are specified uniquely by a pair of partitions $(\rho,\sigma)$. This deformations were studied in depth in \cite{Hwang:2020wpd}, where it is described how to properly follow the RG flow triggered by those deformations. Throughout this paper we will be only interested in the particular cases where one of the two $USp(2N)$ symmetries is broken to $USp(2M) \times USp(2)$, with $M<N$. At the level of the SCI this deformation is implemented as a specialization of the vector of fugacities of the $USp(2N)$ symmetry, let's call it $\vec{x}$, in terms of the fugacities $\vec{y}$ and $v$ of the $USp(2M)$ and $USp(2)$ symmetries respectively:
\begin{align}\label{eq:FE_specialisation}
	& x_i = t^{\frac{N-M+1-2i}{2}}v \qquad \text{for} \quad i=1,\ldots,N-M \,, \nn \\
	& x_i = y_{i-N+M} \qquad \text{for} \quad i=N-M+1,\ldots,N \,.
\end{align}
When such deformation is implemented in an FE theory we depict it as an ``asymmetric" zig-zag line as:
\be\label{fig:FE_asymm_symbol} 
 \begin{tikzpicture}[thick,node distance=3cm,gauge/.style={circle,draw,minimum size=5mm},flavor/.style={rectangle,draw,minimum size=5mm}]
 
	\path (-2,0) node[flavor,blue](x) {$\!2N\!$} -- (1,0) node[flavor,red](y) {$\!2M\!$} -- (1,1) node[flavor](v) {$\!2\!$};
  
	\wigE (x) -- (y);   
	\wigE (v) -- (-0.5,0);
	
\end{tikzpicture}\ee
In the maximal case, $M=0$, therefore breaking $USp(2N)$ completely down to $USp(2)$,
the $FE[USp(2N)]$ theory is dual to $2N \times 2$ fundamental chiral with the addition of extra singlets:
\be\label{fig:FEtoFlav}
 \bpic[thick,node distance=3cm,gauge/.style={circle,draw,minimum size=5mm},flavor/.style={rectangle,draw,minimum size=5mm}]  
	\path (-5,0) node[flavor,blue](x) {$\!2N\!$} -- (-2,0) node[flavor,red](y) {$\!0\!$} -- (-2,1) node[flavor](v) {$\!2\!$};
	\wigE (x) -- (y); \draw (-4,0.4) node {$\Pi$};
	\wigE (v) -- (-3.5,0);
	\draw[-] (x) to[out=120,in=180] (-5,0.6) to[out=0,in=60] (x); \draw (-5,0.7) node[right]{$\color{blue}a$};
	
	\draw (-3.5,-1) node {$\CW = {\color{blue}a}\mathsf{A}_L$};
	
	\draw (-0.5,0) node {$\Longleftrightarrow$};
	
	\path (1,0) node[flavor,blue](x1) {$\!2N\!$} -- (4,0) node[flavor](y1) {$\!2\!$};
	\draw[-] (x1) -- (y1); \draw (2.5,0) node[cross]{}; \draw (2.5,0.4) node {$b$};
	\draw[-] (x1) to[out=120,in=180] (1,0.6) to[out=0,in=60] (x1); \draw (1,0.6) node[cross] {}; \draw (1.1,0.7)[right] node{${\color{blue}a}$};
	
	\path (7,0) node {\begin{tabular}{c|c}
							$\color{blue}a$ & $\tau$ \\
							$\Pi$ & $C$ \\
							$b$ & $\frac{1-N}{2}\tau + C$ 
						\end{tabular}};	
						
	\draw (0.5,-1) node[right] {$\cW = \sum_{j=2}^N Flip[Tr({\color{blue}a}^{j})] + $};
	\draw (1,-1.6) node[right] {$+ \sum_{j=0}^{N-1} Flip[b^2{{\color{blue}a}}^j]$};
							
\epic\ee
As a superconformal index identity we write:
\begin{align}
	& \CI_{FE}^{(N)} ( \vec{x}, \{ t^{\frac{N-1}{2}}y, \ldots, t^{\frac{1-N}{2}}y \}, t, c ) = \nn \\
	& = \prod_{j=1}^N \Ge( t^{\frac{1-N}{2}}\tau c ) \prod_{j=2}^N \Ge( pq t^{-j} ) \prod_{j=0}^{N-1} \Ge( pq t^{N-1-j} c^{-2} ) \,.
\label{mtoid}
\end{align}

\paragraph{Fusion to identity}
An interesting property of the $FE[USp(2N)]$ theory is that gluing together two of them, meaning that we gauge a diagonal subgroup of a $USp(2N)$ symmetry of each theory, triggers an RG flow that leads to a singular delta function theory. This means that the there is a deformed moduli space over which the global $USp(2N)^2$ symmetry is spontaneously broken to its diagonal subgroup. At the level of the SCI this can be written as:
\begin{align}\label{FEdelta}
	\oint d \vec{z}_N \Delta_N(\vec{z},t) \mathcal{I}_{FE}^{(N)} (\vec{x}, \vec{z}, t,c) \mathcal{I}_{FE}^{(N)} (\vec{z}, \vec{y}, t,c^{-1}) = & {}_{\vec{x}}\mathbb{I}_{\vec{y}} (t) \,.
\end{align}
Where the identity operator is defined as in \eqref{eq:4d_identity}.
This property can be demonstrated by iterative applications of the IP duality. Graphically the $\mathbb{I}$-wall is depicted as:
\be\label{FEdelta}
 \bpic[thick,node distance=3cm,gauge/.style={circle,draw,minimum size=5mm},flavor/.style={rectangle,draw,minimum size=5mm}]  
 
	\path (-3,0) node[flavor,blue](x) {$\!2N\!$} -- (-1,0) node[gauge,black](y) {$\!\!\!2N\!\!\!$} -- (1,0) node[flavor,red](z) {$\!2N\!$} 
		--  (2.5,0) node{$\Longleftrightarrow$};
	
	\wigE (x) -- (y); \draw (-2,0.4) node {$\Pi_L$};
	\wigE (z) -- (y); \draw (0,0.4) node {$\Pi_R$};
	\draw (4,0) node {$\mathbb{I}$-wall};
	\draw[-] (y) to[out=60,in=0] (-1,0.6) to[out=180,in=120] (y); \draw (-1,0.7) node[right]{$a$};
	
	\draw (-1,-1) node {$\cW = \CW_{\text{gluing}}$};
	
	\draw (6.5,0) node {\begin{tabular}{c|c}
							$\Pi_L$ & $C$ \\
							$\Pi_R$ & $-C$ \\
							$a$ & $\tau$  
						\end{tabular}};
	
\epic\ee
On the l.h.s. the superpotential $\CW_{\text{gluing}}$ contains the coupling $a(\mathsf{A}_L + \mathsf{A}_R)$, between the antisymmetric chiral $a$ and the antisymmetric operators $\mathsf{A}_L$ and $\mathsf{A}_R$ inside the left and right $FE[USp(2N)]$ theories.
Notice that assigning the R-charge of $A$ to be $\tau$ fixes the R-charge of $\mathsf{A}_L$ and $\mathsf{A}_R$ to be $2-\tau$, as it is in the ``standard" $FE[USp(2N)]$ theory defined in \ref{fig:FE_quiver}. \\
We can also consider the situation where one of the two glued $FE[USp(2N)]$ theories is asymmetric:
\be
 \bpic[thick,node distance=3cm,gauge/.style={circle,draw,minimum size=5mm},flavor/.style={rectangle,draw,minimum size=5mm}]  
 
	\path (-3,0) node[flavor,blue](x) {$\!2M\!$} -- (-1,0) node[gauge,black](y) {$\!\!\!2N\!\!\!$} -- (1,0) node[flavor,red](z) {$\!2N\!$} -- (-3,1) node[flavor](v) {$\!2\!$}
		--  (2.5,0) node{$\Longleftrightarrow$};
	
	\wigE (x) -- (y); \draw (-2,-0.4) node {$\Pi_L$};
	\wigE (v) -- (-2,0);
	\wigE (z) -- (y); \draw (0,-0.4) node {$\Pi_R$};
	\draw (4,0) node {$\mathbb{I}$-wall};
	\draw[-] (y) to[out=60,in=0] (-1,0.6) to[out=180,in=120] (y); \draw (-1,0.7) node[right]{$a$};
	
	\draw (-1,-1) node {$\cW = \CW_{\text{gluing}}$};
	
	\draw (6.5,0) node {\begin{tabular}{c|c}
							$\Pi_L$ & $C$ \\
							$\Pi_R$ & $-C$ \\
							$a$ & $\tau$  
						\end{tabular}};
	
\epic
\label{FEdeltaA}
\ee
In this case we produce an asymmetric $\mathbb{I}$-wall which identifies the Cartans of one $USp(2N)$ with the Cartans of $USp(2M)\times USp(2)$
in the specialization \eqref{eq:FE_specialisation}.
At the level of  SCI we have:
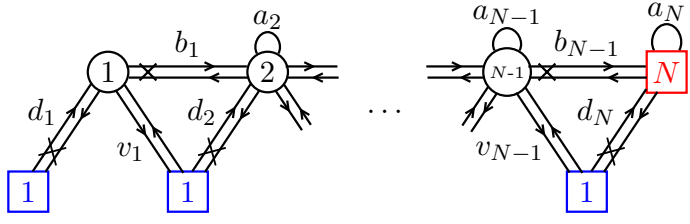
\begin{align}
	\oint d \vec{z}_N \Delta_N(\vec{z},t) & \mathcal{I}_{FE}^{(N)} (\{ \vec{x}, t^{\frac{N-1}{2}}v, \ldots, t^{\frac{1-N}{2}}v \}, \vec{z}, t,c) \mathcal{I}_{FE}^{(N)} (\vec{z}, \vec{y}, t,c^{-1}) = \nn \\
	&= \frac{\prod_{j=1}^{N} 2\pi i y_i}{\Delta_N(\vec{y},t)} \sum_{\sigma \in S_N} \prod_{j=1}^{N} \delta (x_j - y_{\sigma(j)}^\pm) \big\rvert_{x_{M+j}=t^{\frac{N-M+1-2j}{2}v}} \,.
\end{align}

\subsection{$3d$ improved bifundamental: the $FM[U(N)]$ theory}\label{app:FM}
\begin{figure}
\centering
\resizebox{.65\hsize}{!}{
\begin{tikzpicture}[thick,node distance=3cm,gauge/.style={circle,draw,minimum size=5mm},flavor/.style={rectangle,draw,minimum size=5mm}] 
 
	\path (0,0) node[gauge] (g1) {$\!\!\!1\!\!\!$} -- (2,0) node[gauge] (g2)	{$\!\!\!2\!\!\!$} 
		-- (5,0) node[gauge] (g3) {\!\!\tiny{$N$-$1$}\!\!\!} -- (7,0) node[flavor,red] (g4) {$\!N\!$}
		-- (-1,-1.5) node[flavor,blue] (x1) {$\!1\!$} -- (1,-1.5) node[flavor,blue] (x2) {$\!1\!$} -- (6,-1.5) node[flavor,blue] (x3) {$\!1\!$};
		
	\draw[-, shorten >= 6, shorten <= 8, shift={(-0.05,0.07)}, middx arrowsm] (0,0) -- (2,0);
	\draw[-, shorten >= 6, shorten <= 8, shift={(0.05,-0.07)}, midsx arrowsm] (2,0) -- (0,0);
	\draw (0.5,0) node[cross] {}; \draw (1,0.35) node {$b_1$};
	
	\draw[-, shorten >= 5, shorten <= 5, shift={(0.05,0.07)}, mid arrowsm] (2,0) -- (3,0);
	\draw[-, shorten >= 4, shorten <= 6, shift={(0.1,-0.07)}, mid arrowsm] (3,0) -- (2,0);
	
	\draw (3.5,-0.5) node {$\cdots$};	
	
	\draw[-, shorten >= 5.5, shorten <= 2.5, shift={(-0.1,0.07)}, mid arrowsm] (4,0) -- (5,0);
	\draw[-, shorten >= 1.5, shorten <= 7, shift={(-0.05,-0.07)}, mid arrowsm] (5,0) -- (4,0);
	
	\draw[-, shorten >= 6, shorten <= 9, shift={(-0.05,0.07)}, middx arrowsm] (5,0) -- (7,0);
	\draw[-, shorten >= 6.5, shorten <= 8.5, shift={(0.05,-0.07)}, midsx arrowsm] (7,0) -- (5,0);
	\draw (5.5,0) node[cross] {}; \draw (6,0.35) node {$b_{N-1}$};
	
	\draw[-] (g2) to[out=60,in=0] (2,0.5) to[out=180,in=120] (g2); \draw (2,0.65) node {$a_2$};
	\draw[-] (g3) to[out=60,in=0] (5,0.55) to[out=180,in=120] (g3); \draw (5,0.7) node {$a_{N-1}$};
	\draw[-] (g4) to[out=60,in=0] (7,0.6) to[out=180,in=120] (g4); \draw (7,0.75) node {$a_N$};
	
	\draw[-, shorten >= 5.5, shorten <= 8, shift={(-0.1,0.02)}, middx arrowsm] (-1,-1.5) -- (0,0);
	\draw[-, shorten >= 8.5, shorten <= 8, shift={(0.05,0)}, midsx arrowsm] (0,0) -- (-1,-1.5);
	\draw (-0.7,-1) node {\rotatebox{-30}{\LARGE{$\times$}}};
	\draw (-0.5,-0.5) node[left] {$d_1$};
	
	\draw[-, shorten >= 7.5, shorten <= 8.5, shift={(-0.07,0.02)}, mid arrowsm] (0,0) -- (1,-1.5);
	\draw[-, shorten >= 5.5, shorten <= 8, shift={(0.1,0)}, mid arrowsm] (1,-1.5) -- (0,0);
	\draw (0.6,-1) node[left] {$v_1$};
	
	\draw[-, shorten >= 5.5, shorten <= 8, shift={(-0.1,0.02)}, middx arrowsm] (1,-1.5) -- (2,0);
	\draw[-, shorten >= 8.5, shorten <= 8, shift={(0.05,0)}, midsx arrowsm] (2,0) -- (1,-1.5);
	\draw (1.3,-1) node {\rotatebox{-30}{\LARGE{$\times$}}};
	\draw (1.5,-0.5) node[left] {$d_2$};
	
	\draw[-, shorten >= 0, shorten <= 8.5, shift={(-0.07,0.02)}, middx arrowsm] (2,0) -- (2.5,-0.75);
	\draw[-, shorten >= 5.5, shorten <= 3, shift={(0.1,0)}, midsx arrowsm] (2.5,-0.75) -- (2,0);
	
	\draw[-, shorten >= 7, shorten <= 2, shift={(-0.1,0.02)}, midsx arrowsm] (4.5,-0.75) -- (5,0);
	\draw[-, shorten >= 0, shorten <= 9, shift={(0.05,0)}, middx arrowsm] (5,0) -- (4.5,-0.75);
	
	\draw[-, shorten >= 7.5, shorten <= 9.5, shift={(-0.07,0.02)}, mid arrowsm] (5,0) -- (6,-1.5);
	\draw[-, shorten >= 6.5, shorten <= 8, shift={(0.1,0)}, mid arrowsm] (6,-1.5) -- (5,0);
	\draw (5.6,-1) node[left] {$v_{N-1}$};
	
	\draw[-, shorten >= 8, shorten <= 8, shift={(-0.1,0.02)}, middx arrowsm] (6,-1.5) -- (7,0);
	\draw[-, shorten >= 8.5, shorten <= 8.5, shift={(0.05,0)}, midsx arrowsm] (7,0) -- (6,-1.5);
	\draw (6.3,-1) node {\rotatebox{-30}{\LARGE{$\times$}}};
	\draw (6.5,-0.5) node[left] {$d_N$};
	
	\draw (-1,-2.5) node[right] {$\CW = \sum_{j=1}^{N-1} \big[ b_j (a_j + a_{j+1}) \tilde{b}_j + Flip[b_j \tilde{b}_j] \big] +$};
	\draw (-0.5,-3.2) node[right] {$+ \sum_{j=1}^N (\mathfrak{M}^+_j + \mathfrak{M}^-_j) + \sum_{j=1}^{N-1} ( \tilde{v}_j b_j \tilde{d}_{j+1} + v_j \tilde{b}_j d_{j+1} ) $};
	
	
\end{tikzpicture}}
\caption{
Quiver representation of the UV completion of the $FM[U(N)]$ SCFT. Each node, square or round, labeled with a number $n$, represents a gauge or flavor $U(n)$ group, respectively. Each line is a $\CN=2$ chiral in the fundamental/antifundamental representation of the nodes to whom is attached, depending whether the arrow is outgoing or ingoing. Arches denote fields in the traceless adjoint representation. Crosses denote flipping fields. In the superpotential we also have monopoles, we denote by $\mathfrak{M}_i^\pm$ the monopole with charge $\pm1$ under the topological symmetry associated to the $i$-th gauge node.
}
\label{fig:FM_quiver}
\end{figure}

The $FM[U(N)]$ theory is a $3d$ $\CN=2$ SCFT denoted by the following symbol:
\be\label{fig:FM_symbol}
\begin{tikzpicture}[thick,node distance=3cm,gauge/.style={circle,draw,minimum size=5mm},flavor/.style={rectangle,draw,minimum size=5mm}]
	
	\path (-1.5,0) node[flavor,blue](y1) {$\!N\!$} -- (1.5,0) node[flavor,red](y3) {$\!N\!$};
    \wigM (y1) -- (y3); 
	
\end{tikzpicture}\ee
This theory admits a UV Lagrangian description as a quiver of $N-1$ unitary gauge nodes given in figure \ref{fig:FM_quiver}, see also table \ref{tab:FM_fields} for the charges and representation of all the fields.
\begin{table}
\centering
\renewcommand{\arraystretch}{1.15}
\begin{tabular}{|c|c|c|c|c|c|}
	\hline
	 & $U(1)_{R_0}$ & $U(1)_\tau$ & $U(1)_\D$ & $\color{blue}U(1)_{y_j}$ & $\color{red}U(N)$ \\
	\hline
	$b_i, \tilde{b}_i$ &  0 & $1/2$ & 0 & 0 & $\bf 1$ \\
	$b_{N-1}, \tilde{b}_{N-1}$ & 0 & $1/2$ & 0 & 0 & $\bf N,\bar{N}$ \\
	$a_i$ & 2 & $-1$ & 0 & 0 & $\bf 1$ \\
	$a_N$ & 2 & $-1$ & 0 & 0 & $\bf N^2-1$ \\
	$v_i, \tilde{v}_i$ & 2 & $\frac{N-i-2}{2}$ & $-1$ & $\mp \d_{i,j+1}$ & $\bf 1$ \\
	$d_i, \tilde{d}_i$ & 0 & $\frac{i-N}{2}$ & $+1$ & $\pm \d_{i,j}$ & $\bf 1$ \\
	$d_N, \tilde{d}_N$ & 0 & 0 & $+1$ & $\pm \d_{N,j}$ & $\bf \bar{N},N$ \\
	\hline
	
\end{tabular}
\caption{List of abelian charges and representation under the global symmetries of all the fields of the $FM[U(N)]$ theory in figure \ref{fig:FM_quiver}.}
\label{tab:FM_fields}
\end{table}
The $FM$ theory has the UV global symmetry group: 
\begin{align}
	S[{\color{red}U(N)} \times {\color{blue}U(1)}^N] \times U(1)_\tau \times U(1)_\D\,,
\end{align}
in addition to the $U(1)_R$ symmetry. At the IR fixed point, the SCFT is characterized by the enhanced global symmetry:
\begin{align}
	S[{\color{red}U(N)} \times {\color{blue}U(N)}] \times U(1)_\tau \times U(1)_\D \,.
\end{align}
The gauge invariant operators indeed reorganize into representations of the IR symmetry group. The list of the chiral ring generators of the $FM[U(N)]$ SCFT, along with their charges and representations, is given in table \ref{tab:FM_operators}.
\begin{table}[h!]
\centering
\renewcommand{\arraystretch}{1.15}
\begin{tabular}{|c|cc|c|}\hline
{} & ${\color{red}U(N)}$ & ${\color{blue}U(N)}$ & R charge\\ \hline
${\color{red}\mathsf{A}}$ & ${\bf N^2-1}$ & $\bf 1$  & $2-\tau$ \\
${\color{blue}\mathsf{A}}$ & $\bf 1$ & ${\bf N^2-1}$  & $2-\tau$ \\
$\Pi$ & $\bf N$ & $\bf \bar{N}$  & $\D$ \\
$\tilde{\Pi}$ & $\bf \bar{N}$ & $\bf N$ & $\D$ \\
$\mathsf{B}_{n, m}$ & $\bf1$ & $\bf1$  & $2n-2\D+(m-n)\tau$ \\
 \hline
\end{tabular}
\caption{List of all  gauge invariant operators that generate the holomorphic spectrum of the $FM[U(N)]$ SCFT. The R-charge is given as a trial value mixed with the other two abelian symmetries of the theory, $U(1)_\tau$ and $U(1)_\Delta$, whose mixing values are given by the two real variables $\tau$ and $\Delta$
(to avoid clutter we denote the  real mass and  the mixing coefficient by the same letter). 
The $\mathsf{B}_{n,m}$ are ${\color{red}U(N)}\times{\color{blue}U(N)}$ singlets, for $n=1,\ldots,N$ and $m=1,\ldots,N+1-n$, defined by $\mathsf{B}_{1,m} = \CF[d_{N+1-m} \tilde{d}_{N+1-m}]$, $\mathsf{B}_{n>1,m} = v_{N-m} a_{N-m}^{n-2} \tilde{v}_{N-m}$. 
}
\label{tab:FM_operators}
\end{table}
The $S^3_b$ partition function of the $FM$ theory can be defined recursively as:\footnote{Notice that in this work we take all the adjoint fields to be traceless instead of tracefull, differently from the original definition.} 
\begin{align}\label{eq:FM_parfun}
	Z_{FM}^{(N)} (\vec{X},\vec{Y},\tau,\D) = & s_b \big( -i\frac{Q}{2} + 2\D \big) \prod_{j=1}^{N} s_b \big( i\frac{Q}{2} - \D \pm (Y_N - X_j) \big) s_b(-\frac{iQ}{2} + \tau) \nonumber \\
	& s_b(-\frac{iQ}{2} + \tau)^{N-1}  \prod_{j<k}^{N} s_b \big( -i\frac{Q}{2} + \tau \pm (X_j - X_k) \big) \int d\vec{Z}_{N-1} \D_{N-1}(\vec{Z}) \nonumber \\
	& \prod_{j=1}^N\prod_{k=1}^{N-1} s_b \big( \frac{iQ}{2} - \frac{\tau}{2} \pm (X_j - Z_k) \big) \prod_{j=1}^{N-1} s_b \big( -\frac{iQ}{2} + \frac{\tau}{2} + \D \pm (Z_j - Y_N) \big) \nonumber \\
	& Z_{FM}^{(N-1)} \big( \vec{Z},\{ Y_1,\ldots,Y_{N-1} \},\tau,\frac{\tau}{2} + \D \big) \,,
\end{align}
with the basis of the recursion given by:
\begin{align}
	Z_{FM}^{(1)} (X,Y,\tau,\D) = s_b \big( -\frac{iQ}{2} + 2\D \big) s_b \big( \frac{iQ}{2} - \D \pm (X - Y) \big) \,.
\end{align}
The vectors $\vec{X}$ and $\vec{Y}$ are the parameters for the manifest and emergent $U(N)$ symmetries respectively, and $\vec{Z}$ is the set of parameters for the gauge group $U(N-1)$. The convention for the $3d$ partition function is given in appendix \ref{inpaconv}.

\paragraph{$FM[U(N)]$ as $3d$ limit of $FE[USp(2N)]$}
The $FM[U(N)]$ theory can be obtained following a $3d$ limit reduction of the $FE[USp(2N)]$ theory. 
We start from the SCI of the $FE[USp(2N)]$ theory in \eqref{eq:FE_SCI} and define the $3d$ parameters from the $4d$ fugacities as:
\begin{align}
	& x_j = e^{2\pi i r X_j} \quad, \quad y_j = e^{2\pi i r Y_j} \quad, \quad z_j = e^{2\pi i r Z_j} \nn \\
	& t = e^{2\pi i r \tau} \quad, \quad c = e^{2\pi i r \D} \,, \nn \\
	& p = e^{-2rb} \quad, \quad q = e^{2rb^{-1}} \,,
\end{align}
then we perform the limit $r \to 0$ and obtain the following relation:
\begin{align}
	\lim_{r\to 0} I_{FE}^{(N)} (\vec{x},\vec{y},t,c) = C_N Z_{FE^{3d}}^{(N)} (\vec{X},\vec{Y},\tau,\D) \,,
\end{align}
where $C_N$ is a prefactor, which is divergent in the limit $r \to 0$, given as:
\begin{align}\label{eq:FE_div_r}
	C_N = \exp \left[ \frac{i \pi}{12 r} (4\D + (1+2N)(-iQ + 2(N-1)\tau) ) \right] \,.
\end{align}
The $FE^{3d}$ theory is given by the same quiver as in \ref{fig:FE_quiver} where now lines are $3d$ $\CN=2$ chiral multiplets and we also introduce linearly in the superpotential the $USp$ monopole of each gauge group. We then shift the parameters $\vec{X},\vec{Y},\vec{Z}$ by $(+s)$ and perform a real mass deformation sending $s \to + \infty$. This has the effect of Higgsing the gauge symmetries from $USp(2N)$ to $U(N)$, landing finally on the $FM[U(N)]$ theory:
\begin{align}\label{eq:FE3d_limit_to_FM}
	\lim_{s \to + \infty} Z_{FE^{3d}}^{(N)} (\vec{X},\vec{Y},\tau,\D) = K_N e^{i\pi(iQ - 2\D + (N-1)\tau)\sum_{j=1}^N (X_j + Y_j)} 
	Z_{FM}(\vec{X},\vec{Y},\tau,\D) \,,
\end{align}
where $K_N$ is a divergent prefactor:
\begin{align}\label{eq:FE_div_s}
	K_N = \exp \big[ 2i s N \pi ( iQ - 2\D + (N-1)\tau ) \big] \,.
\end{align}

\paragraph{Fusion to identity}
Using the $3d$ limit procedure one can reduce all the identities and properties of the $FE[USp(2N)]$ theory into similar properties for the $FM[U(N)]$ theory. For example performing such limit on the identity wall relation in \eqref{FEdelta} we obtain the following partition function identity:
\begin{align}
	\int d\vec{Z}_N \D_N(\vec{Z},\tau) Z_{FM}^{(N)}(\vec{X},\vec{Z},\tau,\D) Z_{FM}^{(N)}(\vec{Z},\vec{Y},\tau,-\D) = {}_{\vec{X}}\mathbb{I}_{\vec{Y}} (\tau) \,,
\end{align}
where the identity operator is defined as:
\begin{align}
	{}_{\vec{X}}\mathbb{I}_{\vec{Y}} (\tau) = \frac{1}{\D_N(\vec{X},\tau)} \sum_{\s \in S_N} \prod_{j=1}^N \d( X_j - Y_{\s(j)} ) \,.
\end{align}
Which consist in two $FM[U(N)]$ theories glued together with the addition of a monopole superpotential $\CW = \mathfrak{M}^+ + \mathfrak{M}^-$ being dual to an identity operator. We depict this relation as:
\be
 \bpic[thick,node distance=3cm,gauge/.style={circle,draw,minimum size=5mm},flavor/.style={rectangle,draw,minimum size=5mm}]  
 
	\path (-3,0) node[flavor,blue](x) {$\!N\!$} -- (-1,0) node[gauge,black](y) {$\!\!\!N\!\!\!$} -- (1,0) node[flavor,red](z) {$\!N\!$} 
		--  (2.5,0) node{$\Longleftrightarrow$};
	
	\wigM (x) -- (y); \draw (-2,0.4) node {$\Pi_L$};
	\wigM (z) -- (y); \draw (0,0.4) node {$\Pi_R$};
	\draw (4,0) node {$\mathbb{I}$-wall};
	\draw[-] (y) to[out=60,in=0] (-1,0.6) to[out=180,in=120] (y); \draw (-1,0.7) node[right] {$a$};
	
	\draw (-1,-1) node {$\cW = \CW_{\text{gluing}} + \mathfrak{M}^+ + \mathfrak{M}^-$};
	
	\draw (6.5,0) node {\begin{tabular}{c|c}
							$\Pi_L$ & $\D$ \\
							$\Pi_R$ & $-\D$ \\
							$a$ & $\tau$  
						\end{tabular}};
	
\epic
\label{FMdelta}
\ee
The superpotential $\CW_{\text{gluing}}$ contains the coupling between the adjoint $a$ and the two adjoint operators $\mathsf{A}_L$ and $\mathsf{A}_R$ of the left and right $FM[U(N)]$ theories.

\paragraph{Mirror self-duality}
We can also reduce the mirror self-duality of the $FE[USp(2N)]$ theory in \eqref{eq:FE_mirror} to obtain a mirror sel-duality for the $FM[U(N)]$ theory which is:
\begin{align}
	Z_{FM}^{(N)} (\vec{X},\vec{Y},\tau,\D) = Z_{FM}^{(N)} (\vec{Y},\vec{X},\tau,\D) \,.
\end{align}
Where the two non-abelian global symmetries are swapped, meaning that we have exchanged the manifest and emergent $U(N)$ symmetries in the UV representation \ref{fig:FM_quiver}.

\paragraph{Interesting deformations}
Along the lines traced for the $FE[USp(2N)]$ theory, we present two types of deformation also for the $FM[U(N)]$ theory.
The first deformation is realized by adding the linear superpotential term $\d \CW = \mathsf{B}_{1,1}$, which has the effect of breaking completely $U(1)_\D$ while it preserves $U(1)_\tau$. This consist in the specialization $\D = 0$ which reduces the $FM[U(N)]$ theory to the $\mathbb{I}$-wall theory:
\begin{align}
	\mathcal{I}_{FM}^{(N)} (\vec{x}, \vec{y}, \tau , \D = 0) = {}_{\vec{X}}\mathbb{I}_{\vec{Y}} (\tau) \,,
	\label{FMid}
\end{align}
Another interesting case is the specialization obtained by the linear superpotential $\d \CW = \mathsf{B}_{1,2}$. This has the effect of breaking $U(1)_\D \times U(1)_\tau$ down to a $U(1)$ subgroup defined by the constraint $\D = 1 - \tau/2$. The $FM[U(N)]$ theory is deformed into a bifundamental hyper multiplet coupled to adjoint singlets:
\be\label{fig:FM_c=t/2}
 \bpic[thick,node distance=3cm,gauge/.style={circle,draw,minimum size=5mm},flavor/.style={rectangle,draw,minimum size=5mm}]  
 
	\path (-4,0) node[flavor,blue](x) {$\!N\!$} -- (-1,0) node[flavor,red](y) {$\!N\!$};
	
	\wigM (x) -- (y); \draw (-2.5,0.4) node {$\Pi$};
	
	\draw (-2.5,-1) node {$\CW = \mathsf{B}_{1,2}$};
	
	\draw (0.5,0) node {$\Longleftrightarrow$};
	
	\path (2,0) node[flavor,blue](x1) {$\!N\!$} -- (5,0) node[flavor,red](y1) {$\!N\!$};
	
	\draw[-, shorten >= 4, shorten <= 10, shift={(-0.1,0.05)}, middx arrowsm] (2,0) -- (5,0);
	\draw[-, shorten >= 4, shorten <= 10, shift={(0.1,-0.05)}, midsx arrowsm] (5,0) -- (2,0);
	
	\draw[-] (x1) to[out=120,in=180] (2,0.6) to[out=0,in=60] (x1);
	\draw[-] (y1) to[out=120,in=180] (5,0.6) to[out=0,in=60] (y1);
	\draw (2.75,0) node[cross] {}; \draw (3.5,0.4) node {$b$}; \draw (2,0.7) node[right] {$\color{blue}a$}; \draw (5,0.7) node[right] {$\color{red}a$};
	
	\draw (3.5,-1) node {$\CW = b({\color{blue}a} + {\color{red}a})\tilde{b} + $};
	\draw (3.5,-1.6) node {$ + Flip[b \tilde{b}]$};
	
	\path (8,0) node {\begin{tabular}{c|c}
							$\Pi$ & $\tau/2$ \\
							${\color{blue}a}$ & $2-\tau$ \\
							${\color{red}a}$ & $2-\tau$ \\
							$b$ & $\tau/2$ 
						\end{tabular}};	
	
\epic
\label{ironduality}
\ee
As a partition function identity this translate into:
\begin{align}
	Z_{FM}^{(N)} \big( \vec{X},\vec{Y},\tau,\D = \frac{\tau}{2} \big) = 
	& s_b( -\frac{iQ}{2} +\tau)^{2N-1} \prod_{j<k=1}^N \big[ s_b \big( -\frac{iQ}{2} + \tau \pm (X_j - X_k) \big)  \nn \\
	& s_b \big( -\frac{iQ}{2} + \tau \pm (Y_j - Y_k) \big) \big] s_b \big( \frac{iQ}{2} - \frac{\tau}{2} \pm (X_j - Y_k) \big) \,.  
\end{align}
One can also iron a $FM$ theory into a bifundamental hypermultiplet also using the deformation $\d \CW = \mathsf{B}_{2,1}$. This has the effect of breaking $U(1)_\D \times U(1)_\tau$ down to a $U(1)$ subgroup defined by the constraint $\D = \frac{iQ-\tau}{2}$. We have the following property:
\begin{align}
	Z_{FM}^{(N)} \big( \vec{X},\vec{Y},\tau,\D = \frac{iQ - \tau}{2} \big) = s_b \big( \frac{iQ}{2} - \tau \big) \prod_{j,k=1}^N s_b \big( \tau \pm (X_j - Y_k) \big) \,.
\end{align}
Graphically we have:
\be
 \bpic[thick,node distance=3cm,gauge/.style={circle,draw,minimum size=5mm},flavor/.style={rectangle,draw,minimum size=5mm}]  
	
	\path (-4,0) node[flavor,blue](x) {$\!N\!$} -- (-1,0) node[flavor,red](y) {$\!N\!$};
	\wigM (x) -- (y); \draw (-2.5,0.4) node {$\Pi$};
	\draw (-2.5,-1) node {$\CW = \mathsf{B}_{2,1}$};
	
	\draw (0.5,0) node {$\Longleftrightarrow$};
	
	\path (2,0) node[flavor,blue](x1) {$\!N\!$} -- (5,0) node[flavor,red](y1) {$\!N\!$};
	
	\draw[-, shorten >= 4, shorten <= 10, shift={(-0.1,0.05)}, middx arrowsm] (2,0) -- (5,0);
	\draw[-, shorten >= 4, shorten <= 10, shift={(0.1,-0.05)}, midsx arrowsm] (5,0) -- (2,0);
	\draw (2.75,0) node[cross] {}; \draw (3.5,0.4) node {$b$};
	\draw (3.5,-1) node {$\cW = Flip[b \tilde{b}]$};
	
	\path (8,0) node {\begin{tabular}{c|c}
							$\Pi$ & $1 - \tau/2$ \\
							$b$ & $1 - \tau/2$ 
						\end{tabular}};	
	
\epic
\label{fig:FM_c=1-t/2}
\ee
The second category of deformations is given by $U(N)$ breaking superpotential terms. This can be obtained by giving VEVs to any of the two adjoint operators. We consider the case of a VEV such that it breaks one of the global $U(N)$ symmetries down to $U(M) \times U(1)$. Suppose that $\vec{X},\vec{Y}$ are the set of mass parameters respectively for $U(N)$ and $U(M)$ and $V$ is that of $U(1)$, the specialization is as follows:
\begin{align}
	& X_i = \frac{N-M+1-2j}{2}\tau + V \qquad \text{for} \quad i=1,\ldots,N-M \,, \nn \\
	& X_i = Y_{i-N+M} \qquad \text{for} \quad i=N-M+1,\ldots,N \,.
\end{align}
We depict the resulting theory as an ``asymmetric" bifundamental:
\be\label{fig:FM_asymm_symbol} 
 \begin{tikzpicture}[thick,node distance=3cm,gauge/.style={circle,draw,minimum size=5mm},flavor/.style={rectangle,draw,minimum size=5mm}]
 
	\path (-1.5,0) node[flavor,blue](y1) {$\!N\!$} -- (1.5,0) node[flavor,red](y3) {$\!M\!$} -- (1.5,1) node[flavor] (v) {$\!1\!$};
  
	\wigM (y1) -- (y3);   \draw (-0.5,0.3) node{$\Pi$};
	\wigM (v) -- (0,0);
	
\end{tikzpicture}\ee
The case $M=0$ enjoys a duality with a flipped fundamental flavor as:
\be
 \bpic[thick,node distance=3cm,gauge/.style={circle,draw,minimum size=5mm},flavor/.style={rectangle,draw,minimum size=5mm}]  
 
	\path (-5,0) node[flavor,blue](x) {$\!N\!$} -- (-2,0) node[flavor,red](y) {$\!0\!$} -- (-2,1) node[flavor](v) {$\!1\!$};
	
	\wigM (x) -- (y); \draw (-4,0.4) node {$\Pi$};
	\wigM (v) -- (-3.5,0);
	\draw[-] (x) to[out=120,in=180] (-5,0.6) to[out=0,in=60] (x); \draw (-5,0.7) node[right]{$\color{blue}a$};
	
	\draw (-3.5,-1) node {$\CW = 0 $};
	
	\draw (-0.5,0) node {$\Longleftrightarrow$};
	
	\path (1,0) node[flavor,blue](x1) {$\!N\!$} -- (4,0) node[flavor](y1) {$\!1\!$};
	
	\draw[-, shorten >= 4, shorten <= 10, shift={(-0.1,0.07)}, middx arrowsm] (1,0) -- (4,0);
	\draw[-, shorten >= 4, shorten <= 10, shift={(0.1,-0.07)}, midsx arrowsm] (4,0) -- (1,0);
	\draw (2.5,0.4) node {$b$};
	\draw (1,0.6) node[cross]{};
	\draw (2,0) node[cross]{};
	\draw[-] (x1) to[out=120,in=180] (1,0.6) to[out=0,in=60] (x1); \draw (1.1,0.8)[right] node{$\color{blue}a$};
	
	\draw (0.5,-1) node[right] {$\CW = \sum_{j=0}^{N-1} Flip[b{{\color{blue}a}}^j\tilde{b}] + $};
	\draw (1,-1.6) node[right] {$ + \sum_{j=2}^N Flip[Tr{{\color{blue}a}}^j] $};
	
	\path (8,0) node {\begin{tabular}{c|c}
							${\color{blue}a}$ & $\tau$ \\
							$\Pi$ & $\D$ \\
							$b$ & $\frac{(1-N)}{2}\tau + \D$ 
						\end{tabular}};	
							
\epic
\label{fig:FMtoFlav}
\ee
As a identity between partition function we have:
\begin{align}\label{eq:FMtoFlav}
	& Z_{FM}^{(N)} \big( \vec{X}, \big\{ \frac{N-1}{2}\tau + V, \ldots, \frac{1-N}{2}\tau + V \big\}, \tau, \D \big) = \nn \\
	& = \prod_{j=2}^N s_b \big( -\frac{iQ}{2} + j \tau \big)
	\prod_{j=1}^N \big[ s_b \big( \frac{iQ}{2} - \frac{1-N}{2}\tau - \D \pm (X_j - V) \big) 
	s_b \big( -\frac{iQ}{2} + (j-N)\tau + 2\D \big) \big] \,.
\end{align}

\subsection{$3d$ $\CS$-wall: the $FT[U(N)]$ theory}\label{app:FT}
The $FT[U(N)]$ theory is a $3d$ $\CN=4$ SCFT denoted by the following symbol:
\be\label{fig:FT_symbol} 
 \begin{tikzpicture}[thick,node distance=3cm,gauge/.style={circle,draw,minimum size=5mm},flavor/.style={rectangle,draw,minimum size=5mm}]
 
	\path (-1.5,0) node[flavor,blue](y1) {$\!N\!$} -- (1.5,0) node[flavor,red](y3) {$\!N\!$};
  
	\wigT (y1) -- (y3);
	
\end{tikzpicture}\ee
The $FT[U(N)]$ theory  has the following quiver description:
\be\label{fig:FT_UV}
\begin{tikzpicture}[thick,node distance=3cm,gauge/.style={circle,draw,minimum size=5mm},flavor/.style={rectangle,draw,minimum size=5mm}] 
	
	\path (0,0) node[gauge] (g1) {$\!\!\!1\!\!\!$} -- (1.5,0) node[gauge] (g2)	{$\!\!\!2\!\!\!$} 
		-- (4,0) node[gauge] (g3) {\!\!\tiny{$N$-$1$}\!\!\!} -- (5.5,0) node[flavor,red] (g4) {$\!N\!$};
		
	\draw[-, shorten >= 6, shorten <= 8, shift={(-0.05,0.07)}, mid arrowsm] (0,0) -- (1.5,0);
	\draw[-, shorten >= 6, shorten <= 8, shift={(0.05,-0.07)}, mid arrowsm] (1.5,0) -- (0,0);
	\draw (0.75,0.35) node {$b_1$};
	
	\draw[-, shorten >= 5, shorten <= 5, shift={(0.05,0.07)}, mid arrowsm] (1.5,0) -- (2.5,0);
	\draw[-, shorten >= 4, shorten <= 6, shift={(0.1,-0.07)}, mid arrowsm] (2.5,0) -- (1.5,0);
	
	\draw (2.73,0) node {$\cdots$};	
	
	\draw[-, shorten >= 5.5, shorten <= 2.5, shift={(-0.1,0.07)}, mid arrowsm] (3,0) -- (4,0);
	\draw[-, shorten >= 1.5, shorten <= 7, shift={(-0.05,-0.07)}, mid arrowsm] (4,0) -- (3,0);
	
	\draw[-, shorten >= 6, shorten <= 9, shift={(-0.05,0.07)}, mid arrowsm] (4,0) -- (5.5,0);
	\draw[-, shorten >= 6.5, shorten <= 8.5, shift={(0.05,-0.07)}, mid arrowsm] (5.5,0) -- (4,0);
	\draw (4.75,0.35) node {$b_{\text{$N$-$1$}}$};
	
	\draw[-] (g1) to[out=60,in=0] (0,0.5) to[out=180,in=120] (g1); \draw (0,0.7) node {$a_1$};
	\draw[-] (g2) to[out=60,in=0] (1.5,0.5) to[out=180,in=120] (g2); \draw (1.5,0.7) node {$a_2$};
	\draw[-] (g3) to[out=60,in=0] (4,0.55) to[out=180,in=120] (g3); \draw (4,0.75) node {$a_{\text{$N$-$1$}}$};
	\draw[-] (g4) to[out=60,in=0] (5.5,0.6) to[out=180,in=120] (g4); \draw (5.5,0.8) node {$a_N$};
	
	\draw (2.75,-1) node{$\cW = \sum_{i=1}^{N-1} b_i (a_i + a_{i+1})\tilde{b}_i $};
	
	\path (7,-0.25) node[right] {\begin{tabular}{c|c}
							$b_i,\tilde{b}_i$ & $\tau/2$ \\
							$a_i$ & $2-\tau$
						\end{tabular}};
	
\end{tikzpicture}
\ee
Notice that in the picture above all the adjoint chirals $a_j$, for $j=1,\ldots,N-1$, are traceful, while $a_N$ is traceless. \\
The UV global symmetry is ${\color{red}SU(N)} \times {\color{blue}U(1)}^{N-1} \times U(1)_\tau$ which enhances in the IR to ${\color{red}SU(N)}\times{\color{blue}SU(N)} \times U(1)_\tau$. However we will work with an ``off-shell" parameterization so that the manifest symmetry is actually ${\color{red}U(N)}\times{\color{blue}U(N)}$, this will be useful since we want to perform $U(N)$ gaugings of these symmetries. Also we work in the  $\CN=2^\star$ language, where $U(1)_\tau$ is the 
antidiagonal combination of the $U(1)_C \times U(1)_H$ subgroup of the $\CN=4$ non-abelian R-symmetry. \\
The IR spectrum of the theory is given by the two moment maps $\color{red}\mathsf{A}$ and $\color{blue}\mathsf{A}$, that are adjoint for the two $U(N)$ global symmetries and carry R-charge $2-\tau$.

%
%
%
%
%

\paragraph{Asymmetric $\CS$-wall}
Starting from the $FT[U(N)]$ theory is possible to perform a deformation with the effect of breaking the two $U(N)$ global symmetries. To do this we give a VEV to the moment maps in form of Jordan-block matrices. This VEVs are uniquely specified by two partitions $(\r,\s)$ of $N$. In this work we are interested only in the case where one of the two partitions is trivial and the other is such that the $U(N)$ symmetry is broken down to $U(M)\times U(1)$, with $M < N$. Let us consider $\vec{X}$ to be the set of mass parameters for the unbroken $U(N)$ group and $\vec{Y}$ those of $U(M)$ and $v$ for $U(1)$, then this deformation is implemented at the level of the partition function by the specialization:
\begin{align}
	&X_j = \frac{N-M+1-2j}{2}\tau + v \quad \text{for} \quad j=1,\ldots,N-M \,, \nn \\
	&X_j = Y_{j-N+M} \quad \text{for} \quad j=N-M+1, \ldots, N  \,.
\end{align}
The resulting theory is depicted as an asymmetric $FT[U(N)]$ theory as:
\be\label{FTasymm}
 \bpic[thick,node distance=3cm,gauge/.style={circle,draw,minimum size=5mm},flavor/.style={rectangle,draw,minimum size=5mm}]  
	\path (-4,0) node[flavor,blue](x) {$\!N\!$} -- (-1,0) node[flavor,red](y) {$\!M\!$} -- (-1,1) node[flavor](v) {$\!1\!$};
	
	\wigT (x) -- (y);
	\wigT (v) -- (-2.5,0);
	
\epic\ee

\paragraph{Fusion to identity}

Gluing together two $FT[U(N)]$ theories with an extra adjoint chiral that comes couples to the moment maps charged under the gauge group gives an $\mathbb{I}$-wall as:
\begin{align}\label{FTdelta}
	\int d \vec{Z}_N \Delta_N(\vec{Z},\tau) Z_{FT}^{(N)} (\vec{X}, \vec{Z},\tau) Z_{FT}^{(N)} (\vec{Z}, \pm \vec{Y}, \tau) = {}_{\vec{X}}\mathbb{I}_{\mp \vec{Y}}(\tau)\,.
\end{align}
Graphically we can write this identity as:
\be\label{fig:FTdelta}
 \bpic[thick,node distance=3cm,gauge/.style={circle,draw,minimum size=5mm},flavor/.style={rectangle,draw,minimum size=5mm}]  
	\path (-3,0) node[flavor,blue](x) {$\!N\!$} -- (-1,0) node[gauge,black](y) {$\!\!\!N\!\!\!$} -- (1,0) node[flavor,red](z) {$\!N\!$} --  (2.5,0) node{$\Longleftrightarrow$};
	
	\wigT (x) -- (y); \draw (-2,0.3) node {$-$};
	\wigT (z) -- (y); \draw (0,0.3) node {$\pm$};
	\draw (4,0) node {$\mathbb{I}$-wall};
	\draw[-] (y) to[out=60,in=0] (-1,0.6) to[out=180,in=120] (y); 
	
	\draw (-1,-1) node {$\cW = \CW_{\text{gluing}}$};
	
\epic\ee
Where the adjoint chiral, with R-charge $\tau$, is coupled to the gauge charged moment maps of the two $FT[U(N)]$ theories respectively. \\
We can also construct asymmetric $\mathbb{I}$-walls by breaking one of the two global $U(N)$ symmetries as in \ref{FTasymm}. The result is the following:
\be\label{fig:FTdeltaasymm}
 \bpic[thick,node distance=3cm,gauge/.style={circle,draw,minimum size=5mm},flavor/.style={rectangle,draw,minimum size=5mm}]  
	
	\path (-3,0) node[flavor,blue](x) {$\!N\!$} -- (-1,0) node[gauge,black](y) {$\!\!\!N\!\!\!$} -- (1,0) node[flavor,red](z) {$\!M\!$} 
		-- (1,1) node[flavor](v) {$\!1\!$} --  (2.5,0) node{$\Longleftrightarrow$};
	
	\wigT (x) -- (y); \draw (-2,0.3) node {$-$};
	\wigT (z) -- (y); \draw (-0.2,0.3) node {$\pm$};
	\wigT (v) -- (0,0);
	\draw (4,0) node {$\mathbb{I}$-wall};
	\draw[-] (y) to[out=60,in=0] (-1,0.6) to[out=180,in=120] (y); 
	
	\draw (-1,-1) node {$\cW = \CW_{\text{gluing}}$};
	
\epic\ee
Again, the adjoint chiral is coupled to the moment maps of the two $FT$ theories. This identity corresponds to the partition function identity \eqref{FTdelta} with the specialization:
\begin{align}
	&Y_j = \frac{N-M+1-2j}{2}\tau + V \quad \text{for} j=1,\ldots,N-M \nn \,, \\
	&Y_j = W_{j-N+M} \quad \text{for} j=N-M+1, \ldots, N \,,
\end{align}
where $\vec{W}$ is the set of parameters of the $U(M)$ global symmetry, while $V$ is the parameter of the $U(1)$ symmetry.

\section{Star-Triangle dualities}\label{app:star_triangle}
This appendix is a collection of the {\it star-triangle} dualities used throughout the work. All the presented identities descend from the $4d$ braid duality, which was first introduced in \cite{Rains_2018} and then studied in \cite{Pasquetti:2019hxf}. In \cite{BCP1} it was shown that the $4d$ braid duality can be proved iterating IP duality. It was also discussed the operator map, deformations and $3d$ reduction.

\subsection{$4d$ braid duality}\label{app:4d_braid}
The braid duality two $FE[USp(2N)]$ theories, glued with the addition of a flavor, to a single $FE[USp(2N)]$ theory with singlets. 
Graphically it can be depicted as:

\be\label{fig:4d_braid} 
\begin{tikzpicture}[thick,node distance=3cm,gauge/.style={circle,draw,minimum size=5mm},flavor/.style={rectangle,draw,minimum size=5mm}]

	\path (0,0) node[gauge](g) {$\!\!\!2N\!\!\!$} -- (-1.5,-1) node[flavor,blue](x) {$\!2N\!$} -- (1.5,-1) node[flavor,red](y) {$\!2N\!$} 
		-- (0,1.5) node[flavor] (v) {$\!2\!$} -- (3,0) node {$\Longleftrightarrow$};
	
	\wigE (x) -- (g); \draw (-0.75,-0.35) node[left] {$\Pi_L$};
	\wigE (y) -- (g); \draw (0.75,-0.35) node[right] {$\Pi_R$};
	\draw[-] (g) -- (v); \draw (0,0.75) node[right] {$f$};
	\draw[-] (g) to[out=110,in=50] (-0.55,0.45) to[out=230,in=170] (g); \draw (-0.5,0.5) node[left] {$a$};
	
	\draw (0,-2) node {$\cW = a(\mathsf{A}_L + \mathsf{A}_R)$};
	
	\path (0,-3.5) node {\begin{tabular}{c|c}
							$\Pi_L$ & $\pi_L$ \\
							$\Pi_R$ & $\pi_R$ \\
							$f$ & $1 - \pi_L - \pi_R$ \\
							$a$ & $\tau$
						\end{tabular}};
      
    \path (5,1) node[flavor,blue](x) {$\!2N\!$} -- (7,1) node[flavor,red](y) {$\!2N\!$} -- (6,-0.5) node[flavor](v) {$\!2\!$}; 
   
    \wigE (x) -- (y);  \draw (6,1.4) node{$\Pi$}; 
    \draw[-] (x) -- (v);  \draw (5.2,0) node {$l$};
    \draw[-] (y) -- (v);   \draw (6.8,0) node {$r$}; 
	
	\draw (6,-2) node {$\cW = l \Pi r$};
	
	\path (6,-3.5) node {\begin{tabular}{c|c}
							$\Pi$ & $\pi_L + \pi_R$ \\
							$l$ & $1 - \pi_R$ \\
							$r$ & $1 - \pi_L$
						\end{tabular}}; 
	
\end{tikzpicture}\ee
The associated SCI is:
\begin{align}
	\oint d\vec{z}_N \D_N(\vec{z},t) & \prod_{j=1}^N \Ge(\sqrt{pq}(c_L c_R)^{-1} z_j^\pm v^\pm) \CI_{FE}^{(N)} (\vec{x},\vec{z},t,c_L) \CI_{FE}^{(N)} (\vec{z},\vec{y},t,c_R) = \nn \\
	= & \CI_{FE}^{(N)} (\vec{x},\vec{y},t,c_L c_R) \prod_{j=1}^N \big( \Ge(\sqrt{pq} c_R^{-1} x_j^\pm v^\pm) \Ge(\sqrt{pq} c_L^{-1} y_j^\pm v^\pm) \big) \,.
\end{align}
Where the fugacity appearing in the above identity are defined from the R-charge mixings written in \eqref{fig:4d_braid} as:
\begin{align}
	c_L = (pq)^{\pi_L/2} \qquad , \qquad c_R = (pq)^{\pi_R/2} \qquad , \qquad t = (pq)^{\tau/2} \,.
\end{align}
and $\vec{x},\vec{y},\vec{z}$ are the fugacities for the blue,red and gauge symmetries respectively, $v$ is associated to the $USp(2)$ global symmetry.

\subsection{$3d$ braid duality and its deformations}\label{app:3d_braid}

Starting from the $4d$ braid duality and performing the $3d$ reduction combined with suitable real mass deformations we can generate a series of 
$3d$ dualities. Below we collect the dualities relevant for this work.\\

The $3d$ braid duality relates two $FM[U(N)]$ theories glued with the addition of a flavor, with a single $FM[U(N)]$ theory with singlets:
\be\label{fig:3dBraid}
\begin{tikzpicture}[thick,node distance=3cm,gauge/.style={circle,draw,minimum size=5mm},flavor/.style={rectangle,draw,minimum size=5mm}]
	
	\path (0,0) node[gauge] (g1) {$\!\!\!N\!\!\!$} -- (-1.5,-1) node[flavor,blue] (f1) {$\!N\!$} -- (1.5,-1) node[flavor,red] (f2) {$\!N\!$} -- (0,1.5) node[flavor] (f3) {$\!1\!$} -- (3,0) node {$\Longleftrightarrow$};
	
	\wigM (g1) -- (f1); \draw (-0.75,-0.3) node[left] {$\Pi_L$};
	\wigM (g1) -- (f2); \draw (0.75,-0.3) node[right] {$\Pi_R$};
	
	\draw[-, shorten >= 6, shorten <= 9, shift={(-0.07,0.05)}, mid arrowsm] (0,1.5) -- (0,0);
	\draw[-, shorten >= 6, shorten <= 9, shift={(0.07,-0.05)}, mid arrowsm] (0,0) -- (0,1.5);
	\draw (0,0.75) node [right] {$f$};
	
	\draw[-] (g1) to[out=110,in=50] (-0.55,0.45) to[out=230,in=170] (g1);  \draw (-0.5,0.5) node[left] {$a$};
	
	\draw (0,-2) node {$\cW = a(\mathsf{A}_L + \mathsf{A}_R) + \mathfrak{M}^+ + \mathfrak{M}^-$};
	
	\path (0,-4) node {\begin{tabular}{c|c}
							$\Pi_L, \tilde{\Pi}_L$ & $\D_L$ \\
							$\Pi_R, \tilde{\Pi}_R$ & $\D_R$ \\
							$f,\tilde{f}$ & $1 - \D_L - \D_R \mp V$ \\
							$a$ & $\tau$
						\end{tabular}};
	
	\path (5,1) node[flavor,blue] (f1) {$\!N\!$} -- (7,1) node[flavor,red] (f2) {$\!N\!$} -- (6,-0.5) node[flavor] (f3) {$\!1\!$};
	 
	\wigM (f1) -- (f2); \draw (6,1.4) node {$\Pi$};
	
	\draw[-, shorten >= 7.5, shorten <= 9.5, shift={(-0.07,0.02)}, mid arrowsm] (5,1) -- (6,-0.5);
	\draw[-, shorten >= 7.5, shorten <= 8.5, shift={(0.1,0)}, mid arrowsm] (6,-0.5) -- (5,1);
	\draw (5.5,0.15) node[left] {$l$};
	
	\draw[-, shorten >= 7.5, shorten <= 8, shift={(-0.1,0.02)}, mid arrowsm] (6,-0.5) -- (7,1);
	\draw[-, shorten >= 8.5, shorten <= 9, shift={(0.05,0)}, mid arrowsm] (7,1) -- (6,-0.5);
	\draw (6.5,0.15) node[right] {$r$};
	
	\draw (6,-2) node {$\cW = l \Pi \tilde{r} + \tilde{l} \tilde{\Pi}r$};
	
	\path (6,-4) node {\begin{tabular}{c|c}
							$\Pi, \tilde{\Pi}$ & $\D_R + \D_R$ \\
							$l,\tilde{l}$ & $1 - \D_R \mp V$ \\
							$r,\tilde{r}$ & $1 - \D_L \mp V$ \\
						\end{tabular}};
	
\end{tikzpicture}\ee
The associated partition function identity is:
\begin{align}
	\int d\vec{Z}_N \D_N(\vec{Z},\tau) & Z_{FM}^{(N)} (\vec{X},\vec{Z},\tau,\D_L) Z_{FM}^{(N)} (\vec{Z},\vec{Y},\tau,\D_R) \prod_{j=1}^N s_b( \D_L + \D_R \pm (Z_j - V) ) = \nn \\
	& = Z_{FM}^{(N)}(\vec{X},\vec{Y},\tau,\D_L+\D_R) \prod_{j=1}^N \big( s_b(\D_R \pm (X_j - V)) s_b(\D_L \pm (Y_j - V)) \big) \,.
\end{align}\\

If we perform a real mass for $U(1)_V$, sending $V \to +\infty$, we land on the duality:
\be\label{fig:FM_fusion}
\begin{tikzpicture}[thick,node distance=3cm,gauge/.style={circle,draw,minimum size=5mm},flavor/.style={rectangle,draw,minimum size=5mm}]
	
	\path (0,0) node[gauge] (g1) {$\!\!\!N\!\!\!$} -- (-2,0) node[flavor,blue] (f1) {$\!N\!$} -- (2,0) node[flavor,red] (f2) {$\!N\!$} 
		 -- (3.5,0) node {$\Longleftrightarrow$};
	
	\wigM (g1) -- (f1); \draw (-1,0.4) node {$\Pi_L$};
	\wigM (g1) -- (f2); \draw (1,0.4) node {$\Pi_R$};
	\draw (g1) to[out=60,in=0] (0,0.6) to[out=180,in=120] (g1); \draw (0,0.8) node {$a$};
	
	\draw (0,-1) node {$\cW = a(\mathsf{A}_L + \mathsf{A}_R) + \mathfrak{M}^+ $};
	
	\path (0,-2.5) node {\begin{tabular}{c|c}
							$\Pi_L, \tilde{\Pi}_L$ & $\D_L$ \\
							$\Pi_R, \tilde{\Pi}_R$ & $\D_R$ \\
							$a$ & $\tau$
						\end{tabular}};
	
	\path (5,0) node[flavor,blue] (f1) {$\!N\!$} -- (7,0) node[flavor,red] (f2) {$\!N\!$};
	 
	\wigM (f1) -- (f2); \draw (6,0.4) node {$\Pi$};
	
	\draw (6,-1) node {$\cW = 0$};
	
	\path (6,-2) node {\begin{tabular}{c|c}
							$\Pi, \tilde{\Pi}$ & $\D_R + \D_R$ \\
						\end{tabular}};
	
\end{tikzpicture}\ee
The corresponding partition function identity is:
\begin{align}
	\int d\vec{Z}_N & \D_N(\vec{Z},\tau) e^{2\pi i (\D_L+\D_R) \sum_{j=1}^N Z_j}
	Z_{FM}^{(N)} (\vec{X},\vec{Z},\tau,\D_L) Z_{FM}^{(N)} (\vec{Z},\vec{Y},\tau,\D_R) = \nn \\
	& = e^{-2\pi i( \D_R \sum_{j=1}^N X_j + \D_L \sum_{j=1}^N Y_j )} Z_{FM}^{(N)}(\vec{X},\vec{Y},\tau,\D_L+\D_R) \,.
\end{align}\\

There is another interesting deformation  to consider. Starting from \eqref{fig:3dBraid} we first activate the nilpotent VEV deformation 
studied in \eqref{fig:FMtoFlav} which on the star side confines the left $FM[U(N)]$ theory to a flavor of charge $\frac{1-N}{2}\tau + \D_L$, plus singlets. Similarly on the triangle side the effect of this deformation is to confine the $FM[U(N)]$ theory to a flavor of charge  $\frac{1-N}{2}\tau + \D_L + \D_R$, plus singlets. In order to write a consistent duality we add an extra adjoint singlet charged under the $\color{red}U(N)$ symmetry which is coupled to ${\color{red}\mathsf{A}}_R$
\be
\begin{tikzpicture}[thick,node distance=3cm,gauge/.style={circle,draw,minimum size=5mm},flavor/.style={rectangle,draw,minimum size=5mm}]
	
	\path (0,0) node[gauge] (g1) {$\!\!\!N\!\!\!$} -- (-1.5,-1) node[flavor,blue] (f1) {$\!1\!$} -- (1.5,-1) node[flavor,red] (f2) {$\!N\!$} 
		-- (0,1.5) node[flavor] (f3) {$\!1\!$} -- (4,0) node {$\Longleftrightarrow$};
	
	\draw[-, shorten >= 7.5, shorten <= 9, shift={(0,0.1)}, midsx arrowsm] (0,0) -- (-1.5,-1);
	\draw[-, shorten >= 5, shorten <= 10, shift={(-0.05,-0.1)}, middx arrowsm] (-1.5,-1) -- (0,0);
	\draw (-1,-0.66) node {\rotatebox{-45}{\LARGE{$\times$}}};
	\draw (-0.75,-0.3) node[left] {$g$};
	
	\wigM (g1) -- (f2); \draw (0.75,-0.3) node[right] {$\Pi_R$};
	
	\draw[-, shorten >= 6, shorten <= 9, shift={(-0.07,0.05)}, mid arrowsm] (0,1.5) -- (0,0);
	\draw[-, shorten >= 6, shorten <= 9, shift={(0.07,-0.05)}, mid arrowsm] (0,0) -- (0,1.5); 
	\draw (0,0.75) node [right] {$f$};
	
	\draw[-] (g1) to[out=110,in=50] (-0.55,0.45) to[out=230,in=170] (g1); \draw (-0.5,0.5) node[left] {$a$};
	\draw (f2) to[out=30,in=90] (2.1,-1) to[out=-90,in=-30] (f2); \draw (2,-1) node[right] {$\color{red}a$};
	
	\draw (-2.5,-2) node[right] {$\cW = a \mathsf{A}_R + {\color{red}a \mathsf{A}}_R + \mathfrak{M}^+ + \mathfrak{M}^- +$};
	\draw (-2,-2.7) node[right] {$ + \sum_{j=0}^{N-1} Flip[g a^j \tilde{g}] $};
	
	\path (0,-4.5) node {\begin{tabular}{c|c}
							$g,\tilde{g}$ & $\frac{1-N}{2}\tau + \D_L \mp X$ \\
							$\Pi_R, \tilde{\Pi}_R$ & $\D_R$ \\
							$f,\tilde{f}$ & $1 - \D_L - \D_R \mp V$ \\
							$a, {\color{red}a}$ & $\tau$
						\end{tabular}};
	
\begin{scope}[shift={(1,0)}]
	\path (5,1) node[flavor,blue] (f1) {$\!1\!$} -- (7,1) node[flavor,red] (f2) {$\!N\!$} -- (6,-0.5) node[flavor] (f3) {$\!1\!$};
	 
	\draw[-, shorten >= 6, shorten <= 9, shift={(-0.05,0.07)}, midsx arrowsm] (5,1) -- (7,1);
	\draw[-, shorten >= 6, shorten <= 9, shift={(0.05,-0.07)}, middx arrowsm] (7,1) -- (5,1);
	\draw (6,1.4) node {$b$};
	\draw (6.35,1) node[cross] {};
	
	\draw[-, shorten >= 7.5, shorten <= 9.5, shift={(-0.07,0.02)}, mid arrowsm] (5,1) -- (6,-0.5);
	\draw[-, shorten >= 7.5, shorten <= 8.5, shift={(0.1,0)}, mid arrowsm] (6,-0.5) -- (5,1);
	\draw (5.5,0.15) node[left] {$l_j$};
	
	\draw[-, shorten >= 7.5, shorten <= 8, shift={(-0.1,0.02)}, mid arrowsm] (6,-0.5) -- (7,1);
	\draw[-, shorten >= 8.5, shorten <= 9, shift={(0.05,0)}, mid arrowsm] (7,1) -- (6,-0.5);
	\draw (6.5,0.15) node[right] {$r$};
	
	\draw (f2) to[out=60,in=0] (7,1.6) to[out=180,in=120] (f2);
	\draw (7,1.8) node[right] {${\color{red}a}$};
\end{scope}
	
	\draw (4.5,-2) node[right] {$\cW = \sum_{j=1}^{N} ( l_j b {{\color{red}a}}^{j-1} \tilde{r} + \tilde{l} \tilde{b} {{\color{red}a}}^{j-1} r ) + $};
	\draw (5,-2.7) node[right] {$+ \sum_{j=0}^{N-1} Flip[b {{\color{red}a}}^j \tilde{b}] $};
	
	\path (8,-4.5) node {\begin{tabular}{c|c}
							$b,\tilde{b}$ & $\frac{1-N}{2}\tau + \D_R + \D_R \pm X$ \\
							$l_j,\tilde{l}_j$ & $1 + \frac{N+1-2j}{2}\tau - \D_R \mp (V-X)$ \\
							$r,\tilde{r}$ & $1 - \D_L \mp V$ \\
							${\color{red}a}$ & $\tau$
						\end{tabular}};
	
\end{tikzpicture}\ee
Notice on the r.h.s. this deformation has made the original chirals $l,\tilde{l}$ in the fundamental/antifundamental of $U(N)$ into $2N$ chirals.\\
We then perform a  real mass deformation for $U(1)_{\D_L}$,  this has the effect of integrating out the two  flavors in the electric theory (no CS level is generated). On the dual theory this deformation gives mass to the horizontal $b,\tilde{b}$ and the right diagonal $r,\tilde{r}$ flavors. After flipping some singlets on both sides we are left with:
\be
\begin{tikzpicture}[thick,node distance=3cm,gauge/.style={circle,draw,minimum size=5mm},flavor/.style={rectangle,draw,minimum size=5mm}] 
	 
	\path (0,0) node[gauge](g) {$\!\!\!N\!\!\!$} -- (2,0) node[flavor,red](x) {$\!N\!$} -- (3,0) node {$\Longleftrightarrow$};
	 
	\wigM (g) -- (x); \draw (1,0.4) node {$\Pi$};
	\draw[-] (g) to[out=60,in=0] (0,0.6) to[out=180,in=120] (g); \draw (0,0.75) node[left] {$a$};
	
	\draw (1,-1) node {$\cW = a \mathsf{A}$};
	
	\path (4,0) node[flavor](x1) {$\!1\!$} -- (6,0) node[flavor](x2) {$\!1\!$};
	
	\draw[-, shorten >= 6, shorten <= 9, shift={(-0.05,-0.07)}, mid arrowsm] (4,0) -- (6,0);
	\draw[-, shorten >= 6, shorten <= 9, shift={(0.05,0.07)}, mid arrowsm] (6,0) -- (4,0);
	\draw (5,0.4) node{$q_j$};
	
	\draw (5,-1) node {$\CW = 0 $};
	
	\draw (10,0) node {\begin{tabular}{c|c}
		$\Pi$ & $\D$ \\
		$a$ & $\tau$ \\
		$q_j, \tilde{q}_j$ & $1 + \frac{N+1-2j}{2}\tau - \D \pm V$	
	\end{tabular}};
	 
\end{tikzpicture}
\label{fmconf}
\ee
On the l.h.s. we have an $FM[U(N)]$ theory with one node gauged while on the r.h.s. we have $N$ free chirals. At the level of the partition function this reads:
\begin{align}
	\int d\vec{Z}_N \D_N(\vec{Z},\tau) & e^{2 \pi i V \sum_{j=1}^N Z_j} Z_{FM}^{(N)}(\vec{Z},\vec{Y},\tau,\D) = \nn \\
	& = e^{2 \pi i V \sum_{j=1}^N Y_j}  \prod_{j=1}^N s_b(-\frac{N+1-2j}{2}\tau + \D \pm V ) \,.
\end{align}
Notice that the chirals appear with ``wrong" R-charge and with only few Cartans of the flavor symmetry visible. In the IR the emergent symmetry rotating the chirals mixes with the R-charge so that all chirals have the free $R=1/2$ and then we have $N$ free hypers.\\

We can also perform a real mass deformation for the $U(1)_{\D_L}$ and $U(1)_{\D_R}$ symmetries in \eqref{fig:3dBraid}, taking the limit: $\D_L \to -\infty$ and $\D_R \to +\infty$ such that the sum is kept finite: $\D_L + \D_R = \D$, to land on the duality:
\be\label{fig:3d_basicmove}
\begin{tikzpicture}[thick,node distance=3cm,gauge/.style={circle,draw,minimum size=5mm},flavor/.style={rectangle,draw,minimum size=5mm}]
	
	\path (0,0) node[gauge] (g1) {$\!\!\!N\!\!\!$} -- (-1.5,-1) node[flavor,blue] (f1) {$\!N\!$} -- (1.5,-1) node[flavor,red] (f2) {$\!N\!$} 
		-- (0,1.5) node[flavor] (f3) {$\!1\!$} -- (3,0) node {$\Longleftrightarrow$};
	
	\wigT (g1) -- (f1); \draw (-0.75,-0.4) node[left] {$+$};
	\wigT (g1) -- (f2); \draw (0.75,-0.4) node[right] {$-$};
	
	\draw[-, shorten >= 6, shorten <= 9, shift={(-0.07,0.05)}, mid arrowsm] (0,1.5) -- (0,0);
	\draw[-, shorten >= 6, shorten <= 9, shift={(0.07,-0.05)}, mid arrowsm] (0,0) -- (0,1.5);
	\draw (0,0.75) node[right] {$f$};
	
	\draw[-] (g1) to[out=110,in=50] (-0.55,0.45) to[out=230,in=170] (g1); \draw (-0.5,0.5) node[left] {$a$};
	
	\draw (0,-2) node {$\cW = a(\mathsf{A}_L + \mathsf{A}_R)$};
	
	\path (0,-3) node {\begin{tabular}{c|c}
							$f, \tilde{f}$ & $1 - \D$ \\
							$a$ & $\tau$
						\end{tabular}};
	
	\path (5,0) node[flavor,blue] (f1) {$\!N\!$} -- (7,0) node[flavor,red] (f2) {$\!N\!$};
	
	\wigM (f1) -- (f2); \draw (6,0.4) node {$\Pi$};
	
	\draw (6,-2) node {$\cW = 0$};
	
	\path (6,-3) node {\begin{tabular}{c|c}
							$\Pi, \tilde{\Pi}$ & $\D$ 
						\end{tabular}};
	
\end{tikzpicture}\ee
Which we claim to be the $3d$ $\cN=2$ basic $\CS$-duality move. As a partition function identity we have:
\begin{align}
	\int d\vec{Z}_N \D_N(\vec{Z},\tau) & Z_{FT}^{(N)} (\vec{X},\vec{Z},\tau) Z_{FT}^{(N)} (\vec{Z},-\vec{Y},\tau) \prod_{j=1}^N s_b( \D_L + \D_R \pm (Z_j - V) ) = \nn \\
	& = e^{2 \pi i V \sum_{j=1}^N ( Y_j - X_j )} Z_{FM}^{(N)}(\vec{X},\vec{Y},\tau,\D_L+\D_R) \,.
\end{align}
\section{Star-Star dualities}\label{app:star_star}
This appendix is a collection of the {\it star-star} dualities used throughout the work. All the identities descend from the $4d$ generalized star-star duality which was first introduced in \cite{Rains_2018} and later studied in \cite{Bottini:2021vms} and \cite{BCP1}.

\subsection{$4d$ dualities}\label{app:4d_starstar}
The $4d$ generalized star-star duality is a self-duality modulo singlets for two $FE[USp(2N)]$ theories glued with the addition of two flavors. 
We have the following duality:
\be \label{fig:4d_starstar}
 \bpic[thick,node distance=3cm,gauge/.style={circle,draw,minimum size=5mm},flavor/.style={rectangle,draw,minimum size=5mm}]  
    
    \path (-6,0) node[flavor,orange](x1) {$\!2N\!$} -- (-4,0) node[gauge](x2) {$\!\!\!2N\!\!\!$} -- (-2,0) node[flavor,blue](x3) {$\!2N\!$} 
    	-- (-5,-1.5) node[flavor,violet](fU) {$2$} -- (-3,-1.5) node[flavor,red](fD) {$2$} --  (0,0) node{$\Longleftrightarrow$};
    
    \wigE (x1) -- (x2); \draw (-5,0.3) node {$\Pi_L$};
    \wigE (x3) -- (x2); \draw (-3,0.3) node {$\Pi_R$};
    \draw[-] (x2) -- (fU); \draw (-4.5,-0.7) node[left] {$p$};
    \draw[-] (x2) -- (fD); \draw (-3.5,-0.8) node[left] {$q$};
    \draw[-] (x1) -- (fU); \draw (-5.5,-0.8) node[left] {$r$};
    \draw[-] (x3) -- (fD); \draw (-2.5,-0.7) node[left] {$s$};
    \draw [-] (x2) to[out=60, in=0] (-4,0.6) to[out=180,in=120] (x2); \draw (-4,0.7) node[right] {$a$};  
    
    \draw (-6,-2.5) node[right]{$\cW = a ( \mathsf{A}_L + \mathsf{A}_R ) + $};
    \draw (-5.5,-3.1) node[right]{$ + r \Pi_L p + s \Pi_R q $};
	
	\path (-4,-5.5) node {\begin{tabular}{c|c}
							$\Pi_L$ & $\pi_L$ \\
							$\Pi_R$ & $\pi_R$ \\
							$p,q$ & $1 - \frac{\pi_L + \pi_R}{2} \mp \phi$ \\
							$r$ & $1 - \frac{\pi_L - \pi_R}{2} + \phi$ \\
							$s$ & $1 + \frac{\pi_L - \pi_R}{2} - \phi$ \\
							$a$ & $\tau$ 
						\end{tabular}};
    
    \path (2,0) node[flavor,orange](x1) {$\!2N\!$} -- (4,0) node[gauge](x2) {$\!\!\!2N\!\!\!$} -- (6,0) node[flavor,blue](x3) {$\!2N\!$} 
    	-- (3,-1.5) node[flavor,red](fU) {$2$} -- (5,-1.5) node[flavor,violet](fD) {$2$};
    
    \wigE (x1) -- (x2); \draw (3,0.35) node {$\Pi'_L$};
    \wigE (x3) -- (x2); \draw (5,0.35) node {$\Pi'_R$};
    \draw[-] (x2) -- (fU); \draw (3.5,-0.7) node[left] {$p'$};
    \draw[-] (x2) -- (fD); \draw (4.5,-0.8) node[left] {$q'$};
    \draw[-] (x1) -- (fU); \draw (2.5,-0.8) node[left] {$r'$};
    \draw[-] (x3) -- (fD); \draw (5.5,-0.7) node[left] {$s'$};
    \draw [-] (x2) to[out=60, in=0] (4,0.6) to[out=180,in=120] (x2); \draw (4,0.75) node[right] {$a'$};     
    
    \draw (2,-2.5) node[right]{$\cW =  a' (\mathsf{A}'_L + \mathsf{A}'_R) + $};
    \draw (2.5,-3.1) node[right]{$ + r' \Pi'_L p' + r' \Pi'_R q' $};

	\path (4,-5.5) node {\begin{tabular}{c|c}
							$\Pi'_L$ & $\pi_R$ \\
							$\Pi'_R$ & $\pi_L$ \\
							$p',q'$ & $1 - \frac{\pi_L + \pi_R}{2} \pm \phi$ \\
							$r'$ & $1 + \frac{\pi_L - \pi_R}{2} - \phi$ \\
							$s'$ & $1 - \frac{\pi_L - \pi_R}{2} + \phi$ \\
							$a'$ & $\tau$ 
						\end{tabular}};
						
\epic\ee
This duality consist in the following SCI identity;
\begin{align}
	& \oint d\vec{z}_N \D_N(\vec{z},t) \CI_{FE}^{(N)}(\vec{x},\vec{z},t,c_L) \CI_{FE}^{(N)}(\vec{y},\vec{z},t,c_R) \nn \\
	& \prod_{j=1}^N \big[ \Ge((pq)^{1/2} (c_L c_R)^{-1/2} f^{-1} z_j^\pm v_1^\pm) \Ge((pq)^{1/2} (c_L c_R)^{-1/2} f z_j^\pm v_2^\pm ) \nn \\
	& \Ge((pq)^{1/2} (c_L/c_R)^{-1/2} f x_j^\pm v_1^\pm ) \Ge((pq)^{1/2} (\pi_L/\pi_R)^{1/2} f^{-1} y_j^\pm v_2^\pm ) \big]  = \nn \\
	= & \oint d\vec{z}_N \D_N(\vec{z},t) \CI_{FE}^{(N)}(\vec{x},\vec{z},t,c_R) \CI_{FE}^{(N)}(\vec{y},\vec{z},t,c_L) \nn \\
	& \prod_{j=1}^N \big[ \Ge((pq)^{1/2} (c_L c_R)^{-1/2} f z_j^\pm v_2^\pm) \Ge((pq)^{1/2} (c_L c_R)^{-1/2} f^{-1} z_j^\pm v_1^\pm ) \nn \\
	& \Ge((pq)^{1/2} (c_L/c_R)^{1/2} f x_j^\pm v_2^\pm ) \Ge((pq)^{1/2} (c_L/c_R)^{-1/2} f^{-1} y_j^\pm v_1^\pm ) \big] \,.
\end{align}
Starting from this duality we can consider various deformations. In the following paper we are only interested in one case of deformations, which is given by nilpotent VEVs for one of the two $USp(2N)$ symmetries, let us take the blue one for simplicity, with the effect of breaking it down to $USp(2)$. Using the identity \eqref{fig:FEtoFlav}, we obtain the new duality:
\be\label{fig:4d_starstar_nilmass}
 \bpic[thick,node distance=3cm,gauge/.style={circle,draw,minimum size=5mm},flavor/.style={rectangle,draw,minimum size=5mm}]  
    
    \path (-6,0) node[flavor,orange](x1) {$\!2N\!$} -- (-4,0) node[gauge](x2) {$\!\!\!2N\!\!\!$} -- (-2,0) node[flavor,blue](x3) {$\!2\!$} 
    	-- (-5,-1.5) node[flavor,violet](fU) {$2$} -- (-3,-1.5) node[flavor,red](fD) {$2$} --  (0,0) node{$\Longleftrightarrow$};
    
    \wigE (x1) -- (x2); \draw (-5,0.3) node {$\Pi_L$};
    \draw[-] (x3) -- (x2); \draw (-3,0) node[cross] {}; \draw (-3,0.3) node {$b_R$};
    \draw[-] (x2) -- (fU); \draw (-4.5,-0.7) node[left] {$p$};
    \draw[-] (x2) -- (fD); \draw (-3.5,-0.8) node[left] {$q$};
    \draw[-] (x1) -- (fU); \draw (-5.5,-0.8) node[left] {$r$};
    \draw[-] (x3) -- (fD); \draw (-2.5,-0.7) node[left] {$s_j$};
    \draw [-] (x2) to[out=60, in=0] (-4,0.6) to[out=180,in=120] (x2); \draw (-4,0.7) node[right] {$a$};  
    
    \draw (-7,-2.5) node[right]{$\cW = a \mathsf{A}_L + \sum_{j=0}^N Flip[b_R^2 a^j] +$};
    \draw (-6.5,-3.1) node[right]{$ + r \Pi_L p + \sum_{j=1}^N s_j b_R a^{j-1} q $};
	
	\path (-4,-5.5) node {\begin{tabular}{c|c}
							$\Pi_L$ & $\pi_L$ \\
							$b_R$ & $\frac{1-N}{2}\tau + \pi_R$ \\
							$p,q$ & $1 - \frac{\pi_L + \pi_R}{2} \mp \phi$ \\
							$r$ & $1 - \frac{\pi_L - \pi_R}{2} + \phi$ \\
							$s_j$ & $1 + \frac{N-1+2j}{2}\tau + \frac{\pi_L - \pi_R}{2} - \phi$ \\
							$a$ & $\tau$ 
						\end{tabular}};
    
    \path (2,0) node[flavor,orange](x1) {$\!2N\!$} -- (4,0) node[gauge](x2) {$\!\!\!2N\!\!\!$} -- (6,0) node[flavor,blue](x3) {$\!2\!$} 
    	-- (3,-1.5) node[flavor,violet](fU) {$2$} -- (5,-1.5) node[flavor,red](fD) {$2$};
    
    \wigE (x1) -- (x2); \draw (3,0.35) node {$\Pi'_L$};
    \draw[-] (x3) -- (x2); \draw (5,0) node[cross] {}; \draw (5,0.35) node {$b'_R$};
    \draw[-] (x2) -- (fU); \draw (3.5,-0.7) node[left] {$p'$};
    \draw[-] (x2) -- (fD); \draw (4.5,-0.8) node[left] {$q'$};
    \draw[-] (x1) -- (fU); \draw (2.5,-0.8) node[left] {$r'$};
    \draw[-] (x3) -- (fD); \draw (5.5,-0.7) node[left] {$s'_j$};
    \draw [-] (x2) to[out=60, in=0] (4,0.6) to[out=180,in=120] (x2); \draw (4,0.75) node[right] {$a'$};     
    
    \draw (1,-2.5) node[right]{$\cW = a' \mathsf{A}'_L + \sum_{j=0}^N Flip[{b'_R}^2 a^j] + $};
    \draw (1.5,-3.1) node[right]{$ + r' \Pi_L p' + \sum_{j=1}^N s'_j b'_R a^{j-1} q'$};

	\path (4,-5.5) node {\begin{tabular}{c|c}
							$\Pi'_L$ & $\pi_R$ \\
							$b'_R$ & $\frac{1-N}{2}\tau + \pi_L$ \\
							$p',q'$ & $1 - \frac{\pi_L + \pi_R}{2} \pm \phi$ \\
							$r'$ & $1 + \frac{\pi_L - \pi_R}{2} - \phi$ \\
							$s'_j$ & $1 + \frac{N+1-2j}{2}\tau - \frac{\pi_L - \pi_R}{2} + \phi$ \\
							$a'$ & $\tau$ 
						\end{tabular}};
						
\epic\ee
The associated SCI identity is:
\begin{align}
	& \oint d\vec{z}_N \D_N(\vec{z},t) \CI_{FE}^{(N)}(\vec{x},\vec{z},t,c_L) \prod_{j=1}^N \big[ \Ge( t^{\frac{1-N}{2}} c_R z_j^\pm y^\pm ) \nn \\
	& \Ge((pq)^{1/2} (c_L c_R)^{-1/2} f^{-1} z_j^\pm v_1^\pm) \Ge((pq)^{1/2} (c_L c_R)^{-1/2} f z_j^\pm v_2^\pm \nn \\
	& \Ge((pq)^{1/2} (c_L/c_R)^{-1/2} f x_j^\pm v_1^\pm ) \prod_ {k=1}^N \Ge((pq)^{1/2} t^{\frac{N+1-2k}{2}} (\pi_L/\pi_R)^{1/2} f^{-1} y_j^\pm v_2^\pm ) \big]  = \nn \\
	= & \oint d\vec{z}_N \D_N(\vec{z},t) \CI_{FE}^{(N)}(\vec{x},\vec{z},t,c_R) \big[ \Ge( t^{\frac{1-N}{2}} c_L z_j^\pm y^\pm ) \nn \\
	& \Ge((pq)^{1/2} (c_L c_R)^{-1/2} f z_j^\pm v_2^\pm) \Ge((pq)^{1/2} (c_L c_R)^{-1/2} f^{-1} z_j^\pm v_1^\pm \nn \\
	& \Ge((pq)^{1/2} (c_L/c_R)^{1/2} f^{-1} x_j^\pm v_2^\pm ) \prod_ {k=1}^N \Ge((pq)^{1/2} t^{\frac{N+1-2k}{2}} (\pi_R/\pi_L)^{1/2} f y_j^\pm v_1^\pm ) \big] \,.
\end{align}

\subsection{$3d$ dualities}\label{app:3d_starstar}

Starting from the $4d$  star-star duality in \eqref{app:4d_starstar} we can perform a circle compactification followed by  a series of suitable real mass deformations
(along the lines of the discussion in Section \ref{3dred}) to obtain the $3d$ swapping duality:
\be \label{fig:3d_starstar}
 \bpic[thick,node distance=3cm,gauge/.style={circle,draw,minimum size=5mm},flavor/.style={rectangle,draw,minimum size=5mm}]  
    
    \path (-6,0) node[flavor,orange](x1) {$\!N\!$} -- (-4,0) node[gauge](x2) {$\!\!\!N\!\!\!$} -- (-2,0) node[flavor,blue](x3) {$\!N\!$} 
    	 --  (0,0) node{$\Longleftrightarrow$};
    
    \wigM (x1) -- (x2); \draw (-5,0.4) node {$\Pi_L$};
    \wigM (x3) -- (x2); \draw (-3,0.4) node {$\Pi_R$};
    \draw [-] (x2) to[out=60, in=0] (-4,0.6) to[out=180,in=120] (x2); \draw (-4,0.7) node[right] {$a$};
    
    \draw (-4,-1) node{$\cW = a (\mathsf{A}_L + \mathsf{A}_R )$};
	
	\path (-4,-2.5) node {\begin{tabular}{c|c}
							$\Pi_L, \tilde{\Pi}_L$ & $\D_L$ \\
							$\Pi_R, \tilde{\Pi}_R$ & $\D_R$ \\
							$a$ & $\tau$ 
						\end{tabular}};
    
    \path (2,0) node[flavor,orange](x1) {$\!N\!$} -- (4,0) node[gauge](x2) {$\!\!\!N\!\!\!$} -- (6,0) node[flavor,blue](x3) {$\!N\!$};
    	
    \wigM (x1) -- (x2); \draw (3,0.4) node {$\Pi'_L$};
    \wigM (x3) -- (x2); \draw (5,0.4) node {$\Pi'_R$};
    \draw [-] (x2) to[out=60, in=0] (4,0.6) to[out=180,in=120] (x2); \draw (4,0.75) node[right] {$a'$};
  
    \draw (4,-1) node {$\cW = a'( \mathsf{A}'_L + \mathsf{A}'_R ) $};
	
	\path (4,-2.5) node {\begin{tabular}{c|c}
							$\Pi'_L, \tilde{\Pi}'_L$ & $\D_R$ \\
							$\Pi'_R, \tilde{\Pi}'_R$ & $\D_L$ \\
							$a$ & $\tau$ 
						\end{tabular}};
						
\epic\ee
The associated identity between partition functions is:
\begin{align}
	&\int d\vec{Z}_N \D_N( \vec{Z},\tau ) e^{2 \pi i W \sum_{j=1}^N Z_j} Z_{FM}^{(N)}(\vec{X},\vec{Z},\tau, \D_L) Z_{FM}^{(N)}(\vec{Z},\vec{Y},\tau,\D_R) = \nn \\
	= & e^{2\pi i W \sum_{j=1}^N ( X_j + Y_j) } \int d\vec{Z}_N \D_N( \vec{Z},\tau ) e^{-2 \pi i W \sum_{j=1}^N Z_j} Z_{FM}^{(N)}(\vec{X},\vec{Z},\tau, \D_R) Z_{FM}^{(N)}(\vec{Z},\vec{Y},\tau,\D_L) \,.
\end{align}
By breaking one of the two $U(N)$ symmetries down to $U(1)$ in the previous duality we get:
\be\label{fig:3d_starstar_specialised}
 \bpic[thick,node distance=3cm,gauge/.style={circle,draw,minimum size=5mm},flavor/.style={rectangle,draw,minimum size=5mm}]  
    
    \path (-6,0) node[flavor,orange](x1) {$\!N\!$} -- (-4,0) node[gauge](x2) {$\!\!\!N\!\!\!$} -- (-2,0) node[flavor,blue](x3) {$\!1\!$} 
    	 --  (0,0) node{$\Longleftrightarrow$};
    
    \wigM (x1) -- (x2); \draw (-5,0.4) node {$\Pi_L$};
    
    \draw[-, shorten >= 6, shorten <= 9, shift={(-0.05,0.07)}, midsx arrowsm] (-4,0) -- (-2,0);
	\draw[-, shorten >= 6, shorten <= 9, shift={(0.05,-0.07)}, middx arrowsm] (-2,0) -- (-4,0); 
	\draw (-2.65,0) node[cross] {}; \draw (-3,0.4) node {$b_R$};
    
    \draw [-] (x2) to[out=60, in=0] (-4,0.6) to[out=180,in=120] (x2); \draw (-4,0.7) node[right] {$a$};
    
    \draw (-4,-1) node {$\cW = a \mathsf{A}_L + \sum_{j=0}^{N-1} Flip[b_R a^j \tilde{b}_R] $};
	
	\path (-4,-2.5) node {\begin{tabular}{c|c}
							$\Pi_L, \tilde{\Pi}_L$ & $\D_L$ \\
							$b_R, \tilde{b}_R$ & $\frac{1-N}{2}\tau +  \D_R \mp Y$ \\
							$a$ & $\tau$ 
						\end{tabular}};
    
    \path (2,0) node[flavor,orange](x1) {$\!N\!$} -- (4,0) node[gauge](x2) {$\!\!\!N\!\!\!$} -- (6,0) node[flavor,blue](x3) {$\!1\!$};
    	
    \wigM (x1) -- (x2); \draw (3,0.4) node {$\Pi'_L$};
    
    \draw[-, shorten >= 6, shorten <= 9, shift={(-0.05,0.07)}, midsx arrowsm] (4,0) -- (6,0);
	\draw[-, shorten >= 6, shorten <= 9, shift={(0.05,-0.07)}, middx arrowsm] (6,0) -- (4,0);
	\draw (5.35,0) node[cross] {}; \draw (5,0.4) node {$b'_R$};
	
    \draw [-] (x2) to[out=60, in=0] (4,0.6) to[out=180,in=120] (x2); \draw (4,0.75) node[right] {$a'$};
  
    \draw (4,-1) node {$\cW = a' \mathsf{A}'_L + \sum_{j=0}^{N-1} Flip[b'_R {a'}^j \tilde{b}'_R] $};
	
	\path (4,-2.5) node {\begin{tabular}{c|c}
							$\Pi'_L, \tilde{\Pi}'_L$ & $\D_R$ \\
							$b'_R, \tilde{b}'_R$ & $\frac{1-N}{2}\tau + \D_L \mp Y$ \\
							$a'$ & $\tau$ 
						\end{tabular}};
						
\epic\ee
The associated identity between partition functions is:
\begin{align}
	&\int d\vec{Z}_N \D_N( \vec{Z},\tau ) e^{2 \pi i W \sum_{j=1}^N Z_j} Z_{FM}^{(N)}(\vec{X},\vec{Z},\tau, \D_L) \nn \\
	& \prod_{j=1}^N \big[ s_b( \frac{iQ}{2} - \frac{1-N}{2}\tau - \D_R \pm (Z_j - Y)) s_b( -\frac{iQ}{2} + (j-N)\tau + 2\D_R ) \big] = \nn \\
	= & e^{2\pi i W (\sum_{j=1}^N X_j + N Y_j) } \int d\vec{Z}_N \D_N( \vec{Z},\tau ) e^{-2 \pi i W \sum_{j=1}^N Z_j} 
	Z_{FM}^{(N)}(\vec{X},\vec{Z},\tau, \D_R) \nn \\
	& \prod_{j=1}^N \big[ s_b( \frac{iQ}{2} - \frac{1-N}{2}\tau - \D_L \pm (Z_j - Y))  s_b( -\frac{iQ}{2} + (j-N)\tau + 2\D_L ) \big] \,.
\end{align}\\

It is also possible to prove the following duality:
\be\label{fig:FM_dressedmonopole}
 \bpic[thick,node distance=3cm,gauge/.style={circle,draw,minimum size=5mm},flavor/.style={rectangle,draw,minimum size=5mm}]  
    
    \path (-6,0) node[flavor,orange](x1) {$\!N\!$} -- (-4,0) node[gauge](x2) {$\!\!\!N\!\!\!$} -- (-2,0) node[flavor,blue](x3) {$\!N\!$} 
    	 --  (0,0) node{$\Longleftrightarrow$};
    
    \wigM (x1) -- (x2); \draw (-5,0.4) node {$\Pi_L$};
    \wigM (x3) -- (x2); \draw (-3,0.4) node {$\Pi_R$};
    \draw [-] (x2) to[out=60, in=0] (-4,0.6) to[out=180,in=120] (x2); \draw (-4,0.7) node[right] {$a$};
    
    \draw (-4,-1) node {$\cW = a ( \mathsf{A}_L + \mathsf{A}_R ) + \mathfrak{M}_a^+$};
	
	\path (-4,-2.5) node {\begin{tabular}{c|c}
							$\Pi, \tilde{\Pi}$ & $\D_L/2 + \tau/4 + \phi$ \\
							$\Pi, \tilde{\Pi}$ & $\D_R/2 + \tau/4 - \phi$ \\
							$a$ & $\tau$ 
						\end{tabular}};
    
    \path (2,0) node[flavor,orange](x1) {$\!N\!$} -- (4,0) node[gauge](x2) {$\!\!\!N\!\!\!$} -- (6,0) node[flavor,blue](x3) {$\!N\!$};
    
    \draw[-, shorten >= 6, shorten <= 9, shift={(-0.05,-0.07)}, middx arrowsm] (2,0) -- (4,0);
	\draw[-, shorten >= 6, shorten <= 9, shift={(0.05,0.07)}, midsx arrowsm] (4,0) -- (2,0); 
    \draw (2.5,0) node[cross] {}; \draw (3,0.4) node {$b'_L$};
    
    \wigM (x3) -- (x2); \draw (5,0.4) node {$\Pi'_R$};
    \draw [-] (x1) to[out=60, in=0] (2,0.6) to[out=180,in=120] (x1); \draw (2,0.75) node[right] {${\color{orange}a}'$};
    
    \draw (4,-1) node {$\cW = b'_L ({\color{orange}a}' + \mathsf{A}_R) \tilde{b}'_L + Flip[b'_L \tilde{b}'_L] $};
    
    \path (4,-2.5) node {\begin{tabular}{c|c}
							$b'_L, \tilde{b}'_L$ & $\tau/2$ \\
							$\Pi'_R, \tilde{\Pi}'_R$ & $\D$ \\
							${\color{orange}a}'$ & $2-\tau$  
						\end{tabular}};
						
\epic\ee
The associated identity between partition functions is:
\begin{align}
	& \oint d\vec{Z}_N \D_N(\vec{Z},\tau) e^{2 \pi i (\D - \tau/2) \sum_{j=1}^N Z_j } 
	Z_{FE}^{(N)}(\vec{X},\vec{Z},\tau,\D/2 + \tau/4 + \phi) \nn \\
	& Z_{FE}^{(N)}(\vec{Z},\vec{Y},\tau,\D/2 + \tau/4 - \phi) = \nn \\
	= & e^{2\pi i (\frac{\tau}{4} - \frac{\D}{2} - \phi) \sum_{j=1}^N( X_j + Y_j) } \oint d\vec{Z}_N \D_N(\vec{Z},\tau) 
	e^{4 \pi i \phi  \sum_{j=1}^N Z_j } Z_{FE}^{(N)}(\vec{Z},\vec{Y},\tau,\D) \nn \\ 
	& s_b(-\frac{iQ}{2} + \tau) \prod_{j,k=1}^N s_b( \frac{iQ}{2} - \frac{\tau}{2} \pm (X_j - Z_k) )  \,.
\end{align}


\section{Monopole R-charge}\label{app:FMmonopoles}

In this section we discuss the monopoles in the SQCD mirror.

Let's first focus on monopoles with unit magnetic flux $\mathfrak{M}^{\pm1,0,\cdots,0}$, charged under the $U(1)_{X_2-X_1}$ topological symmetry.  In this case we can easily calculate its R-charge by considering the Lagrangian description of the improved bifundamental  assuming that we are gauging its manifest symmetry.
By doing so we find:
\begin{align}
R[\mathfrak{M}^{(\pm1,0,\cdots,0)}]&=(N-1)(1-\tau/2)+(1-B_2)+(N-1)(1-2+\tau)+ \nn\\
&+(1-B_1-(N-1)\tau/2) +(N-1)(1-\tau) -(N-1)=
2-B_1-B_2 \,,
\end{align}
in the first line we have the improved bifundamental contribution given by the contribution 
 $(N-1)$ flavors with R-charge $\tau/2$, one flavor with charge $B_2$ and one adjoint chiral with charge $2-\tau$.
 We then have the contribution of the $V_1\tilde V_1$ flavor, the adjoint $a$ and the vector multiplet.
As expected this matches the corresponding meson R-charge $R[Q_1\tilde Q_2]=R[Q_2\tilde Q_1]=2-B_1-B_2$.

We can then consider  monopoles with magnetic flux $\mathfrak{M}^{(0,\cdots,\pm1,0,\cdots,0)}$, charged under the $U(1)_{X_i-X_{i-1}}$ topological symmetry. 
In this case we have to take into account the contribution of the improved bifundamentals on the left and on the right of the node. Thanks to the self-mirror property of the improved bifundamentals we can always calculate this contribution assuming that we are gauging the manifest symmetries of the two improved bifundamentals. So we have:
 
\begin{align}
R[\mathfrak{M}^{(0,\cdots,\pm1,0,\cdots,0)}]&=(N-1)(1-\tau/2)+(1-B_i)+(N-1)(1-2+\tau)+ \nn\\
&+(N-1)(1-\tau/2)+(1-B_{i+1})+(N-1)(1-2+\tau)+ \nn\\
&+(N-1)(1-\tau) -(N-1)=2-B_i-B_{i+1} \,,
\end{align}
in the first two lines we have the contributions of the left and right improved bifundamentals
in the last line, the contribution of the gluing adjoing $a$ and of the vector.

As expected this matches the corresponding meson R-charge $R[Q_i\tilde Q_{i+1}]=R[Q_{i+1}\tilde Q_i]=2-B_i-B_{i+1}$.

To calculate the R-charge of the other monopoles with magnetic flux given by strings of consecutive $\pm1$ we need the contribution of the genearalised bifundamental when we simultaneously gauge its manifest and emergent symmetry which we can't directly calculate 
from the Lagrangian.

For example,  the R-charge of the  monopole charged under the second and third gauge node is given by:
\begin{align}
R[\mathfrak{M}^{(0,\pm1,\pm1,0,\cdots,0)}]&=(N-1)(1-\tau/2)+(1-B_1)+(N-1)(1-2+\tau)+ \nn\\
&+(N-1)(1-\tau/2)+(1-B_{4})+(N-1)(1-2+\tau)+ \nn\\
&+2(N-1)(1-\tau) -2(N-1)+GB[\pm1,\pm1]= \nn \\
&= 2-B_2-B_{4} \,,
\end{align}
in the first two lines we have the contribution of the first and fourth genearlised bifundamental which we can calculate using the 
Lagrangian description. In the third line we have the contribution of the adjoints and vector mutiplets at the gauged nodes
and the contribution to the third improved bifundamental which we conjecture to be:
\be
\label{assumption}
GB[\pm1,\pm1]=(N-1) (-\tau) \,,
\ee
as the contribution of an ordinary bifundamental flavor of charge $1-\tau/2$. 
Assuming (\ref{assumption}) the charged of the monoples match those of the electric $Q_i\tilde Q_{i+k}$ mesons, we have:
\begin{align}
R[\mathfrak{M}^{(0,\ldots,0,\pm1,\ldots,\pm1,0,\ldots,0)}] &= (N-1)(1-\tau/2)+(1-B_1)+(N-1)(1-2+\tau)+ \nn\\
&+(N-1)(1-\tau/2)+(1-B_{4})+(N-1)(1-2+\tau)+ \nn\\
&+(k+1)(N-1)(1-\tau) -(k+1)(N-1)+ k GB[\pm1,\pm1] \nn\\
&=2-B_j-B_{j+k+1} \,.
\end{align}

We checked this assumption with the index where we can see that it gives the correct R-charge of monopoles
visibile in the expansion at low $N_c$ and $N_f$,
In particular in  the abelian case the R-charge of the monopoles can be computed exactly since the improved bifundamental reduces to just a standard one and we can verify that the assumption is correct in this case.

One can also play a similar game in the $SU(N)$ SQCD mirror (see section \ref{sec:SUN_SQCD}) to establish a map for the baryons. In this case the problem is more complicated and we do not have a complete closed formula for the contribution of an improved bifundamental to the R-charge of a monopole. However we observed empirically that the baryon map can be established by assuming the following formulae:
\begin{align}
	& GB[\pm m, \pm (m-1)] = \begin{cases}
	m(N-m)\tau + (1-\D)	\qquad \text{for} \quad 1 < m < N \\
	(N-1)\tau + (1-\D) \qquad \text{for} \quad m=1,N \\
	\end{cases} \,, \\
	& GB[\pm m, \pm m] = m(N-m)\tau \,.
\end{align}
The subcase $m=1$ of these assumptions indeed coincide with the result found for the map of the mesons. A more generic formula could be provided by the understanding of the operator map for the dressed baryons in the $SU(N)$ SQCD, however we do not have a clear solution to this problem and we address this to a future work.

\section{Quiver mirror pair via the dualization algorithm}\label{app:Quiver_algorithm}
In the following section we present how the duality proposed in \ref{fig:MoreNodes_ex} can be derived using the dualization algorithm. \\
We start from the electric theory as parameterized in figure \eqref{fig:MoreNodes_Elec_manifest}. We then cut the theory into $\CN=2$ QFT blocks as defined in section \ref{sec:3d_algorithm}, we obtain:
\be 
 \begin{tikzpicture}[thick,node distance=3cm,gauge/.style={circle,draw,minimum size=5mm},flavor/.style={rectangle,draw,minimum size=5mm}]
 
	\path (0,0) node[flavor](g1) {$\!0\!$} -- (1.5,0) node[flavor](g2) {$\!N\!$} 
		-- (2.5,0) node[flavor](g3) {$\!N\!$} -- (2,1.5) node[flavor](x1) {$\!1\!$} -- (3,1.5) node[flavor](x2) {$\!1\!$}
		-- (4,0) node[flavor](g4) {$\!N\!$} -- (5.5,0) node[gauge](g5) {$\!\!\!N\!\!\!$} -- (7.5,0) node[flavor](g6) {$\!N\!$}
		-- (8.5,0) node[flavor](g7) {$\!N\!$} -- (8,1.5) node[flavor](y1) {$\!1\!$} -- (9,1.5) node[flavor](y2) {$\!1\!$} 
		-- (10,0) node[flavor](g8) {$\!N\!$} -- (11.5,0) node[flavor](g9) {$\!0\!$};
		
	\draw[-, shorten >= 6, shorten <= 9, shift={(-0.05,-0.07)}, mid arrowsm] (0,0) -- (1.5,0);
	\draw[-, shorten >= 6, shorten <= 9, shift={(0.05,0.07)}, mid arrowsm] (1.5,0) -- (0,0); 
	\draw[blue] (1.5,-0.5) node {\scriptsize{$(W_L)$}};
	
	\draw[-, shorten >= 6, shorten <= 10, shift={(-0.05,-0.07)}, mid arrowsm] (2.5,0) -- (2,1.5);
	\draw[-, shorten >= 6, shorten <= 10, shift={(0.05,0.07)}, mid arrowsm] (2,1.5) -- (2.5,0); 
	\draw[-, shorten >= 10, shorten <= 6, shift={(-0.05,0.07)}, mid arrowsm] (2.5,0) -- (3,1.5);
	\draw[-, shorten >= 10, shorten <= 6, shift={(0.05,-0.07)}, mid arrowsm] (3,1.5) -- (2.5,0); 
	\draw[blue] (2.25,0.75) node[left] {\scriptsize{$1-B_1$}};
	\draw[blue] (2.8,0.75) node[right] {\scriptsize{$1-B_{F_1}$}};
	\draw (2.8,0) node[right] {\large{$\mathbb{I}(\tau)$}};
	
	\wigM (g4) -- (g5); \draw[blue] (4.75,0.35) node {\scriptsize{$D_1$}};
	\wigM (g5) -- (6.25,0);
	\wigM (g6) -- (6.75,0);
	\draw[-] (g5) to[out=60,in=0] (5.5,0.6) to[out=180,in=120] (g5); \draw[blue] (5.5,0.8) node {\scriptsize{$\tau$}};
	\draw[blue] (4,-0.5) node {\scriptsize{$(-W_1)$}}; \draw[blue] (5.5,-0.5) node {\scriptsize{$(W_1-W_2)$}}; 
	\draw[blue] (7.5,-0.5) node {\scriptsize{$(W_K)$}};
	
	\draw[-, shorten >= 6, shorten <= 10, shift={(-0.05,-0.07)}, mid arrowsm] (8.5,0) -- (8,1.5);
	\draw[-, shorten >= 6, shorten <= 10, shift={(0.05,0.07)}, mid arrowsm] (8,1.5) -- (8.5,0); 
	\draw[-, shorten >= 10, shorten <= 6, shift={(-0.05,0.07)}, mid arrowsm] (8.5,0) -- (9,1.5);
	\draw[-, shorten >= 10, shorten <= 6, shift={(0.05,-0.07)}, mid arrowsm] (9,1.5) -- (8.5,0);
	\draw[blue] (8.25,0.75) node[left] {\scriptsize{$1-C_1$}};
	\draw[blue] (8.8,0.75) node[right] {\scriptsize{$1-C_{F_2}$}};
	\draw (8.8,0) node[right] {\large{$\mathbb{I}(\tau)$}};
	
	\draw[-, shorten >= 6, shorten <= 9, shift={(-0.05,-0.07)}, mid arrowsm] (10,0) -- (11.5,0);
	\draw[-, shorten >= 6, shorten <= 9, shift={(0.05,0.07)}, mid arrowsm] (11.5,0) -- (10,0);
	\draw[blue] (10,-0.5) node {\scriptsize{$(-W_R)$}};
									  
\end{tikzpicture}\ee
This consist in the following partition function identity:
\begin{align}\label{eq:quiverdual_step1}
	& \int \prod_{a=1}^{K+1} \big( d\vec{Z}^{(a)}_N \D_N (\vec{Z}^{(a)},\tau) \big) e^{2\pi i W_L \sum_{j=1}^N Z_j^{(a)} } 
	\prod_{j=1}^N \prod_{a=1}^{F_1} s_b( B_a \pm (Z^{(1)}_j - X_a) ) \nn \\
	& \prod_{a=1}^{K} Z_{NS}^{(N)}(\vec{Z}^{(a)},\vec{Z}^{(a+1)}, \tau, B_a,  -W_a) 
	\prod_{j=1}^N \prod_{a=1}^{F_2} s_b( C_a \pm (Z^{(K+1)}_j - Y_a) ) e^{-2\pi i W_R \sum_{j=1}^N Z^{(K+1)}_j } = \nn \\
	= & \int \prod_{a=1}^{K+1} \big( d\vec{Z}^{(a)}_N \D_N (\vec{Z}^{(a)},\tau) \big) \prod_{a=1}^2 \big( d\vec{W}^{(a)}_N \D_N (\vec{W}^{(a)}, \tau) \big)
	e^{2\pi i W_L \sum_{j=1}^N Z_j^{(a)} }  \nn \\
	& \prod_{j=1}^N \prod_{a=1}^{F_1} s_b( B_a \pm (Z^{(1)}_j - X_a) ) {}_{\vec{Z}^{(1)}}\mathbb{I}_{\vec{W}^{(1)}} (\tau) 
	\prod_{a=1}^{K} Z_{NS}^{(N)}(\vec{Z}^{(a)},\vec{Z}^{(a+1)}, \tau, D_a,  -W_a) \nn \\
	& \prod_{j=1}^N \prod_{a=1}^{F_2} s_b( C_a \pm (Z^{(K+1)}_j - Y_a) ) {}_{\vec{Z}^{(K+1)}}\mathbb{I}_{\vec{W}^{(2)}} (\tau) 
	e^{-2\pi i W_R \sum_{j=1}^N W^{(2)}_j } \,,
\end{align}
which is a trivial identity after we use the two identity operators to cancel the two extra integrations, taking into account that:
\begin{align}
	\int d\vec{W} \D_N (\vec{W},\tau) {}_{\vec{Z}}\mathbb{I}_{\vec{W}} (\tau) = 1 \,.
\end{align}

Then dualize each block using one of the basic duality moves \ref{fig:3d_flavors_basicmove} and \ref{fig:3d_asymm_basicmove}. Performing this dualization and then gluing back all the results we obtain: 
\be
\resizebox{.95\hsize}{!}{
\begin{tikzpicture}[thick,node distance=3cm,gauge/.style={circle,draw,minimum size=5mm},flavor/.style={rectangle,draw,minimum size=5mm}]
	
	\path (0,0) node[flavor] (f1) {$\!0\!$} -- (1.5,0) node[gauge] (g1) {$\!\!\!0\!\!\!$} -- (1.5,1.5) node[flavor] (x1) {$\!1\!$} 
		-- (3,0) node[gauge] (g2) {$\!\!\!N\!\!\!$} -- (4.5,0) node[gauge] (g3) {$\!\!\!N\!\!\!$} -- (6,0) node[gauge] (g4) {$\!\!\!N\!\!\!$} 
		-- (8,0) node[gauge] (g5) {$\!\!\!N\!\!\!$} -- (7.5,1.5) node[flavor](y1) {$\!1\!$} -- (8.5,1.5) node[flavor](y2) {$\!1\!$}
		-- (10,0) node[gauge] (g6) {$\!\!\!N\!\!\!$} -- (11.5,0) node[gauge] (g7) {$\!\!\!N\!\!\!$} 
		-- (13,0) node[gauge] (g8) {$\!\!\!N\!\!\!$} -- (14.5,0) node[gauge] (g9) {$\!\!\!0\!\!\!$}
		-- (14.5,1.5) node[flavor] (x2) {$\!1\!$} -- (16,0) node[flavor] (f2) {$\!0\!$};
						
	\wigT (f1) -- (g1); \draw (0.75,-0.3) node {$+$};
	
	\draw[-, shorten >= 6, shorten <= 9, shift={(-0.07,-0.05)}, mid arrowsm] (1.5,0) -- (1.5,1.5);
	\draw[-, shorten >= 6, shorten <= 9, shift={(0.07,0.05)}, mid arrowsm] (1.5,1.5) -- (1.5,0);
	\draw[blue] (1.5,2) node {\scriptsize{$W_L$}};
	 
	\wigT (g1) -- (g2); \draw (2.25,-0.3) node {$-$};
	\wigT (x1) -- (2.25,0);
	\wigT (g2) -- (g3); \draw (3.75,-0.3) node {$+$};
	\wigM (g3) -- (g4); \draw[blue] (5.25,0.35) node {\scriptsize{$B_1$}};
	\wigM (g4) -- (6.75,0); 
	\draw (7,0) node {$\cdots$};
	\wigM (g5) -- (7.25,0);
	
	\draw[-, shorten >= 6, shorten <= 10, shift={(-0.05,-0.07)}, mid arrowsm] (8,0) -- (7.5,1.5);
	\draw[-, shorten >= 6, shorten <= 10, shift={(0.05,0.07)}, mid arrowsm] (7.5,1.5) -- (8,0); 
	\draw (8.02,1.5) node {\small{$\cdots$}};
	\draw[-, shorten >= 6, shorten <= 10, shift={(-0.05,0.07)}, mid arrowsm] (8.5,1.5) -- (8,0);
	\draw[-, shorten >= 6, shorten <= 10, shift={(0.05,-0.07)}, mid arrowsm] (8,0) -- (8.5,1.5);
	\draw[blue] (7.5,2) node {\scriptsize{$W_1$}}; \draw[blue] (8.5,2) node {\scriptsize{$W_K$}};
	\draw[blue] (7.75,0.75) node[left] {\scriptsize{$1-D_1$}}; \draw[blue] (8.25,0.75) node[right] {\scriptsize{$1-D_K$}};
	
	\wigM (g5) -- (8.75,0);
	\draw (9,0) node {$\cdots$};
	\wigM (g6) -- (9.25,0);
	\wigM (g6) -- (g7); \draw[blue] (10.75,0.35) node {\scriptsize{$C_{F_2}$}};
	\wigT (g7) -- (g8); \draw (12.25,-0.3) node {$-$};
	\wigT (g8) -- (g9); \draw (13.75,-0.3) node {$+$};
	\wigT (x2) -- (13.75,0);
	
	\draw[-, shorten >= 6, shorten <= 9, shift={(-0.07,-0.05)}, mid arrowsm] (14.5,0) -- (14.5,1.5);
	\draw[-, shorten >= 6, shorten <= 9, shift={(0.07,0.05)}, mid arrowsm] (14.5,1.5) -- (14.5,0);
	\draw[blue] (14.5,2) node {\scriptsize{$W_R$}};
	
	\wigT (g9) -- (f2); \draw (15.25,-0.3) node {$-$};
	
	\draw (g1) to[out=-60,in=0] (1.5,-0.6) to[out=180,in=-120] (g1);
	\draw (g2) to[out=60,in=0] (3,0.6) to[out=180,in=120] (g2); \draw[blue] (3,0.75) node {\scriptsize{$\tau$}};
	\draw (g3) to[out=60,in=0] (4.5,0.6) to[out=180,in=120] (g3); \draw[blue] (4.5,0.75) node {\scriptsize{$\tau$}};
	\draw (g4) to[out=60,in=0] (6,0.6) to[out=180,in=120] (g4); \draw[blue] (6,0.75) node {\scriptsize{$\tau$}};
	\draw (g5) to[out=-60,in=0] (8,-0.6) to[out=180,in=-120] (g5); \draw[blue] (8.1,-0.6) node[right] {\scriptsize{$\tau$}};
	\draw (g6) to[out=60,in=0] (10,0.6) to[out=180,in=120] (g6); \draw[blue] (10,0.75) node {\scriptsize{$\tau$}};
	\draw (g7) to[out=60,in=0] (11.5,0.6) to[out=180,in=120] (g7); \draw[blue] (11.5,0.75) node {\scriptsize{$\tau$}};
	\draw (g8) to[out=60,in=0] (13,0.6) to[out=180,in=120] (g8); \draw[blue] (13,0.75) node {\scriptsize{$\tau$}};
	\draw (g9) to[out=-60,in=0] (14.5,-0.6) to[out=180,in=-120] (g9);
	
	\draw[blue] (4.5,-0.5) node {\scriptsize{$(X_1)$}}; \draw[blue] (6,-0.5) node {\scriptsize{$(X_2-X_1)$}}; \draw[blue] (8,-1) node {\scriptsize{$(-X_{F_1}+Y_{1})$}}; \draw[blue] (10,-0.5) node {\scriptsize{$(Y_{F_2}$-$Y_{\text{$F_2$-$1$}})$}}; \draw[blue] (11.5,-0.5) node {\scriptsize{$(-Y_{F_2})$}};
	
\end{tikzpicture}}
\ee
In the figure we do not give all the singlets produced from the dualization to avoid cluttering. This step consist in starting from \eqref{eq:quiverdual_step1} and using the basic moves \eqref{eq:basic_moves_F} and \eqref{eq:basic_moves_asymm}, obtaining:
\begin{align}\label{eq:quiverdual_step2}
	& \int \prod_{a=1}^{F_1+4} \big( d\vec{Z}^{(a)}_N \D_N (\vec{Z}^{(a)},\tau) \big) \prod_{a=1}^{F_2+3} \big( d\vec{M}^{(a)}_N \D_N (\vec{M}^{(a)}, \tau) \big) \prod_{j=2}^N s_b( \frac{iQ}{2} - j\tau )^2 \nn \\
	& Z_{\CS^{-1}}^{(N)} ( \{ \frac{N-1}{2}\tau + W_L, \ldots, \frac{1-N}{2}\tau + W_L \}, \vec{Z}^{(1)}, \tau) 
	Z_{\CS}^{(N)} (\vec{Z}^{(1)},\vec{Z}^{(2)},\tau)  \nn \\
	& \prod_{a=1}^{F_1} Z_{NS}^{(N)}(\vec{Z}^{(a+1)}, \vec{Z}^{(a+2)}, \tau, B_a, -X_a)
	Z_{\CS^{-1}}^{(N)} ( \vec{Z}^{(F_1+2)}, \vec{Z}^{(F_1+3)}, \tau) Z_{\CS}^{(N)} (\vec{Z}^{(F_1+3)}, \vec{Z}^{(F_1+4)}, \tau) \nn \\
	& \prod_{j=1}^N \prod_{a=1}^K s_b ( D_a \pm ( Z^{(F_1+4)}_j - W_a ) )
	Z_{\CS^{-1}}^{(N)} ( \vec{Z}^{(F_1+4)}, \vec{M}^{(1)}, \tau) Z_{\CS}^{(N)}( \vec{M}^{(1)}, \vec{M}^{(2)}, \tau) \nn \\
	& \prod_{a=1}^{F_2} Z_{NS}^{(N)}(\vec{M}^{(a+1)}, \vec{M}^{(a+2)}, \tau, C_a, -Y_a) 
	Z_{\CS^{-1}}^{(N)} (\vec{M}^{(F_2+2)}, \vec{M}^{(F_2+3)},\tau) \nn \\
	& Z_{\CS}^{(N)} ( \vec{M}^{(F_2+3)}, \{ \frac{N-1}{2}\tau + W_R, \ldots, \frac{1-N}{2}\tau + W_R \},  \tau) \,.
\end{align}

We now get rid of the identity walls. We recall that the effect of the asymmetric $0-N$ identity wall is to break the first (and similarly the last) $U(N)$ gauge symmetry down to $U(1)$, the effect of such deformation in an improved bifundamental is to make it into a flavor using the duality \eqref{fig:FMtoFlav}. We then  get: 
\be
\resizebox{.95\hsize}{!}{
\begin{tikzpicture}[thick,node distance=3cm,gauge/.style={circle,draw,minimum size=5mm},flavor/.style={rectangle,draw,minimum size=5mm}]
	\path (0,0) node[gauge](g1) {$\!\!\!N\!\!\!$} -- (2,0) node[gauge](g2) {$\!\!\!N\!\!\!$} -- (5,0) node[gauge] (g3) {$\!\!\!N\!\!\!$} 
	 		-- (7,0) node[gauge] (g4) {$\!\!\!N\!\!\!$} -- (9,0) node[gauge] (g5) {$\!\!\!N\!\!\!$} -- (12,0) node[gauge] (g6) {$\!\!\!N\!\!\!$} 
	 		-- (14,0) node[gauge] (g7) {$\!\!\!N\!\!\!$} -- (0,1.5) node[flavor] (x1) {$\!1\!$} -- (6.25,1.5) node[flavor] (x2) {$\!1\!$}
	 		-- (7.75,1.5) node[flavor] (x22) {$\!1\!$} -- (14,1.5) node[flavor] (x3) {$\!1\!$};
 
	\wigM (g1) -- (g2); \draw[blue] (1,0.3) node {\scriptsize{$B_2$}};
	\wigM (g2) -- (3,0); 
	\draw (3.5,0) node {$\cdots$};
	\wigM (g3) -- (4,0);
	\wigM (g3) -- (g4); \draw[blue] (6,0.3) node {\scriptsize{$B_{F_1}$}};
	\wigM (g4) -- (g5); \draw[blue] (8,0.3) node {\scriptsize{$C_1$}};
	\wigM (g5) -- (10,0);
	\draw (10.5,0) node {$\cdots$};
	\wigM (g6) -- (11,0);
	\wigM (g6) -- (g7); \draw[blue] (13,0.3) node {\scriptsize{$C_{F_2-1}$}};
	
	\draw[-, shorten >= 6, shorten <= 9, shift={(-0.07,-0.05)}, midsx arrowsm] (0,0) -- (0,1.5);
	\draw[-, shorten >= 6, shorten <= 9, shift={(0.07,0.05)}, middx arrowsm] (0,1.5) -- (0,0); 
	\draw (0,0.9) node[cross] {}; \draw[blue] (0,0.75) node [left] {\scriptsize{$\frac{1-N}{2}\tau + B_1$}}; \draw[blue] (0,2) node {\scriptsize{$W_L$}};
	
	\draw[-, shorten >= 6, shorten <= 10, shift={(-0.05,-0.07)}, mid arrowsm] (7,0) -- (6.25,1.5);
	\draw[-, shorten >= 6, shorten <= 10, shift={(0.05,0.07)}, mid arrowsm] (6.25,1.5) -- (7,0); 
	\draw[-, shorten >= 6, shorten <= 10, shift={(-0.05,0.07)}, mid arrowsm] (7.75,1.5) -- (7,0);
	\draw[-, shorten >= 6, shorten <= 10, shift={(0.05,-0.07)}, mid arrowsm] (7,0) -- (7.75,1.5);
	\draw[blue] (6.625,0.75) node[left] {\scriptsize{$1-D_1$}}; \draw[blue] (6.25,2) node {\scriptsize{$W_1$}};
	\draw (7,1.5) node {$\cdots$};
	\draw[blue] (7.375,0.75) node[right] {\scriptsize{$1-D_K$}}; \draw[blue] (7.75,2) node {\scriptsize{$W_K$}};
	
	\draw[-, shorten >= 6, shorten <= 9, shift={(-0.07,-0.05)}, midsx arrowsm] (14,0) -- (14,1.5);
	\draw[-, shorten >= 6, shorten <= 9, shift={(0.07,0.05)}, middx arrowsm] (14,1.5) -- (14,0); 
	\draw (14,0.9) node[cross] {}; \draw[blue] (14,0.75) node [right] {\scriptsize{$\frac{1-N}{2}\tau + C_{F_2}$}}; \draw[blue] (14,2) node {\scriptsize{$W_R$}};
	
	\draw[blue] (0,-0.5) node {\scriptsize{$(X_2 - X_1)$}}; \draw[blue] (2,-0.5) node {\scriptsize{$(X_3 - X_2)$}};
	\draw[blue] (5,-0.5) node {\scriptsize{$(X_{F_1} - X_{F_1-1})$}}; \draw[blue] (7,-1) node {\scriptsize{$(Y_1 - X_{F_1})$}};
	\draw[blue] (9,-0.5) node {\scriptsize{$(Y_2 - Y_1)$}}; \draw[blue] (11.9,-0.5) node {\scriptsize{$(Y_{F_2-1} - Y_{F_2-2})$}};
	\draw[blue] (14.1,-0.5) node {\scriptsize{$(Y_{F_2} - Y_{F_2-1})$}};
	\draw[-] (g1) to[out=150,in=90] (-0.6,0) to[out=-90,in=-150] (g1); \draw[blue] (-0.6,0) node[left] {\scriptsize{$\tau$}};
	\draw[-] (g2) to[out=60,in=0] (2,0.6) to[out=180,in=120] (g2); \draw[blue] (2,0.7) node[right] {\scriptsize{$\tau$}};
	\draw[-] (g3) to[out=60,in=0] (5,0.6) to[out=180,in=120] (g3); \draw[blue] (5,0.7) node[right] {\scriptsize{$\tau$}};
	\draw[-] (g4) to[out=-60,in=0] (7,-0.6) to[out=180,in=-120] (g4); \draw[blue] (7,-0.7) node[right] {\scriptsize{$\tau$}};
	\draw[-] (g5) to[out=60,in=0] (9,0.6) to[out=180,in=120] (g5); \draw[blue] (9,0.7) node[right] {\scriptsize{$\tau$}};
	\draw[-] (g6) to[out=60,in=0] (12,0.6) to[out=180,in=120] (g6); \draw[blue] (12,0.7) node[right] {\scriptsize{$\tau$}};
	\draw[-] (g7) to[out=30,in=90] (14.6,0) to[out=-90,in=-30] (g7); \draw[blue] (14.6,0) node[right] {\scriptsize{$\tau$}};

\end{tikzpicture}}
\ee
Which is the result depicted in \ref{fig:MoreNodes_Magn_manifest}. Indeed, evaluating the result for $K=1$ gives the mirror dual in \eqref{fig:TwoNodes_Magn}. This step consist in starting from \eqref{eq:quiverdual_step2} and using the formula for the $\mathbb{I}$-walls \eqref{FTdelta} and then the duality for the asymmetric $FM[U(N)]$ to a flavor \eqref{eq:FMtoFlav}. We then get the final result:
\begin{align}
	& \int \prod_{a=1}^{F_1+F_2+1} \big( d\vec{Z}^{(a)}_N \D_N (\vec{Z}^{(a)},\tau) \big) 
	e^{2 \pi i ( -X_1 \sum_{j=1}^N Z^{(1)}_j +  Y_{F_2} \sum_{j=1}^N Z^{(F_1+F_2+1)}_j )} \nn \\
	& \prod_{a=2}^{F_1} Z_{NS}^{(N)}(\vec{Z}^{(a-1)}, \vec{Z}^{(a)}, \tau, B_a, -X_a)
	\prod_{j=1}^N \prod_{a=1}^K s_b ( D_a \pm ( Z^{(F_1)}_j - W_a ) )  \nn \\
	& \prod_{a=1}^{F_2} Z_{NS}^{(N)}(\vec{Z}^{(F_1+a-1)}, \vec{M}^{(F_1+a)}, \tau, C_a, -Y_a) \nn \\
	& \prod_{j=1}^N \big( s_b ( \frac{iQ}{2} - \frac{1-N}{2}\tau - B_1 \pm ( Z^{(1)}_j - W_L ) ) s_b ( -\frac{iQ}{2} + (j-N)\tau + 2 B_1 ) \big) \nn \\
	& \prod_{j=1}^N \big( s_b ( \frac{iQ}{2} - \frac{1-N}{2}\tau - C_{F_2} \pm ( Z^{(F_1+F_2+1)}_j - W_R ) ) s_b ( -\frac{iQ}{2} + (j-N)\tau + 2 C_{F_2} ) \big) \,.
\end{align}
Which matches with the partition function of the magnetic theory in figure \eqref{fig:MoreNodes_Elec_manifest}.

\bibliographystyle{JHEP}
\bibliography{ref}

\end{document}